\documentclass{aa}  

\usepackage{graphicx}
\usepackage{txfonts}
\usepackage{subfigure}
\usepackage{longtable}
\usepackage{color}
\usepackage{icomma}
\usepackage{bm}
\usepackage{marginnote}
\usepackage{textcomp}

\usepackage{amstext}
\usepackage[normalem]{ulem}

\begin{document} 

   \title{A study of accretion and disk diagnostics \\ in the NGC 2264 cluster \thanks{Tables 2 and 3 are only available in electronic form
at the CDS via anonymous ftp to cdsarc.u-strasbg.fr (130.79.128.5) or via http://cdsweb.u-strasbg.fr/cgi-bin/qcat?J/A+A/} \thanks{This  work  is  based  on  observations  made with FLAMES instrument on the Very Large Telescope under  program  ID  088.C-0239.} }


   \author{Alana P. Sousa
          \inst{1,2}
          \and
          Silvia H. P. Alencar\inst{1}%
           \and
          Luisa M. Rebull\inst{3}
           \and
          Catherine C. Espaillat\inst{4}
           \and
          Nuria Calvet\inst{5} 
           \and
          Paula S. Teixeira\inst{6}
          }

   \institute{Departamento de F\'isica-Icex-UFMG 
               Ant\^onio Carlos, 6627, 31270-901. Belo Horizonte, MG, Brazil\\
              \email{alana@fisica.ufmg.br}
     \and          
     Université Grenoble Alpes, IPAG, F-38000 Grenoble, France     
     \and
     Infrared Science Archive (IRSA), IPAC, 1200 E.\ California Blvd., California Institute of Technology, Pasadena, CA 91125, USA
     \and
     Department of Astronomy, Boston University, 725 Commonwealth Avenue, Boston, MA 02215, USA
     \and
     Department of Astronomy, University of Michigan, 830 Dennison Building, 500 Church Street, Ann Arbor, MI 48109, USA
     \and
     Scottish Universities Physics Alliance (SUPA), School of Physics and Astronomy, University of St. Andrews, North Haugh, Fife, KY16 9SS, St. Andrews, UK
    }
    
   \date{Received May 26, 2015; accepted September 04, 2015}

 
  \abstract
   {
   Understanding disk dissipation is essential for studying how planets form. Disk 
gaps and holes, which almost correspond to dust-free regions, are inferred from infrared 
observations of T Tauri stars (TTS), indicating the existence of a transitional phase 
between thick accreting disks and debris disks. Transition disks  are usually 
referred to as candidates for newly formed planets.
   We searched for transition disk candidates belonging to NGC 2264. Using 
stellar and disk parameters obtained in the observational multiwavelength 
campaign CSI2264, we characterized accretion, disk, and stellar properties of transition
disk candidates and compared them to systems with a full disk and diskless stars
We modeled the spectral energy distribution (SED) of a sample of $401$ TTS, 
   observed with both CFHT equipped with  MegaCam and IRAC instrument on the Spitzer, with Hyperion SED 
fitting code using photometric data from the U band ($0.3\,\mu\mathrm{m}$) to the 
Spitzer/MIPS $24\,\mu\mathrm{m}$ band. We used the SED modeling to distinguish transition 
disk candidates, full disk systems, and diskless stars.
 We classified  $\sim52\,\%$ of the sample as full disk systems, $\sim41\,\%$ as diskless 
stars, and $\sim7\,\%$ of the systems as transition disk candidates, among which seven systems 
are new transition disk candidates belonging to the NGC 2264 cluster. The sample 
of transition disk candidates present dust in the inner disk similar to anemic disks, 
according to the $\alpha_\mathrm{IRAC}$ classification, which shows that anemic disk 
systems can be candidate transition disks. We show that the presence of a 
dust hole in the inner disk does not stop the accretion process since $82\,\%$ of 
transition disk candidates accrete and show $\mathrm{H}\alpha$, UV excess, 
and mass accretion rates at the same level as full disk systems. We estimate 
the inner hole sizes, ranging from $0.1$ to $78\,\mathrm{AU}$, for the sample of  
transition disk candidates. In only $\sim18\,\%$ of the transition disk candidates, 
the hole size could be explained by X-ray photoevaporation from stellar radiation. }

   \maketitle
   
\newcommand \ang{\AA\,}
\newcommand \angn{\AA\,}
\newcommand \ha{$\mathrm{H}\alpha$\,}
\newcommand \hb{$\mathrm{H}\beta$\,}
\newcommand \hem{$\mathrm{He}$\,I\ (5876\ang)\,}
\newcommand \he{$\mathrm{He}$\,I\ (6678 \ang)\,}
\newcommand \He{$\mathrm{He}$\,I\,}
\newcommand \dn{Na\,{\sc i} D\,}
\newcommand \fc{$F_{\mbox{\small c}}$\,}
\newcommand \kms{$\mathrm{km s}^{-1}$\,}
\newcommand \oi{[O\,{\sc i}]\,}
\newcommand \sii{[S\,{\sc ii}]\,}
\newcommand \ms{$M_{\odot}$\,} 
   
\section{Introduction}

Circumstellar disks are ubiquitous around young stars and are the 
sites of planet formation. The analysis of disk structure and evolution 
is therefore an essential step in understanding the formation of planets.
There are still many questions as to how the gas and dust in
circumstellar disks evolve into planetary systems. Infrared (IR) observations of 
young low mass stars show that the number of circumstellar disks 
decreases with age with a typical timescale
for dissipation around $\sim 6\,\mathrm{Myr}$ \citep[e.g.,][]{2001ApJ...553L.153H,2007ApJ...671.1784H,2014A&A...561A..54R}. 

Several mechanisms of disk dissipation have been proposed in the literature. 
The inner disk is partly dissipated by accretion onto the star 
with a typical rate of about $10^{-8}\,\mathrm{M}_\sun\mathrm{yr}^{-1}$ \citep{1998ApJ...495..385H} in the T Tauri phase. 
Disks can be photoevaporated by the central star's high-energy radiation, 
which is a very efficient mechanism for gas dispersal, once the mass accretion rate drops
below the photoevaporation mass loss rate \citep{2001MNRAS.328..485C,2004ApJ...611..360A,2011MNRAS.412...13O,2013MNRAS.430.1392R,2014prpl.conf..475A}. 
The disk material may be driven out of the system through disk winds and 
jets \citep{konigl1989self, pelletier1992hydromagnetic, safier1993centrifugally, shu1994magnetocentrifugally}. 
The disk can also be consumed in the coagulation of grains and planet formation 
\citep{papaloizou1999critical,2005A&A...434..971D,2005ApJ...631.1180H,2011ApJ...729...47Z}.

Circumstellar disks are often detected due to IR excess 
with respect to the stellar photosphere. Dust in the disk is heated by 
the central star and accretion and it then reemits IR wavelengths. The spectral 
energy distribution (SED) of a star with an optically thick disk, therefore, presents 
an IR excess added to the stellar photospheric contribution.
As the disk disperses, the SED gradually shows less IR excess 
\citep{lada1987star}. Furthermore, regions of the disk that are almost dust-free, 
that correspond to gaps and holes, are inferred from IR observations 
of T Tauri stars (TTS). Inner disk hole SEDs are characterized by a lack of emission 
excess in the near-IR but normal thick disk IR emission at mid-IR
wavelengths \citep[e.g.,][]{2013MNRAS.428.3327K,2014prpl.conf..497E}.  
Inner disk clearing was confirmed with resolved millimeter continuum images 
\citep{2009ApJ...698..131H,2011ApJ...732...42A}. ALMA later revealed 
inner disk substructures such as rings and thin gaps less than 1 AU wide   
\citep[e.g.,][]{2016ApJ...820L..40A,2018ApJ...869L..41A}.

The formation of holes or gaps in the disk can be explained by dust coagulation due to Brownian motion of
dust grains \citep{2007A&A...461..215O,2013ApJ...764..146G}, or the presence of a newly 
formed planet that dynamically clears a specific region of the disk 
\citep{2002ApJ...568.1008C,2011ApJ...729...47Z,2016PASA...33....5O}. Photoevaporation may also clear a hole in the 
circumstellar disk but only in the final stages of disk accretion \citep{2011MNRAS.412...13O}. Most systems with an 
inner disk hole, however, still accrete at typical T Tauri rates 
\citep[e.g.,][]{2010ApJ...718.1200M,2012ApJ...747..103E,2014A&A...568A..18M}. These disks are called transition disks.
Transition disks are not common in star forming regions,
representing only about $10\,\%$ of young star-disk systems \citep{2016PASA...33....5O,2016ApJ...828...46A}. 
This indicates that disk dispersal is rapid compared to disk lifetime.
 
We searched for transition disk candidates in the young ($\sim3\,\mathrm{Myr}$) stellar cluster
NGC 2264, located at a distance of $\sim760\,\mathrm{pc}$ from the Sun. 
We used data from the Coordinated Synoptic Investigation of NGC 2264 (CSI2264)
observational campaign \citep{Cody2013} to characterize the transition disk candidates 
in terms of various accretion diagnostics ($\mathrm{H}\alpha$  
line emission, ultraviolet excess), disk parameters (IR excess), and stellar parameters
(mass, effective temperature, radius) by
comparing them to the characteristics of systems with full disks and diskless stars.

In Section \ref{sec:data} we describe the data used in this work, in Section \ref{sec:sample}
we explain the sample selection criteria, and in Section \ref{sec:result} we present the determination of 
accretion and disk characteristics of our sample. The discussion and analysis of inner hole properties 
are in Section \ref{sec:discuss}, and in Section \ref{sec:concl} we present our conclusions.

\section{Observations}\label{sec:data}
We used data from the CSI2264, an international campaign 
that included simultaneous photometric and spectroscopic 
observations of the young cluster NGC 2264 with satellites and ground-based 
telescopes at various wavelengths 
\citep[see][for more details about the campaign]{Cody2013}. 
During the CSI2264 campaign, NGC 2264 was simultaneously observed in the optical with the CoRoT satellite for 40 days
and with Spitzer satellite \citep{werner04} at $3.6$ and  $4.5\,\mu \mathrm{m}$ for 30 days.
We obtained 20 epochs of observations 
on the Very Large Telescope (VLT) equipped with the Fiber Large 
Array Multi Element Spectrograph (FLAMES), four to six observations simultaneously with 
the CoRoT data, depending on the target. Photometry in the $\it{u}$ and $\it{r}$ bands was also obtained 
with MegaCam (Canada-France-Hawaii Telescope-CFHT) for $14$ nights about one month 
after the end of the CoRoT observations.

Additionally to the CSI2264 data, to construct SEDs in Section \ref{sec:disk}, we also used 
data from catalog surveys, such as $UBVR_cI_c$ optical photometry from \cite{2002AJ....123.1528R}, 
$ugriz$ optical photometry from SDSS \citep{1998AJ....116.3040G}, 
near-IR photometry $JHK_s$ from 2MASS, IRAC \citep{2004ApJS..154...10F}, and MIPS \citep{2004ApJS..154...25R} data 
from Spitzer satellite, and observations from the Wide-field Infrared Survey 
Explorer (WISE) at $3.4$, $4.6$, $12.0$, and $22\,\mu m$ \citep{2010AJ....140.1868W}. 

\section{Sample of stars} \label{sec:sample}
There is no consensus in the literature on how to classify a star-disk system as a transition disk. 
In the simplest definition, transition disks are systems with little near-IR 
excess and strong mid-IR excess. 
In Section \ref{sec:disk}, we discuss some different ways from the literature
to define a transition disk.

We selected inner disks with holes and gaps based on the flux 
observed in the near and mid-IR, and on the fitting of SEDs,
that is, the flux emitted by the sources in various photometric bands vs. their 
central wavelengths. 
As we intend to analyze the accretion and disk properties of transition disk systems, 
our sample of stars only contains TTS that were observed with IRAC instrument on the Spitzer, which had 
an $\alpha_{IRAC}$ index (the slope of the SED between $3.6\,\mu\mathrm{m}$ and $8\,\mu\mathrm{m}$) 
measured by \cite{2012A&A...540A..83T} and were also observed in \textit{u}, \textit{g}, \textit{r}, and \textit{i} 
bands at CFHT with  MegaCam \citep{2014A&A...570A..82V}.  The IRAC completeness limits, from 
\cite{2012A&A...540A..83T}, are $13.25\,\mathrm{mag}$ for the $3.6\,\mu\mathrm{m}$, $4.5\,\mu\mathrm{m}$, and $5.8\,\mu\mathrm{m}$ 
IRAC bands and $12.75\,\mathrm{mag}$ for the $8\,\mu\mathrm{m}$ IRAC band. The IRAC sample from  \cite{2012A&A...540A..83T} 
was also limited to magnitude uncertainties smaller than $0.1\,\mathrm{mag}$ for all IRAC bands. Consequently, 
some IRAC sources belonging to NGC 2264 could be missing compared to other literature works that used IRAC observations.
Our sample is composed of $401$ TTS 
that were observed with both  MegaCam and IRAC and appeared in \cite{2012A&A...540A..83T}, to select the systems with
the largest number of measured stellar, accretion, and disk parameters. As the sample is not complete, disk frequencies inferred 
from these data are not reliable.

\subsection{SED model}\label{sec:sed}
We calculated synthetic SEDs for all the objects in our sample, using the
Python-based fitting code sedfitter\footnote{In this work, we used version v1.0 
of the code, available to download at https://doi.org/10.5281/zenodo.235786.} 
\citep{2017A&A...600A..11R} based on Hyperion, an open-source parallelized 
three-dimensional dust continuum radiative transfer code by \cite{2011A&A...536A..79R}. 
This fitting model consists of modular sets with different components 
that can include a stellar photosphere, a disk, an envelope, and 
ambipolar cavities. 
The code allows the user to choose the best set of models for their 
sample of stars and eliminate unphysical models. We tested 
three sets of models\footnote{All the modular sets of synthetic spectral energy distributions 
were downloaded from http://doi.org/10.5281/zenodo.166732, using version v1.1 of the models.}: model 1 - only a central star; 
model 2 - a central star and a passive disk,  which is a disk that absorbs the stellar radiation and re-radiates it 
in the IR, without taking into account accretion radiation;  and 
model 3 - a central star, a passive disk, and an inner disk hole.

\begin{figure*} 
 \centering
\subfigure[]{\includegraphics[scale=0.60]{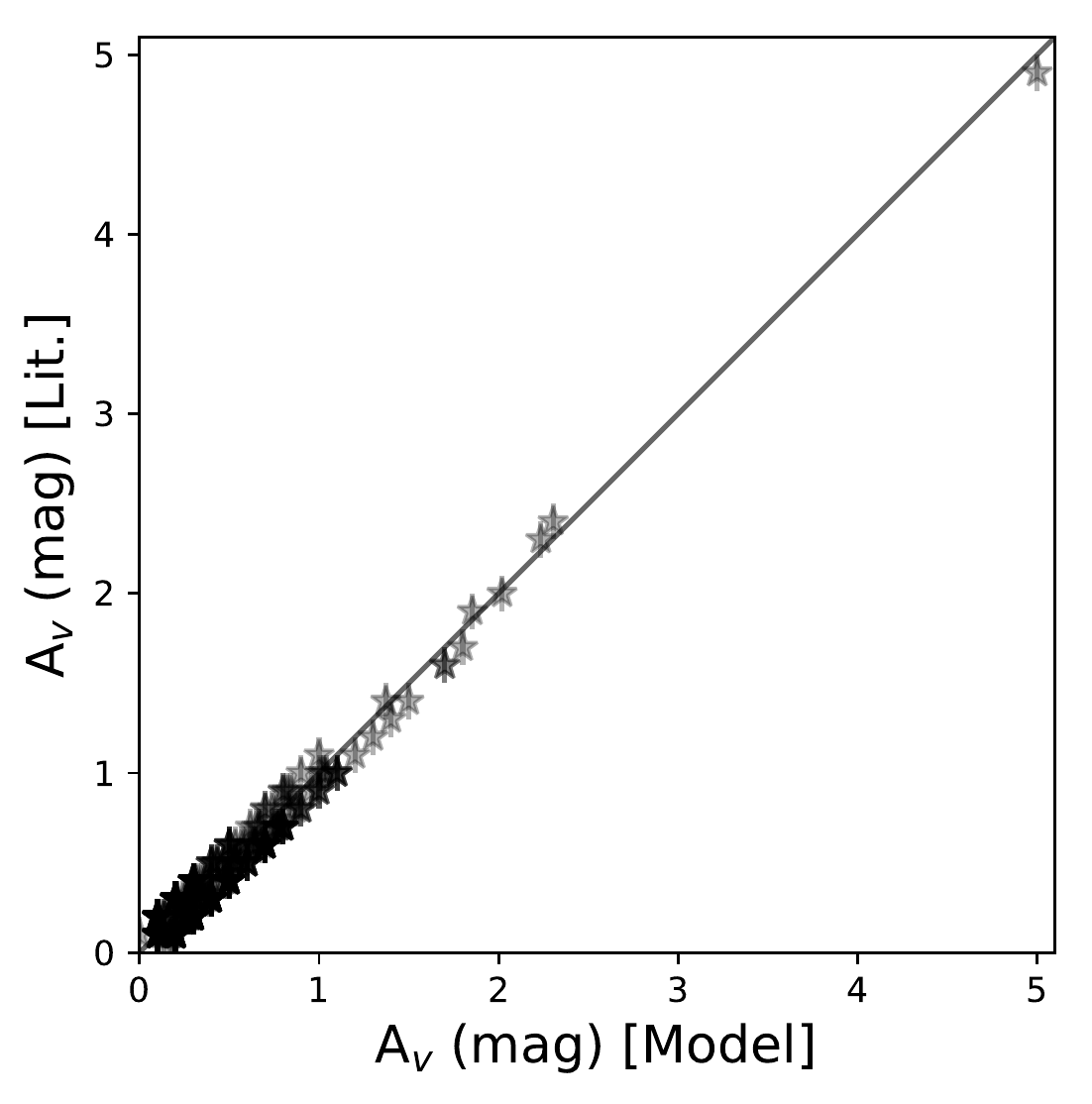}}
\subfigure[]{\includegraphics[scale=0.60]{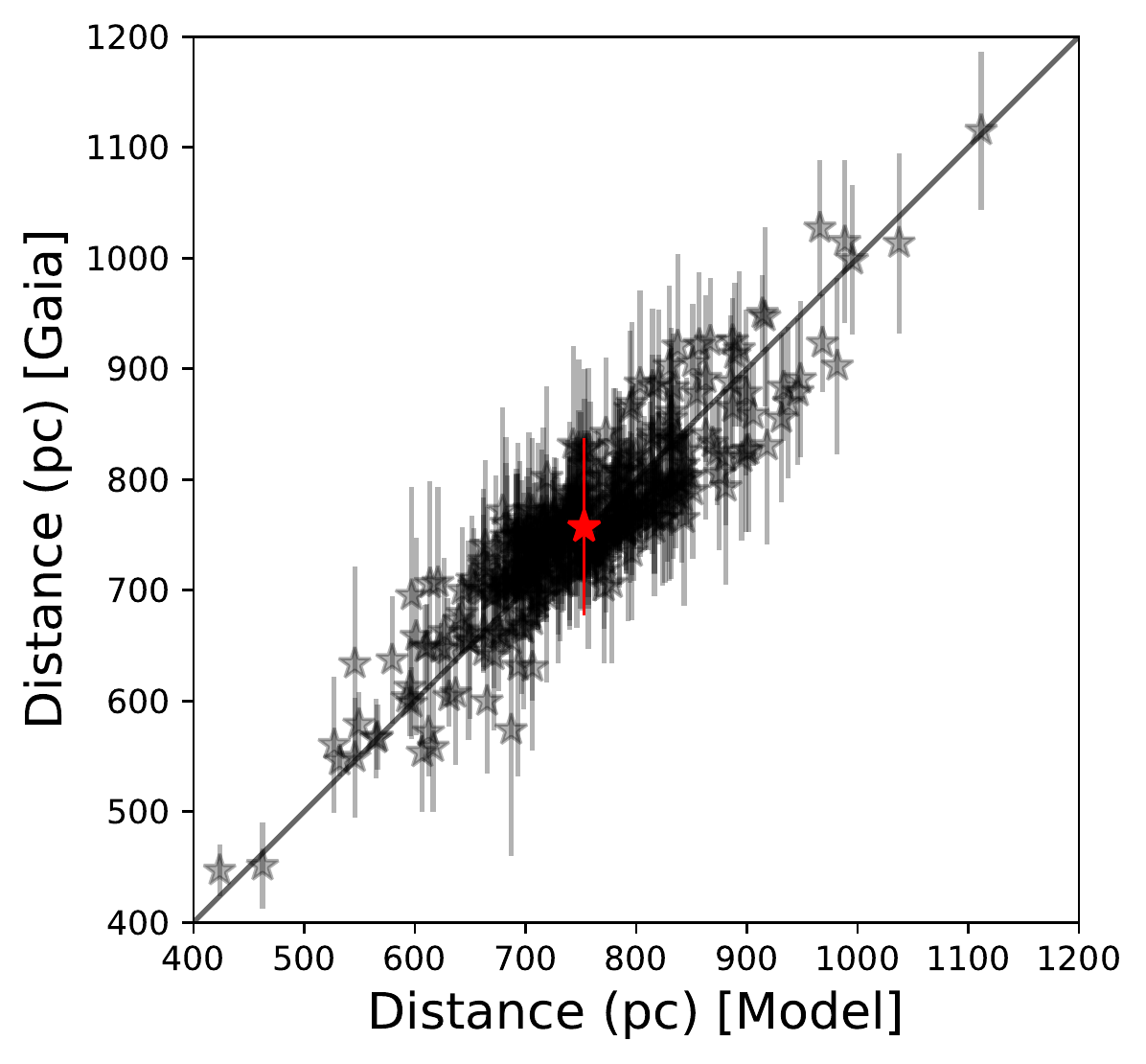}}\\
\subfigure[]{\includegraphics[scale=0.60]{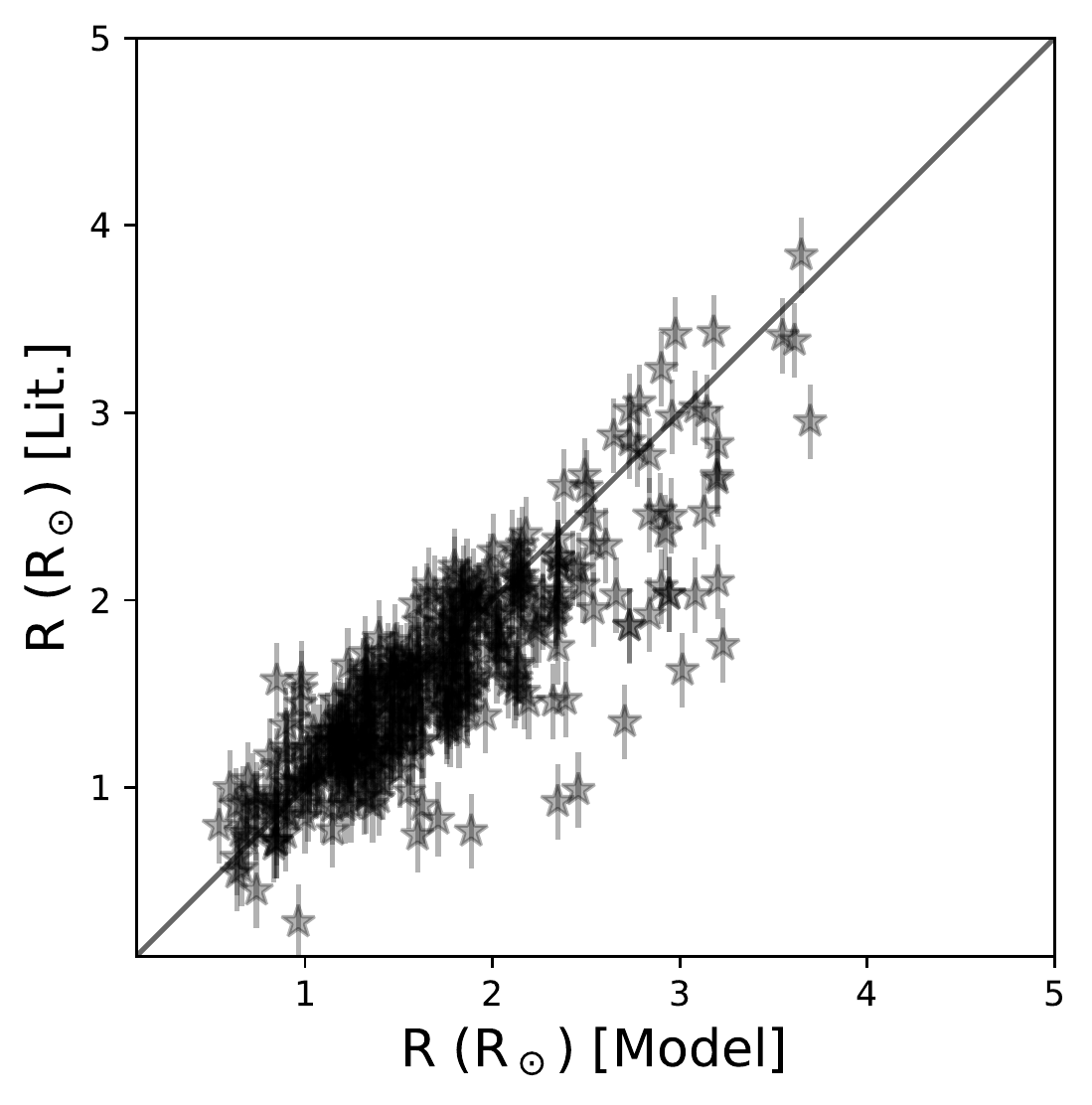}}
\subfigure[]{\includegraphics[scale=0.60]{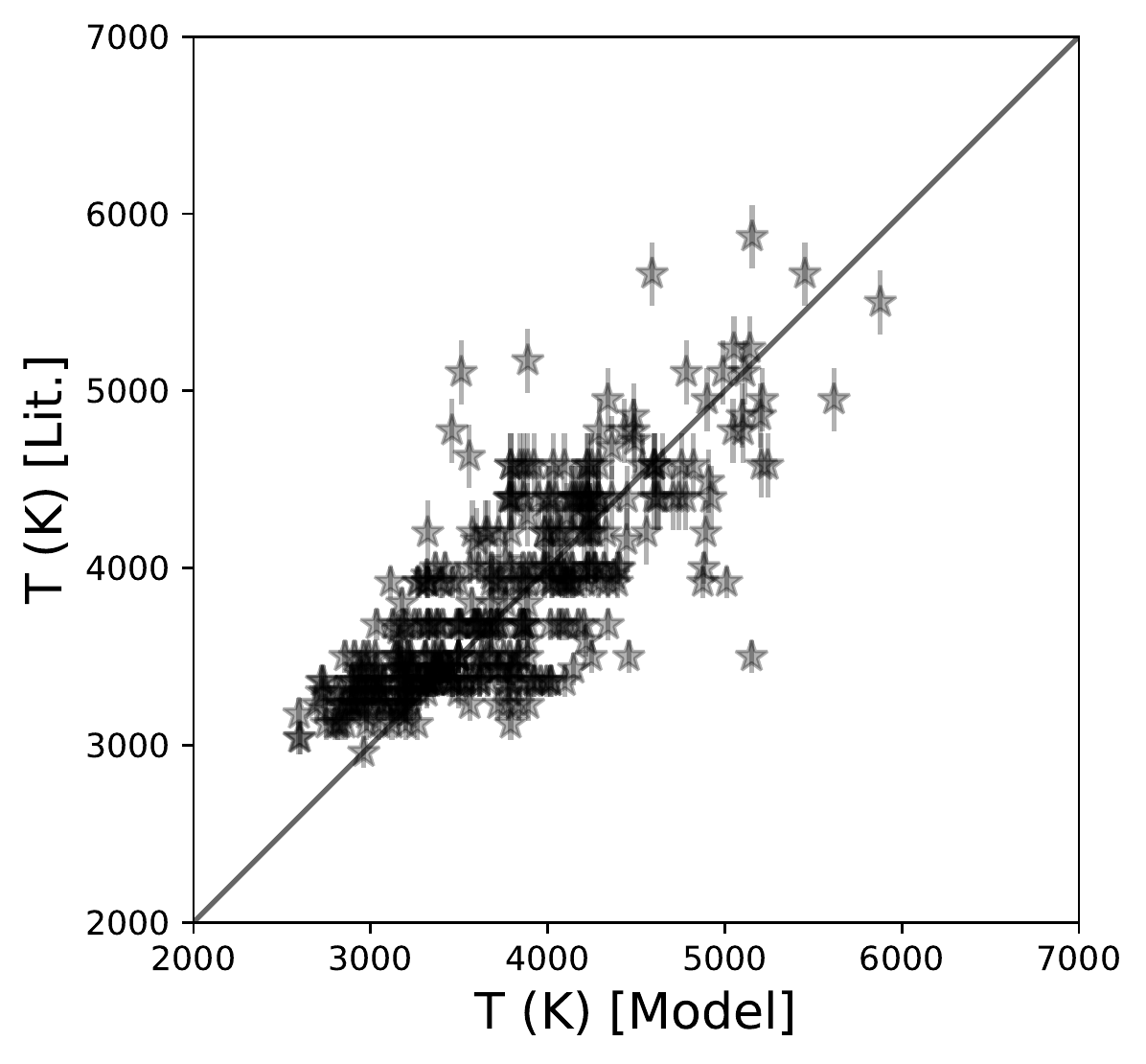}}
\caption{\label{fig:Parm_comp} Stellar parameters modeled with Hyperion SED 
fitting code \citep{2017A&A...600A..11R} compared to stellar parameters 
obtained by \cite{2014A&A...570A..82V}. a) $A_V$, b) distance from the Sun. The red point and error bars represent the mean and standard deviation of the distance of the cluster members studied in this work ($d=757\pm80 \,\mathrm{pc}$), c) stellar radii, d) stellar temperatures. $A_V$ and distances are input parameters of the code, while stellar radius 
and temperature are output parameters of the Hyperion SED Model. The distances were obtained from the 
 Gaia parallax data \citep{2018A&A...616A...9L}. The solid line shows a slope equal to $1$. Based on these plots, we conclude that our modeling is doing at least as well as the prior \cite{2014A&A...570A..82V} modeling.} 
\end{figure*}

The SEDs have been constructed for the $401$ stars in our sample that were 
modeled with the three model sets. 
The observational fluxes, from the U band to the $24\,\mu\mathrm{m}$ MIPS band 
taken from the literature, were not dereddened. 
The input parameters of the models are the fluxes and photometric apertures, 
a range of $A_V$, and the distance of the star to the Sun. We used the individual 
value of $A_V$ to each star from \cite{2014A&A...570A..82V}. We used 
the individual distance values estimated from parallax data obtained from the 
Gaia mission second release \citep{2016A&A...595A...1G,2018A&A...616A...1G}. 
The distance values were calculated using Bayesian methods, following \cite{2018A&A...616A...9L}.  
The individual distance values for our sample of stars 
are listed in Tables \ref{tab:TransDisk}, \ref{tab:Fulldisk}, and \ref{tab:Diskless}. 
The $A_v$ and the distance uncertainty of each star were used to define the range of $A_v$ and distances to be searched for during the 
SED modeling. Within the allowed ranges, the SED program can choose the $A_V$ and distance of each system that correspond to the best fit to the observed flux data. 
For the systems that did not have Av and/or distance ( Gaia parallax) available, 
we chose the values that represented a best fit to the data.  We used the same extinction law described in \cite{2007ApJS..169..328R}.
In the model calculations, the inner disk radius ($\mathrm{R_{in}}$) is set to
the dust sublimation radius ($\mathrm{R_{sub}}$) in disks without an inner hole (model 2) 
or varied from $1$ to $1000$ $\mathrm{R_{sub}}$ \citep{2017A&A...600A..11R} to
reach a best fit, when an inner disk hole is present (model 3). 

\begin{figure*}[htb!] 
 \centering
\includegraphics[scale=0.74]{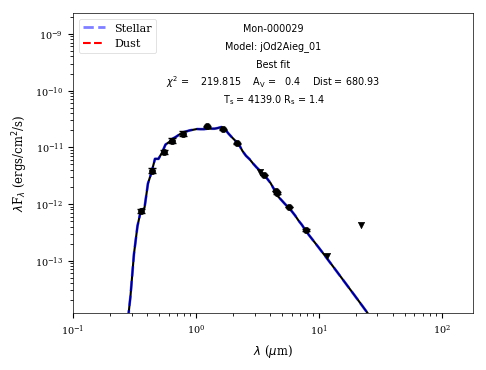}
\includegraphics[scale=0.74]{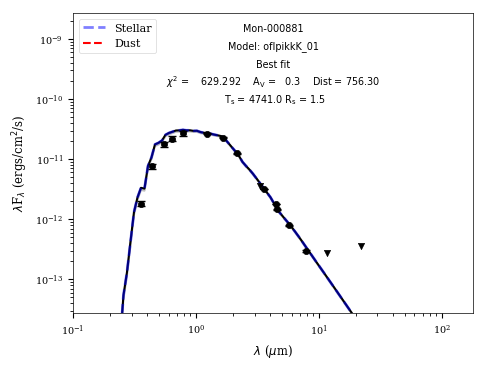}\\
\includegraphics[scale=0.74]{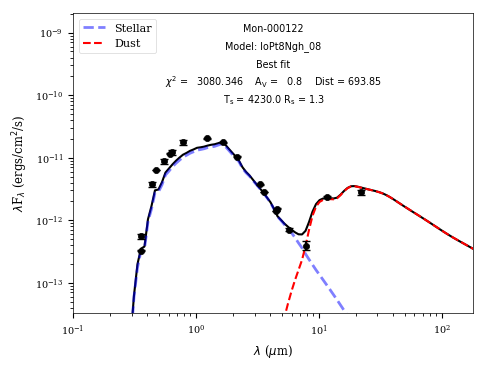}
\includegraphics[scale=0.74]{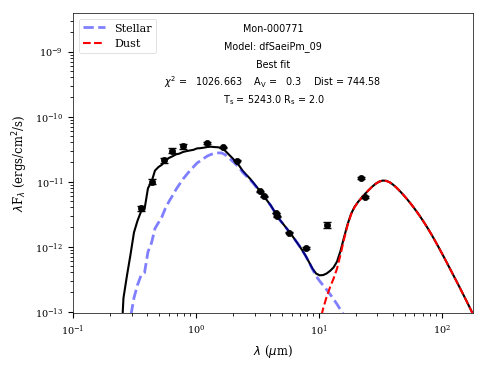}\\
\includegraphics[scale=0.74]{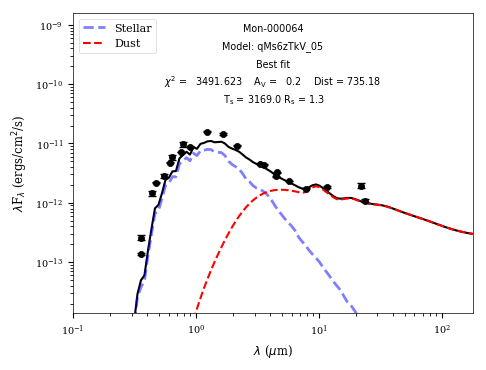}
\includegraphics[scale=0.74]{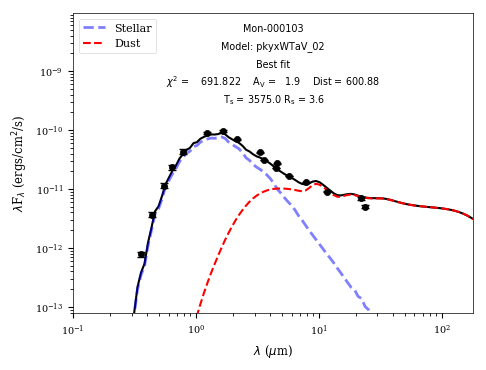}\\
\caption{\label{fig:SEDs} Examples of synthetic SED fitting, using Hyperion SED 
code \citep{2017A&A...600A..11R},  for our sample of systems 
classified as diskless stars (top), transition disk candidates (middle) and full disk systems (bottom). 
The circles show data from the U band ($0.3\,\mu\mathrm{m}$) to the MIPS $24\,\mu\mathrm{m}$ band. 
Triangles, when present, represent the upper limits. The black solid 
line is the best fit to the SED \citep{2017A&A...600A..11R} and 
the dashed lines correspond to the stellar (blue) and dust (red) emission components 
\citep{2011A&A...536A..79R}. We found in our sample $7\%$ transition disk candidates, 
$52\%$ systems with full disks, and $41\%$ diskless stars.} 
\end{figure*}

Among the $401$ TTS in our sample, only $152$ have detections 
available at $22\,\mu\mathrm{m}$ (WISE) or $24\,\mu\mathrm{m}$ (MIPS), 
which are important for constraining the outer disk. Without these data, 
the SED fitting is highly degenerate in the outer disk region. 
The MIPS observations targeted preferentially stars that 
presented IR excess; we therefore expect that stars which were not detected in these bands 
to have little or no excess at these wavelengths \citep{2012A&A...540A..83T}.  

The SED models were calculated for each source and the different output models were 
compared with the observational data through a $\chi^2$ test.  
The best model was the one presenting the minimum $\chi^2$ value. For the $152$ stars that 
had $22\,\mu\mathrm{m}$ or $24\,\mu\mathrm{m}$ detections, we calculated models 2 and 3, 
choosing the best model that fit the data and showed output stellar 
parameters (temperature and radii) from the fitting procedure close to the values available
in the literature.
A total of $249$ stars did not have $24\,\mu\mathrm{m}$ and/or $22\,\mu\mathrm{m}$
measurements or have only an upper limit at these wavelengths. Among those
systems, we fit models 2 and 3 to the $85$ stars that presented excess in the inner disk 
(based in the $\alpha_\mathrm{IRAC}$ index) or signs of accretion  (based in the \ha line analyze \citep{2016A&A...586A..47S} or UV excess \citep{2014A&A...570A..82V}). The $164$ stars 
without IR excess and presenting no accretion signature were fit with model 1, composed
only of a stellar photosphere. The sample of diskless stars was used to compare its characteristics 
with the sample of transition disk candidates. Among the $401$ TTS in our sample,
the SEDs of $7$ stars could not be fit by any reasonable physical model,
sometimes due to a large discrepancy between the temperature or radius of the best model 
compared to the literature values or the best model did not produce a good fit to the observational data.

In Fig. \ref{fig:Parm_comp} we show a comparison of the $A_V$, distance, stellar radii, and stellar temperatures 
found by Hyperion SED fitting model with the values by \cite{2014A&A...570A..82V} and the distances calculated using 
Gaia parallax data following \cite{2018A&A...616A...9L}. As $A_V$ and the distances are input parameters of the model, the values returned 
by the model agree with the literature value. The stellar temperature and radius are output parameters and 
the best-fit values are not always accurate, but for most sources, the deviations from the literature values 
are small. 

The SEDs have been extensively used in the literature to identify transition disks 
\citep[e.g.,][]{2005ApJ...630L.185C,2010ApJ...718.1200M,2016PASA...33....5O,2018ApJ...863...13G}. 
We used the SED to measure the near-IR and mid-IR flux excess \citep{1989AJ.....97.1451S} and the model 
including an inner hole to confirm or refute the presence of a hole in the inner disk 
\citep[e.g.,][]{2010ApJ...718.1200M}. 
In this work, we selected as transition disk candidates stars with inner-disk holes according 
to the SED fitting (systems with $\mathrm{R_{in}}>\mathrm{R_{sub}}$). We also only selected systems that have 
$24\,\mu\mathrm{m}$ flux detected above the photospheric level larger than the $10\,\mu\mathrm{m}$ flux.
We found $28$ stars that fulfill all these conditions in our sample. Among the remaining
systems, $209$ stars were modeled with a full disk and $164$ as diskless. 
In Fig. \ref{fig:SEDs} we show some examples of SEDs of diskless stars, transition 
disk candidates, and full disk systems. The black solid line is the best fit (based on $\chi^2$) 
that represents the total flux of the system. The dashed red and blue lines are the components 
of emission from dust and the central star, respectively. 
The dust and stellar emission components were computed using the Hyperion 
code\footnote{In this work, we used version 0.9.8 of the Hyperion code available 
to download from http://www.hyperion-rt.org/ and model sets from https://doi.org/10.5281/zenodo.572233.} 
\citep{2011A&A...536A..79R}, after determining the best the SED model for each system \citep{2017A&A...600A..11R}.
In Sect. \ref{sec:hole} we analyze the disk and the inner-disk hole
determined values. From here on, the results will be discussed in terms of three groups: diskless stars, transition disk 
candidates, and full disk systems.  In Table \ref{tab:TransDisk}, \ref{tab:Fulldisk}, and \ref{tab:Diskless} we show the observational parameters 
of the transition disk candidates, full disk systems, and diskless stars, respectively.

\begin{table*}[htb!]
\tiny
\addtolength{\tabcolsep}{-3.5pt}  
\caption{\label{tab:TransDisk} Observational parameters of our sample of transition disk candidates.}
\begin{center}
\begin{tabular}{llllllllllllll}
  \hline\hline 
Mon ID\tablefootmark{a} & TwoMass ID\tablefootmark{b} & SpT\tablefootmark{c} & $\mathrm{EW}_{\mathrm{H}\alpha}$\tablefootmark{d} & $\mathrm{W}10\%_{\mathrm{H}\alpha}$\tablefootmark{d} & $\mathrm{EW}_{\mathrm{H}\alpha}$\tablefootmark{e} & $Class$\tablefootmark{f} & $\alpha_{\mathrm{IRAC}}$ \tablefootmark{g}  & $R_\star$\tablefootmark{h} & $T\star$\tablefootmark{h} & $UV_{Exc.}$\tablefootmark{h} & 
$\dot{M}_{UV_{Exc.}}$\tablefootmark{h} & $\dot{M}_{H\alpha}$\tablefootmark{d} & Dist\tablefootmark{i} \\
 & &   & $(\mathring{\mathrm{A}})$ & $(\mathrm{km}\mathrm{s}^{-1})$  &$(\mathring{\mathrm{A}})$ & & & $(R_\odot)$ & $(K)$ & $(mag)$ & $(M_\odot yr^{-1})$ &  $(M_\odot yr^{-1})$ & ($\mathrm{Pc}$) \\
\hline
Mon-000040  &  06410892+0956476  &  M4.5   &         &       &         &  c  &  -2.29  &    0.85  &    3175.  &       -0.340  &      4.3E-10  &               &    664.7  \\
Mon-000120  &  06411911+0926294  &  M3     &         &       &  155.9  &  c  &  -2.12  &    2.09  &    3360.  &       -1.930  &      1.7E-07  &               &    749.3  \\
Mon-000122  &  06414711+0938047  &  K6     &         &       &    1.4  &  w  &  -2.84  &    1.57  &    4200.  &       -0.190  &               &               &    714.3  \\
Mon-000177  &  06410620+0936229  &  G5     &   11.1  &  412  &   10.0  &  c  &  -2.03  &    2.03  &    5660.  &        0.067  &               &      9.3E-09  &    703.0  \\
Mon-000273  &  06411837+0939411  &  M1     &         &       &  123.5  &  c  &  -2.28  &    1.60  &    3680.  &       -1.050  &      2.1E-08  &               &    560.4  \\
Mon-000280  &  06404100+0927543  &  K4     &   12.8  &  486  &   13.2  &  c  &  -1.97  &    1.07  &    4580.  &       -0.090  &      8.9E-10  &      5.9E-10  &    741.1  \\
Mon-000296  &  06405059+0954573  &  K2     &   10.6  &  375  &   11.2  &  c  &  -2.14  &    1.71  &    4950.  &       -0.084  &               &      1.7E-09  &    741.7  \\
Mon-000314  &  06404459+0932261  &  M3     &   53.2  &  287  &   60.0  &  c  &  -2.61  &  3 1.35  &    3360.  &       -1.230  &      9.5E-09  &      1.1E-08  &    751.7  \\
Mon-000328  &  06412700+0930131  &  M1     &   31.9  &  231  &   25.8  &  c  &  -1.88  &    1.34  &    3680.  &       -0.340  &      2.4E-09  &      3.9E-09  &    755.4  \\
Mon-000342  &  06405573+0946456  &  M4     &         &       &   21.1  &  c  &  -1.79  &    1.57  &    3230.  &       -1.270  &      1.7E-08  &               &    946.9  \\
Mon-000433  &  06410111+0934522  &  M1     &         &       &    7.0  &  c  &  -2.24  &  3 0.99  &    3680.  &       -0.100  &      1.4E-09  &               &    766.3  \\
Mon-000452  &  06410160+0937287  &  M2.5   &         &       &  154.8  &  c  &  -1.71  &    0.92  &    3430.  &       -2.340  &      1.9E-08  &               &    829.2  \\
Mon-000502  &  06405777+0930502  &  K7     &         &       &    7.3  &  w  &  -2.37  &    2.03  &    4000.  &       -0.033  &               &               &    663.7  \\
Mon-000637  &  06404921+0957387  &  M1     &         &       &   50.8  &  c  &  -2.67  &    1.64  &    3680.  &       -0.420  &      3.2E-09  &               &    720.1  \\
Mon-000676  &  06414780+0934096  &  K5     &         &       &         &  w  &  -1.53  &    2.32  &    4395.  &        0.517  &               &               &    728.8  \\
Mon-000771  &  06411827+0933535  &  K4     &         &       &   28.9  &  c  &  -2.39  &    1.83  &    4580.  &       -0.270  &               &               &    762.9  \\
Mon-000824  &  06410183+0938411  &  K4     &   -0.5  &  493  &    1.5  &  c  &  -2.41  &    2.23  &    4580.  &       -0.068  &               &      3.3E-09  &    606.9  \\
Mon-000860  &  06415492+0942527  &  M2.5   &         &       &  261.0  &  c  &  -3.23  &    1.18  &    3430.  &       -0.900  &      5.6E-09  &               &    708.7  \\
Mon-000879  &  06410338+0940448  &  M1     &         &       &   16.5  &  c  &  -2.69  &    1.99  &    3680.  &       -0.130  &      1.2E-08  &               &   1027.0  \\
Mon-000937  &  06405255+0952059  &  K7     &    8.5  &  371  &    6.9  &  c  &  -2.40  &    2.10  &    4000.  &       -0.010  &      1.0E-08  &      7.2E-10  &    890.7  \\
Mon-000961  &  06411247+0908509  &  M3.5   &         &       &         &  w  &  -1.20  &    0.90  &    3300.  &        0.010  &      7.9E-11  &               &    706.1  \\
Mon-000965  &  06404600+0917582  &  M2     &   14.4  &  305  &         &  c  &  -2.51  &    1.72  &    3500.  &       -0.540  &      1.0E-08  &      1.8E-09  &    886.9  \\
Mon-000997  &  06403077+0948076  &  M1     &         &       &   72.8  &  c  &  -4.22  &    0.85  &    3680.  &       -0.120  &      2.8E-10  &               &           \\
Mon-001033  &  06404102+0947577  &  K7     &    6.3  &  434  &    7.5  &  c  &  -1.97  &    1.67  &    4000.  &       -0.700  &      2.0E-08  &      4.7E-09  &    737.2  \\
Mon-001094  &  06403164+0948233  &  K5     &         &       &   54.6  &  c  &  -1.96  &    1.49  &    4395.  &       -0.670  &      1.6E-08  &               &    750.2  \\
Mon-001229  &  06403697+0939097  &  M1     &         &       &    5.0  &  w  &  -1.92  &    1.46  &    3680.  &       -0.248  &               &               &    753.1  \\
Mon-001287  &  06401110+0921270  &  M3     &  126.0  &  315  &  163.1  &  c  &  -1.62  &    1.08  &    3360.  &       -1.930  &      1.9E-08  &               &    795.2  \\
Mon-001308  &  06395924+0927245  &  M0     &         &       &         &  c  &  -1.87  &    1.59  &    3920.  &       -0.800  &      1.9E-08  &               &    744.9  \\
\hline
\end{tabular}
\end{center}
 \tablefoot{This table is ordered according to the Mon ID.}\\
 \tablefoottext{a}{CSIMon is an internal identification of the CSI 2264 campaign. Here, CSI was omitted for brevity.}
 \tablefoottext{b}{2Mass identification.}
 \tablefoottext{c}{Spectral type obtained by \cite{2014A&A...570A..82V}.}
 \tablefoottext{d}{$\mathrm{H}\alpha$ parameters obtained using FLAMES spectra by 
\cite{2016A&A...586A..47S}. They used the convention that positive $\mathrm{H}\alpha$ 
equivalent width indicates $\mathrm{H}\alpha$ in emission, and negative values 
correspond to $\mathrm{H}\alpha$ in absorption.} 
 \tablefoottext{e}{$\mathrm{H}\alpha$ equivalent width obtained by \cite{2005AJ....129..829D}.}
 \tablefoottext{f}{Classification as CTTS (c) and WTTS (w) by \cite{2016A&A...586A..47S} and \cite{2014A&A...570A..82V}.}
 \tablefoottext{g}{$\alpha_{IRAC}$ is the slope of the SED 
between $3.6\,\mu\mathrm{m}$ and $8\,\mu\mathrm{m}$ obtained by \cite{2012A&A...540A..83T}.}
 \tablefoottext{h}{Parameters obtained using CFHT data by \cite{2014A&A...570A..82V}.}
 \tablefoottext{i}{Distance from the Sun which was obtained from the Gaia parallax data, following \citep{2018A&A...616A...9L}.}
\end{table*}
  
\begin{longtab}
\tiny
\addtolength{\tabcolsep}{-3.5pt}  
\begin{center}
\begin{longtable}{llllllllllllll}
\caption{\label{tab:Fulldisk} Observational parameters of our sample of full disk systems.$^*$}\\
\hline\hline 
Mon ID\tablefootmark{a} & TwoMass ID\tablefootmark{b} & SpT\tablefootmark{c} & $\mathrm{EW}_{\mathrm{H}\alpha}$\tablefootmark{d} & $\mathrm{W}10\%_{\mathrm{H}\alpha}$\tablefootmark{d} & $\mathrm{EW}_{\mathrm{H}\alpha}$\tablefootmark{e} & $Class$\tablefootmark{f} & $\alpha_{\mathrm{IRAC}}$ \tablefootmark{g}  & $R_\star$\tablefootmark{h} & $T\star$\tablefootmark{h} & $UV_{Exc.}$\tablefootmark{h} & 
$\dot{M}_{UV_{Exc.}}$\tablefootmark{h} & $\dot{M}_{H\alpha}$\tablefootmark{d} & Dist\tablefootmark{i}  \\
 & &   & $(\mathring{\mathrm{A}})$ & $(\mathrm{km}\mathrm{s}^{-1})$  &$(\mathring{\mathrm{A}})$ & & & $(R_\odot)$ & $(K)$ & $(mag)$ & $(M_\odot yr^{-1})$ &  $(M_\odot yr^{-1})$ & ($\mathrm{Pc}$)\\
\hline
\endfirsthead
 \caption{Continued.} \\
\hline\hline
Mon ID\tablefootmark{a} & TwoMass ID\tablefootmark{b} & SpT\tablefootmark{c} & $\mathrm{EW}_{\mathrm{H}\alpha}$\tablefootmark{d} & $\mathrm{W}10\%_{\mathrm{H}\alpha}$\tablefootmark{d} & $\mathrm{EW}_{\mathrm{H}\alpha}$\tablefootmark{e} & $Class$\tablefootmark{f} & $\alpha_{\mathrm{IRAC}}$ \tablefootmark{g} & $R_\star$\tablefootmark{h} & $T\star$\tablefootmark{h} & $UV_{Exc.}$\tablefootmark{h} & 
$\dot{M}_{UV_{Exc.}}$\tablefootmark{h} & $\dot{M}_{H\alpha}$\tablefootmark{d} & Dist\tablefootmark{i}  \\
 & &   & $(\mathring{\mathrm{A}})$ & $(\mathrm{km}\mathrm{s}^{-1})$  &$(\mathring{\mathrm{A}})$ & & & $(R_\odot)$ & $(K)$ & $(mag)$ & $(M_\odot yr^{-1})$ &  $(M_\odot yr^{-1})$ & ($\mathrm{Pc}$) \\
\hline
\endhead
\hline
\endfoot
\hline
\endlastfoot
Mon-000007  &  06415304+0958028  &  K7     &         &       &         &  c  &  -1.47  &          &    4000.  &       -0.650  &      6.0E-08  &               &           \\
Mon-000017  &  06413199+1000244  &  K5     &         &       &   13.1  &  c  &  -2.69  &    1.45  &    4395.  &       -0.270  &      5.8E-09  &               &    694.0  \\
Mon-000028  &  06410511+0958461  &  M3.5   &         &       &  294.5  &  c  &  -1.64  &    0.57  &    3300.  &       -1.400  &      7.4E-10  &               &    705.8  \\
Mon-000056  &  06415315+0950474  &  K5     &         &       &    1.8  &  c  &  -1.36  &    1.52  &    4395.  &        0.010  &      5.8E-09  &               &    756.8  \\
Mon-000059  &  06410334+1000478  &  M4     &         &       &         &  c  &  -1.44  &    0.76  &    3230.  &       -1.110  &      1.2E-09  &               &    788.8  \\
Mon-000063  &  06411193+0959412  &  M2.5   &         &       &   19.4  &  c  &  -1.37  &    1.19  &    3430.  &       -1.030  &      6.2E-09  &               &    748.9  \\
Mon-000064  &  06411070+0957424  &  M1.5   &         &       &    3.5  &  w  &  -1.15  &    1.62  &    3590.  &       -0.320  &      4.7E-09  &               &    736.4  \\
Mon-000080  &  06411795+1004021  &  M3     &         &       &         &  w  &  -1.12  &    1.17  &    3360.  &       -0.460  &      1.9E-09  &               &    723.2  \\
Mon-000081  &  06405978+1002126  &  M2     &         &       &  155.0  &  c  &  -1.01  &    1.00  &    3500.  &       -1.340  &      5.4E-09  &               &    667.7  \\
Mon-000090  &  06410896+0933460  &  M3     &         &       &   51.0  &  c  &  -0.88  &    2.26  &    3360.  &       -0.550  &      1.1E-08  &               &    786.6  \\
Mon-000099  &  06411610+0926435  &  M4.5   &         &       &         &  w  &  -1.44  &    1.16  &    3175.  &        0.360  &      1.1E-10  &               &    743.3  \\
Mon-000103  &  06405954+0935109  &  K6     &   15.6  &  401  &    6.4  &  c  &  -1.14  &    3.84  &    4200.  &        0.591  &               &      9.6E-09  &    658.2  \\
Mon-000106  &  06405272+0928437  &  M0.5   &         &       &         &  c  &  -1.40  &    0.74  &    3800.  &       -1.260  &      1.3E-09  &               &    764.8  \\
Mon-000113  &  06410028+0928341  &  K6     &         &       &  314.6  &  c  &  -1.25  &    0.71  &    4200.  &       -2.380  &      6.3E-09  &               &    672.4  \\
Mon-000117  &  06405413+0948434  &  M2.5   &         &       &  353.0  &  c  &  -2.34  &    1.21  &    3430.  &       -2.090  &      3.4E-08  &               &    752.9  \\
Mon-000119  &  06412100+0933361  &  K6     &   10.5  &  466  &   10.6  &  c  &  -1.44  &    2.04  &    4200.  &       -0.147  &               &      3.3E-09  &           \\
Mon-000126  &  06405783+0941201  &  M0     &         &       &   26.4  &  c  &  -1.09  &    2.02  &    3920.  &       -0.540  &      2.1E-08  &               &    573.7  \\
Mon-000136  &  06411599+0926094  &  M0     &         &       &  174.1  &  c  &  -0.53  &    1.99  &    3920.  &       -1.150  &      1.6E-08  &               &    792.4  \\
Mon-000137  &  06404712+0942077  &  M2     &         &       &   11.4  &  c  &  -1.54  &    1.36  &    3500.  &       -0.360  &      2.9E-09  &               &    557.9  \\
Mon-000145  &  06412346+0945586  &  K7     &         &       &    4.0  &  w  &  -2.46  &    1.39  &    4000.  &        0.050  &               &               &    820.3  \\
Mon-000153  &  06405990+0947044  &  M3     &         &       &   39.9  &  c  &  -1.71  &    1.35  &    3360.  &       -1.220  &      9.1E-09  &               &    791.7  \\
Mon-000165  &  06410481+0944333  &  K4.5   &         &       &   92.9  &  c  &  -0.62  &    0.76  &    4490.  &       -1.580  &      1.5E-09  &               &    743.8  \\
Mon-000168  &  06414287+0925084  &  K5:M0  &   44.9  &  390  &   86.0  &  c  &  -1.38  &    1.20  &    4157.  &       -1.450  &      1.1E-08  &      3.8E-09  &    776.0  \\
\hline
\end{longtable}
\end{center}
\vspace{-0.5cm}
 \tablefoot{Only a portion of this table is shown here. A full version is available at the CDS. This table is ordered according to the Mon ID.}
 \tablefoottext{a}{CSIMon is an internal identification of the CSI 2264 campaign. Here, CSI was omitted for brevity.}
 \tablefoottext{b}{2Mass identification.}
 \tablefoottext{c}{Spectral type obtained by \cite{2014A&A...570A..82V}.}
 \tablefoottext{d}{$\mathrm{H}\alpha$ parameters obtained using FLAMES spectra by 
\cite{2016A&A...586A..47S}. They used the convention that positive $\mathrm{H}\alpha$ 
equivalent width indicates $\mathrm{H}\alpha$ in emission, and negative values 
correspond to $\mathrm{H}\alpha$ in absorption.} 
 \tablefoottext{e}{$\mathrm{H}\alpha$ equivalent width obtained by \cite{2005AJ....129..829D}.}
 \tablefoottext{f}{Classification as CTTS (c) and WTTS (w) by \cite{2016A&A...586A..47S} and \cite{2014A&A...570A..82V}.}
 \tablefoottext{g}{$\alpha_{IRAC}$ is the slope of the SED 
between $3.6\,\mu\mathrm{m}$ and $8\,\mu\mathrm{m}$ obtained by \cite{2012A&A...540A..83T}.}
 \tablefoottext{h}{Parameters obtained using CFHT data by \cite{2014A&A...570A..82V}.}
 \tablefoottext{i}{Distance from the Sun which was obtained from the  Gaia parallax data, following \citep{2018A&A...616A...9L}.
 (*) The systems Mon-000168, Mon-000358, Mon-000361, Mon-000687, Mon-000753 could be classified as transition disk using transition disk select criteria from \cite{2018ApJ...863...13G}, see text. }
 \end{longtab}

 \begin{longtab}
\tiny
\addtolength{\tabcolsep}{-4pt}  
\begin{center}
\begin{longtable}{llllllllllllll}
\caption{\label{tab:Diskless} Observational parameters of our sample of diskless stars.}\\
\hline\hline 
Mon ID\tablefootmark{a} & TwoMass ID\tablefootmark{b} & SpT\tablefootmark{c} & $\mathrm{EW}_{\mathrm{H}\alpha}$\tablefootmark{d} & $\mathrm{W}10\%_{\mathrm{H}\alpha}$\tablefootmark{d} & $\mathrm{EW}_{\mathrm{H}\alpha}$\tablefootmark{e} & $Class$\tablefootmark{f} & $\alpha_{\mathrm{IRAC}}$ \tablefootmark{g}  & $R_\star$\tablefootmark{h} & $T\star$\tablefootmark{h} & $UV_{Exc.}$\tablefootmark{h} & 
$\dot{M}_{H\alpha}$\tablefootmark{d} & Dist\tablefootmark{i}  \\
 & &   & $(\mathring{\mathrm{A}})$ & $(\mathrm{km}\mathrm{s}^{-1})$  &$(\mathring{\mathrm{A}})$ & &  & $(R_\odot)$ & $(K)$ & $(mag)$ &  $(M_\odot yr^{-1})$ & ($\mathrm{Pc}$) \\
\hline
\endfirsthead
 \caption{Continued.} \\
\hline\hline
Mon ID\tablefootmark{a} & TwoMass ID\tablefootmark{b} & SpT\tablefootmark{c} & $\mathrm{EW}_{\mathrm{H}\alpha}$\tablefootmark{d} & $\mathrm{W}10\%_{\mathrm{H}\alpha}$\tablefootmark{d} & $\mathrm{EW}_{\mathrm{H}\alpha}$\tablefootmark{e} & $Class$\tablefootmark{f} & $\alpha_{\mathrm{IRAC}}$ \tablefootmark{g}  & $R_\star$\tablefootmark{h} & $T\star$\tablefootmark{h} & $UV_{Exc.}$\tablefootmark{h}  & $\dot{M}_{H\alpha}$\tablefootmark{d}  & Dist\tablefootmark{i}\\
 & &   & $(\mathring{\mathrm{A}})$ & $(\mathrm{km}\mathrm{s}^{-1})$  &$(\mathring{\mathrm{A}})$ & &  & $(R_\odot)$ & $(K)$ & $(mag)$ &  $(M_\odot yr^{-1})$ & ($\mathrm{Pc}$) \\
\hline
\endhead
\hline
\endfoot
\hline
\endlastfoot
Mon-000018  &  06411322+0955086  &  K3:K4  &         &       &    3.0  &  w  &  -2.73  &    2.03  &    4630.  &        0.216   &               &    812.9  \\
Mon-000023  &  06411242+0955001  &  M3     &         &       &         &  w  &  -2.74  &    0.98  &    3360.  &       -0.011   &               &    699.7  \\
Mon-000029  &  06410328+0957549  &  K7     &         &       &    1.5  &  w  &  -2.85  &    1.68  &    4000.  &       -0.040   &               &    706.5  \\
Mon-000033  &  06410726+0958311  &  K5     &         &       &    1.4  &  w  &  -2.78  &    1.87  &    4395.  &       -0.046   &               &    723.1  \\
Mon-000050  &  06410153+1000365  &  K4     &         &       &    1.1  &  w  &  -2.88  &    1.47  &    4580.  &       -0.081   &               &    746.5  \\
Mon-000055  &  06413491+1001472  &  M2     &         &       &         &  w  &  -2.80  &    1.53  &    3500.  &       -0.111   &               &    711.3  \\
Mon-000057  &  06410393+0958094  &  M3     &         &       &    4.1  &  w  &  -2.74  &    2.16  &    3360.  &        0.172   &               &    736.6  \\
Mon-000060  &  06411532+0954509  &  M3     &         &       &    4.0  &  w  &  -2.92  &    1.25  &    3360.  &        0.016   &               &    998.3  \\
Mon-000066  &  06410357+1000353  &  M1     &         &       &    6.9  &  w  &  -2.66  &    1.53  &    3680.  &        0.027   &               &    760.0  \\
Mon-000075  &  06411159+1002235  &  K4     &         &       &         &  w  &  -2.84  &    2.18  &    4580.  &        0.079   &               &    762.9  \\
Mon-000086  &  06404645+0959463  &  M0     &         &       &         &  w  &  -3.04  &    1.45  &    3920.  &        0.221   &               &    717.2  \\
Mon-000104  &  06410417+0952020  &  K6     &         &       &    1.3  &  w  &  -2.79  &    1.87  &    4200.  &        0.010   &               &    763.1  \\
Mon-000108  &  06411484+0932358  &  M3     &         &       &    2.2  &  w  &  -2.85  &    3.00  &    3360.  &        0.235   &               &    883.5  \\
Mon-000121  &  06414901+0941061  &  M0     &         &       &    1.7  &  w  &  -2.72  &    1.95  &    3920.  &        0.200   &               &    923.2  \\
Mon-000135  &  06405999+0928500  &  K7     &         &       &    1.7  &  w  &  -2.59  &    1.64  &    4000.  &        0.009   &               &    752.9  \\
Mon-000139  &  06405367+0958000  &  M1     &         &       &    3.2  &  w  &  -3.16  &    1.36  &    3680.  &       -0.145   &               &    709.2  \\
Mon-000142  &  06404480+0949478  &  M2     &         &       &    5.5  &  w  &  -3.01  &    1.13  &    3500.  &        0.100   &               &    726.1  \\
Mon-000143  &  06410454+0926092  &  M3     &         &       &    3.8  &  w  &  -2.80  &    1.03  &    3360.  &       -0.000   &               &    772.1  \\
Mon-000149  &  06411330+0951544  &  M1     &         &       &    2.0  &  w  &  -2.69  &    1.25  &    3680.  &       -0.081   &               &    733.5  \\
Mon-000151  &  06411829+0928330  &  M0     &         &       &    6.2  &  w  &  -2.66  &    0.94  &    3920.  &        0.030   &               &    759.8  \\
Mon-000158  &  06404484+0946384  &  K0.5   &         &       &         &  w  &  -3.80  &    1.05  &    5170.  &       -0.626   &               &    660.7  \\
Mon-000159  &  06405146+0937144  &  M1     &         &       &    3.4  &  w  &  -4.65  &    1.38  &    3680.  &        0.055   &               &    701.9  \\
\hline
\end{longtable}
\end{center}
 \tablefoot{Only a portion of this table is shown here. A full version is available at the CDS. This table is ordered according to the Mon ID.}\\
 \tablefoottext{a}{CSIMon is an internal identification of the CSI 2264 campaign. Here, CSI was omitted for brevity.}
 \tablefoottext{b}{2Mass identification.}
 \tablefoottext{c}{Spectral type obtained by \cite{2014A&A...570A..82V}.}
 \tablefoottext{d}{$\mathrm{H}\alpha$ parameters obtained using FLAMES spectra by 
\cite{2016A&A...586A..47S}. They used the convention that positive $\mathrm{H}\alpha$ 
equivalent width indicates $\mathrm{H}\alpha$ in emission, and negative values 
correspond to $\mathrm{H}\alpha$ in absorption.  The accretion rate attributed to WTTS, here represented by the diskless sample, 
corresponds to a contribution from nebular and chromospheric \ha line emission.} 
 \tablefoottext{e}{$\mathrm{H}\alpha$ equivalent width obtained by \cite{2005AJ....129..829D}.}
 \tablefoottext{f}{Classification as CTTS (c) and WTTS (w) by \cite{2016A&A...586A..47S} and \cite{2014A&A...570A..82V}.}
 \tablefoottext{g}{$\alpha_{IRAC}$ is the slope of the SED 
between $3.6\,\mu\mathrm{m}$ and $8\,\mu\mathrm{m}$ obtained by \cite{2012A&A...540A..83T}.}
 \tablefoottext{h}{Parameters obtained using CFHT data by \cite{2014A&A...570A..82V}.}
 \tablefoottext{i}{Distance from the Sun which was obtained from the  Gaia parallax data, following \citep{2018A&A...616A...9L}.}
 \end{longtab}

\section{Results} \label{sec:result}

\subsection{Star formation sites in NGC 2264}

Star formation is still ongoing in some regions of NGC 2264
\citep{2004A&A...417..557L,2009AJ....138.1116S,2012A&A...540A..83T}. These regions have the most embedded members ($\mathrm{A_v}>3.0$) in the cluster, as seen in \cite{2012A&A...540A..83T}. 
Since the youngest stars are clustered, we wondered if the transition disks were clustered too. 
We checked if transition disk candidates were randomly located in NGC 2264
or preferentially clustered in some spatial regions. 

\begin{figure}[htb!] 
 \centering
 \includegraphics[width=\hsize]{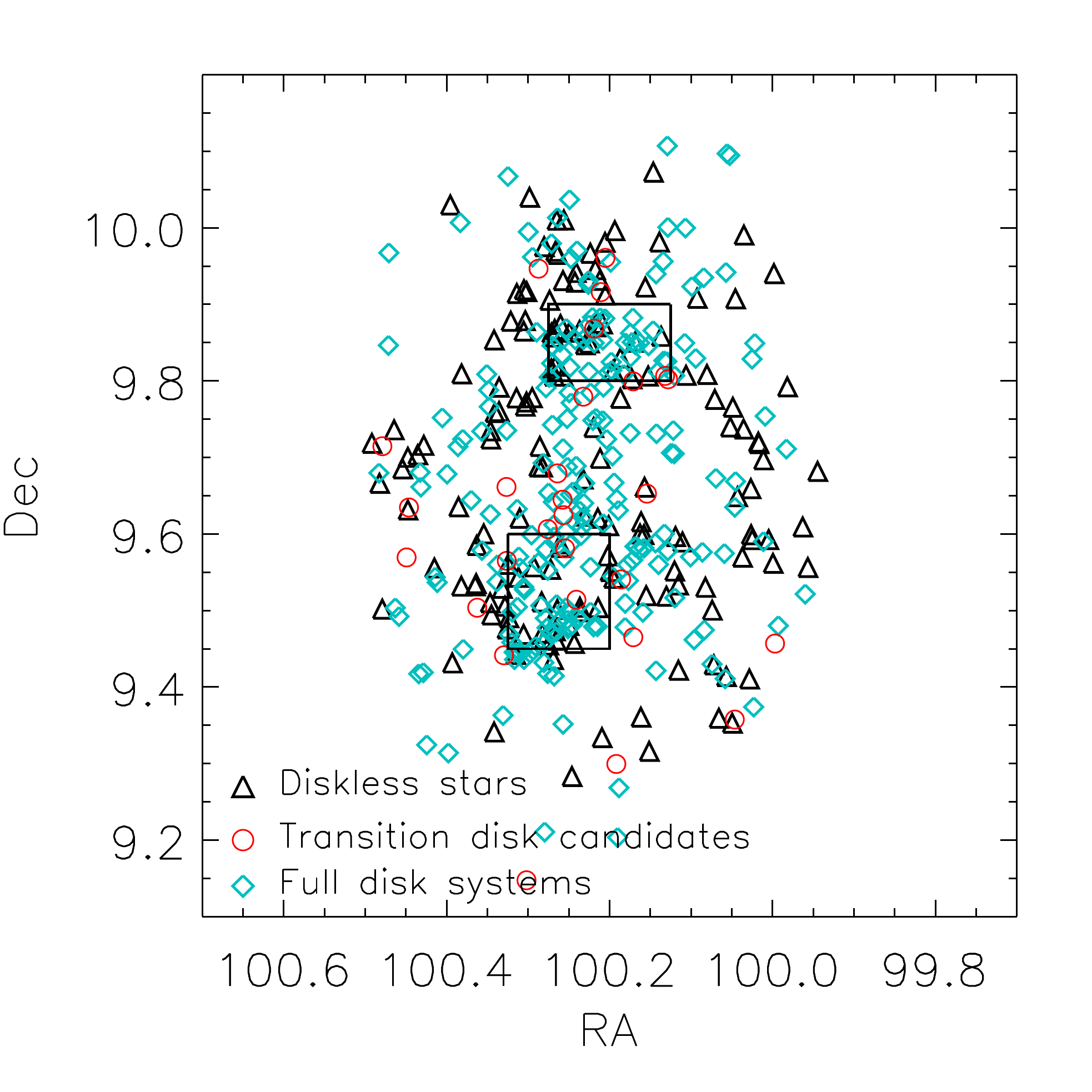}
 \caption{\label{fig:Coord} Spatial distribution of stars belonging to NGC 2264 and analyzed in this work. 
Our classification of the systems is represented with different symbols and colors. The two boxes delimit 
the most active star formation regions, as defined by \cite{2004A&A...417..557L}. The population in these 
boxes are predominantly composed of full disk systems. Transition disk candidates and diskless stars 
are preferentially found outside of the active star formation regions.}  
\end{figure}

We show in Fig. \ref{fig:Coord} 
the spatial distribution of our three groups of stars. The boxes define 
the regions of the most active star formation of the cluster taken from 
\cite{2004A&A...417..557L}. We calculated the fraction of stars of a given group that falls 
in the boxes, that is the number of systems of a given group in the boxes divided by the total number of systems 
in the boxes. To compare with that, we also calculated the fraction of each 
group that falls out of the boxes, that is the number of systems of a given group out 
of the boxes divided by the total number of systems out of the boxes.
We found that in the boxes, $\sim 59\,\%$ of systems are full disks, $\sim 4\,\%$ 
are transition disk candidates, and $\sim 37\,\%$ correspond to diskless stars. 
The fractions of systems that fall out of the boxes are $\sim49\,\%$ full 
disks, $\sim\,8\%$ transition disk candidates, and $\sim43\,\%$ diskless stars. 
The population in the active star formation regions are predominantly composed 
of full disk systems and the fraction of transition disk candidates and diskless stars
in these regions is considerably smaller than outside, which 
may point to an evolutionary scenario for the cluster, as discussed by \cite{2014A&A...570A..82V}.

\subsection{Disk diagnostics}\label{sec:disk}

\begin{figure*}[htb!] 
 \centering
 
\includegraphics[width=17cm]{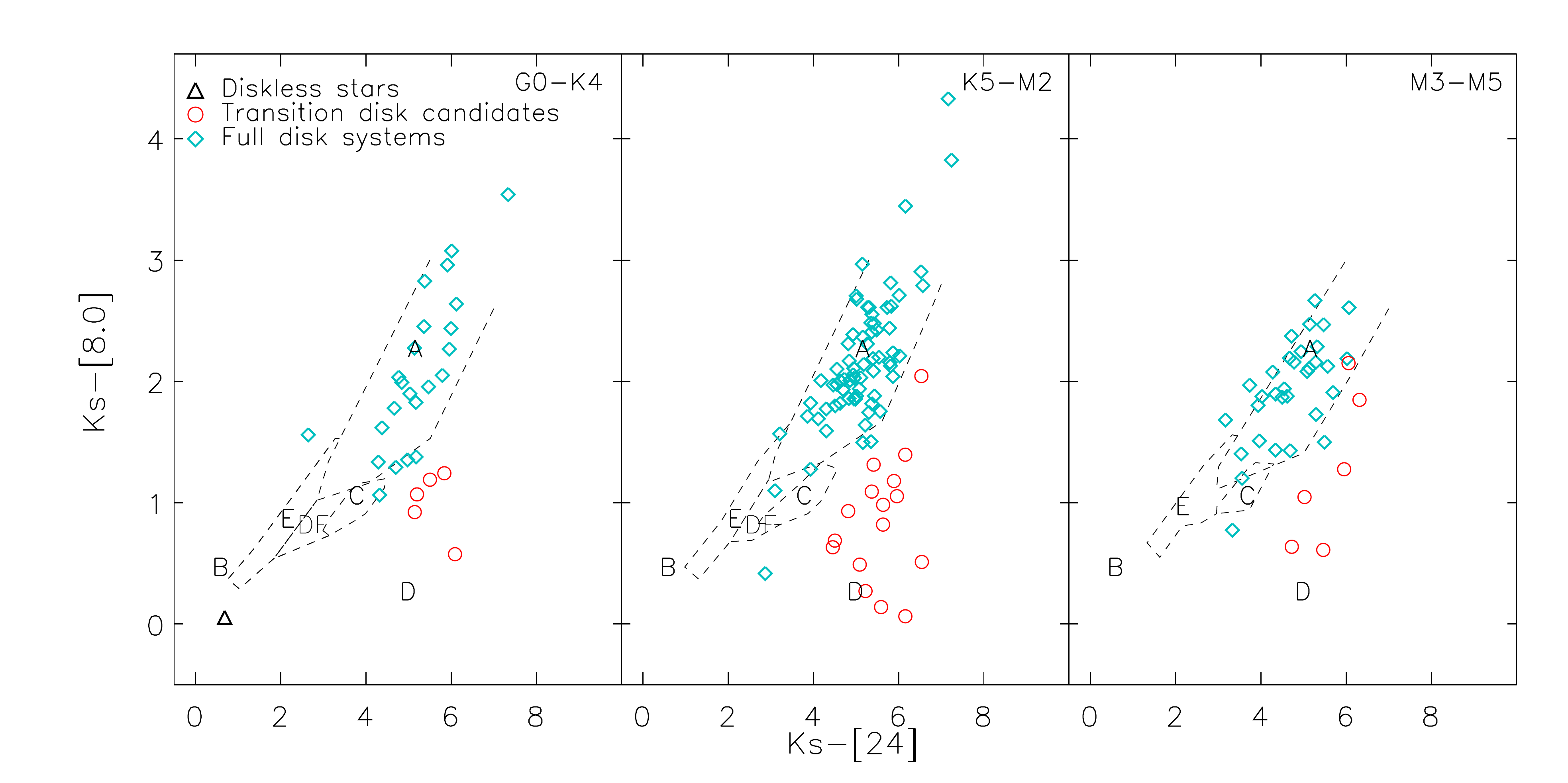}\\
\caption{\label{fig:KvsSpitzer}Near-IR and mid-IR color-color diagram for stars belonging to NGC 2264 
broken up by spectral type ranges into three plots. We can see 
two different populations: stars with full disks present emission excess above the photospheric level 
in the inner and outer parts of the disk, while transition disk candidates present little dust emission 
in the inner disk, and dust emission in the outer disk compatible with full disk systems. 
The dashed lines delimit regions with different disk dispersal mechanisms computed by 
\cite{2013MNRAS.428.3327K}. Region (A) represents thick primordial disks, (B) stellar photospheres, 
(C) ultra-settled primordial disks (flat disks), (D) inside to outside depleted disks, (E) homogeneously 
depleted disks and (DE) is a region that may contain systems with homogeneously depleted 
disks and inside to outside depleted disks. The stars we have selected as full disks and transition disks fall in regions of this diagram that are consistent with that expected from the models.}
\end{figure*}

Transition disks have a dust hole and gap in the inner disk region,  
characterized by a lower quantity of dust compared to the outer disk. Therefore, 
we expect that transition disks will show less excess in near-IR than full disks, while the excess at
longer wavelengths may be compatible with full disk systems. 
In Fig. \ref{fig:KvsSpitzer}, we show 
$K_s-[8.0]$ vs. $K_s-[24]$ diagrams. The stellar magnitudes were dereddened with individual Av value from \cite{2014A&A...570A..82V} and the $A_\lambda/A_v$ relation from the SVO Filter\footnote{The SVO Filter Profile Service. Rodrigo, C., Solano, E., Bayo, A. http://ivoa.net/documents/Notes/SVOFPS/index.html}, which uses the extinction law by \cite{1999PASP..111...63F}, improved by \cite{2005ApJ...619..931I} in the IR. 
We can see that stars with full disks present excess above the photospheric 
emission in the inner and outer parts of the disk, while transition disks present emission in 
the outer disk compatible with a full disk system and lower emission in the near IR than full disks.

As a star moves from a full disk to a diskless system, it moves in different ways in the IR color-color diagrams in Fig. \ref{fig:KvsSpitzer}. 
Disk dispersal can occur from outside to inside due to some external factor like 
photoevaporation by a close high mass star \citep{2001MNRAS.325..449S}. 
These systems will present emission above the photospheric level in the inner disk and 
little emission in the outer disk. The disk can also be dispersed homogeneously throughout 
its radius. In this case, the disk will evolve through a diagonal path in Fig. \ref{fig:KvsSpitzer}, 
the inner and outer disks presenting the same emission above the photospheric level at each epoch.
However, \cite{2013MNRAS.428.3327K} show that this type of disk dispersal is not common. 
Disk dispersal can also proceed from inside to outside due, for example, to 
photoevaporation by high-energy radiation of the central star and planet formation. In Fig. \ref{fig:KvsSpitzer}, 
these systems should present IR excesses in the outer disk, and weak or no IR excesses in the inner disk.

In Fig. \ref{fig:KvsSpitzer}, we plot lines from theoretical models that separate stars with disks in 
different evolutionary stages \citep{2013MNRAS.428.3327K}. Our disk classification is in 
agreement with the theoretical disk evolution obtained by \cite{2013MNRAS.428.3327K}, 
since our systems classified as full disks are predominantly located in the region 
with primordial disks (represented by letter A in the plot) and our transition disk 
candidates are in the region of inside to outside disk evolution (represented by letter D in the plot). We do not have 
a significant number of systems in regions C and E that are expected to contain systems 
with ultra-settled primordial disks, which correspond to flat disks and  homogeneously depleted disks, respectively.  
Unfortunately, we do not have MIPS data for the sample of diskless stars used in this work, which we expect 
to show no excess emission at all wavelengths, as seen, for example, in 
\cite{2016PASA...33....5O}, and should be located in region B in the plot.

In the literature there are several criteria to select transition disk candidates 
\citep[e.g.,][]{2009A&A...504..461F,2013ApJS..207....5F,2010ApJ...708.1107M,2010ApJ...718.1200M,2010ApJ...712..925C,2013A&A...552A.115R}, 
using IR photometric and spectroscopic data. We can compare our selection criteria with those used in some previous works.  

\cite{2010ApJ...718.1200M}, used photometric fluxes in IRAC and MIPS bands to classify
a system as a transition disk.  In a $[3.6]-[8]$ versus $[8]-[24]$ diagram they considered that systems 
with $0>[3.6]-[8]<1.1$ and $3.2<[8]-[24]<5.3$ were transition disks, while systems with $1.1<[3.6]-[8]<1.8$ 
and $3.2<[8]-[24]<5.3$ presented a small excess emission and were classified as pre-transition disk 
by \cite{2013ApJ...769..149K}. Pre-transition disks were defined as systems with a gap between an inner 
and an outer disk \citep{2008ApJ...682L.125E}. 
The selection criteria  adopted by \cite{2013ApJS..207....5F} are based in the $K_s -[5.8]$ versus $[8]-[24]$ diagram. 
Systems are classified as transition disks if  $[8]-[24]\geq2.5$ and if 
$K_s-[5.8]\leq0.56+([8]-[24])\times0.15$.
\cite{2010ApJ...712..925C} proposed a less restrictive transition disk selection criteria, 
using the $[3.6]-[24]$ versus $[3.6]-[4.5]$ diagram, where a system is classified 
as transition disk if $[3.6]-[24] > 1.5$ and $[3.6]-[4.5] < 0.25$.
\cite{2010ApJ...708.1107M} used as transition disk selection criteria the slopes 
in $log(\lambda F_\lambda)$ versus $log(\lambda)$. They proposed the following 
limits to the slopes: for weak emission sources, $\alpha_{3.6-5.8}<1.8$ and $-1.5\leq\alpha_{8-24}\leq0.0$, 
and for transition disks, $\alpha_{3.6-5.8}<1.8$ and $\alpha>0.0$.

\begin{figure*} 
 \centering
\subfigure[]{\includegraphics[scale=0.45]{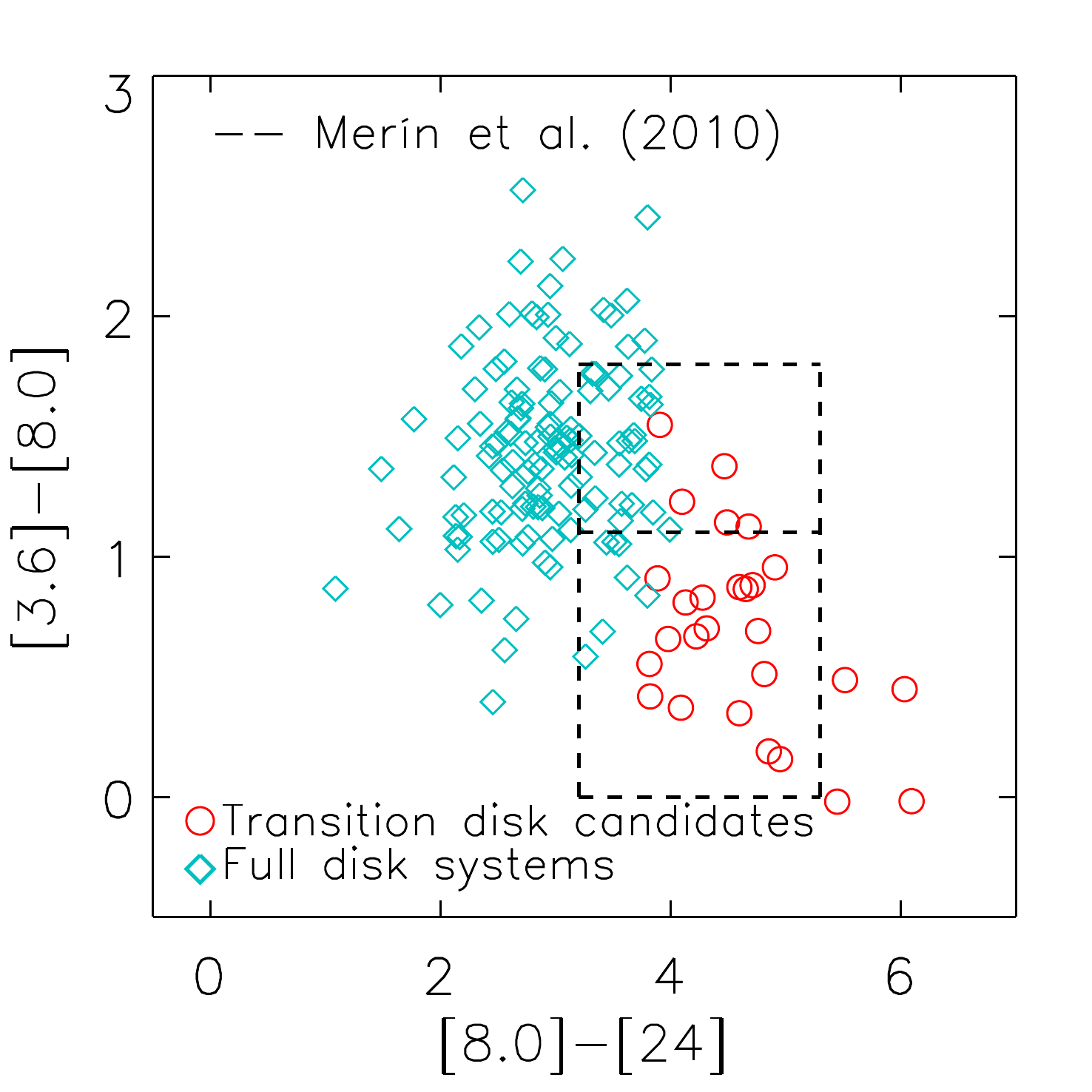}}
\subfigure[]{\includegraphics[scale=0.45]{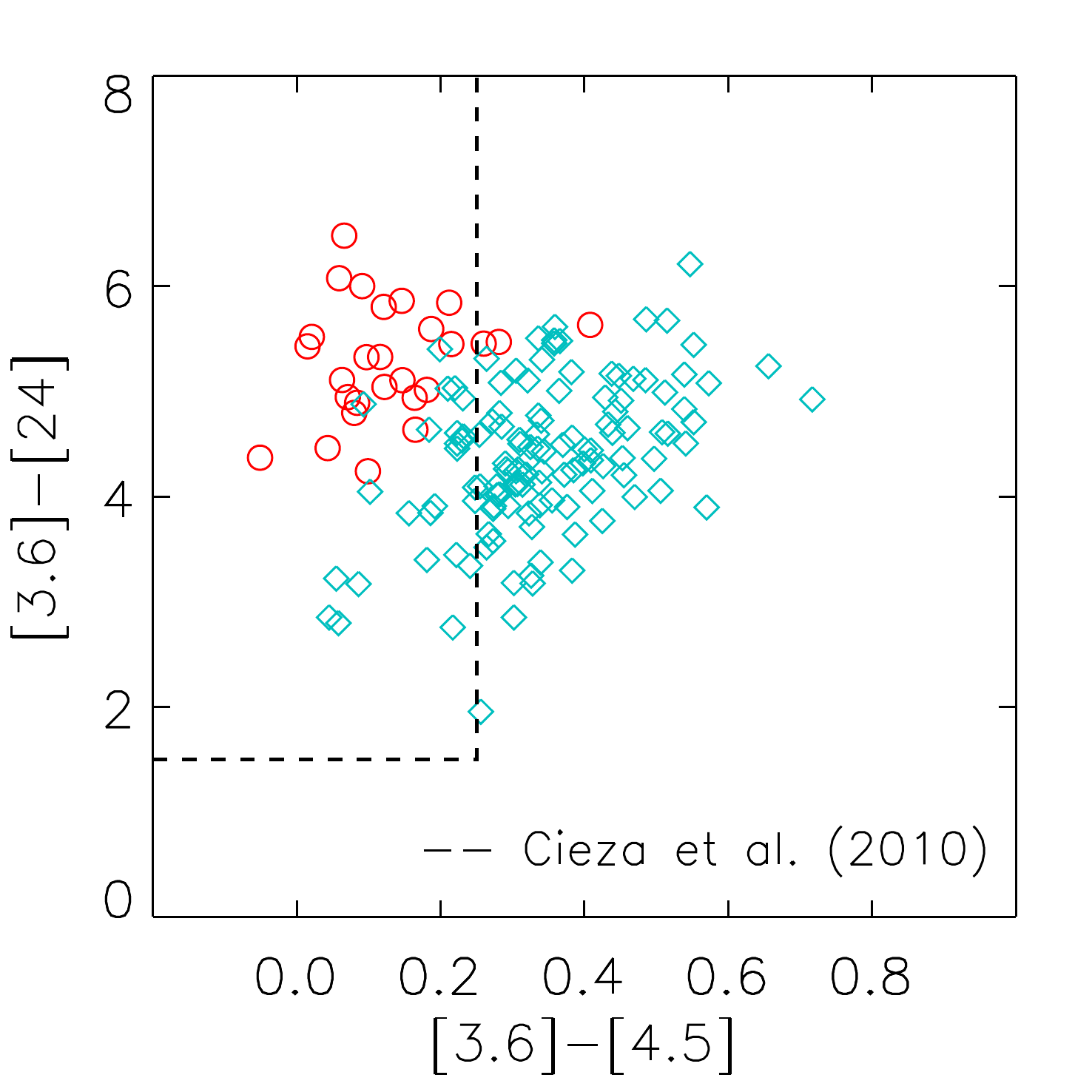}}\\
\subfigure[]{\includegraphics[scale=0.45]{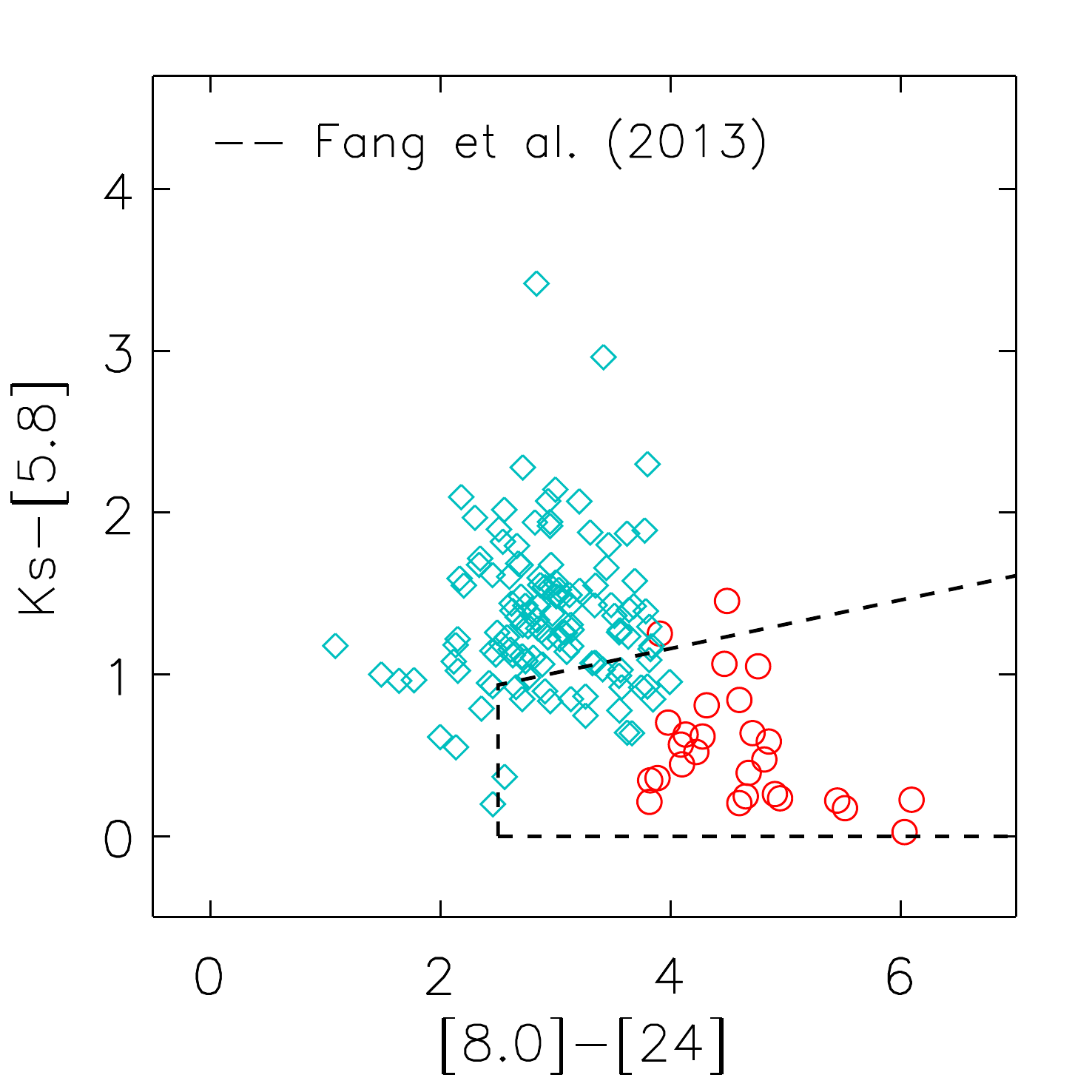}}
\subfigure[]{\includegraphics[scale=0.45]{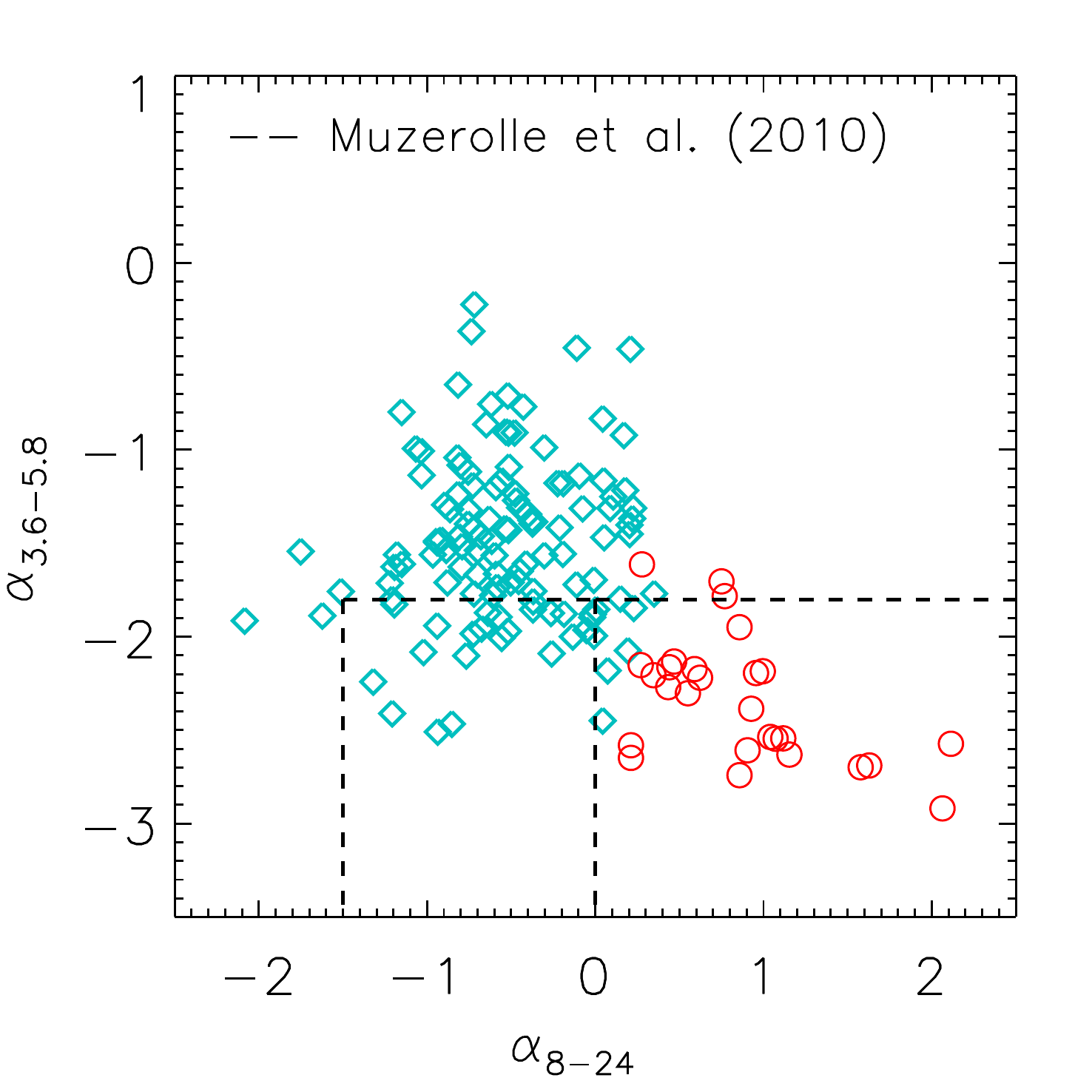}}
\caption{\label{fig:Selec_crit} Different diagrams that are used in previous works to separate 
transition disk systems from the full disk sample. a) Selection criteria that were used by 
\cite{2010ApJ...718.1200M}. Systems that fall in the bottom dashed box are classified 
as transition disks and systems that fall in the top dashed box are systems with small excess 
in the IRAC bands.  b) Selection criteria used by \cite{2010ApJ...712..925C}.  
The dashed lines separate the transition disks. c) Selection criteria from  
\cite{2013ApJS..207....5F}. Systems that fall in the dashed region are transition disks. 
d) Selection criteria from \cite{2010ApJ...708.1107M}. Systems in the left dashed box 
are classified as weak-excess sources and systems that lie in the right dashed box 
are transition disks. Almost all of our transition disk candidates fulfill the 
selection criteria from the literature. } 
\end{figure*}

In Fig. \ref{fig:Selec_crit} we show the selection criteria discussed above and compare them
with our selection criteria.  We can see  that our  transition disk sample is consistent with 
the various selection criteria. But we also see that $52$ systems that we classify as full disks 
agree with the transition disk or pre-transition disk criteria from 
\cite{ 2010ApJ...718.1200M}, \cite{2010ApJ...712..925C}, \cite{2013ApJS..207....5F}, 
and/or \cite{2010ApJ...708.1107M}. Only $15$ of them would be classified as transition disks 
by at least three of these criteria, following the selection criteria from \cite{2018ApJ...863...13G}.  
We checked the SEDs of these systems and most of them are not compatible with a transition disk SED. 
Five systems could be classified as transition disks according to their SEDs, but the 
photometric data are not good enough to confirm such a classification. We marked these five systems 
in Table \ref{tab:Fulldisk}.

The $\alpha_{IRAC}$ index allows a classification of inner disk evolution, as 
proposed by \cite{Lada2006}. They classified star-disk systems as photospheres 
(no dust in the inner disk) when $\alpha_{IRAC}\, < \,-2.56$, anemic disk 
(optically thin disk) if $-2.56\, < \alpha_{IRAC}\, <\,-1.80$, 
optically thick disk if $-1.80 \,< \alpha_{IRAC}\, < \,-0.5$, 
flat spectrum for systems with $-0.5\, < \, \alpha_{IRAC}\, <\, 0.5$, 
and class I sources if $\alpha_{IRAC}\, >\, 0.5$. 

Anemic disk systems fall in the transition region between stars with and 
without dust in the inner disk. This type of disk was also classified as an 
evolved disk by \citet{2007ApJ...671.1784H}.
Using only the $\alpha_{IRAC}$ index that was measured by \cite{2012A&A...540A..83T}, we would have 
classified $179$ ($\sim45\%$) systems as diskless stars, that correspond to photospheres, 
$78$ ($\sim 19\%$) as transition disk candidates, which correspond to the anemic disks, and 
$144$ ($\sim36\%$) as full disk systems, which are the systems classified as flat spectra and thick disks.

\begin{figure}[htb!] 
 \centering
 \includegraphics[width=\hsize]{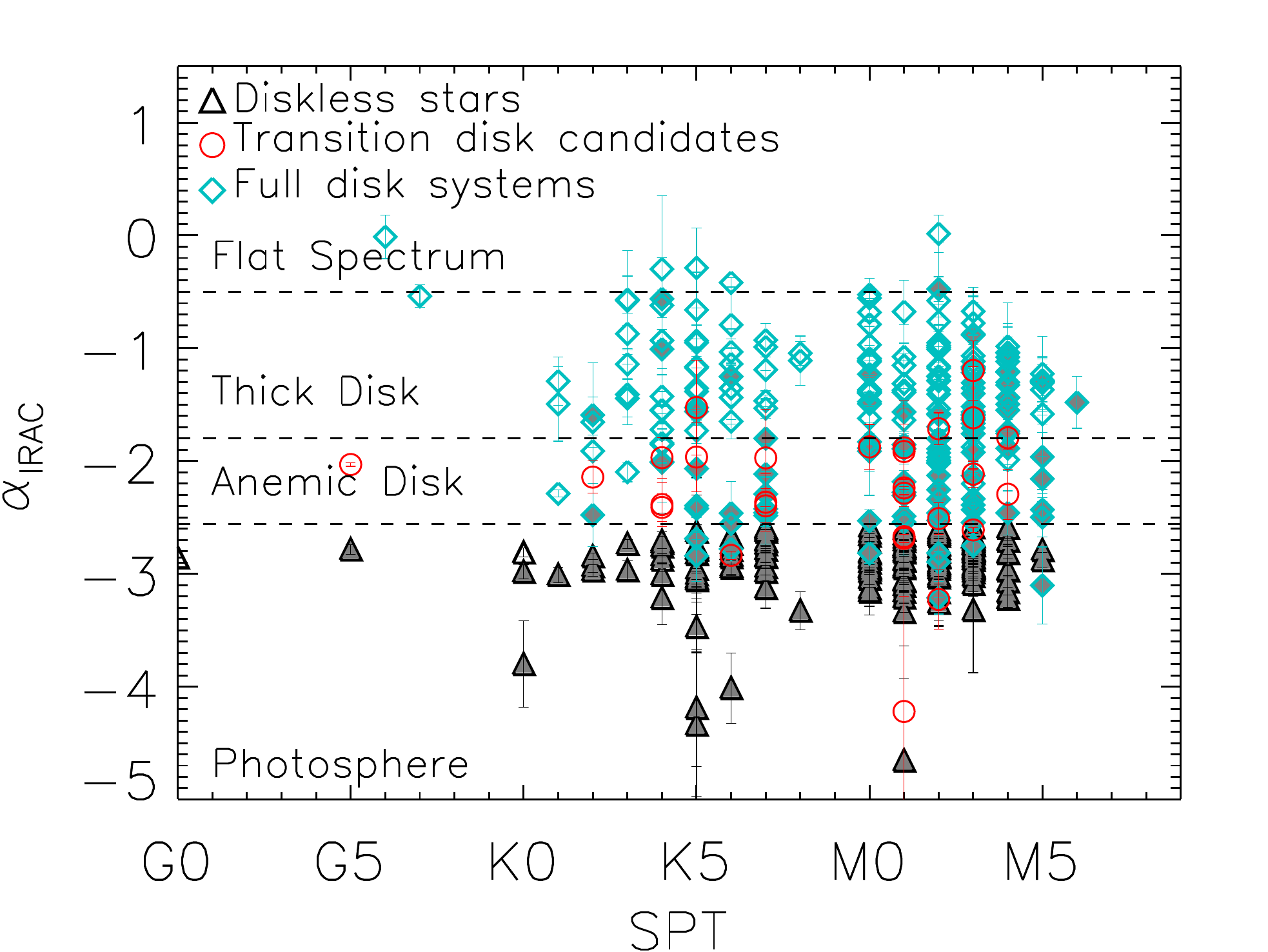}
 \caption{\label{fig:alpha_spt} Slope of the SED from 
$3.6\,\mu\mathrm{m}$ to $8\,\mu\mathrm{m}$ ($\alpha_{IRAC}$) measured by \cite{2012A&A...540A..83T} as a function 
of the spectral type obtained by \cite{2014A&A...570A..82V,2005AJ....129..829D,2002AJ....123.1528R,1956ApJS....2..365W}. 
Symbols indicate our categorization (notation as in Fig. \ref{fig:KvsSpitzer}). 
Our sample of transition disk candidates presents dust in 
the inner disk similar to anemic disks, which shows that anemic disk systems can be good 
candidates to transition disks. Systems for which we do not have data at 
$22/24\,\mu\mathrm{m}$ are identified by gray filled symbols. } 
\end{figure}

Our disk classification based on SED modeling is shown in Fig. \ref{fig:alpha_spt}, where 
we can see the overall good agreement with the $\alpha_{IRAC}$ disk classification. 
All the flat spectra and almost all the thick disk systems are indeed classified as full disks 
according to our SEDs and most of the photospheres correspond to the
diskless SED systems. The transition disk candidates are generally found among 
the anemic disks, but a significant fraction of anemic disks would not be classified 
as transition disk candidates based on the SED analysis. Among the 
$78$ systems classified as anemic disks, only $31$ have available data at 
$22/24\,\mu\mathrm{m}$ and $17$ of these were classified as transition disk candidates 
by the SED fitting. The other $47$ anemic disks, identified in Fig. \ref{fig:alpha_spt}, 
do not have $22/24\,\mu\mathrm{m}$ data, 
and we could not properly model their outer disk emission, as discussed in 
Sect. \ref{sec:sed} and as shown by \cite{2007ApJS..169..328R}. Therefore the number 
of transition disk systems in our sample could be larger than what we could determine. 

\subsection{Accretion diagnostics}\label{sec:accretion}
 
Accreting young low-mass stars are known as classical T Tauri stars (CTTS) 
and are characterized by strong and broad emission lines (e.g., \ha, \hb, \He) 
that vary in intensity and morphology as the star-disk system rotates and accretes, 
and UV excess above the photospheric emission. When the accretion process 
ceases, the young low-mass stars are called weak line T Tauri stars (WTTS).
Surprisingly, most transition disk systems are found to be accreting \citep{2014A&A...568A..18M}. 
It is therefore interesting to analyze how the presence of a hole in the inner disk 
influences accretion to the star in this disk evolutionary phase.

The \ha line is variable, intense and broad in CTTS. This line 
can be formed in different regions of the star-disk system, such as
accretion funnels, winds, and the chromosphere. In typical CTTS, the \ha line 
emission component is thought to come mostly from the accretion funnel, and it 
is used as an accretion diagnostic \citep{2003ApJ...582.1109W,2009A&A...504..461F,2016A&A...586A..47S}. 
The measured equivalent width of \ha depends on the stellar continuum contribution, and
is therefore spectral type dependent.
For a star to be considered a CTTS, the equivalent width of its \ha line 
($\mathrm{EW}_{\mathrm{H}\alpha}$) should be larger than $3 \, \mathring{\mathrm{A}}$ 
for spectral types K0-K5, larger than $10 \, \mathring{\mathrm{A}}$ for K7-M2.5, 
larger than $20 \, \mathring{\mathrm{A}}$ for M3-M5.5, or larger than 
$40 \, \mathring{\mathrm{A}}$ for M6-M7.5 \citep{2003ApJ...582.1109W}.
Stars that show \ha width at $10\%$ of maximum intensity 
($\mathrm{W}10\%_{\mathrm{H}\alpha}$) higher than 270 km/s are also 
considered CTTS \citep{2003ApJ...582.1109W}, 
since gas at such high velocities cannot be explained by the stellar chromosphere alone. 

\begin{figure*}
 \centering
\subfigure[]{\includegraphics[width=8cm]{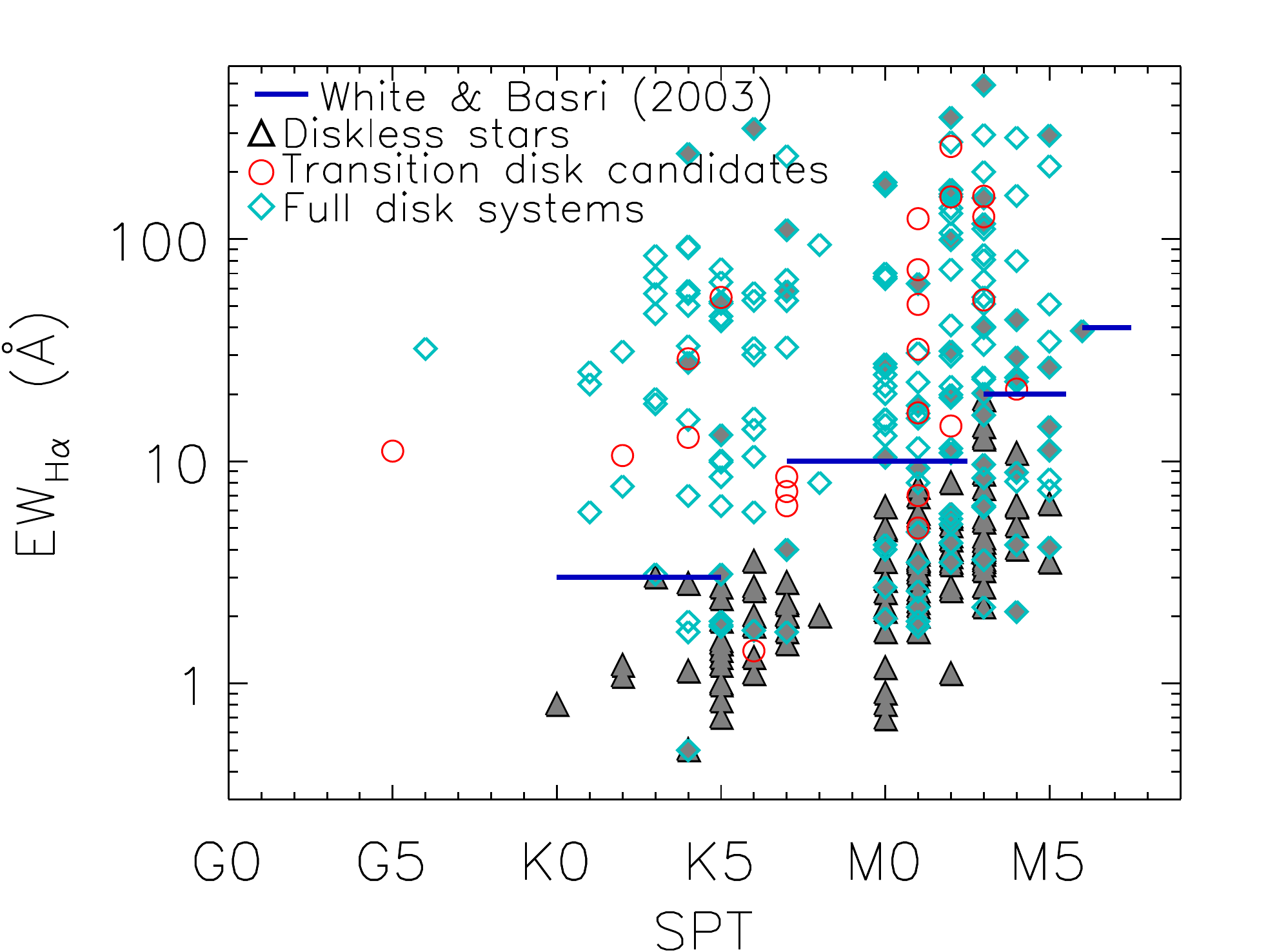}}
\subfigure[]{\includegraphics[width=8cm]{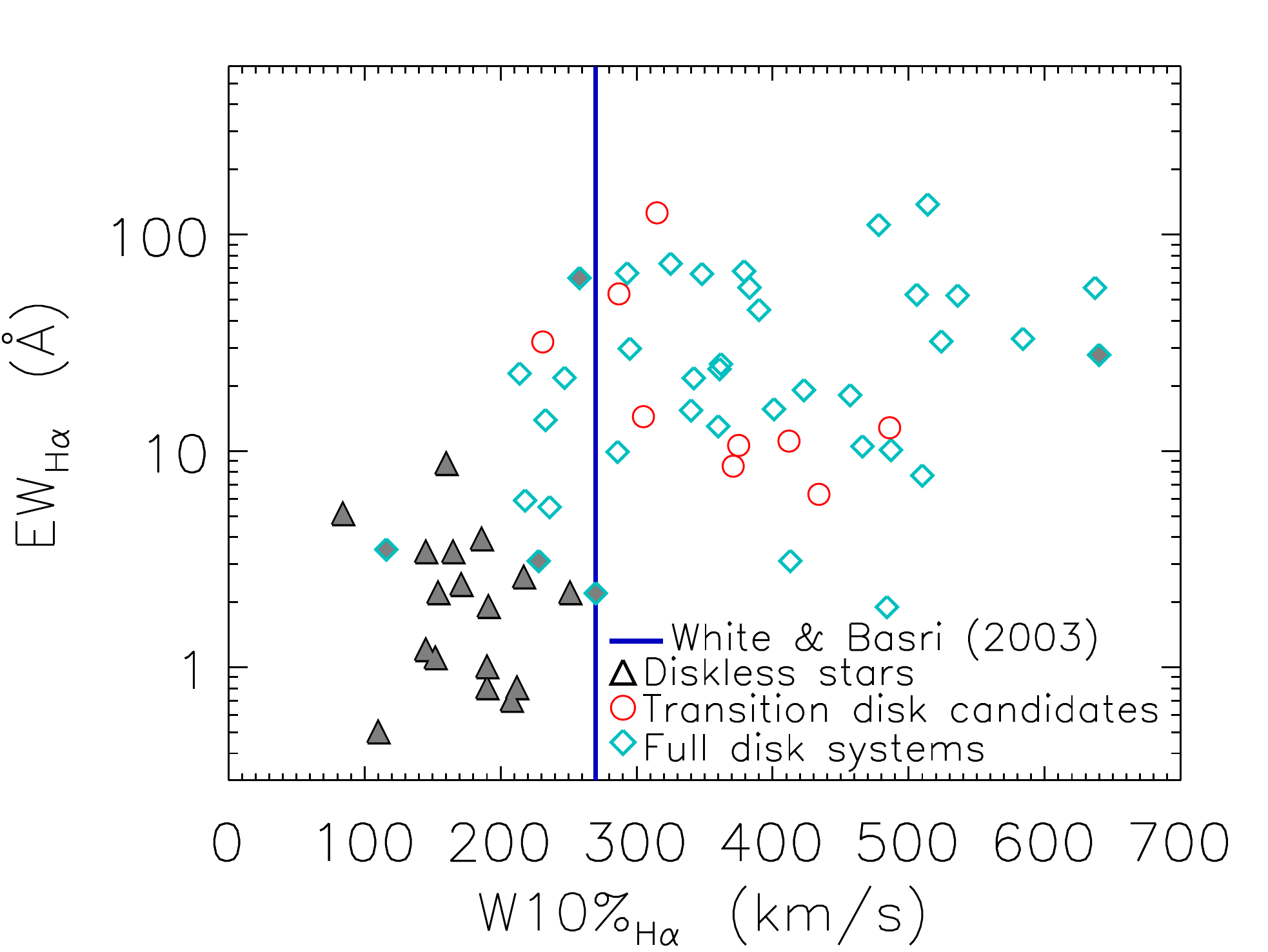}}
 \caption{\label{fig:EWspt} \ha criteria (blue lines) to select CTTS and WTTS \citep{2003ApJ...582.1109W}. 
a) $\mathrm{H}\alpha$ equivalent width vs. spectral type. b) $\mathrm{H}\alpha$ equivalent 
width vs. $\mathrm{H}\alpha$ width at $10\%$ of maximum intensity.  Transition disk candidates are predominantly 
found among accreting systems. Therefore, the presence of a hole in the inner disk does not stop the accretion process, suggesting that gas is flowing through the hole. Systems for which we do not have data at $22/24\,\mu\mathrm{m}$ are identified by gray filled symbols.} 
\end{figure*}

In Fig. \ref{fig:EWspt} shows the $\mathrm{EW}_{\mathrm{H}\alpha}$ as a 
function of spectral type and the 
$\mathrm{W}10\%_{\mathrm{H}\alpha}$ as a function of $\mathrm{EW}_{\mathrm{H}\alpha}$, 
for full disk systems, transition disk candidates, and diskless stars according to our classification. 
The $\mathrm{EW}_{\mathrm{H}\alpha}$ were taken from \cite{2016A&A...586A..47S} 
and \cite{2005AJ....129..829D}, $\mathrm{W}10\%_{\mathrm{H}\alpha}$ comes from \cite{2016A&A...586A..47S}, 
and the spectral types are from \cite{2014A&A...570A..82V}. 
In Table \ref{tab:TransDisk}, \ref{tab:Fulldisk}, and \ref{tab:Diskless}, we show the classification of the stars as CTTS or WTTS for 
all the systems analyzed in this work. In particular, $\sim82\,\%$ of the transition 
disk candidates belonging to NGC 2264 still accrete, suggesting that gas is able to flow through the 
inner disk hole. 

In Fig. \ref{fig:EWspt}a, we see a considerable number of full disk systems that 
fall below the $\mathrm{EW}_{\mathrm{H}\alpha}$ threshold of CTTS (blue lines). 
Among them, $17$ stars were observed at $22/24\,\mu\mathrm{m}$ and 
have their outer disk emission and their classification as full disk systems
well determined. The other $28$ full disk systems that fall below the accretion threshold 
do not have data at $22/24\,\mu\mathrm{m}$ and their outer disk contribution is not
constrained. However, as these systems present emission excess in the inner disk 
(based on the $\alpha_{IRAC}$ index obtained by \cite{2012A&A...540A..83T}), 
we classified them as full disks. Among these $28$ systems, $22$ do not present 
other accretion signatures and are true WTTS, despite having near-IR excess. 

\begin{figure*}
 \centering
\subfigure[]{\includegraphics[scale=0.35]{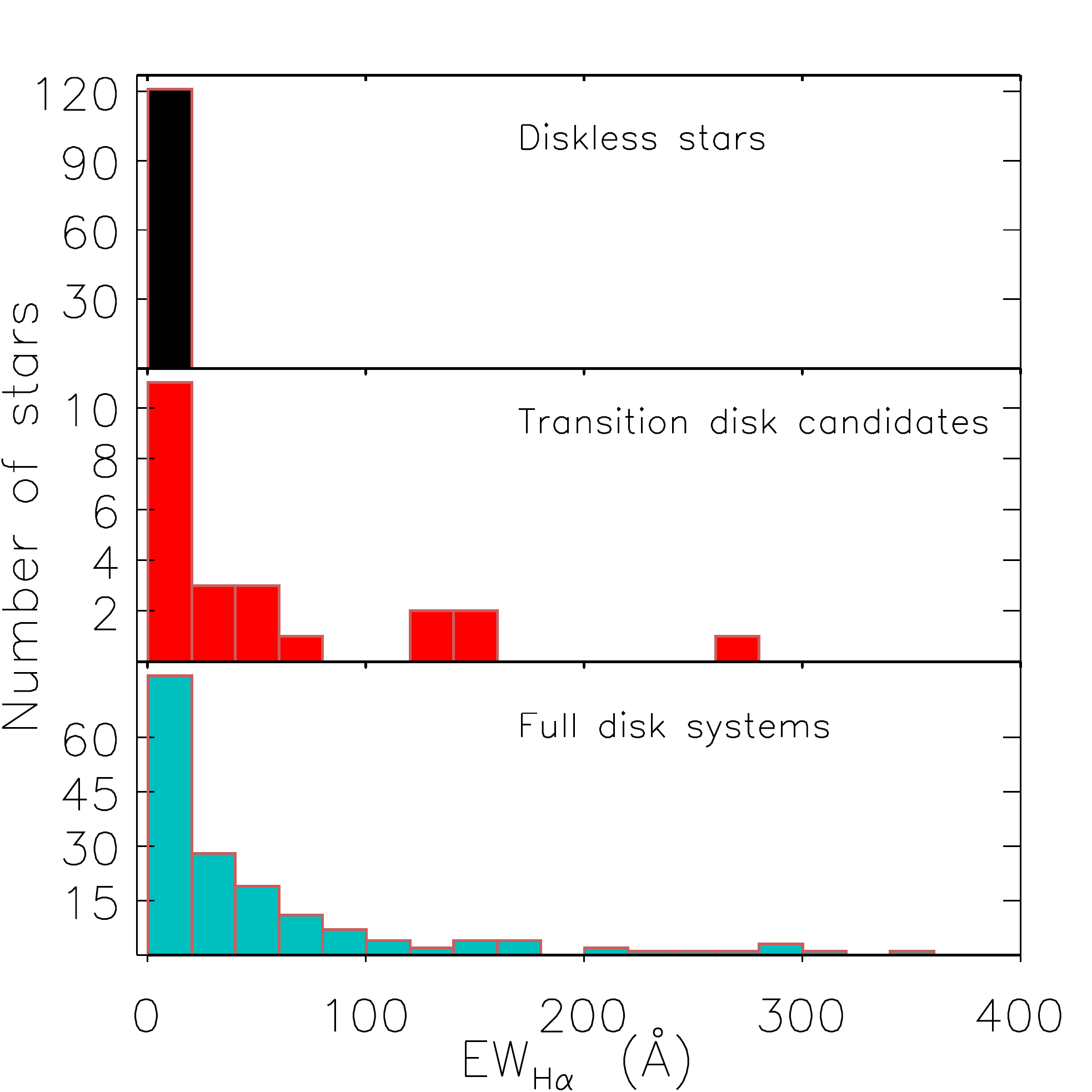}}
\subfigure[]{\includegraphics[scale=0.35]{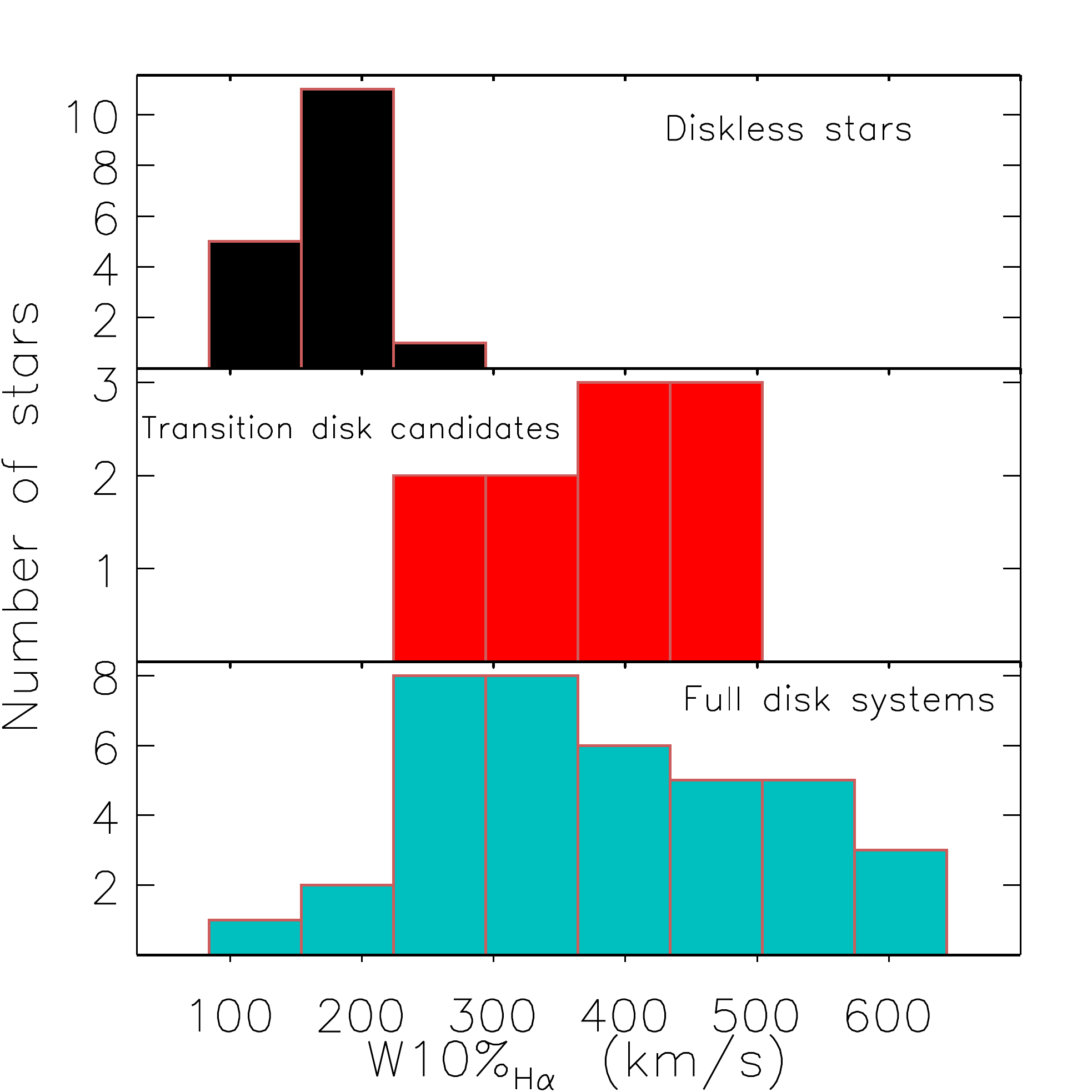}}
 \caption{\label{fig:histEW} Distribution of \ha emission line parameters. (a) \ha equivalent width 
and (b) \ha width at $10\%$ of maximum intensity of our sample of diskless stars (black), 
transition disk candidates (red) and full disk systems (light blue). Transition disk candidates 
distributions are more similar to the full disk systems than diskless stars, see the text.} 
\end{figure*}

In Fig. \ref{fig:histEW} we show the distribution of $\mathrm{EW}_{\mathrm{H}\alpha}$ 
and $\mathrm{W}10\%_{\mathrm{H}\alpha}$ of our sample of stars. The $\mathrm{EW}_{\mathrm{H}\alpha}$ 
mean values are $(3.7\pm0.3)\,\mathring{\mathrm{A}}$ for diskless stars,  
$(51\pm14)\,\mathring{\mathrm{A}}$ for transition disk candidates, and $(59\pm8)\,\mathring{\mathrm{A}}$ 
for full disk systems. The transition disk distribution of $\mathrm{EW}_{\mathrm{H}\alpha}$ is more similar to 
the full disk system distribution than the diskless stars. A two-sided 
Kolmogorov-Smirnov(K-S) test could not prove that the transition disk and the full disk 
distributions of $\mathrm{EW}_{\mathrm{H}\alpha}$ are statistically different, with a probability of $98\%$ to be the same. 
However, a K-S test indicates that the distribution of transition disk candidates and 
diskless stars are statistically different, with probability of less than $1\%$ to be the same.
The mean values of $\mathrm{W}10\%_{\mathrm{H}\alpha}$ are  $(172\pm10)\,\mathring{\mathrm{A}}$ for diskless systems,  
$(371\pm27)\,\mathring{\mathrm{A}}$ for transition disk candidates, and $(383\pm20)\,\mathring{\mathrm{A}}$ 
for full disk systems. A K-S test shows a probability of $83\%$ that the distributions of $\mathrm{W}10\%_{\mathrm{H}\alpha}$ for 
transition disks and full disk stars are equal. Comparing the distribution of 
transition disk candidates with diskless stars, a K-S test shows that they are 
statistically different with a probability of less then $1\%$ to be the same.

\begin{figure}[!htb!] 
 \centering
 \includegraphics[scale=0.4]{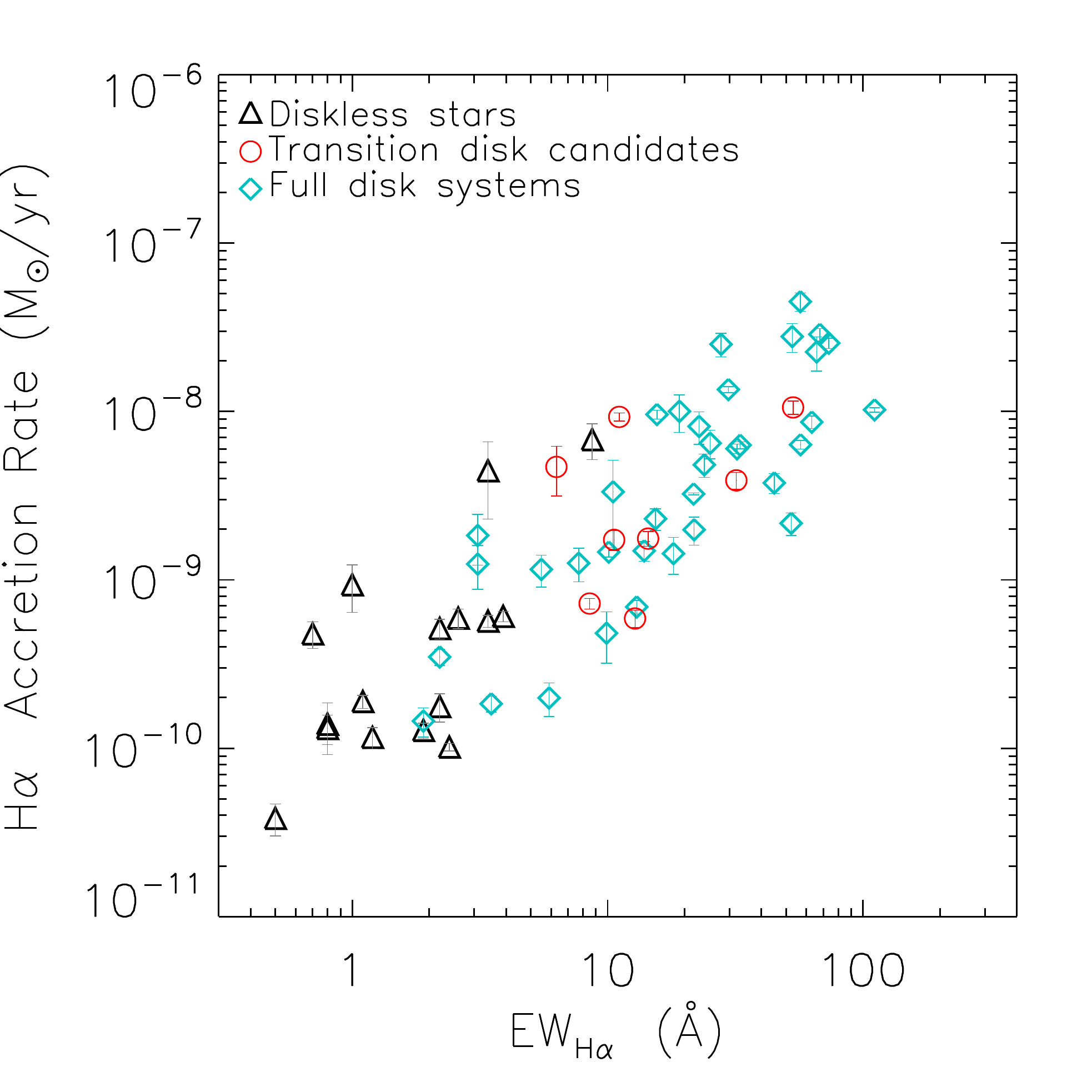}
 \caption{\label{fig:AccRateHa} Mass accretion rates from $\mathrm{H}\alpha$ equivalent width calculated 
by \cite{2016A&A...586A..47S} as a function of $\mathrm{H}\alpha$ equivalent width. The error bar 
represents the night-to-night variability of the accretion rate.} 
\end{figure}

Using $\mathrm{EW}_{\mathrm{H}\alpha}$, \cite{2016A&A...586A..47S} calculated the mass accretion 
rates ($\dot{\mathrm{M}}_{\mathrm{H}\alpha}$) for a sample of stars belonging to NGC 2264 and 
observed with the FLAMES spectrograph on VLT. These accretion rates represent a lower limit, because their
computation did not take into account absorption due to winds that the \ha line may exhibit, which decreases
the total line equivalent width. In Fig. \ref{fig:AccRateHa}, we show 
$\dot{\mathrm{M}}_{\mathrm{H}\alpha}$ from \cite{2016A&A...586A..47S} as a function of $\mathrm{EW}_{\mathrm{H}\alpha}$ for 
full disk systems, transition disk candidates and diskless stars. The $\dot{\mathrm{M}}_{\mathrm{H}\alpha}$ 
for transition disk candidates falls between the values computed for full disk systems and diskless stars. 
The mean $\dot{\mathrm{M}}_{\mathrm{H}\alpha}$ value of each group corroborates this analysis: 
$(9.9\pm3.1)\times10^{-10}\,$\ms yr$^{-1}$ for diskless stars, $(4.1\pm1.2)\times10^{-9}$\ms yr$^{-1}$ for transition disk candidates,
and $(8.2\pm0.5)\times10^{-9}$\ms yr$^{-1}$ for full disk systems. As explained in \cite{2013AA...551A.107M} 
and \cite{2016A&A...586A..47S}, the accretion rate attributed to WTTS, here represented by the diskless sample, 
corresponds to a contribution from nebular and chromospheric \ha line emission.

In CTTS, the stellar magnetic field is expected to be strong enough to truncate 
the disk at a few stellar radii from the star. 
The gas accreting from the disk follows the stellar magnetic field and hits the star at high 
latitudes, creating hot spots at temperatures higher than the stellar photospheric temperature. 
The UV excess in CTTS comes only from hot spots \citep[e.g.,][]{1998ApJ...509..802C,2015A&A...581A..66V,2014A&A...570A..82V}, 
and it is a more direct diagnostic of accretion than \ha line, except for low accretors \citep{2011ApJ...743..105I}.  
UV excesses for the known NGC 2264 members were measured by \cite{2014A&A...570A..82V} with  MegaCam instrument. 
The UV excess was obtained comparing the observed data with a reference non-accreting threshold
defined by the fit of a polynomial function 
to the locus of non-accreting systems on the \textit{r} versus \textit{u-r} color-magnitude diagram. 
The UV excess was defined as $\mathrm{E}(u)=(u-r)_\mathrm{obs}-(u-r)_\mathrm{ref}$, 
where $(u-r)_\mathrm{obs}$ is the observed color of the star and $(u-r)_\mathrm{ref}$ 
is a reference non-accreting color at $r_\mathrm{obs}$ magnitude. The more negative the
$\mathrm{E}(u)$ values, the larger the UV excess. 

\begin{figure*}
 \begin{center}
\subfigure[]{\includegraphics[width=8cm]{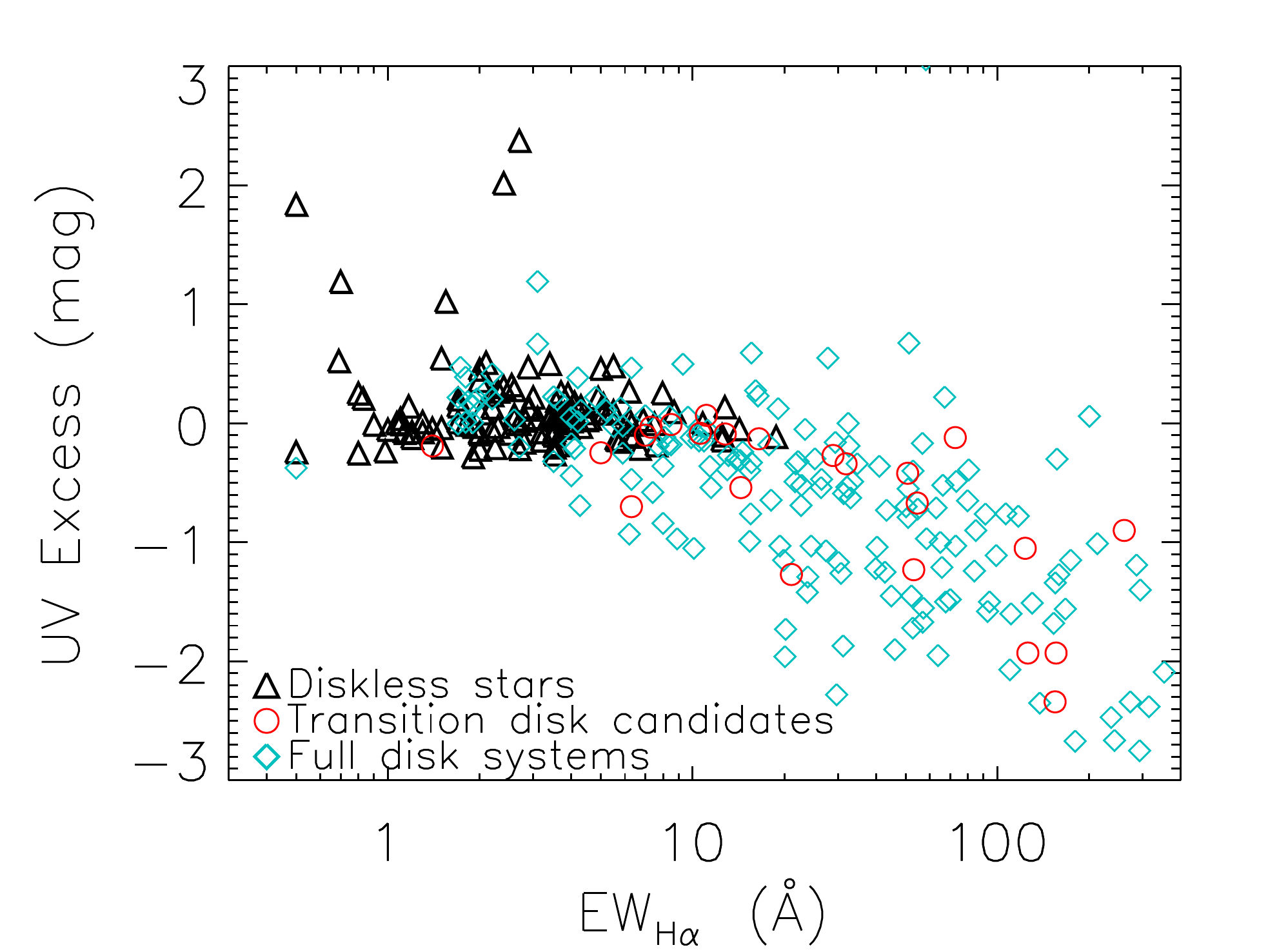}}
\subfigure[]{\includegraphics[width=8cm]{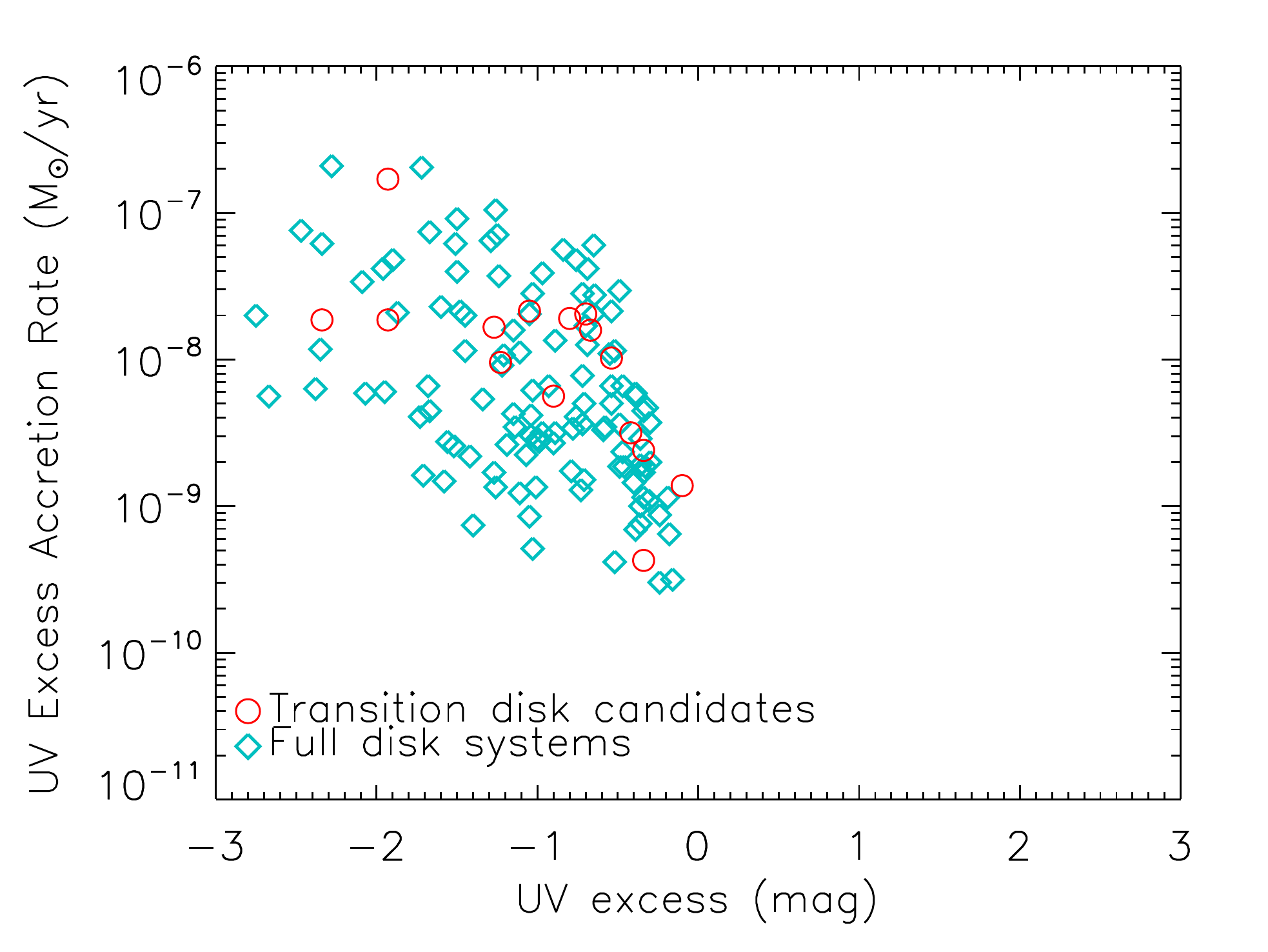}}
 \end{center}
\caption{\label{fig:UV} UV excess as a function of $\mathrm{H}\alpha$ equivalent 
width (a) and mass accretion rate as a function of UV excess (b). More negative values 
indicate larger UV excess. The UV excess and mass accretion rate were calculated by 
\cite{2014A&A...570A..82V} and the $\mathrm{H}\alpha$ equivalent width was measured by
\cite{2016A&A...586A..47S} \& \cite{2005AJ....129..829D}. As expected, diskless stars 
do not present UV excess, while full disk systems do. Transition disk candidates, in general, 
present UV excess similar to stars with full disks, which is consistent with active accretion, as
discussed in the text.} 
\end{figure*}

We show in Fig. \ref{fig:UV}a the UV excess versus $\mathrm{EW}_{\mathrm{H}\alpha}$. 
As expected, most stars classified as diskless do not accrete and do not present 
UV excess. Most full disk systems present UV excess, while the transition disk candidates, 
in general, present UV excess comparable to full disk systems. The mean UV excess 
is $0.10\pm0.03\,\mathrm{mag}$ for diskless stars, $-0.5\pm0.1\,\mathrm{mag}$ for transition disk candidates, 
and $-0.60\pm0.06\,\mathrm{mag}$ for full disk systems.

With the UV excess, \cite{2014A&A...570A..82V} calculated mass accretion rates for 
a sample of TTS belonging to NGC 2264, shown in Fig. \ref{fig:UV}b. 
The mean mass accretion rate values for our sample of stars are $(2\pm1)\times10^{-8}\,\mathrm{M_\sun yr^{-1}}$ 
for transition disk candidates and $(2.1\pm0.4)\times10^{-8}\,\mathrm{M_\sun yr^{-1}}$ for 
full disk systems. The mean accretion rates are different from the 
$\mathrm{H}\alpha$ accretion rates, but also show that the accretion rates 
of the transition disk candidates are similar to the accretion rates of full disk systems. 

\section{Discussion}\label{sec:discuss}

\subsubsection{Photometric period}
The CTTS are contracting and accreting; both physical phenomena should increase the 
stellar angular momentum. However, CTTS are found to be slow rotators, with typical 
periods of about 4 to 8 days in young clusters. Some efficient mechanism of 
angular momentum transfer from the star to the disk or the interstellar medium must then take place.
The accretion process is related to the CTTS braking, possibly through disk locking, since at the end of the accretion phase,
WTTS start their spin-up toward the Main Sequence
\citep{1993A&A...272..176B,2001AJ....121.1676R,2005A&A...430.1005L,2007ApJ...667..308C,2007A&A...463.1081M,2009IAUS..258..363I,2015A&A...578A..89V,2017A&A...603A.106R,2017A&A...599A..23V,2018AJ....155..196R}.

\begin{figure}[htb!] 
 \centering
{\includegraphics[scale=0.4]{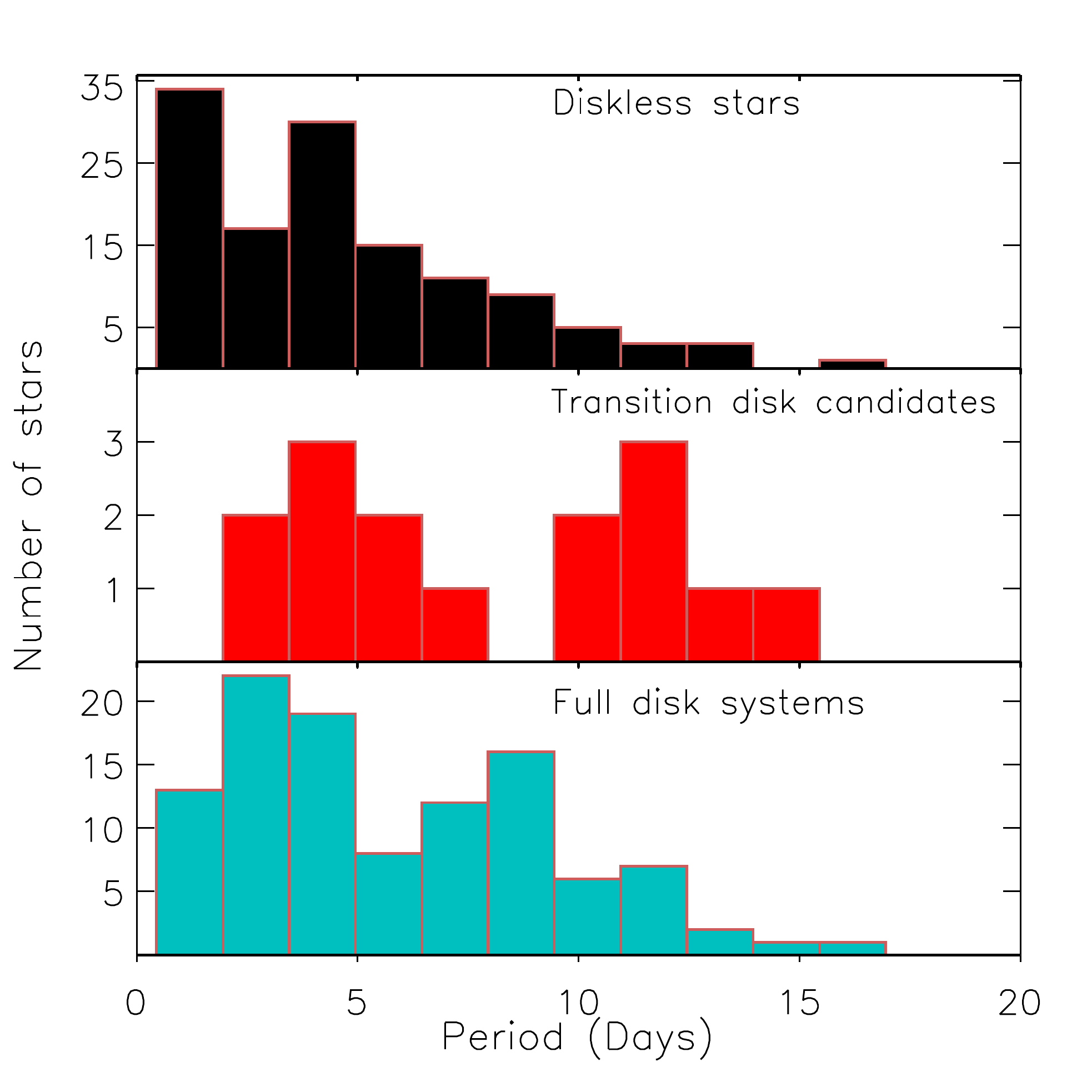}}
 \caption{\label{fig:histPer} Period distribution of the samples of diskless stars (black), transition disk candidates (red), and full disk systems (light blue), that were analyzed in this work. Diskless stars are substantially more rapid rotators than full disk systems, as expected, since most stars with full disks are CTTS and most diskless stars are WTTS.} 
\end{figure}

The period distribution of young low-mass stars belonging to NGC 2264 
shows that WTTS are substantially more rapid rotators than CTTS. 
These results were most recently obtained using CoRoT data from the 2008 \citep{2013MNRAS.430.1433A} 
and 2011 \citep{2017A&A...599A..23V} campaigns.
We analyzed the periodicity of our sample of stars, using the periods obtained 
by \cite{2016A&A...586A..47S} for some CTTS and by \cite{2017A&A...599A..23V} for
some CTTS and the WTTS: both papers use observations from the 2011 CoRoT campaign. 
We complemented our period data with photometric optical periods gathered from the literature in \cite{2006A&A...455..903F}.
The period distribution is shown in Fig. \ref{fig:histPer}. Stars with full disks are slower 
rotators (mean period of $6.0\pm0.4\,\mathrm{days}$) compared to diskless stars 
(mean period of $4.8\pm0.3\,\mathrm{days}$).  This result agrees with the disk locking scenario discussed in the previous 
paragraph, since most full disk systems are CTTS and all diskless stars are WTTS. 
The mean period of the transition disk candidates is $7.8\pm1.0\,\mathrm{days}$ but the sample of periodic transition disk 
candidates is small compared to the diskless and full disk samples.

Following the light curve morphological classification from \cite{2010A&A...519A..88A}, 
the CoRoT light curves of the CTTS were classified by \cite{2016A&A...586A..47S},
as spot-like (sinusoidal variations due stable spots at the stellar surface), 
AA Tau-like (quasi-periodical eclipse like light curves explained as occultation 
of the stellar photosphere by an inner disk warp 
\cite[see][for more details about AA Tau-like light curves]{2010A&A...519A..88A,2014AJ....147...82C,2014A&A...567A..39F,2015A&A...577A..11M},  
and non-periodic light curves, which includes non-periodic occultations of the 
stellar photosphere by the inner disk, accretion bursts \citep{2014AJ....147...83S}, 
and variations not associated with an unique physical phenomenon.  

\begin{figure}[htb!] 
 \centering
{\includegraphics[scale=0.4]{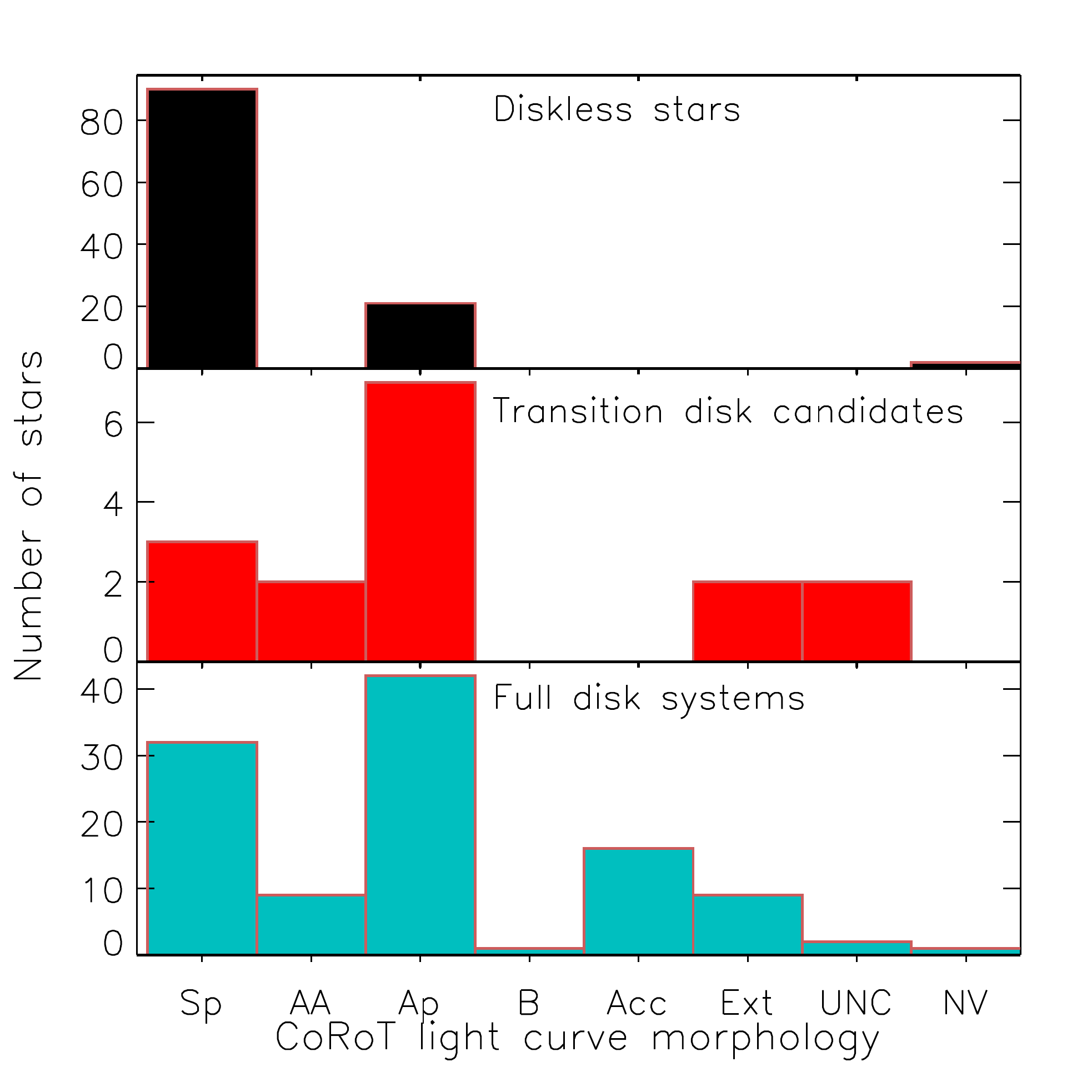}}
 \caption{\label{fig:histLC} Distribution of morphological classification of 2011 CoRoT 
light curve of our sample of TTS. The CTTS morphological classification was taken 
from \cite{2016A&A...586A..47S}, while the  WTTS morphological classification is from this work. 
\textquotedblleft SP\textquotedblright \, corresponds to spot-like, 
\textquotedblleft AA\textquotedblright \,  to AA Tau-like, \textquotedblleft Ap\textquotedblright \, to 
aperiodic variations not related to a unique physical phenomenon, \textquotedblleft B\textquotedblright \, to eclipsing binary, 
\textquotedblleft Acc\textquotedblright \, to accretion bursts, \textquotedblleft Ext\textquotedblright \,
 to non-periodic extinction, \textquotedblleft UNC\textquotedblright \, to unclassified light curves, 
and \textquotedblleft NV\textquotedblright \, to non-variable light curves. }
\end{figure}

We classified all the 2011 CoRoT WTTS light curve using the same nomenclature, 
and we adopted the morphological classification of the CTTS from \cite{2016A&A...586A..47S}. 
The distribution of morphological classifications of the CoRoT light curve is shown 
in Fig. \ref{fig:histLC}. Almost all ($\sim78\%$) of the diskless stars are classified as spot-like. 
As the light curve morphology is related to system evolution \citep{2016A&A...586A..47S}, 
diskless systems, which are no longer accreting, are expected to present spot-like light curves, 
as their photometric variability is mainly due to the cold surface spots.

Full disk systems are mostly CTTS and their light curve variability can be due to 
different physical processes, like accretion, circumstellar occultation, and spots. 
They present a variety of light curve morphologies, $\sim38\%$ being classified as non-periodic, 
probably the result of many physical phenomena acting
together (accretion and spots, for example). The sample of transition disk candidates observed 
by CoRoT is small, but as most of them are also CTTS, we would then 
expect a light curve morphology similar to the full disk systems. It is interesting to note 
that three transition disk candidates are classified as AA Tau-like systems and two 
as due to aperiodic extinction. Extinction systems (periodic or not) are known to have 
dust in the inner disk \citep{2010A&A...519A..88A,2014AJ....147...82C,2014A&A...567A..39F,2015A&A...577A..11M,2016A&A...586A..47S}, 
which is responsible for the occultation of the stellar photosphere, and they are common
among the anemic disks \citep{2010A&A...519A..88A,2016A&A...586A..47S}.

\subsection{Transition disk candidates in NGC 2264 from the literature}\label{sec:Lit}
Systems with near-IR deficits and mid to far IR emission excess were initially
associated with an inner disk hole by \cite{1989AJ.....97.1451S}, studying star-disk systems
in the Taurus-Auriga star formation complex. Over the years, many new objects with inner 
holes were identified and have been studied in detail 
\citep{2005ApJ...630L.185C,2007MNRAS.378..369N,2008ApJ...682L.125E,2010ApJ...708.1107M,2011ApJ...732...42A,2014A&A...568A..18M,2015A&A...578A..23B}. 

\begin{table*}[htb!]
\tiny
\caption{\label{tab:Model_param} Parameters obtained by Hyperion SED fitting code for sample of transition disk candidates.}
\begin{center}
\begin{tabular}{lrrrrrrrrrr}
 \hline\hline
 Mon ID\tablefootmark{a} &Av\tablefootmark{b} & $\mathrm{Dist}$\tablefootmark{b}&$\mathrm{T_\star}$\tablefootmark{b} &$\mathrm{R}_\star$\tablefootmark{b} &$\mathrm{M}_{\mathrm{disk}}$\tablefootmark{b} & $\mathrm{R_H}$\tablefootmark{c}&$\mathrm{R_{in}}$\tablefootmark{b} & $\mathrm{R_{Sub}}$\tablefootmark{c}& $\mathrm{R_{Max}}$\tablefootmark{b}& $i$\tablefootmark{b}\\
&  & $(\mathrm{pc})$ & $(\mathrm{K})$  & $(\mathrm{R_\odot})$  & $(\mathrm{M_\odot})$ & $(\mathrm{AU})$ & $(\mathrm{R_{Sub}})$& $(\mathrm{AU})$ & $(\mathrm{AU})$ &  $(^o)$\\
\hline
Mon-000040* &   0.1  &  698.2  &  2832.0  &   0.91  &   0.0005293  &   5.85  &  419.1  &  0.0140  &   372.7  &  77.2   \\
Mon-000120  &   1.2  &  698.2  &  3775.0  &   1.84  &   0.0001934  &   0.53  &   10.4  &  0.0514  &   472.9  &  21.9   \\
Mon-000122* &   0.8  &  693.4  &  4230.0  &   1.33  &   0.0008640  &   5.84  &  124.4  &  0.0469  &   112.0  &  76.7   \\
Mon-000177  &   0.3  &  690.2  &  4590.0  &   2.66  &    0.002971  &   0.23  &    2.0  &  0.1114  &   951.0  &  55.2   \\
Mon-000273  &   0.2  &  527.2  &  3738.0  &   1.31  &   1.844E-05  &  17.85  &  498.1  &  0.0358  &  1314.0  &  68.1   \\
Mon-000280  &   0.5  &  699.8  &  3922.0  &   1.33  &    0.006200  &   0.09  &    2.2  &  0.0401  &    61.1  &  14.0   \\
Mon-000296  &   0.6  &  758.6  &  5214.0  &   1.60  &   8.402E-06  &   0.12  &    1.4  &  0.0872  &   601.8  &  64.8   \\
Mon-000314  &   0.3  &  701.5  &  2918.0  &   1.76  &    0.004796  &   3.58  &  125.1  &  0.0286  &   186.8  &  76.4   \\
Mon-000328  &   0.2  &  753.4  &  3338.0  &   1.50  &   8.831E-07  &   3.51  &  108.5  &  0.0324  &  1093.0  &   9.3   \\
Mon-000342* &   0.3  &  916.2  &  3836.0  &   2.09  &   4.624E-08  &  18.42  &  306.7  &  0.0601  &   117.9  &  82.4   \\
Mon-000433  &   1.2  &  726.1  &  3506.0  &   2.46  &   2.443E-06  &   7.39  &  126.0  &  0.0587  &   250.9  &  65.4   \\
Mon-000452* &   0.1  &  827.9  &  3605.0  &   2.35  &   3.533E-06  &   6.00  &  101.0  &  0.0594  &  1796.0  &  86.2   \\
Mon-000502  &   1.4  &  662.2  &  4263.0  &   1.74  &   1.757E-05  &   9.43  &  150.7  &  0.0626  &   425.6  &  77.7   \\
Mon-000637  &   1.1  &  756.8  &  3650.0  &   1.82  &   0.0005777  &   8.66  &  183.2  &  0.0473  &  1172.0  &  79.2   \\
Mon-000676  &   0.3  &  712.9  &  3864.0  &   2.35  &   7.821E-05  &   8.50  &  123.6  &  0.0688  &    66.9  &  75.7   \\
Mon-000771  &   0.3  &  744.7  &  5243.0  &   2.02  &   1.408E-06  &  77.93  &  696.8  &  0.1118  &  2331.0  &  83.7   \\
Mon-000824  &   0.9  &  636.8  &  3864.0  &   2.35  &   7.821E-05  &   8.50  &  123.6  &  0.0688  &    66.9  &   3.6   \\
Mon-000860  &   0.4  &  690.2  &  3807.0  &   0.90  &   0.0001289  &   4.34  &  170.2  &  0.0255  &    53.9  &  49.2   \\
Mon-000879  &   0.3  &  966.1  &  3864.0  &   2.35  &   7.821E-05  &   8.50  &  123.6  &  0.0688  &    66.9  &   3.6   \\
Mon-000937* &   0.8  &  948.4  &  3563.0  &   3.20  &   6.025E-08  &  22.31  &  282.1  &  0.0791  &   314.0  &  72.1   \\
Mon-000961  &   0.1  &  682.3  &  2723.0  &   1.15  &   0.0003476  &   0.28  &   17.1  &  0.0162  &  2359.0  &  56.0   \\
Mon-000965  &   0.1  &  803.5  &  3028.0  &   2.05  &    0.001175  &   9.50  &  263.3  &  0.0361  &  1448.0  &   6.4   \\
Mon-000997* &   0.8  &  663.7  &  3374.0  &   1.00  &   2.228E-08  &   2.61  &  118.5  &  0.0220  &   270.4  &  57.9   \\
Mon-001033* &   0.5  &  756.8  &  3999.0  &   1.70  &   0.0009192  &  16.66  &  312.0  &  0.0534  &   221.4  &  65.8   \\
Mon-001094  &   0.3  &  727.8  &  3948.0  &   1.70  &   2.806E-07  &  21.66  &  416.8  &  0.0520  &   112.7  &  19.8   \\
Mon-001229  &   1.3  &  798.0  &  4063.0  &   1.41  &   0.0006760  &   0.59  &   12.9  &  0.0459  &   120.1  &  10.1   \\
Mon-001287  &   1.9  &  814.7  &  3995.0  &   1.06  &   8.285E-08  &   2.64  &   79.8  &  0.0331  &   164.0  &  12.6   \\
Mon-001308  &   0.3  &  774.5  &  5011.0  &   2.13  &   1.981E-06  &  20.18  &  188.7  &  0.1069  &   298.1  &  86.4   \\
\hline
\end{tabular}
\end{center}
\tablefoot{This table is ordered according to the Mon ID}\\
\tablefoottext{a}{CSIMon is an internal identification of the CSI 2264 campaign. Here, CSI was omitted for brevity.}
\tablefoottext{b}{The model parameters of the best SED fitting: reddening ($\mathrm{A_v}$), distance of the star to the Sun (Dist), central star temperature ($\mathrm{T_\star}$) and radius ($\mathrm{R_\star}$), disk mass ($\mathrm{M_{disk}}$), inner disk radius ($\mathrm{R_{in}}$), outer disk radius ($\mathrm{R_{Max}}$) and inclination of the system ($i$).}
\tablefoottext{c}{Parameters determined using parameters from SED fitting model.}
\tablefoottext{*}{New transition disk candidates identified in this work.}
\end{table*}

Systems with inner disk holes have been identified in NGC 2264 by \cite{2009AJ....138.1116S}, 
using Spitzer IRAC and MIPS data.  
The observed star-disk systems, in \cite{2009AJ....138.1116S}, were classified as pre-transition and transition disks, 
following \cite{2008ApJ...682L.125E}. If the SED flux at $8.0\,\mu\mathrm{m}$ was smaller than the flux at $24\,\mu\mathrm{m}$, and if the sign of 
$\alpha_{\mathrm{IRAC}}$ and $\alpha_{\mathrm{IRAC-MIPS}}$ (the SED slope between $8.0\,\mu\mathrm{m}$ and $24\,\mu\mathrm{m}$)
was different, the system was considered a candidate to have a hole/gap in the disk. Additionally, 
these systems were classified as pre-transition disks if $-0.3\geqslant\alpha_{\mathrm{IRAC}}\geqslant-1.8$ 
and as a transition disk if $\alpha_{\mathrm{IRAC}}<-1.8$ \citep{2009AJ....138.1116S}.   

Using these selection criteria, \cite{2009AJ....138.1116S} found $13$ systems 
with pre-transition disks and $24$ systems with transition disks belonging to NGC 2264. 
Among these $37$ systems found by \cite{2009AJ....138.1116S}, only $18$ are part of the sample of 
stars analyzed in this work and only one of them was not classified as a transition disk candidate by us. 
We did not separate transition and pre-transition disks in our sample, 
and two systems (Mon-000177 and Mon-000961) of our transition disk candidates 
were classified as pre-transition disk by \cite{2009AJ....138.1116S}. 

\cite{2012A&A...540A..83T} found only three transition disks in NGC 2264. 
Their selection criteria were very strict and classified as transition disks
only accreting systems that did not present dust in the inner disk 
(photospheres according to the $\alpha_{\mathrm{IRAC}}$ classification) and 
had excess at $24\,\mu\mathrm{m}$, indicating the
existence of a thick outer disk. In our sample, we have only one star classified as a  
photosphere that has available data at $24\,\mu\mathrm{m}$. This star accrete but 
does not present signs of an outer thick disk. 

\cite{2014ApJ...794..124R} also identified transition disks in their sample of stars from NGC 2264 
observed with the Spitzer satellite. Their selection was based on photometric criteria. 
They classified as transition disks those Class III systems that have $[5.8]-[24]>2.5\,\mathrm{mag}$ or 
$[4.5]-[24]>2.5\,\mathrm{mag}$ and $[3.6]<14\,\mathrm{mag}$, which are systems that have 
excess at $24\,\mu\mathrm{m}$. They classified $44$ systems as transition disks using these 
criteria, but only $26$ are part of our sample. Among the $26$ systems we have in common,
our classification agrees with theirs for $17$ transition disk candidates, while the other $9$
systems were classified by us as full disks.
\begin{figure*}[htb!] 
 \centering
 \includegraphics[scale=0.3]{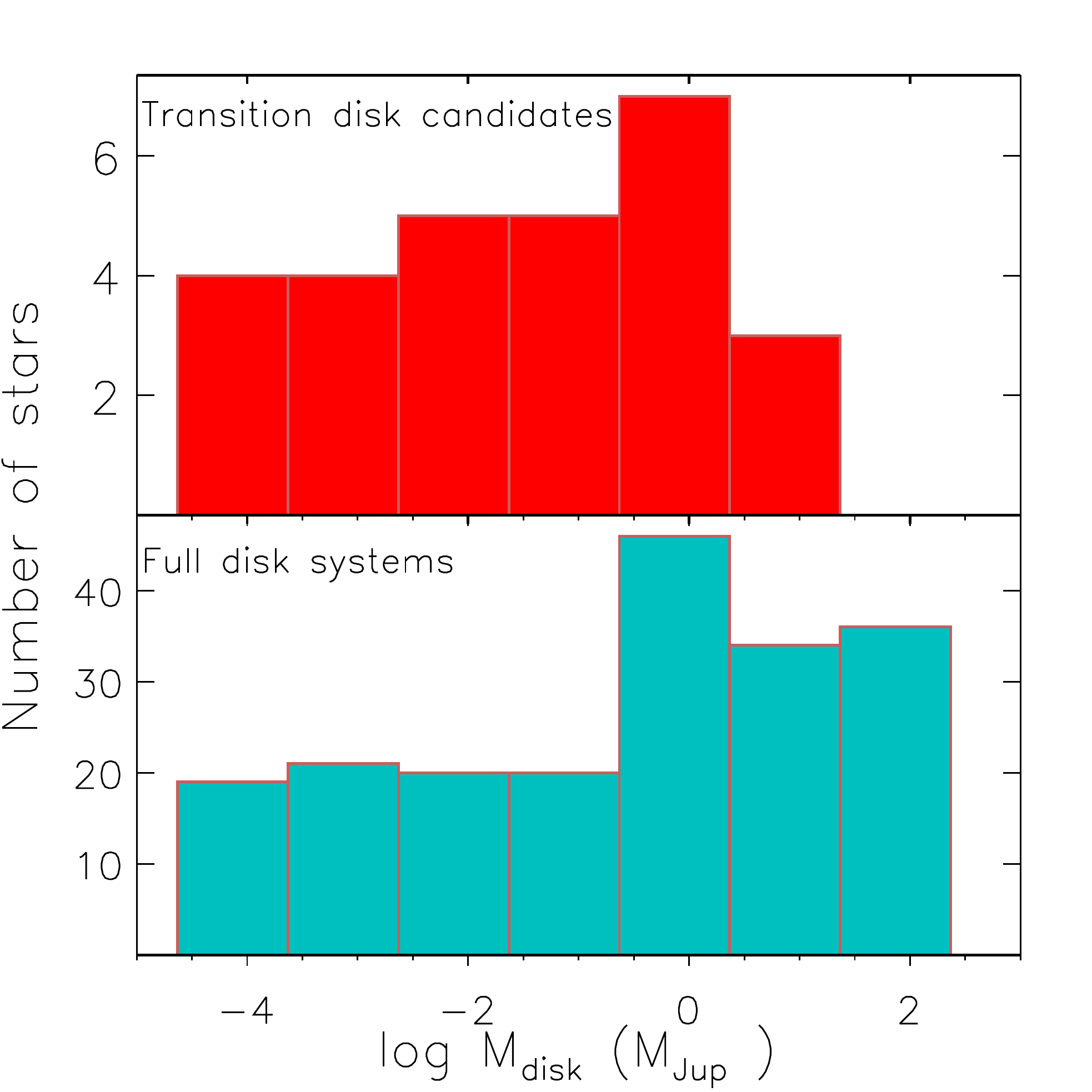}
 \includegraphics[scale=0.3]{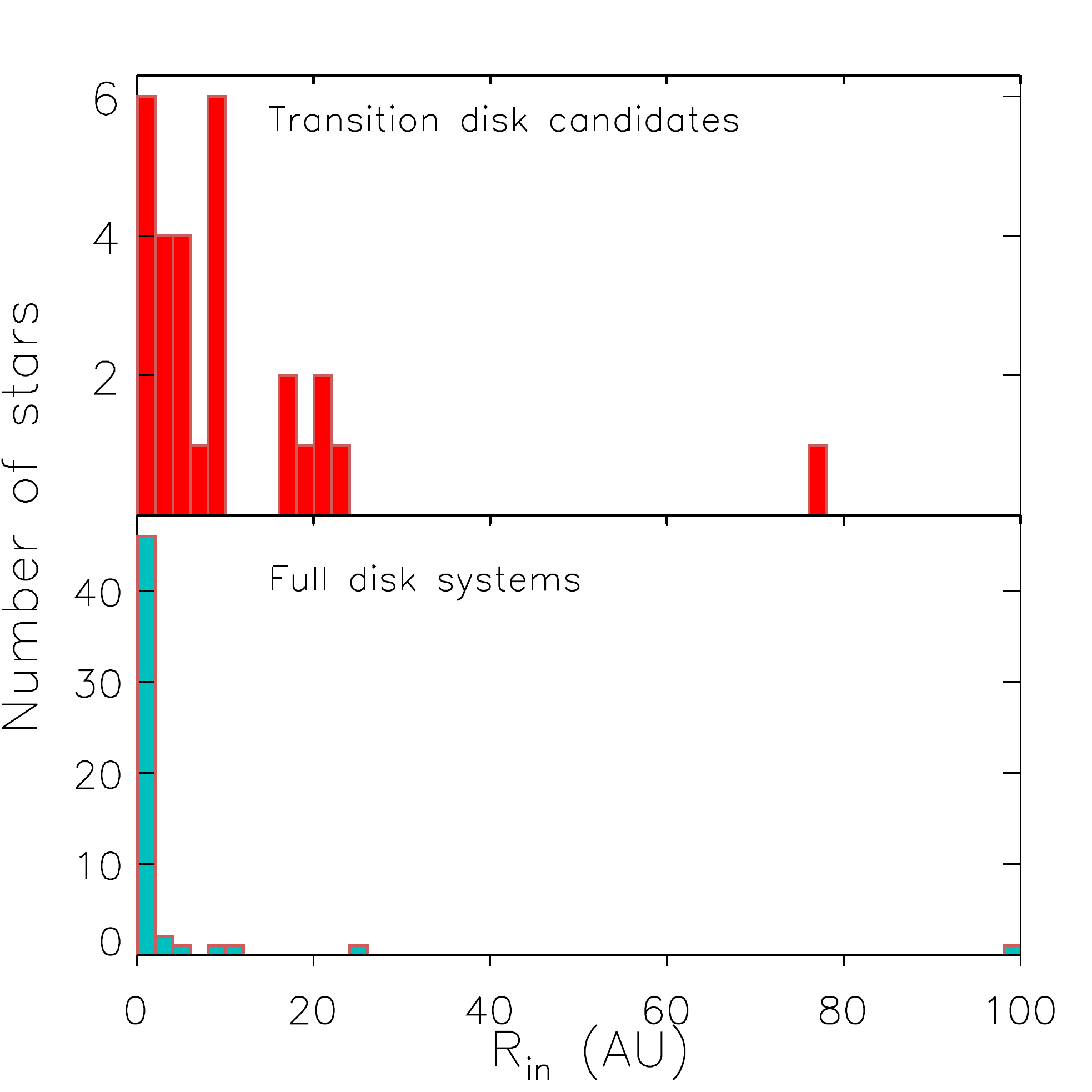}
 \includegraphics[scale=0.3]{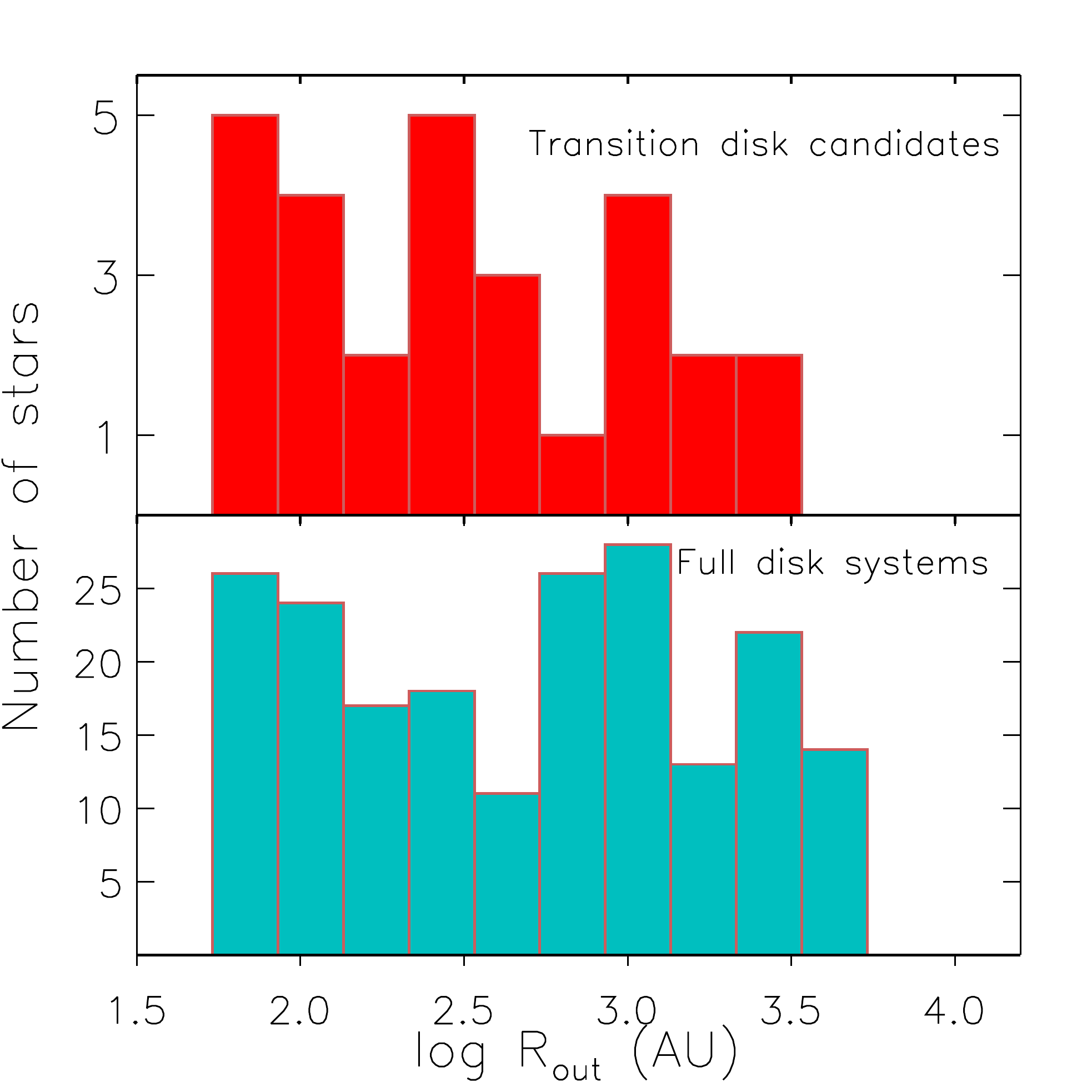}
 \caption{\label{fig:Diskparam} Distribution of disk parameters obtained by the SED model 
fitting of transition disk and full disk
systems. (Left) Disk dust mass. (Middle) Disk inner radius. (Right) Disk outer radius. For the transition disk sample, 
the disk inner radius is the inner hole size and for the sample of full disk systems it corresponds to 
the dust sublimation radius. As the transition disk systems have a disk inner hole, the inner disk 
radius of the transition disk systems tends to be larger than the full disk systems.  The disk mass and the external disk radii are however only estimated values and correspond to lower limits, see text.}
\end{figure*}

In this work we found seven new transition disk candidates belonging to NGC 2264
that had not been classified as transition disks before. These systems 
were classified as Class II or III by \cite{2009AJ....138.1116S} \& \cite{2014ApJ...794..124R}. 
We identified these systems in Table \ref{tab:Model_param} with an asterisk after the CSIMon ID.

\subsection{Disk and inner hole characteristics}\label{sec:hole}
Disk parameters were computed by Hyperion SED fitting code. The size of the inner disk 
hole corresponds to the inner radius of the disk ($\mathrm{R}_\mathrm{in}$), 
which is an output parameter of Hyperion SED fitting code
\citep{2017A&A...600A..11R}. As explained in Sec. \ref{sec:sed}, 
$\mathrm{R}_\mathrm{in}$ is the dust sublimation radius for model 2, 
that is composed of a star and a disk. In model 3, that includes a star 
and a disk with an inner disk hole, $\mathrm{R}_\mathrm{in}$, that can vary 
from $1$ to $1000\,\mathrm{R_{sub}}$. To determine the sublimation radius, 
we used the empirical equation obtained by \cite{2004ApJ...617.1177W} for a fixed type of dust:
\begin{equation}\label{eq:rsub}
 \dfrac{\mathrm{R_{Sub}}}{\mathrm{R_\star}}=\left(\dfrac{\mathrm{T_{Sub}}}{\mathrm{T_\star}}\right)^{-2.085},
\end{equation}
where $\mathrm{T_{Sub}}$ is the dust sublimation temperature ($1600\,\mathrm{K}$) 
and $\mathrm{R_\star}$ and $\mathrm{T_\star}$ are the radius and effective 
temperature of the central star, respectively. In Fig \ref{fig:Diskparam}, 
we show the distributions of dust disk mass, the inner disk radii, and the external 
disk radii.  The disk mass and the external disk radii are however only estimated values and correspond to lower limits, 
since we do not have broad wavelength coverage.

\begin{figure*}[htb!] 
 \centering
\subfigure[]{\includegraphics[scale=0.4]{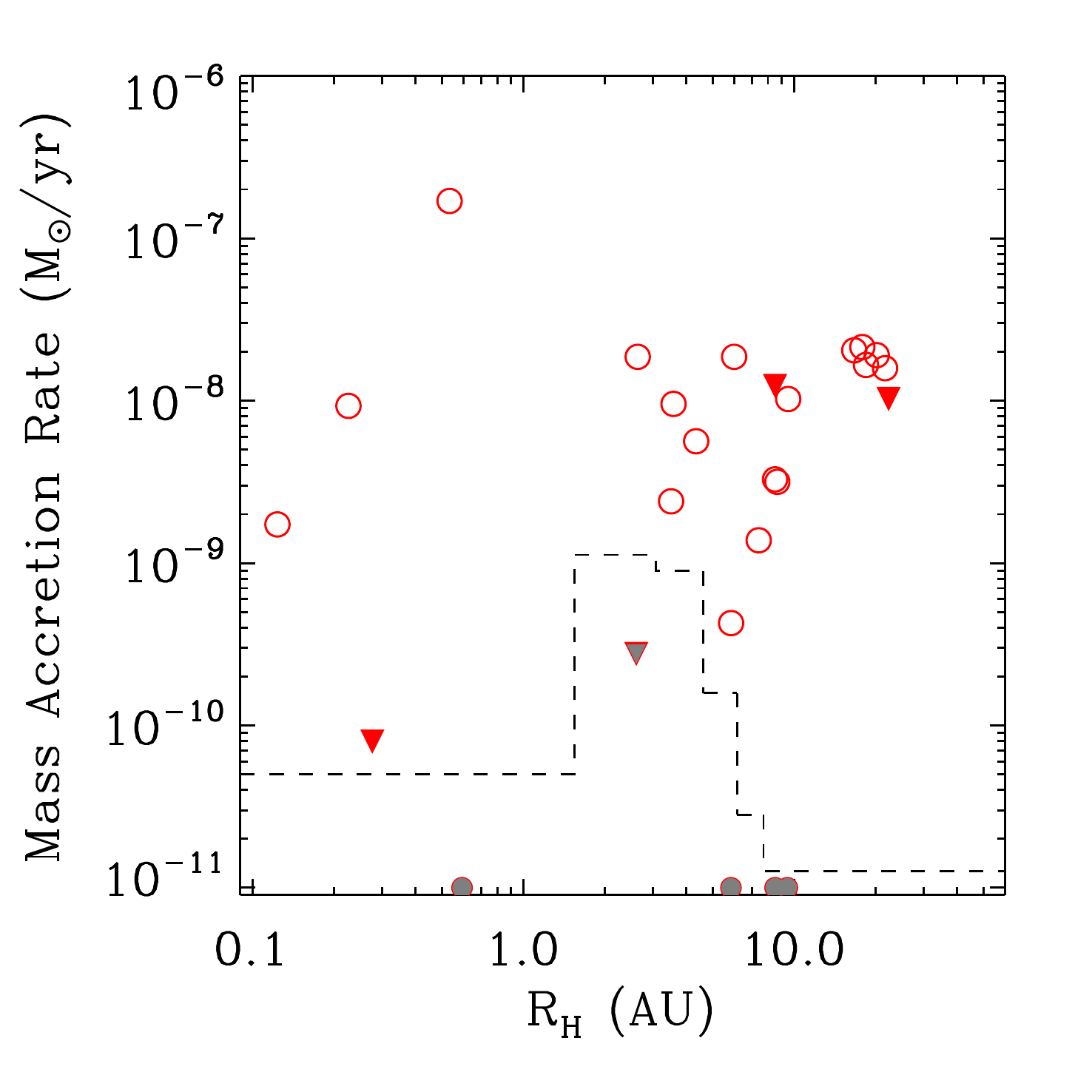}}
\subfigure[]{\includegraphics[scale=0.4]{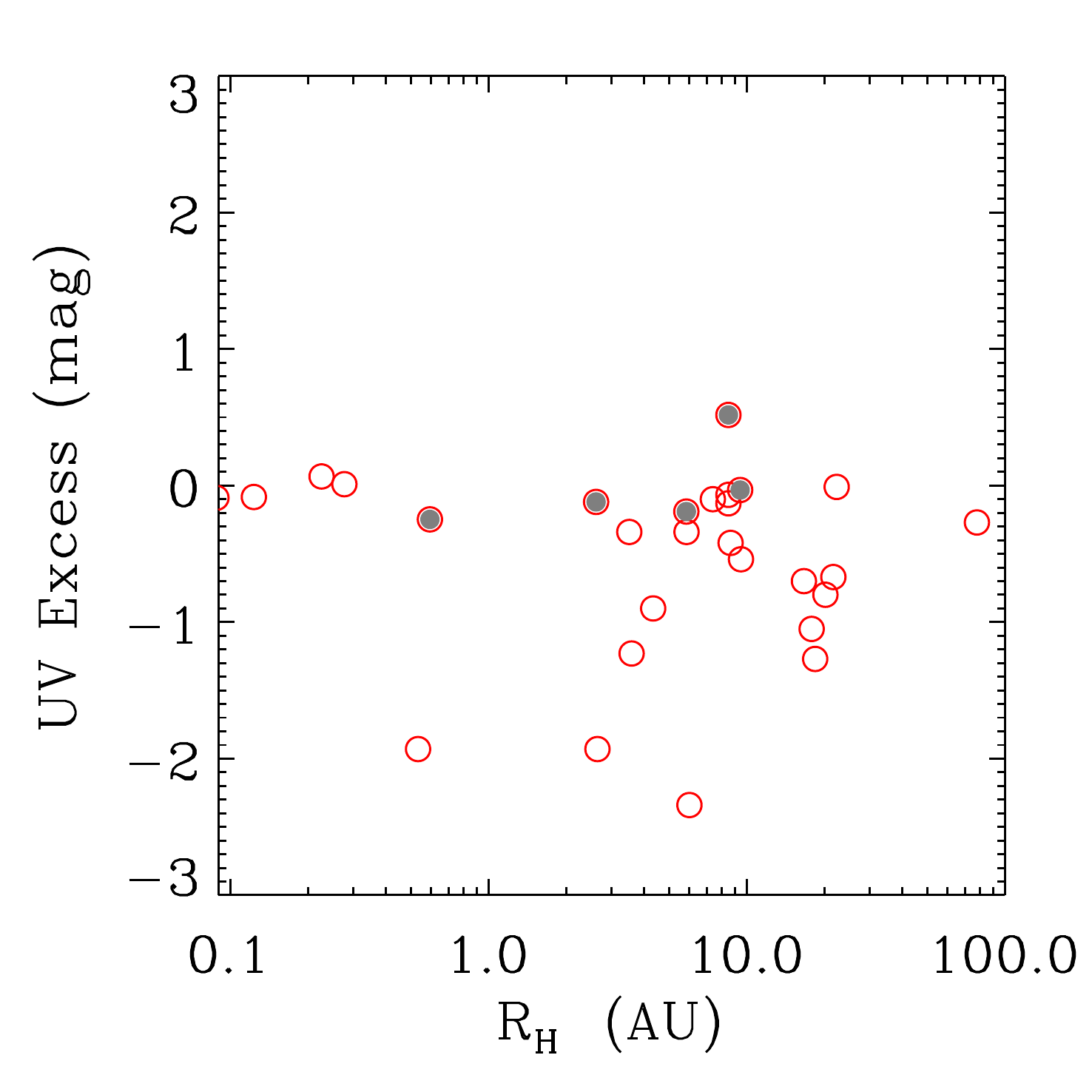}}
\subfigure[]{\includegraphics[scale=0.4]{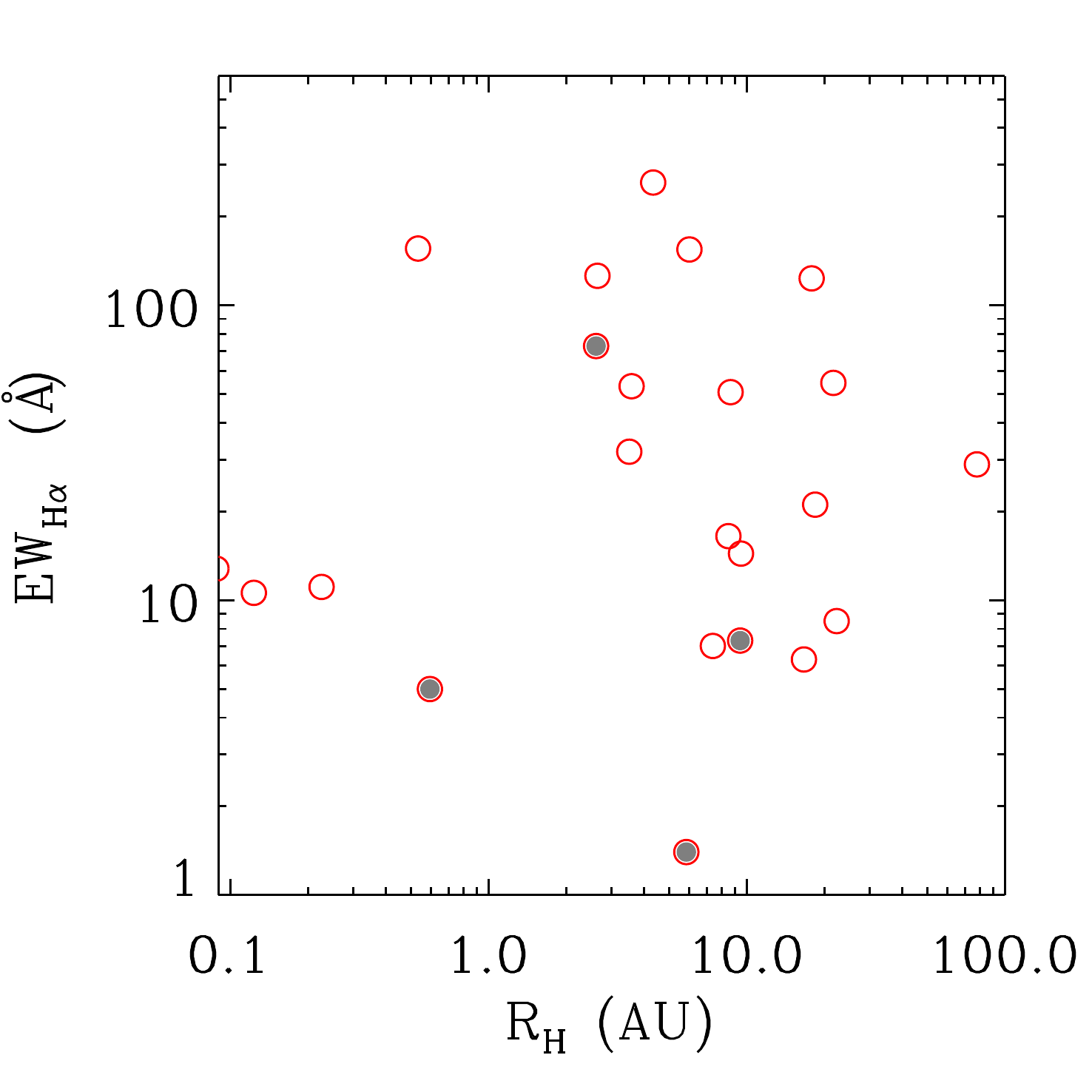}}
 \caption{\label{fig:AccvsRH} Accretion diagnostic as a function of disk hole size for transition disk candidates. 
(a) Mass accretion rates are from UV excess (preferentially) \citep{2014A&A...570A..82V} and 
from $\mathrm{H\alpha}$ equivalent width \citep{2016A&A...586A..47S}. For WTTS, we fix the 
value of the mass accretion rate at $1\times10^{-11}\,\mathrm{M}_\sun\mathrm{yr}^{-1}$. 
The upside down triangles correspond to the upper limits of the mass accretion rate. The dashed line 
represents the region where the inner hole can be explained by X-ray photoevaporation 
\citep{2011MNRAS.412...13O,2017MNRAS.472.2955O}. Only $\sim 18\,\%$ of our sample of transition disk 
candidates can be explained by X-ray photoevaporation of the inner 
disk by stellar radiation. (b) UV emission excess \citep{2014A&A...570A..82V}. 
(c) $\mathrm{H\alpha}$ equivalent width \citep{2016A&A...586A..47S,2005AJ....129..829D}. 
Gray filled symbols identify systems that fall in the region where the inner disk hole 
can be explained by X-ray photoevaporation.} 
\end{figure*}
\begin{figure*}[htb!] 
 \centering
\subfigure[]{\includegraphics[scale=0.4]{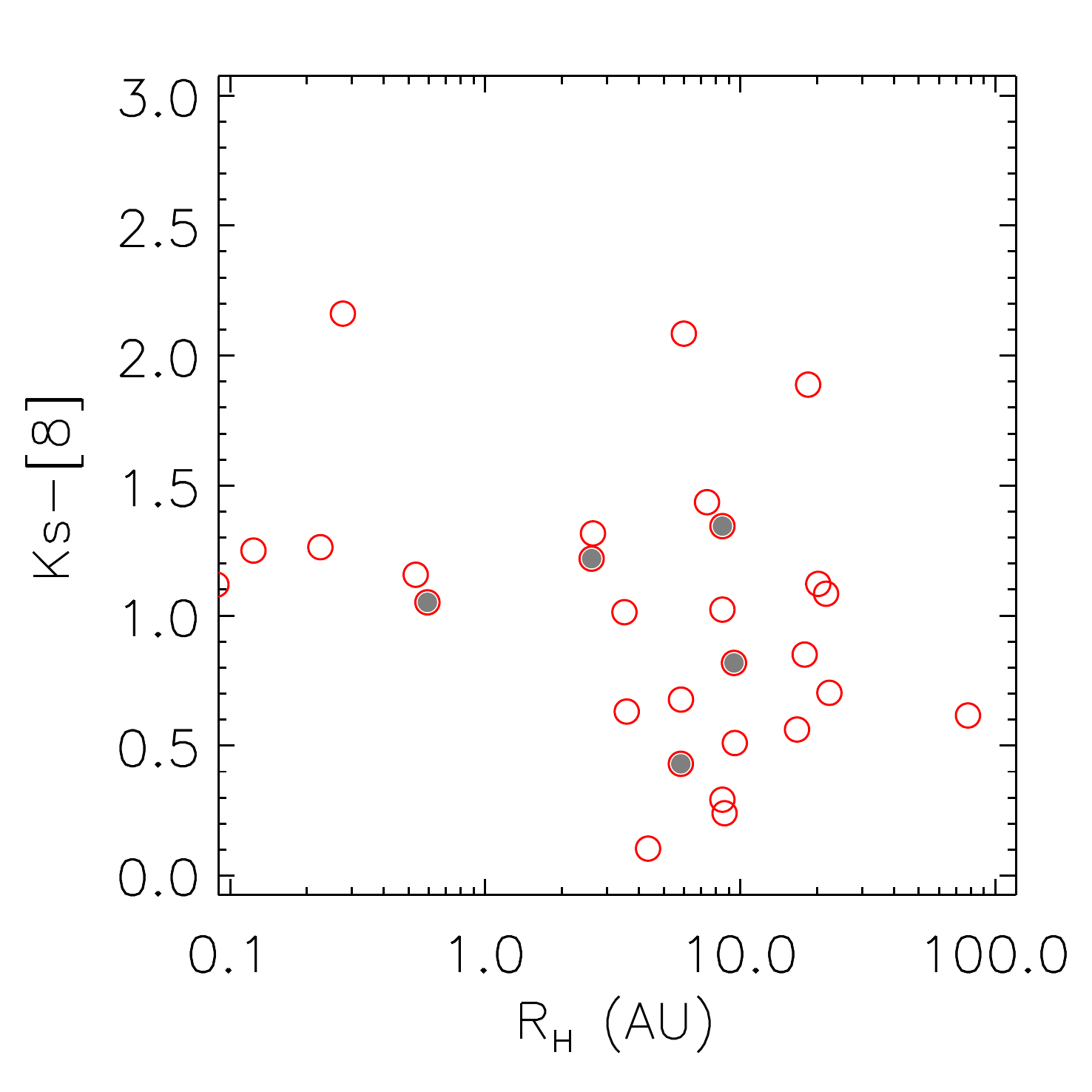}}
\subfigure[]{\includegraphics[scale=0.4]{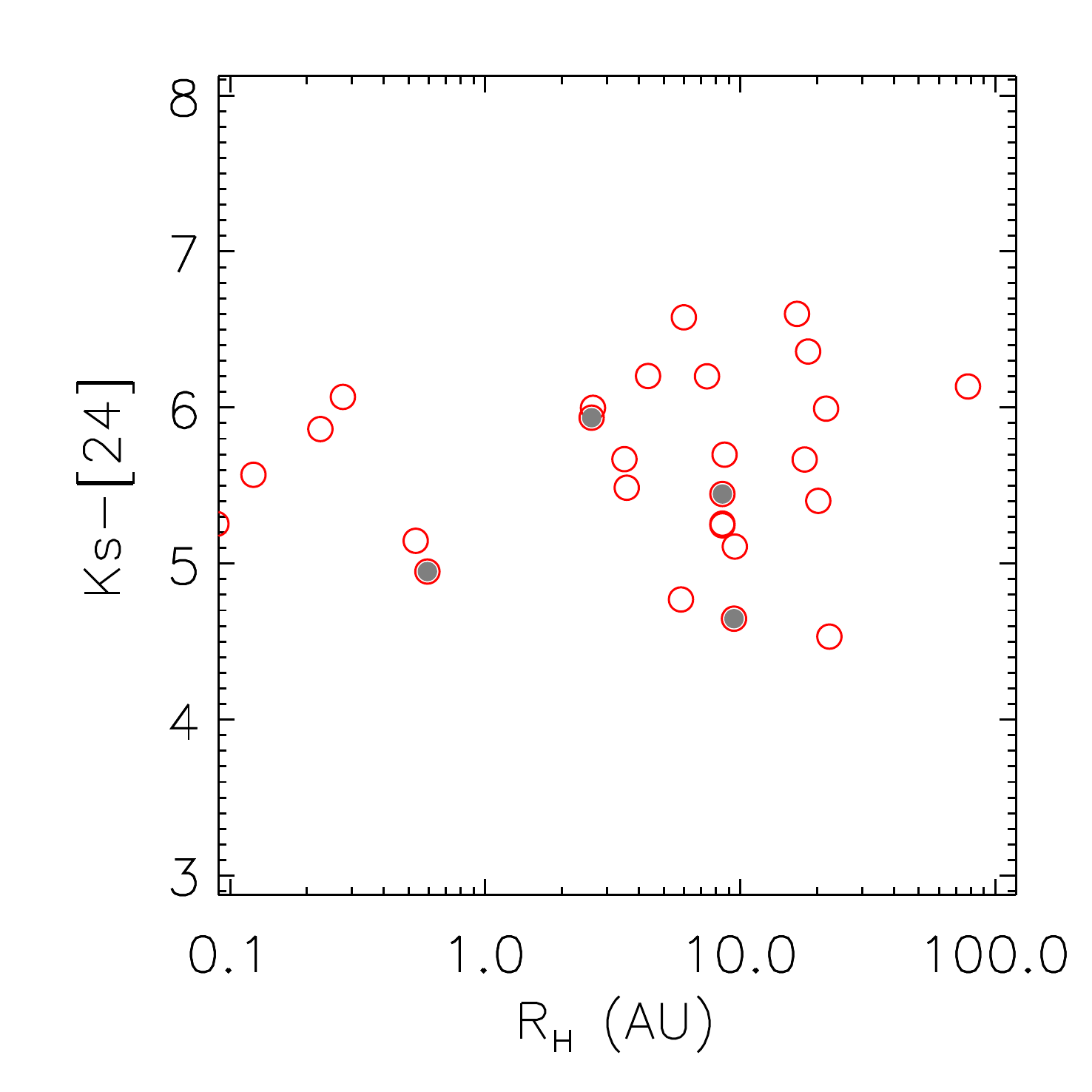}}
\subfigure[]{\includegraphics[scale=0.4]{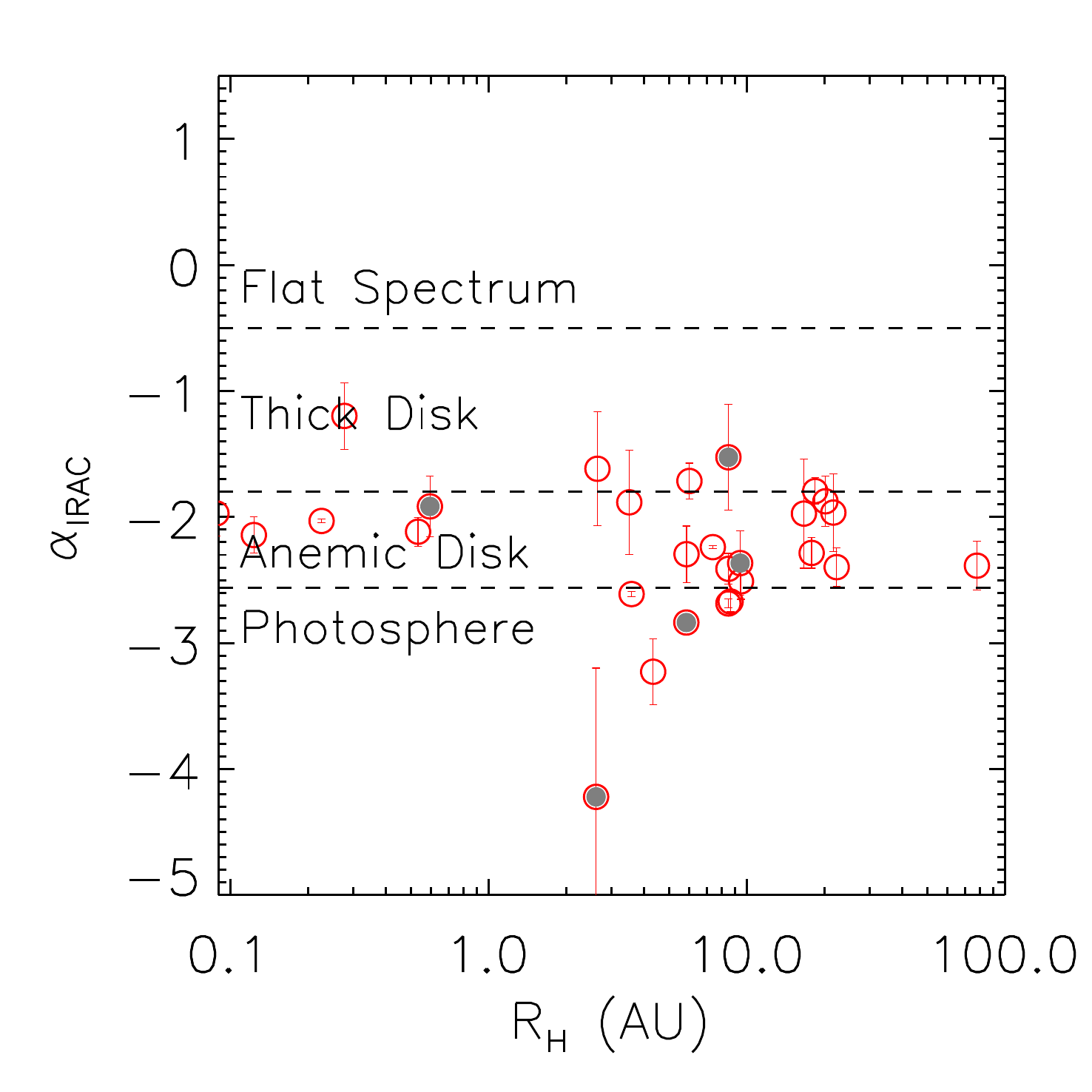}}
 \caption{\label{fig:magvsRH} Disk diagnostics as a function of disk hole size for our sample 
of transition disk candidates. (a) Near-IR color, (b) mid-IR color, and 
(c) $\alpha_{\mathrm{IRAC}}$ index. Gray filled symbols identify systems that fall in the region where the inner disk hole
can be explained by X-ray photoevaporation.} 
\end{figure*}

The size of the inner disk hole ($\mathrm{R_H}$) is given by the inner disk radius when it is 
larger than the sublimation radius. We calculated $\mathrm{R_H}$ for all transition disk candidates. 
Then we compared the inner disk hole size with different characteristics of the star-disk system; 
see Figs. \ref{fig:AccvsRH} and \ref{fig:magvsRH}. In Table \ref{tab:Model_param}, we show the parameters 
obtained by Hyperion SED fitting code for each transition disk candidate and the hole size inferred from the best model parameters.

In our sample of transition disk candidates, the hole size varies from $0.09$ to $78\,\mathrm{AU}$, 
with a mean value of $10.4\pm2.8\,\mathrm{AU}$. These estimated hole sizes are smaller than the 
values usually found in the literature in other star forming regions.
Normally there is a considerable population with hole sizes larger than $10\,\mathrm{AU}$, as seen, 
for example, in \cite{2017RSOS....470114E}. However, the largest hole sizes from models, found in the literature, were 
obtained by SED fitting with additional mid and far IR observational data \citep[e.g.,][]{2016A&A...592A.126V}. As discussed in \cite{2010ApJ...718.1200M}, systems with larger hole sizes are also more easily found with IR spectra than photometric data. 
The other possibility is that the SED model used in this work, which has only a passive disk 
and does not take into account the heating of the disk due to accretion, tends to produce 
smaller disk holes, as reported by \cite{2010ApJ...718.1200M}. 

Different mechanisms can be responsible for creating a hole in the inner disk. 
Photoevaporation of the inner disk by stellar radiation can open holes, 
but, in general, with small radii ($\mathrm{R_H}<10\,\mathrm{AU}$) and for small 
mass accretion rates ($\lesssim10^{-9}\,\mathrm{M}_\sun\mathrm{yr}^{-1}$) 
\citep{2011MNRAS.412...13O,2014A&A...568A..18M,2016PASA...33....5O}.  
Photoevaporation is actually more efficient in non-accreting systems, opening large holes in the disk 
($\gtrsim20\,\mathrm{AU}$) \citep{2016PASA...33....5O}. 

In Fig. \ref{fig:AccvsRH}a we plot the mass accretion rate as a function of the hole size 
and identify the region where the inner hole can be due to photoevaporation alone, according to
the model by \citet{2011MNRAS.412...13O} and \citet{2017MNRAS.472.2955O}.  
As the WTTS do not accrete, we fix the mass accretion rate at 
$1\times10^{-11}\,\mathrm{M_\sun}\mathrm{yr}^{-1}$ for these systems to show the position of these systems on the plot. 
We indicated the five systems that fall in the photoevaporation region in Fig. \ref{fig:AccvsRH}.
We can see that almost all our transition disk candidates are outside of the region where 
photoevaporation can explain the inner disk hole. Our sample of non-accreting transition 
disks is small, and the four systems also present small inner holes ($\mathrm{R_H}<10\,\mathrm{AU}$). 
Observed non-accreting transition disks with the largest holes are indeed rare \citep{2016PASA...33....5O}. 

To analyze the relation of the hole size with the accretion process, 
we also plot the UV excess (Fig. \ref{fig:AccvsRH}b) and the 
$\mathrm{H\alpha}$ equivalent width (Fig. \ref{fig:AccvsRH}c) 
as a function of the hole size. Transition disk systems that present
UV excesses (more negative values in the plot) tend to have larger holes 
compared to transition disk systems without UV excesses.

We analyzed the relation of the hole size with the IR emission in the 
inner and outer disk and with the $\alpha_{\mathrm{IRAC}}$ index (Fig. \ref{fig:magvsRH}). Transition disk candidates 
with lower emission in the inner disk tend to have large holes (Fig. \ref{fig:magvsRH}a). 
Our data do not show a relation between the hole size and the mid-IR emission and the $\alpha_{\mathrm{IRAC}}$ index
(Fig. \ref{fig:magvsRH}bc). 

Planet formation in the disk is one of the most plausible mechanisms to explain transition disk systems
that present signs of accretion, despite the fact that the models do not explain all the 
observational characteristics of transition disks \citep{2016PASA...33....5O,2017RSOS....470114E,2011ApJ...729...47Z}. 
Considering the evolution sequence of a disk with a planet, as discussed by \cite{2016PASA...33....5O}, 
when a newly formed planet is large enough to open a gap in the disk, it effectively traps the mm dust 
in the outer disk. At this point, the inner disk is still not dust free, and the SED looks like a full 
disk. After that, due to some mechanism that is not well 
understood, the inner disk turns almost dust free and the SED is consequently modeled as a disk with an inner 
hole, that correspond to an accreting transition disk system characterized to be mm-bright.  
During this phase, the young planet can then migrate to orbits with 
smaller radii ($\lesssim 10\,\mathrm{AU}$). Consequently the dust from the outer disk can move to inner orbits due to disk viscosity,  
and the inner disk can be photoevaporated by radiation of the central star, forming non-accreting transition disks 
with small holes and bright in the mm or systems with a gap formed by photoevaporation of the outer disk. 
The current disk models with a planet predict more non-accreting systems 
with large holes, and accreting systems with small holes and mm-bright, than found 
observationally \citep[e.g.,][]{2016PASA...33....5O} and
also confirmed with our sample (see Figure \ref{fig:AccvsRH}a). 
The systems that fall out of the region limited by the dashed lines in Figure \ref{fig:AccvsRH}a 
are good candidates to have proto-planets in different stages. Exploring and modeling these systems using data from wavelengths long than $24\,\mathrm{\mu m}$ is beyond the scope of this work. 

In Figure \ref{fig:MassvsRH} we show the mass and temperature of the central star (obtained by\cite{2014A&A...570A..82V})   as a function of the disk hole 
{size for our transition disk candidates. \cite{2009ApJ...700.1017K} found a linear relation 
between the size of the disk hole and the mass and temperature of the central star for a sample 
of transition disks belonging to Chamaeleon I, indicating the dependence of the hole size 
with the central star's properties. 
In our sample of transition disk candidates, this tendency is not well defined in the plot.
In Figure \ref{fig:MassvsRH}, we also overplotted literature data of semimajor axis of the orbit of confirmed 
exoplanets and their host star masses. Despite the fact that the masses and temperatures of our stars are slightly 
lower than the exoplanet host stars, the hole sizes ($R_H$) of our transition disks and the semimajor axis of the 
exoplanet orbits are compatible. This result is intuitively expected, if the planets were responsible for opening the 
gap in the inner disk.

\begin{figure*}[htb!] 
 \centering
\subfigure[]{\includegraphics[scale=0.4]{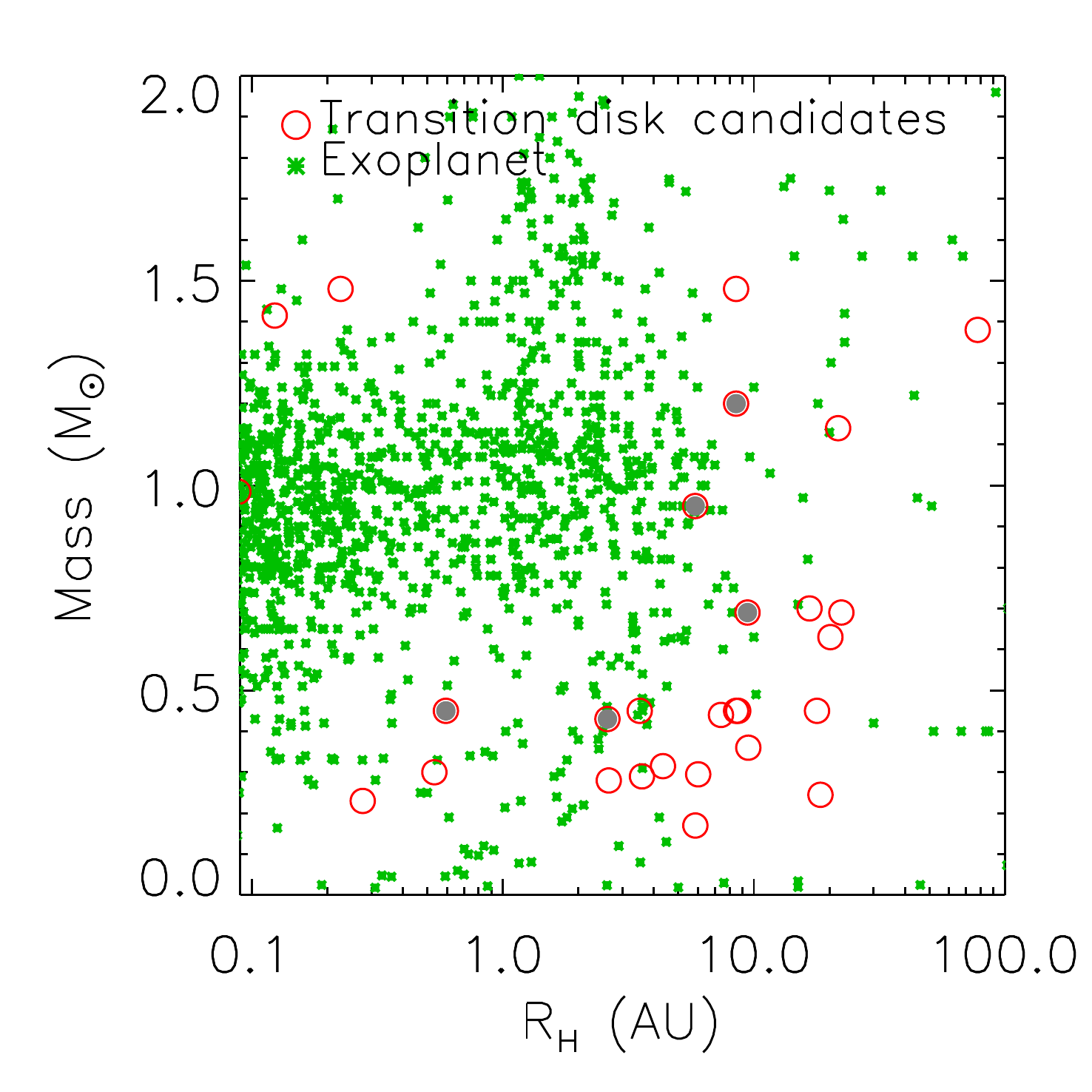}}
\subfigure[]{\includegraphics[scale=0.4]{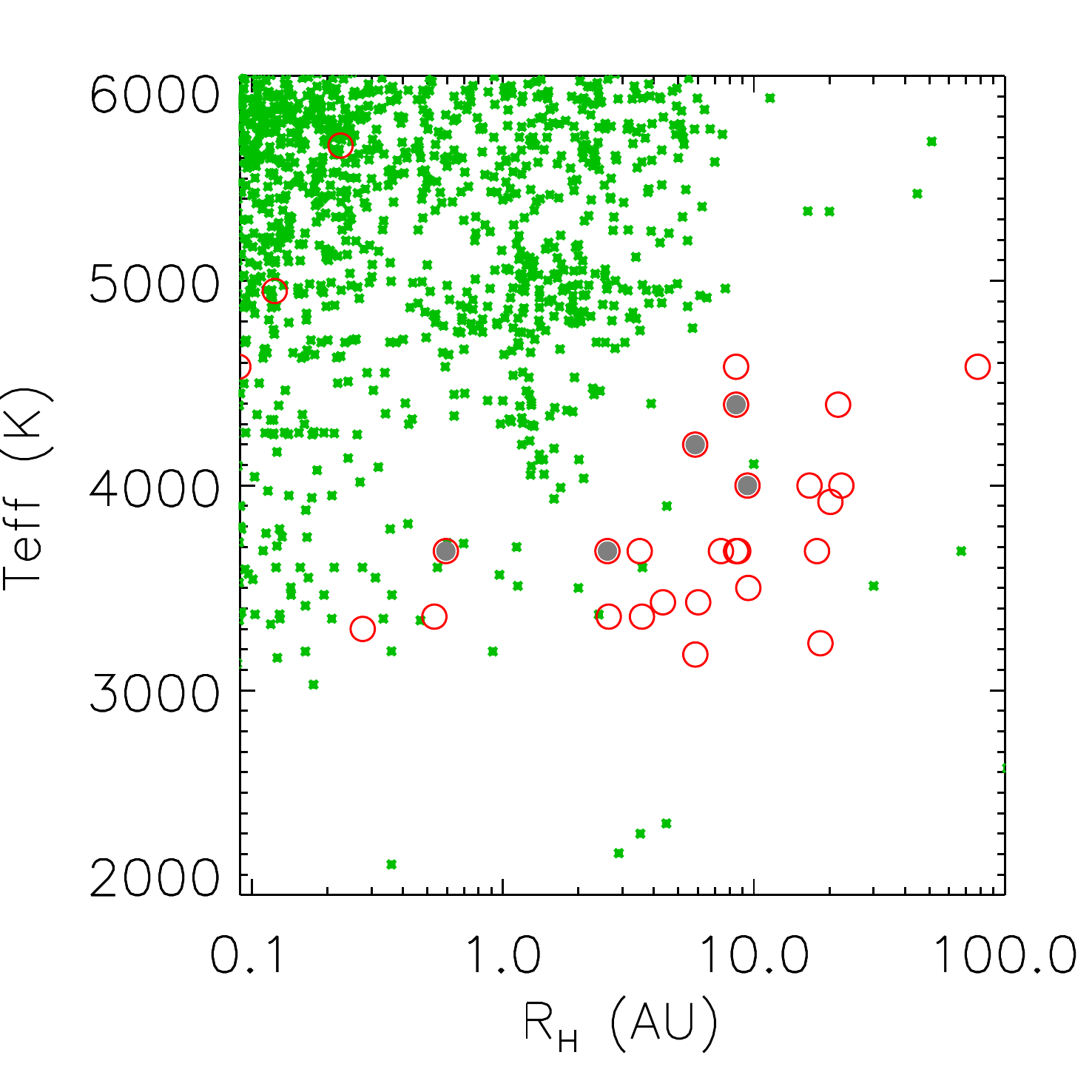}}
 \caption{\label{fig:MassvsRH} Mass and temperature of the central star obtained by \cite{2014A&A...570A..82V} as a function 
of disk hole size. The overplotted data from the literature correspond to exoplanets with confirmed detection
and the vertical axis is plotted as a function of the semimajor axis of the exoplanet orbit. Despite the fact that the masses and temperatures of our stars are slightly lower than the exoplanet host stars, the hole sizes of our transition disks and the semimajor axis of the exoplanet orbits are compatible. The data of the exoplanets and the host star were taken from http://exoplanet.eu. 
Gray filled symbols identify systems that fall in the region where the inner disk hole
can be explained by X-ray photoevaporation. Our data do not show a clear linear tendency between the mass and the temperature of central star with the inner hole size, as seen in literature. } 
\end{figure*}

\section{Conclusions}\label{sec:conclusion}\label{sec:concl}

In this work we searched for transition disk candidates in the young 
cluster NGC 2264. Our sample of $401$ TTS was observed 
with Spitzer equipped with IRAC instrument and MegaCam on CFHT, and corresponds to the
star-disk systems with well defined stellar, disk, and accretion parameters. 
The main results of this work are described below.

 We modeled the SEDs of $401$ TTS with the 
Hyperion SED fitting code, using three different sets of models that 
correspond to a single stellar photosphere, a star with a full passive 
disk and a star with a passive disk and an inner disk hole. 
With the results of the model, we separated the star-disk systems 
into full disk systems ($209$ systems), diskless stars ($164$ systems),  
and transition disk candidates ($28$ systems). The transition disk candidates represent 
$7\,\%$ of our sample and confirm that transition disks are a rapid phase of disk evolution.

 We have shown that  transition disk systems present $\mathrm{H}\alpha$, UV excess, 
and mass accretion rates at the same level as accreting full disk systems. 
It shows that the presence of a hole in the inner disk does not 
stop the accretion process, since $\sim82\,\%$ of our transition 
disk candidates still accrete, suggesting that gas is 
flowing through the disk hole, as found in transition disks from 
other star formation regions.

 Our sample of transition disk candidates have dust in 
the inner disk similar to anemic disks, which indicates that
anemic disks can be in general good candidates to have transition disks.

 In color-color diagrams, we see three different populations. 
Stars with a full disk present IR excess above the photospheric emission, 
in the inner and outer parts of the disk. The transition disk systems 
present weak dust emission in the inner disk, like diskless stars, and dust emission 
in the outer disk like full disk systems, which is consistent with 
disk evolution from inside to outside, associated with
photoevaporation of the inner disk by the stellar radiation or planet formation.
  
 We estimated the hole size of our sample of transition disk 
candidates and found sizes from $0.09$ to $78\,\mathrm{AU}$ with a 
mean value of $10.4\pm2.8\,\mathrm{AU}$. Among our sample of transition disk candidates, only $\sim 18\,\%$ 
have small mass accretion rates ($\lesssim10^{-9}\,\mathrm{M}_\sun\mathrm{yr}^{-1}$) 
and small hole sizes ($\mathrm{R_H}<10\,\mathrm{AU}$) that
can be explained by X-ray photoevaporation of the inner 
disk by stellar radiation. We also show that $\sim82\%$ 
could be explained by planet formation in different evolutionary stages (with small and large hole sizes).

\begin{acknowledgements}
APS and SHPA acknowledge support from CNPq, CAPES and Fapemig. We also thank Laura Venuti and Julia Roquette for helpful discussions.

This publication makes use of data products from the Two Micron All Sky Survey, 
which is a joint project of the University of Massachusetts and the Infrared 
Processing and Analysis Center/California Institute of Technology, funded by 
the National Aeronautics and Space Administration and the National Science Foundation.
This work has made use of data from the European Space Agency (ESA) mission
 Gaia (\url{https://www.cosmos.esa.int/gaia}), processed by the {\it Gaia}
Data Processing and Analysis Consortium (DPAC,
\url{https://www.cosmos.esa.int/web/gaia/dpac/consortium}). Funding for the DPAC
has been provided by national institutions, in particular the institutions
participating in the Gaia Multilateral Agreement. 

This work is based in part on observations made with the Spitzer Space Telescope, which is operated by the Jet Propulsion Laboratory, California Institute of Technology under a contract with NASA.

This publication makes use of data products from the Wide-field Infrared Survey Explorer, which is a joint project of the University of California, Los Angeles, and the Jet Propulsion Laboratory/California Institute of Technology, funded by the National Aeronautics and Space Administration.

Funding for SDSS-III has been provided by the Alfred P. Sloan Foundation, the Participating Institutions, the National Science Foundation, and the U.S. Department of Energy Office of Science. The SDSS-III web site is http://www.sdss3.org/.
SDSS-III is managed by the Astrophysical Research Consortium for the Participating Institutions of the SDSS-III Collaboration including the University of Arizona, the Brazilian Participation Group, Brookhaven National Laboratory, Carnegie Mellon University, University of Florida, the French Participation Group, the German Participation Group, Harvard University, the Instituto de Astrofisica de Canarias, the Michigan State/Notre Dame/JINA Participation Group, Johns Hopkins University, Lawrence Berkeley National Laboratory, Max Planck Institute for Astrophysics, Max Planck Institute for Extraterrestrial Physics, New Mexico State University, New York University, Ohio State University, Pennsylvania State University, University of Portsmouth, Princeton University, the Spanish Participation Group, University of Tokyo, University of Utah, Vanderbilt University, University of Virginia, University of Washington, and Yale University. 
\end{acknowledgements}


\bibliographystyle{aa}   
\bibliography{ref}

\begin{appendix}\label{}
\section{Spectral energy distribution of all systems modeled with Hyperion model fitting} 

\begin{figure*}
\includegraphics[scale=0.29]{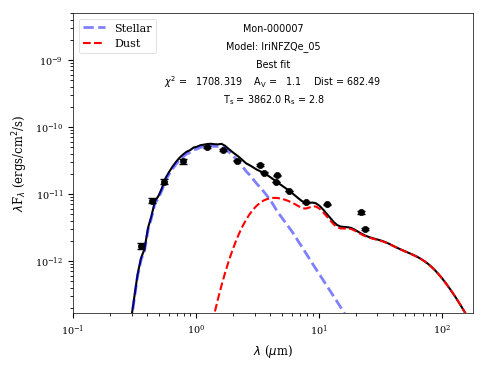}
\includegraphics[scale=0.29]{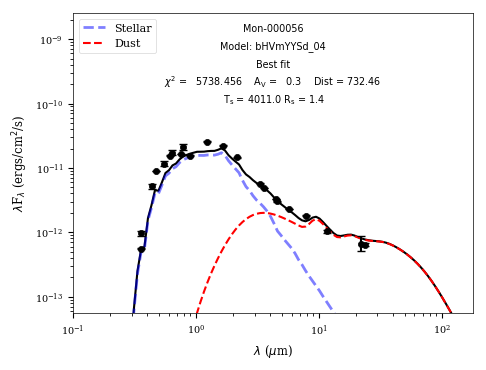}
\includegraphics[scale=0.29]{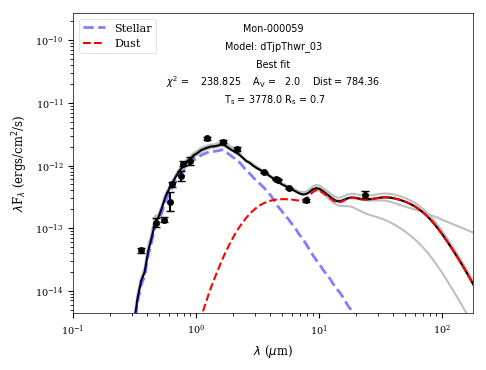}
\includegraphics[scale=0.29]{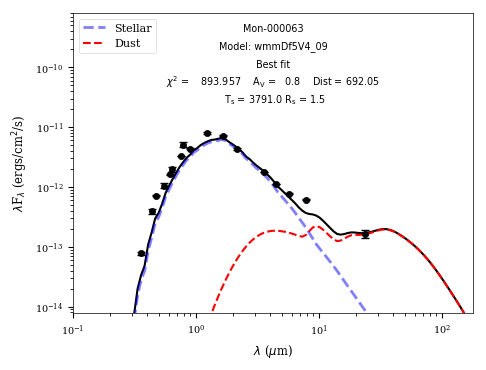}
\includegraphics[scale=0.29]{all_Mon-000064}
\includegraphics[scale=0.29]{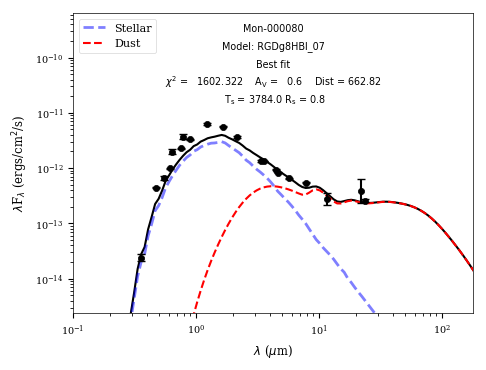}
\includegraphics[scale=0.29]{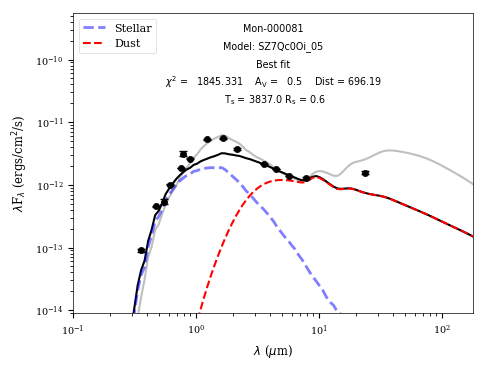}
\includegraphics[scale=0.29]{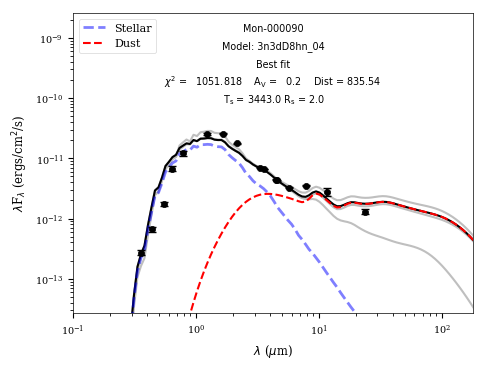}
\includegraphics[scale=0.29]{all_Mon-000103}
\includegraphics[scale=0.29]{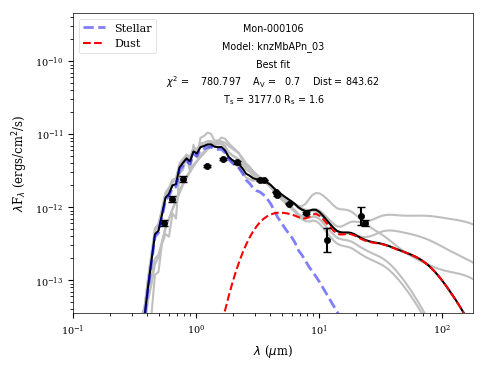}
\includegraphics[scale=0.29]{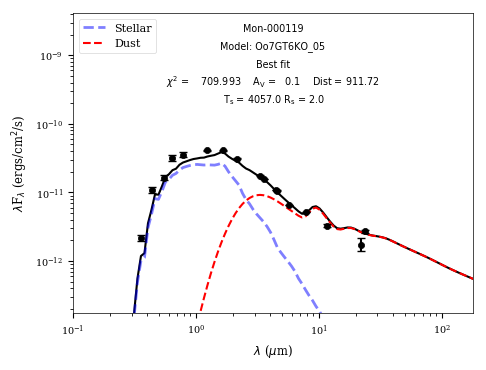}
\includegraphics[scale=0.29]{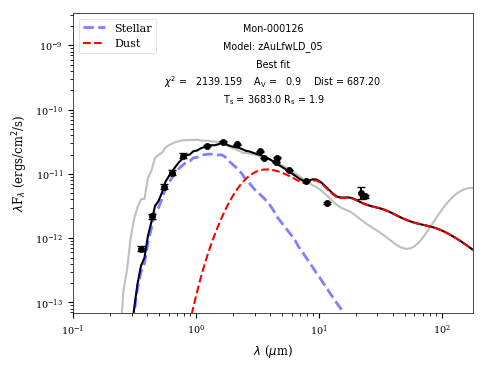}
\includegraphics[scale=0.29]{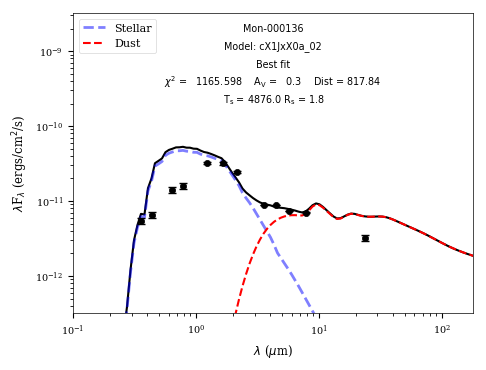}
\includegraphics[scale=0.29]{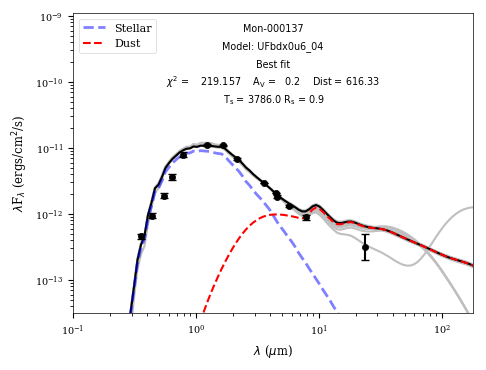}
\includegraphics[scale=0.29]{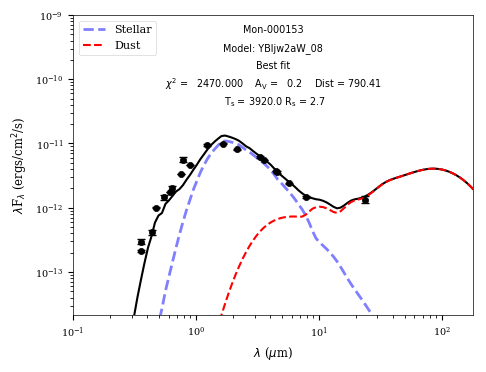}
\includegraphics[scale=0.29]{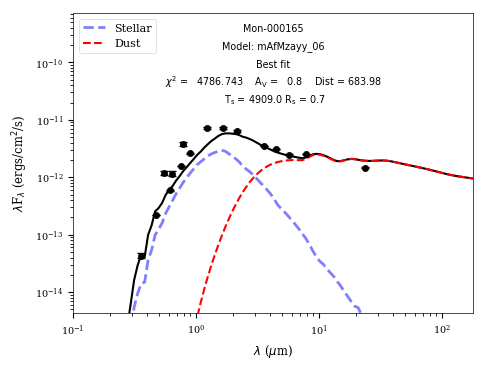}
\includegraphics[scale=0.29]{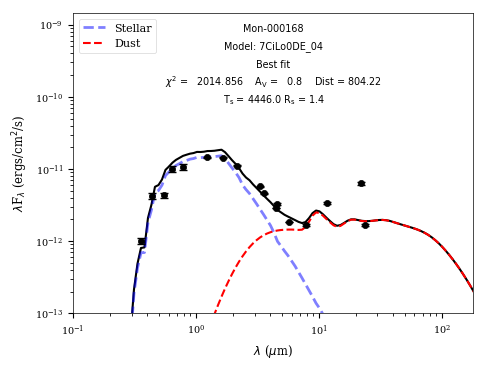}
\includegraphics[scale=0.29]{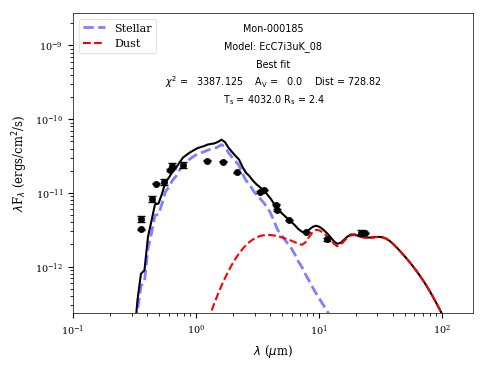}
\includegraphics[scale=0.29]{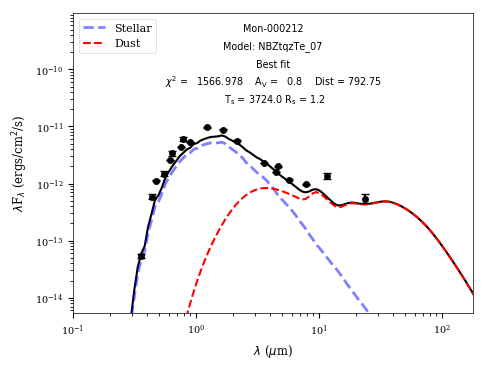}
\includegraphics[scale=0.29]{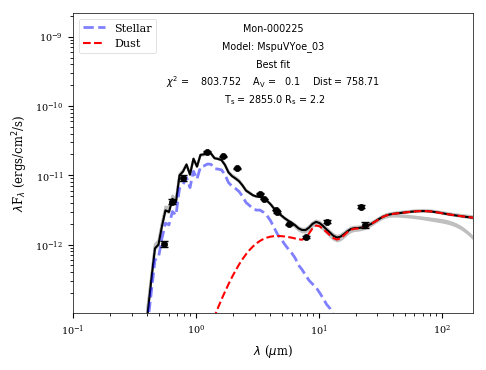}
\includegraphics[scale=0.29]{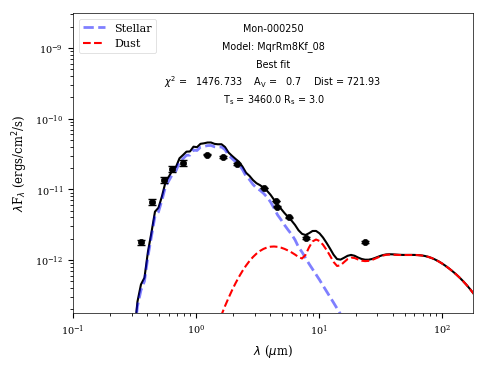}
\includegraphics[scale=0.29]{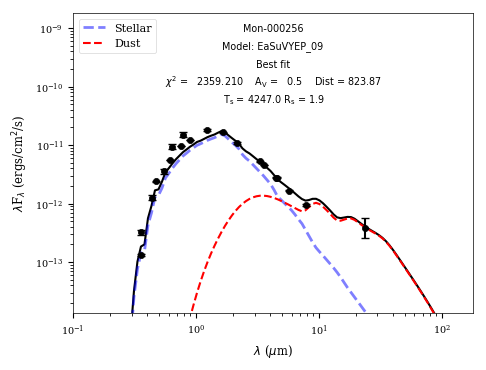}
\includegraphics[scale=0.29]{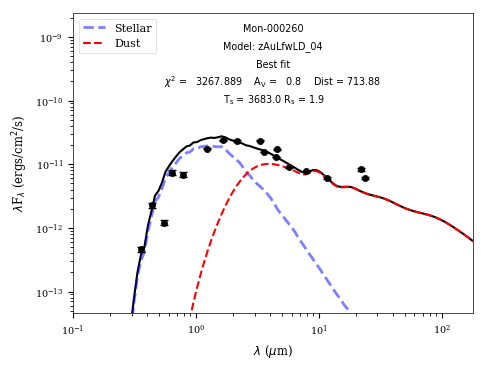}
\includegraphics[scale=0.29]{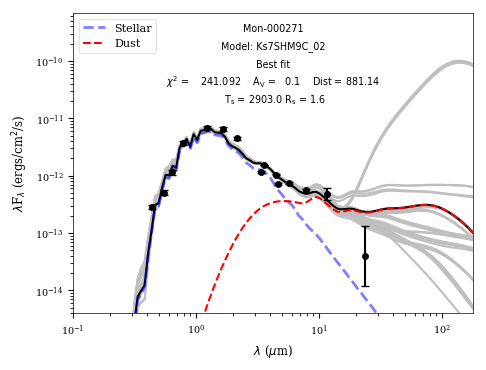}
\includegraphics[scale=0.29]{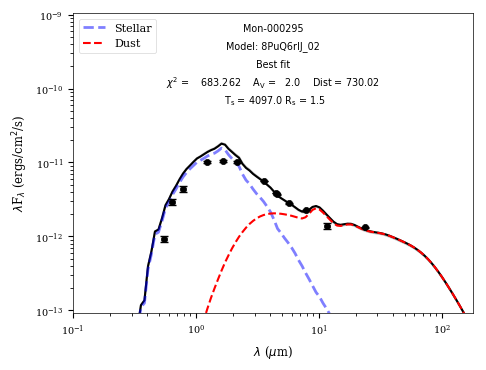}
\includegraphics[scale=0.29]{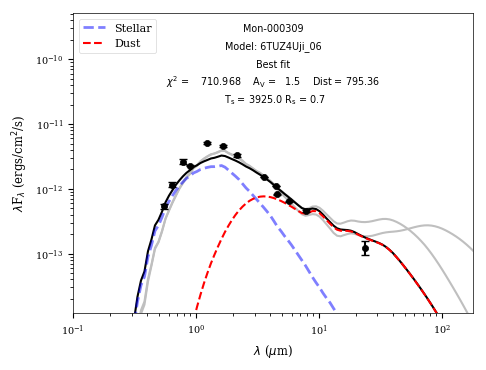}
\includegraphics[scale=0.29]{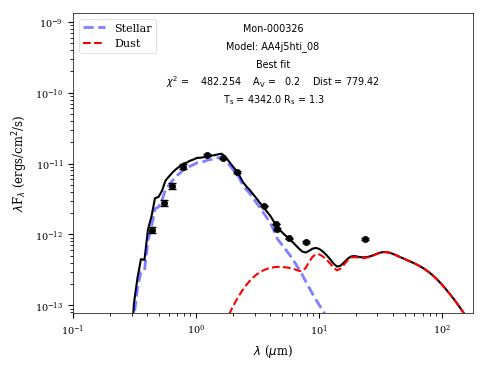}
\includegraphics[scale=0.29]{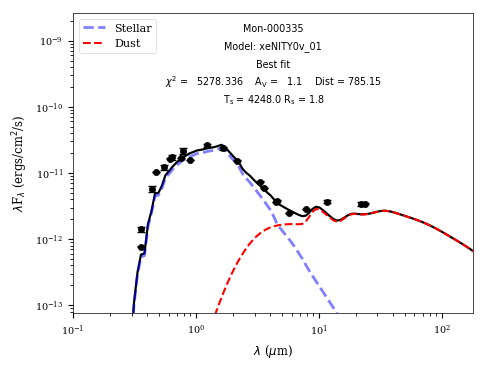}
\includegraphics[scale=0.29]{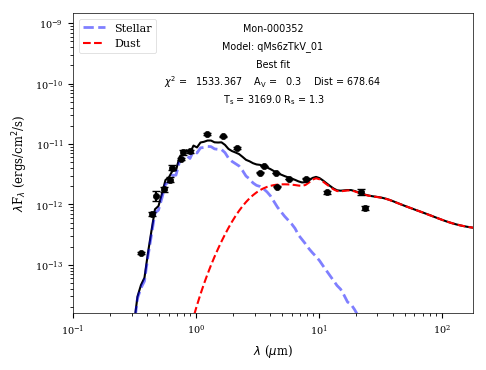}
\includegraphics[scale=0.29]{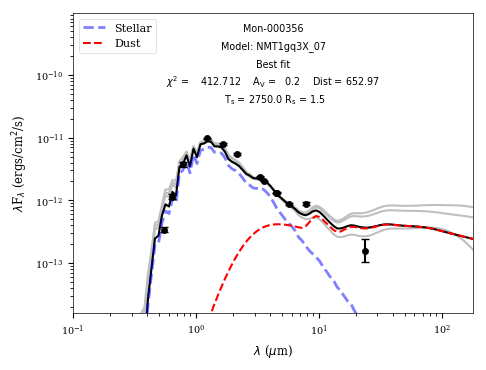}
\includegraphics[scale=0.29]{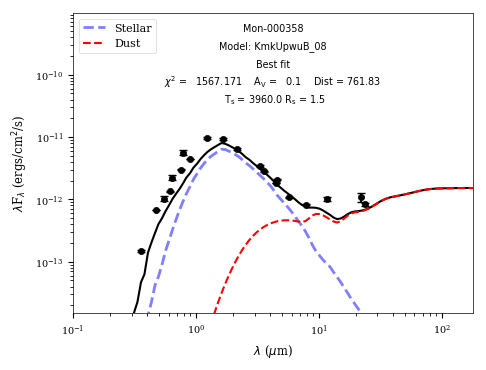}
\includegraphics[scale=0.29]{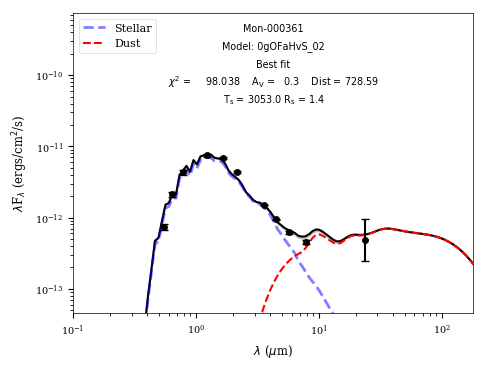}
\includegraphics[scale=0.29]{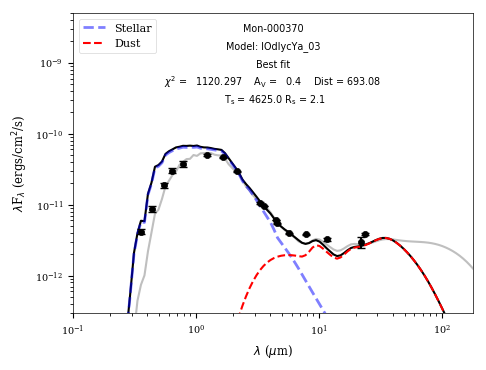}
\includegraphics[scale=0.29]{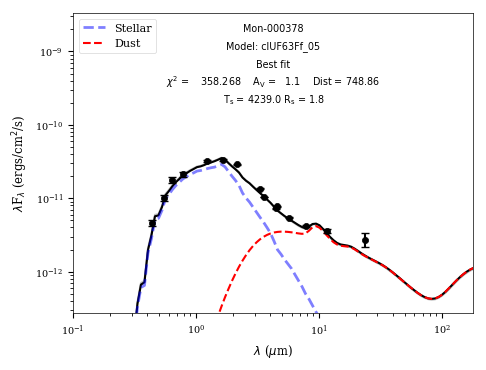}
\includegraphics[scale=0.29]{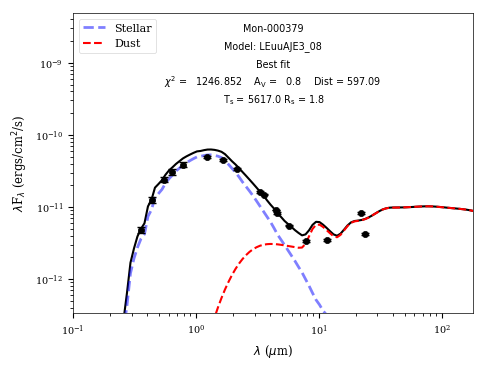}
\includegraphics[scale=0.29]{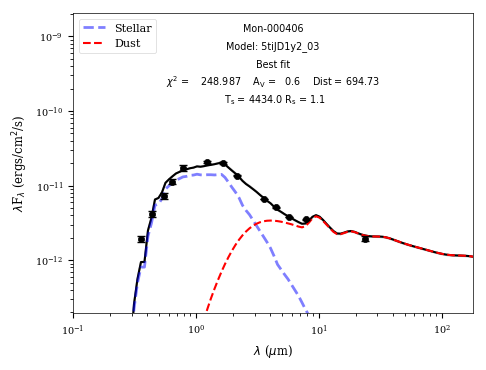}
\includegraphics[scale=0.29]{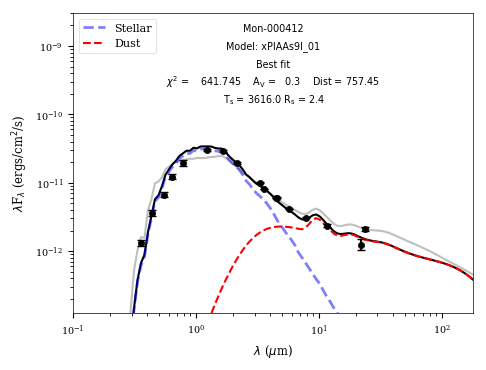}
\includegraphics[scale=0.29]{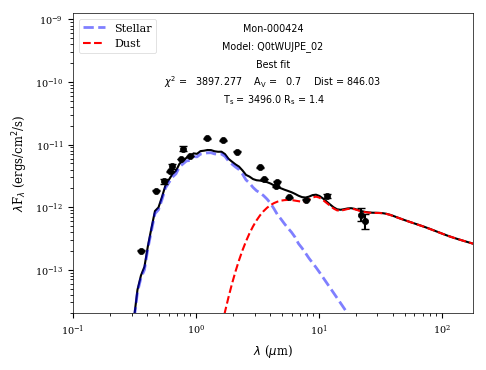}
\includegraphics[scale=0.29]{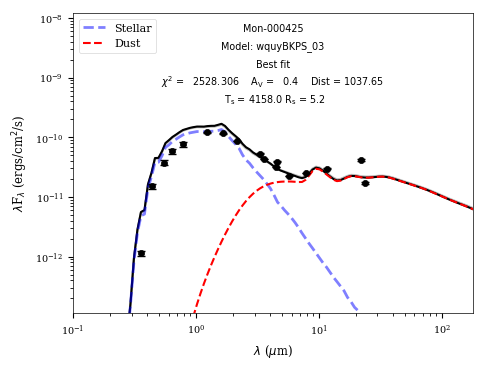}
\includegraphics[scale=0.29]{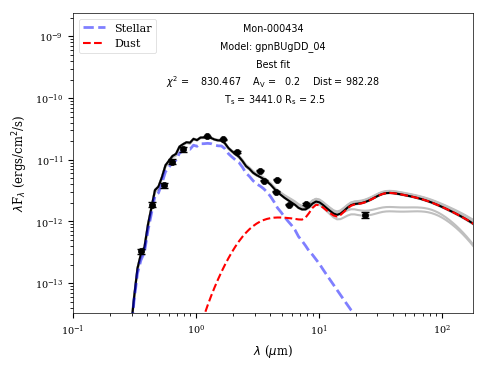}
\caption{\label{fig:Disk1} Spectral energy distribution of systems classified as full disks. 
The circles show data from the U band ($0.3\,\mu\mathrm{m}$) to the MIPS $24\,\mu\mathrm{m}$ band. 
The black solid line is the best fit to the SED \citep{2017A&A...600A..11R} and the 
dashed lines correspond to the stellar (blue) and dust (red) emission components 
\citep{2011A&A...536A..79R}. The gray solid lines are all the model with 
$(\chi^2 - \chi^2_{best})>3\mathrm{n_{data}}$, where $\mathrm{n_{data}}$ is 
the number of available data points from the literature, following 
\cite{2017A&A...600A..11R}. All these systems were observed by  Spitzer 
at $24\,\mu\mathrm{m}$ and/or WISE at $22\,\mu\mathrm{m}$.} 
\end{figure*}

\begin{figure*}
\includegraphics[scale=0.29]{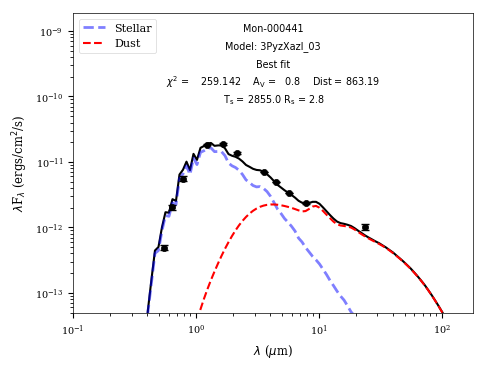}
\includegraphics[scale=0.29]{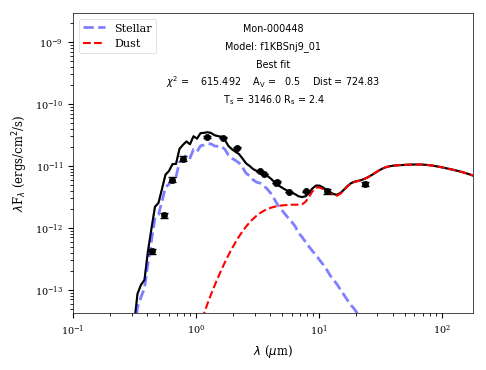}
\includegraphics[scale=0.29]{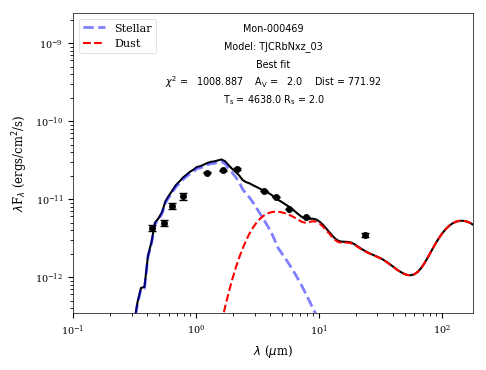}
\includegraphics[scale=0.29]{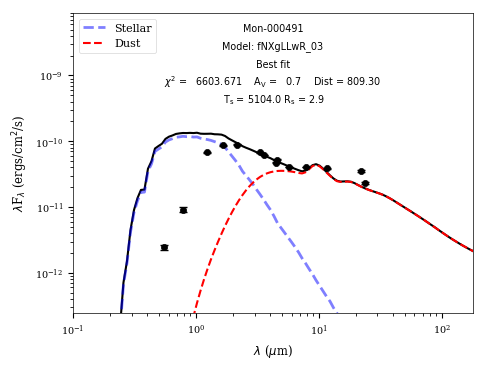}
\includegraphics[scale=0.29]{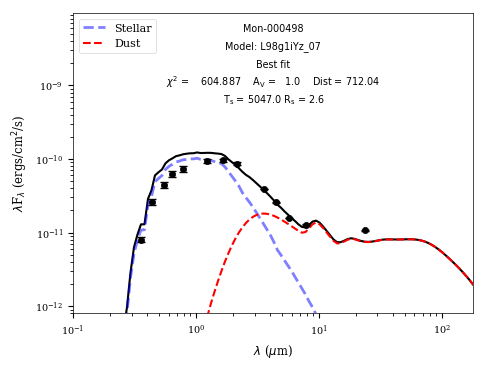}
\includegraphics[scale=0.29]{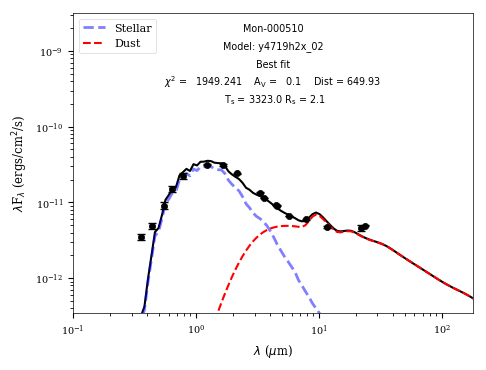}
\includegraphics[scale=0.29]{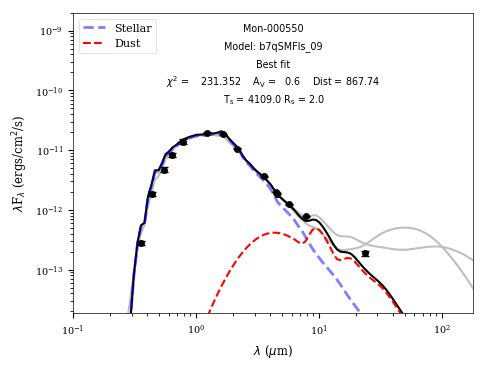}
\includegraphics[scale=0.29]{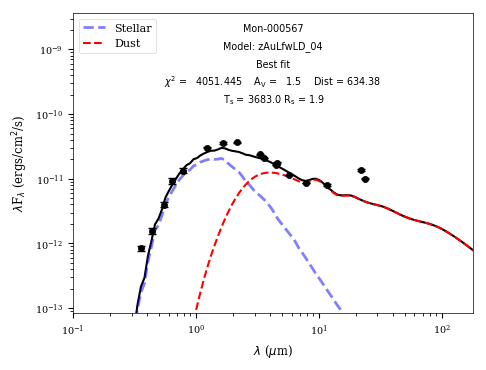}
\includegraphics[scale=0.29]{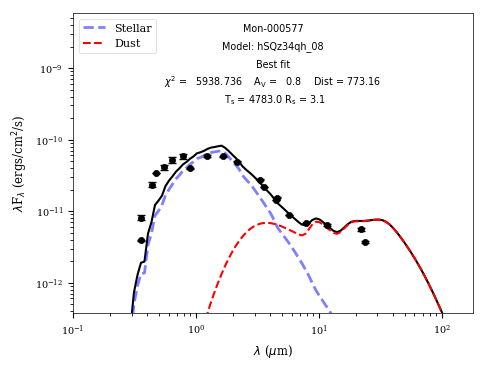}
\includegraphics[scale=0.29]{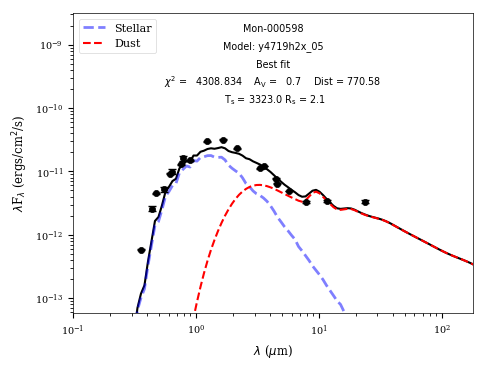}
\includegraphics[scale=0.29]{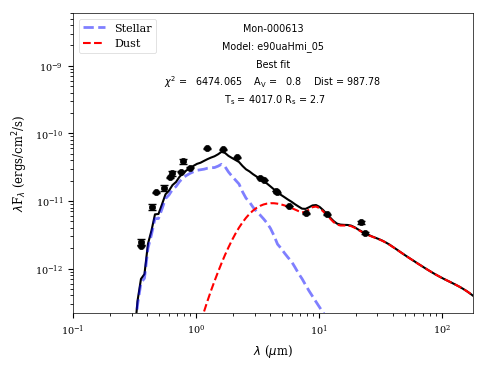}
\includegraphics[scale=0.29]{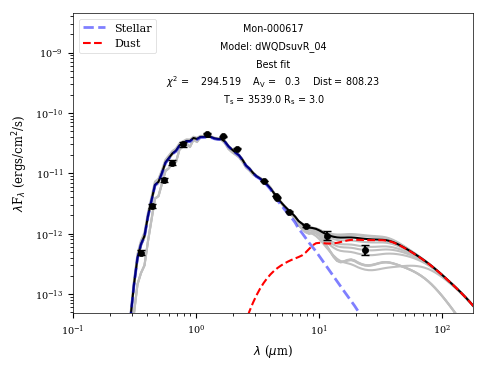}
\includegraphics[scale=0.29]{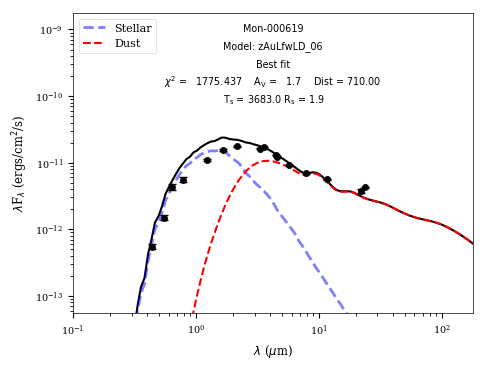}
\includegraphics[scale=0.29]{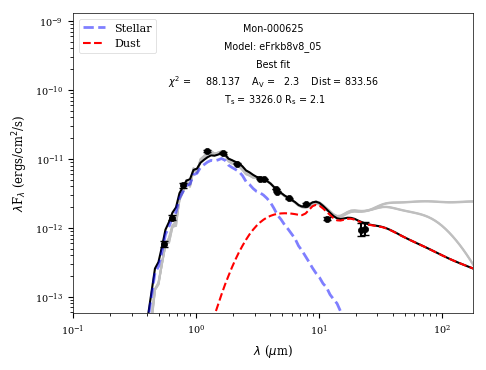}
\includegraphics[scale=0.29]{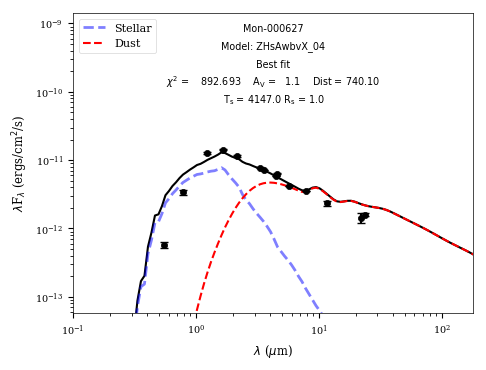}
\includegraphics[scale=0.29]{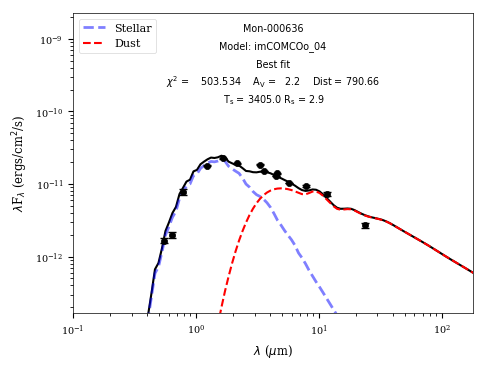}
\includegraphics[scale=0.29]{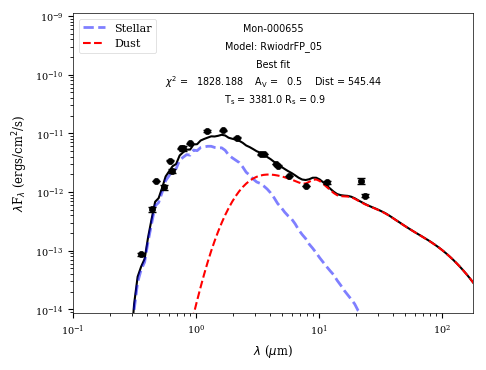}
\includegraphics[scale=0.29]{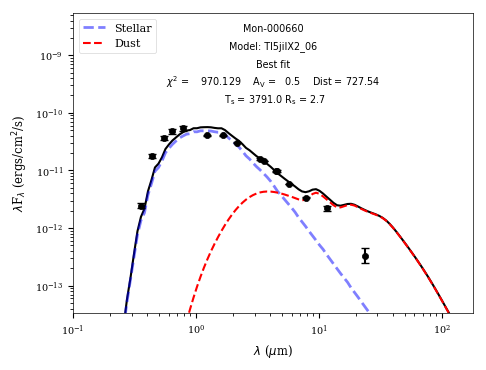}
\includegraphics[scale=0.29]{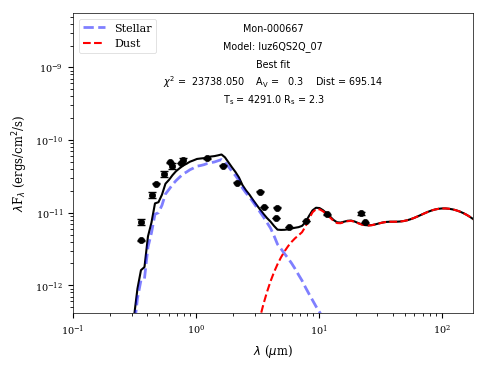}
\includegraphics[scale=0.29]{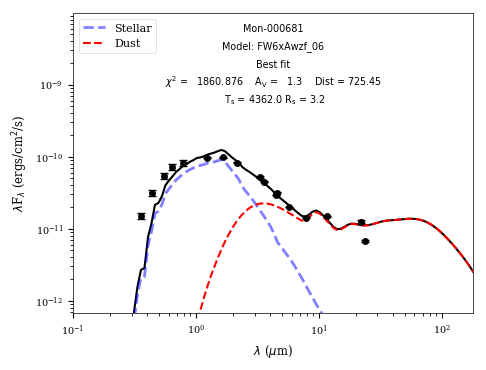}
\includegraphics[scale=0.29]{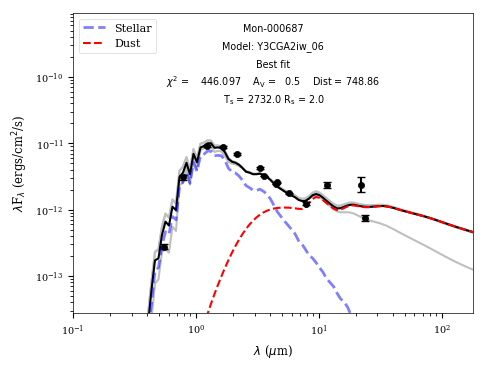}
\includegraphics[scale=0.29]{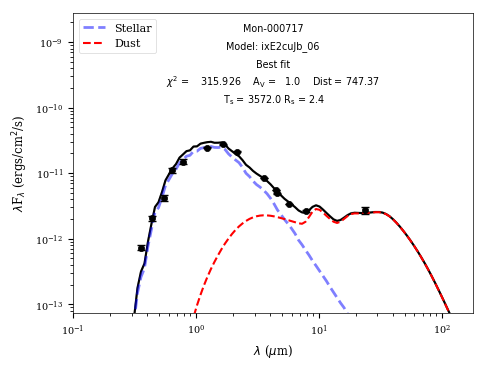}
\includegraphics[scale=0.29]{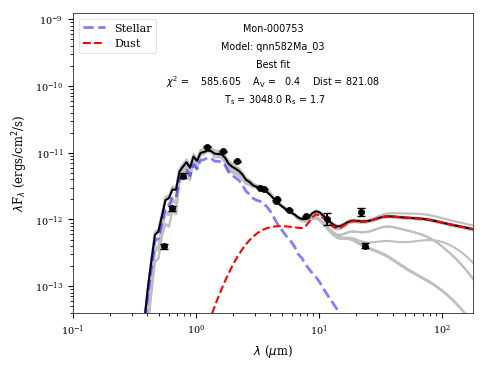}
\includegraphics[scale=0.29]{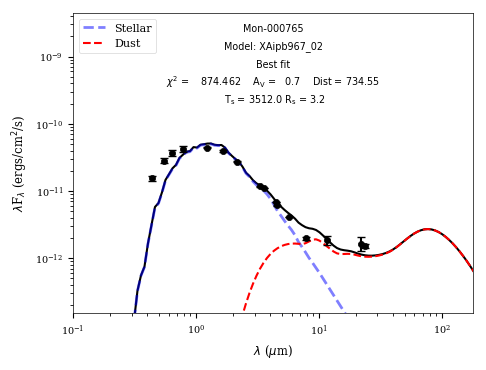}
\includegraphics[scale=0.29]{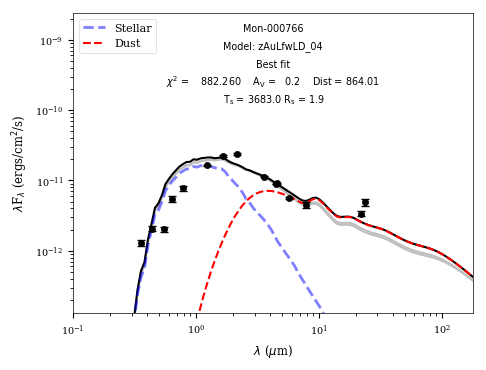}
\includegraphics[scale=0.29]{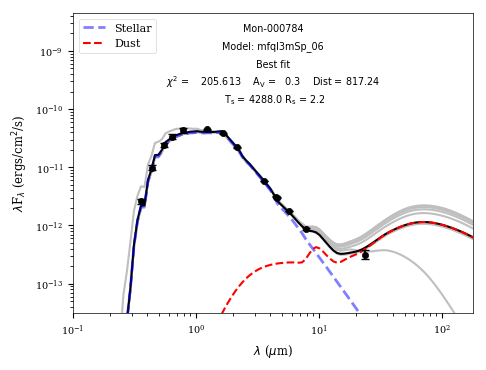}
\includegraphics[scale=0.29]{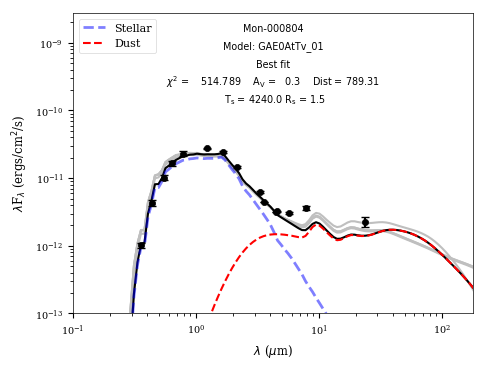}
\includegraphics[scale=0.29]{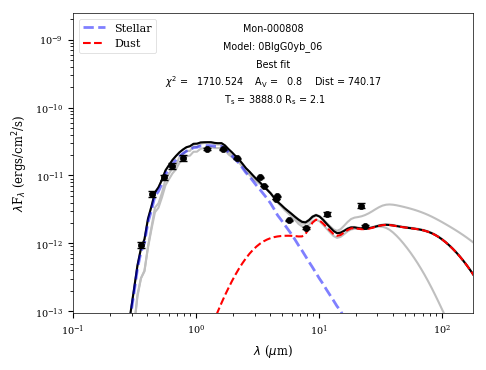}
\includegraphics[scale=0.29]{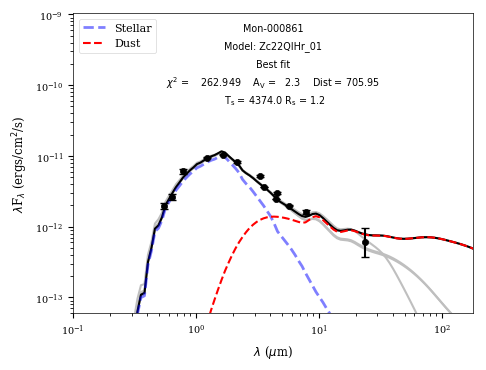}
\includegraphics[scale=0.29]{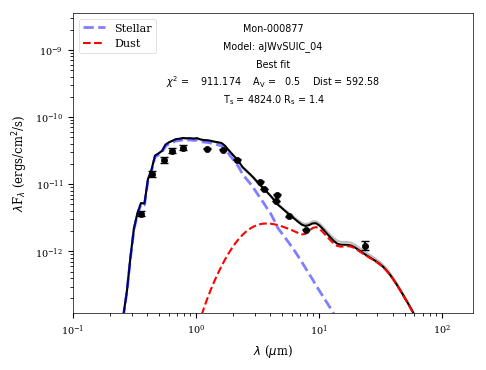}
\includegraphics[scale=0.29]{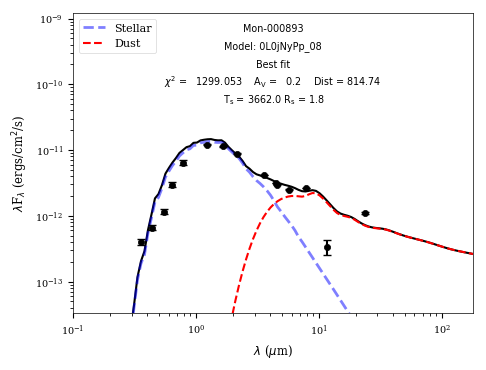}
\includegraphics[scale=0.29]{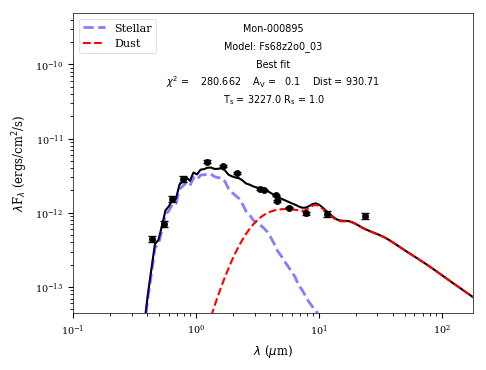}
\includegraphics[scale=0.29]{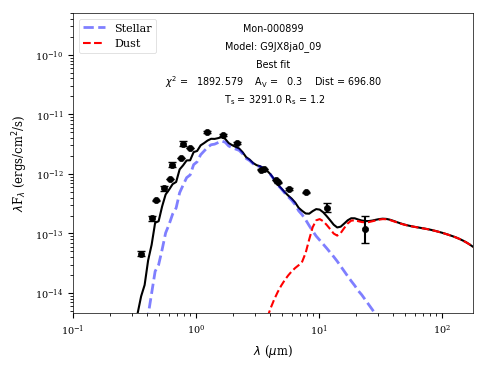}
\includegraphics[scale=0.29]{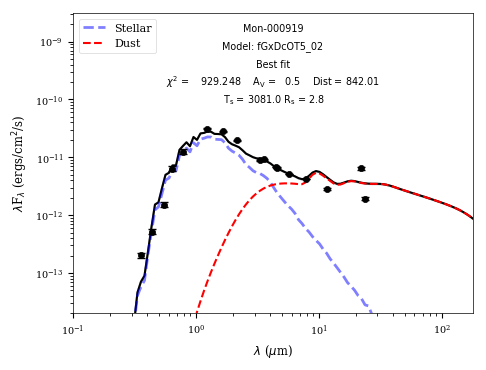}
\includegraphics[scale=0.29]{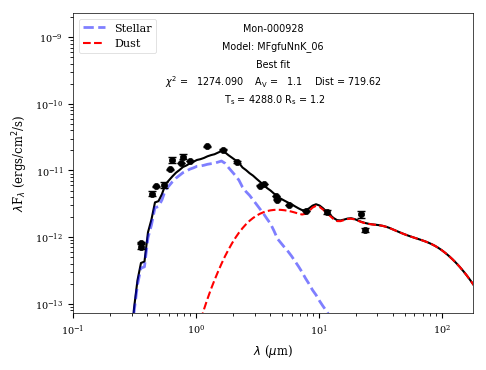}
\includegraphics[scale=0.29]{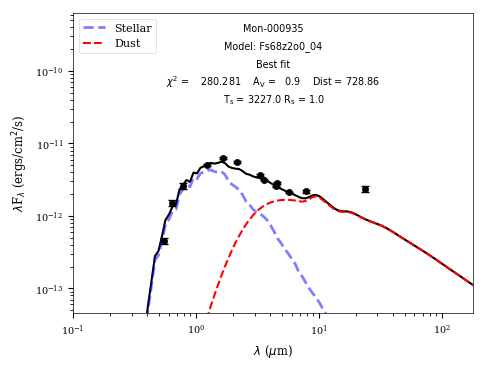}
\includegraphics[scale=0.29]{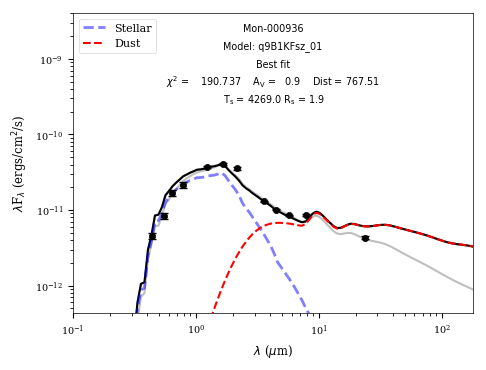}
\includegraphics[scale=0.29]{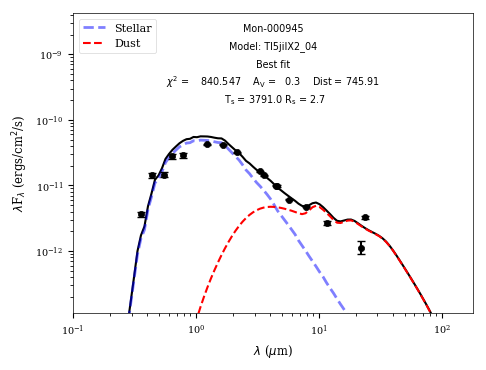}
\includegraphics[scale=0.29]{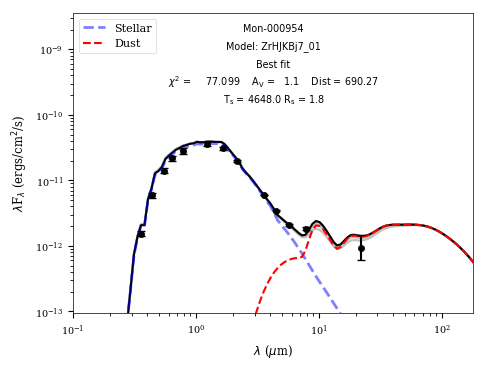}
\includegraphics[scale=0.29]{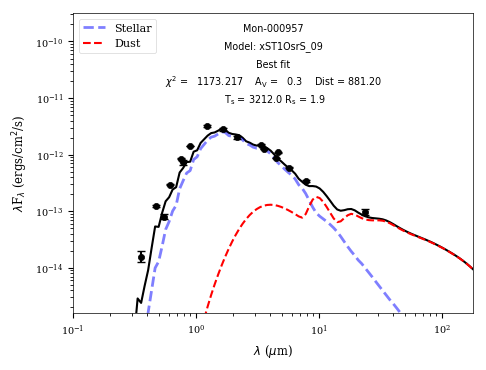}
\caption{\label{fig:Disk2} The same as Fig. \ref{fig:Disk1}.}
\end{figure*}
\begin{figure*}
\includegraphics[scale=0.29]{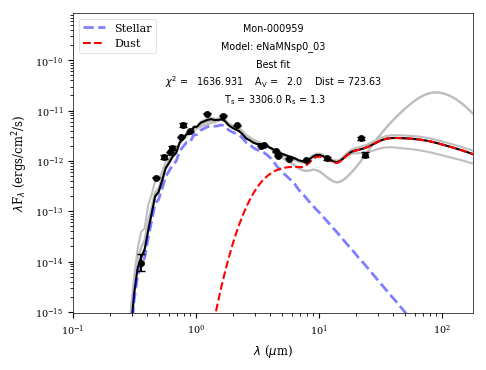}
\includegraphics[scale=0.29]{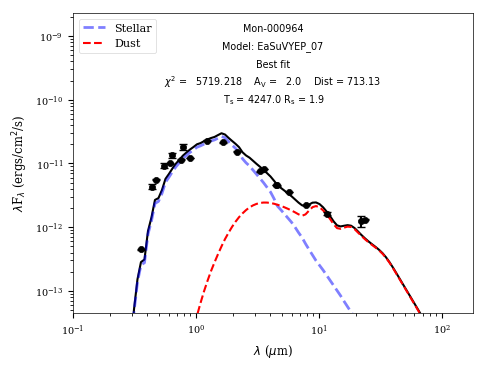}
\includegraphics[scale=0.29]{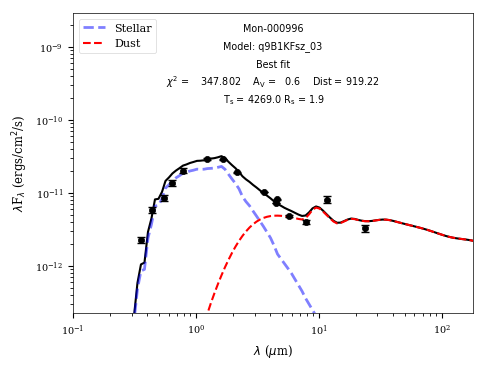}
\includegraphics[scale=0.29]{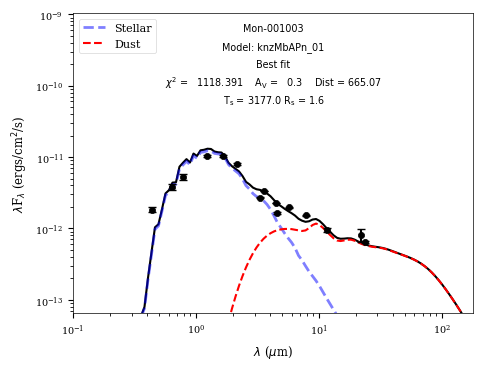}
\includegraphics[scale=0.29]{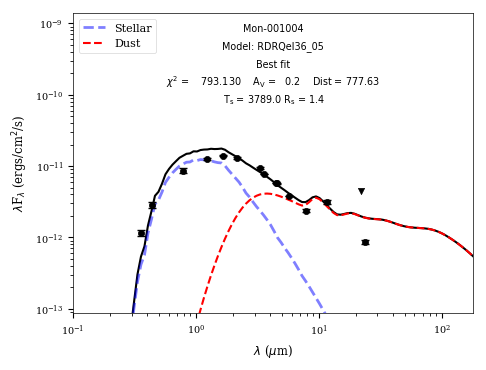}
\includegraphics[scale=0.29]{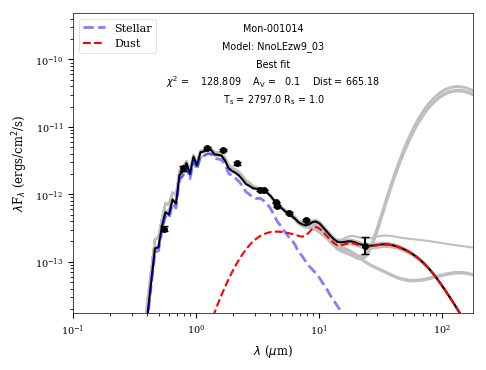}
\includegraphics[scale=0.29]{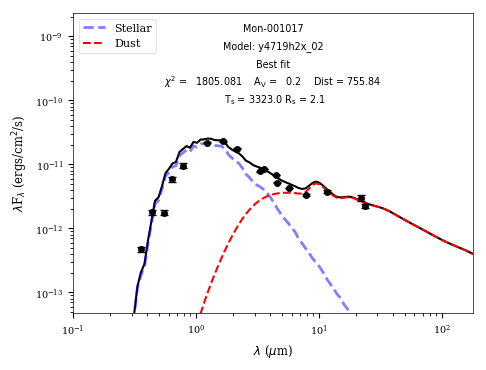}
\includegraphics[scale=0.29]{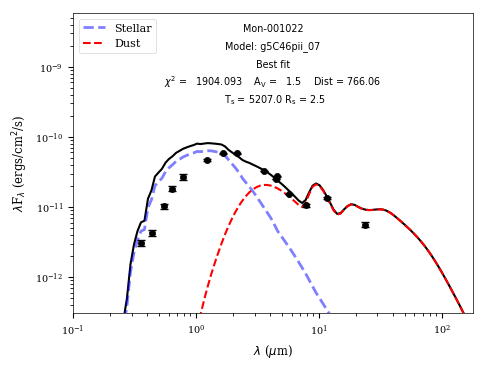}
\includegraphics[scale=0.29]{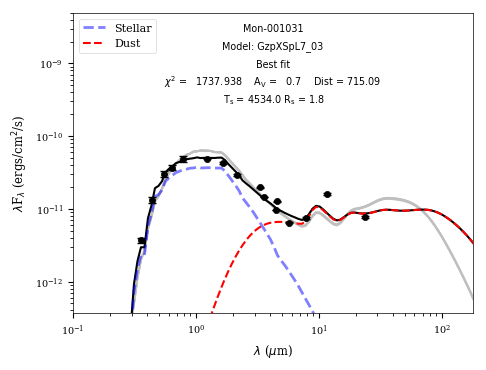}
\includegraphics[scale=0.29]{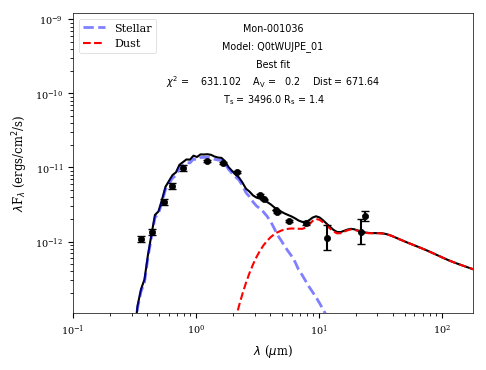}
\includegraphics[scale=0.29]{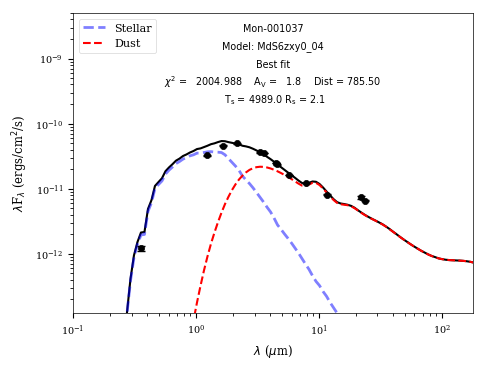}
\includegraphics[scale=0.29]{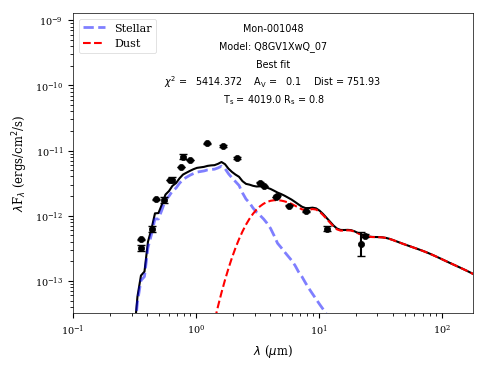}
\includegraphics[scale=0.29]{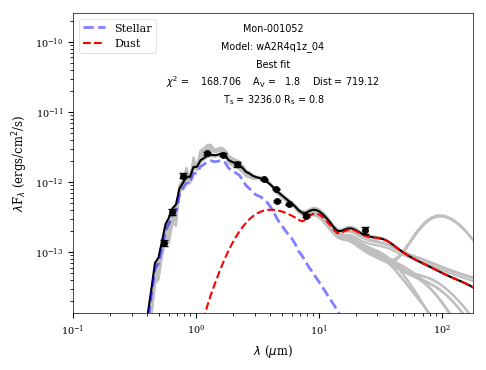}
\includegraphics[scale=0.29]{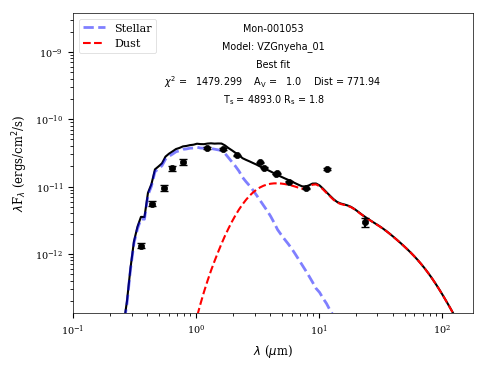}
\includegraphics[scale=0.29]{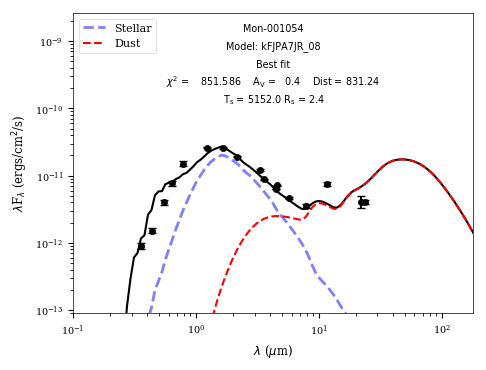}
\includegraphics[scale=0.29]{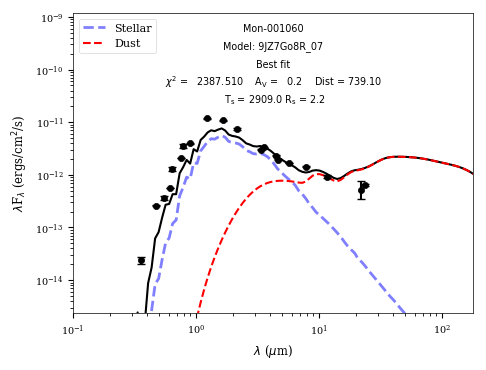}
\includegraphics[scale=0.29]{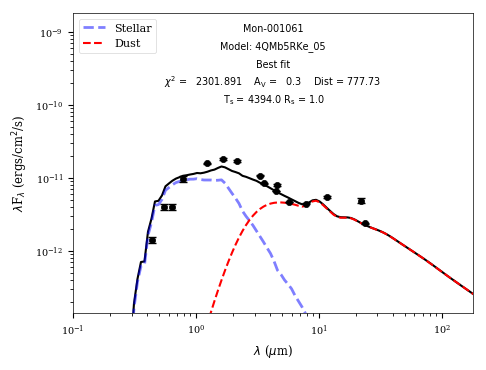}
\includegraphics[scale=0.29]{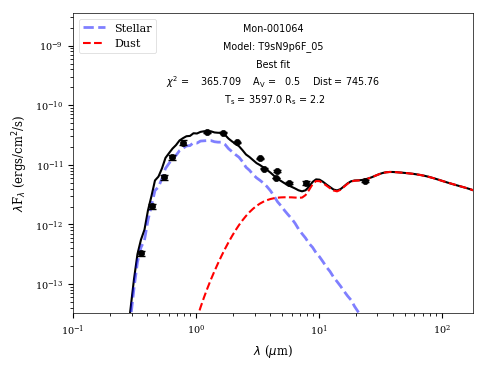}
\includegraphics[scale=0.29]{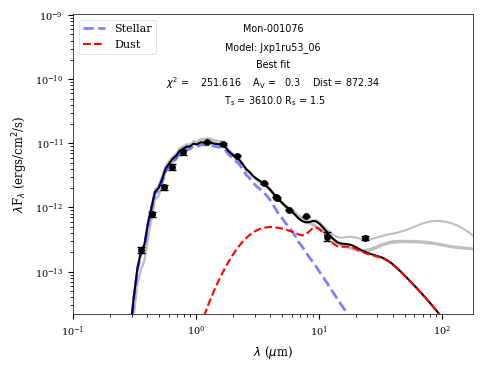}
\includegraphics[scale=0.29]{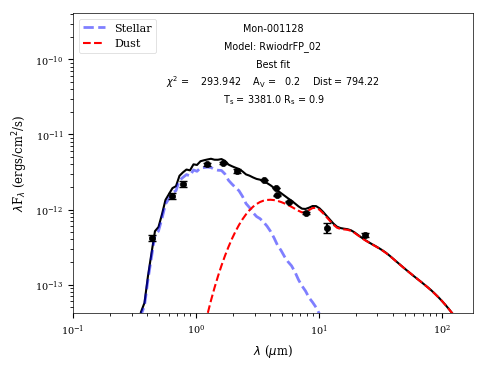}
\includegraphics[scale=0.29]{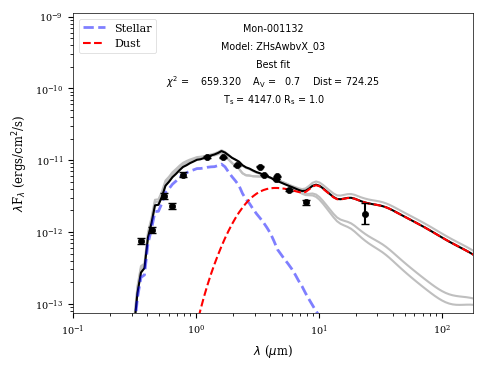}
\includegraphics[scale=0.29]{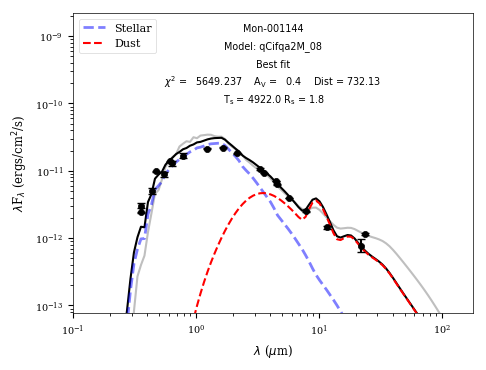}
\includegraphics[scale=0.29]{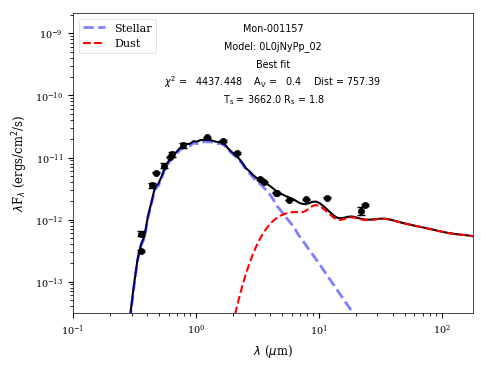}
\includegraphics[scale=0.29]{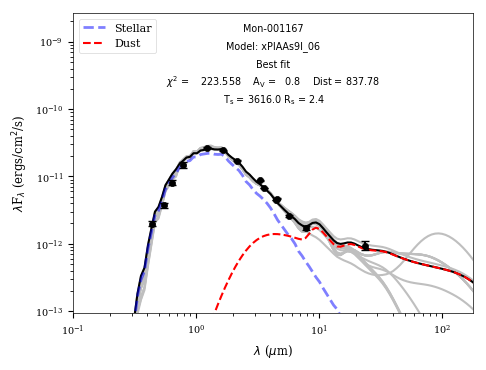}
\includegraphics[scale=0.29]{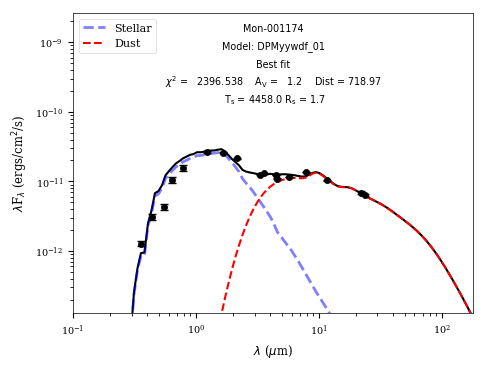}
\includegraphics[scale=0.29]{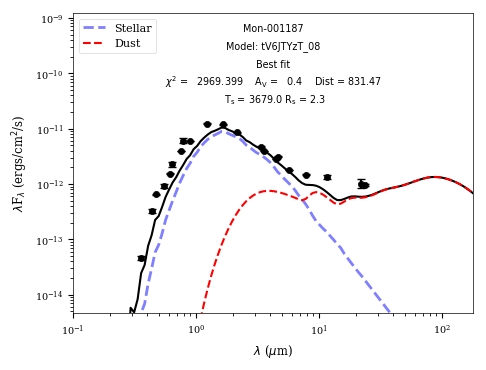}
\includegraphics[scale=0.29]{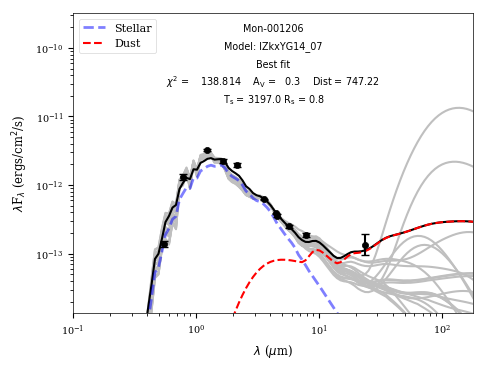}
\includegraphics[scale=0.29]{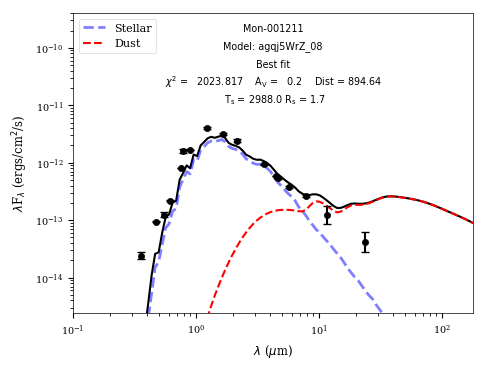}
\includegraphics[scale=0.29]{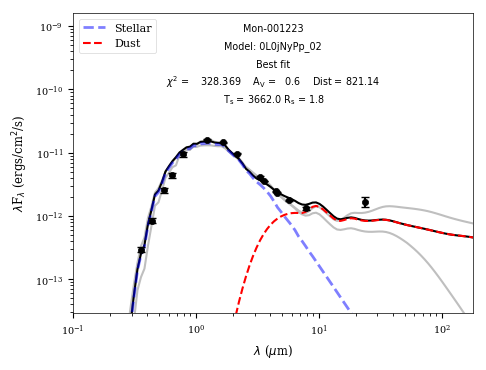}
\includegraphics[scale=0.29]{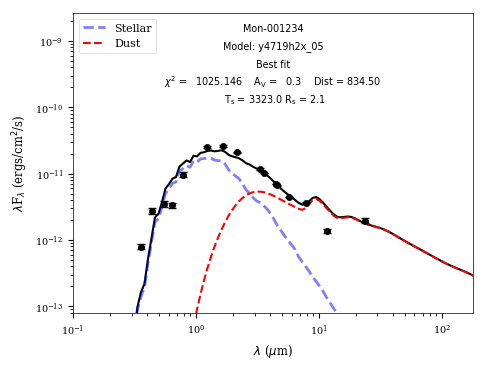}
\includegraphics[scale=0.29]{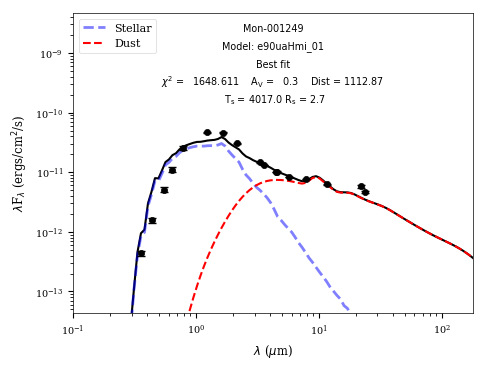}
\includegraphics[scale=0.29]{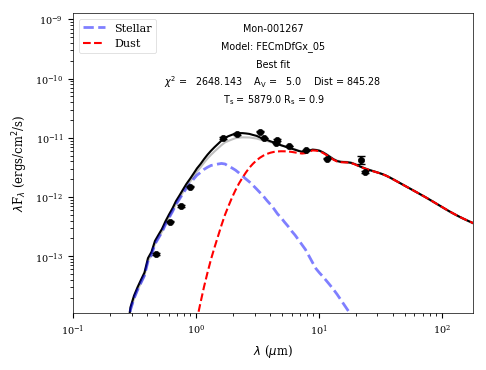}
\includegraphics[scale=0.29]{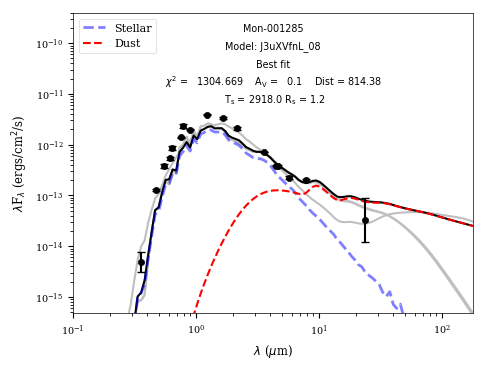}
\includegraphics[scale=0.29]{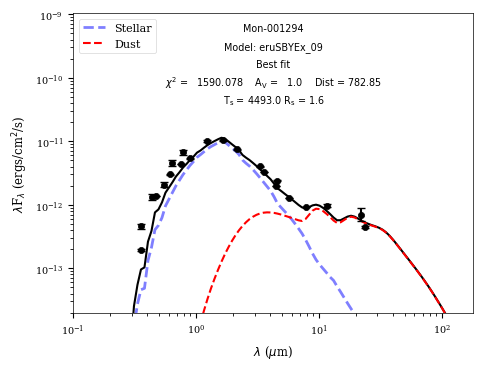}
\includegraphics[scale=0.29]{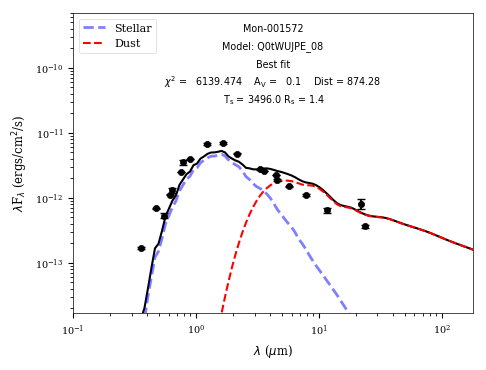}
\includegraphics[scale=0.29]{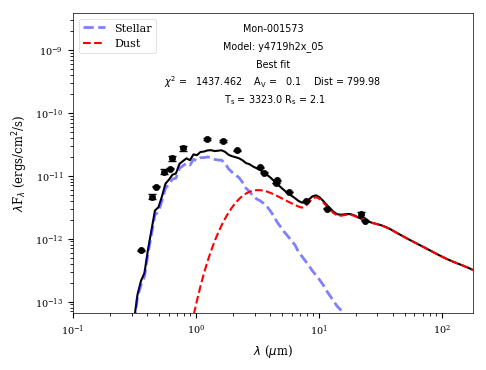}
\includegraphics[scale=0.29]{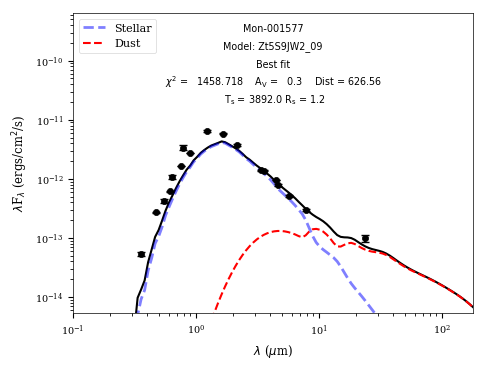}
\includegraphics[scale=0.29]{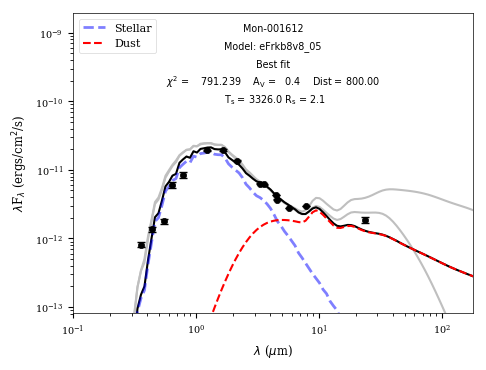}
\includegraphics[scale=0.29]{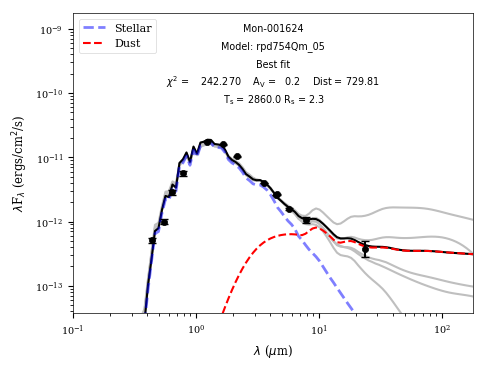}
\caption{\label{fig:Disk3} The same as Fig. \ref{fig:Disk1}. }
\end{figure*}

\begin{figure*}
\includegraphics[scale=0.29]{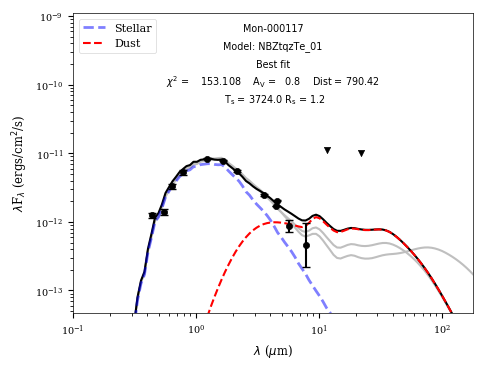}
\includegraphics[scale=0.29]{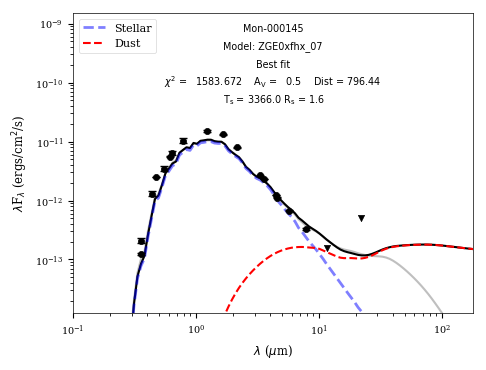}
\includegraphics[scale=0.29]{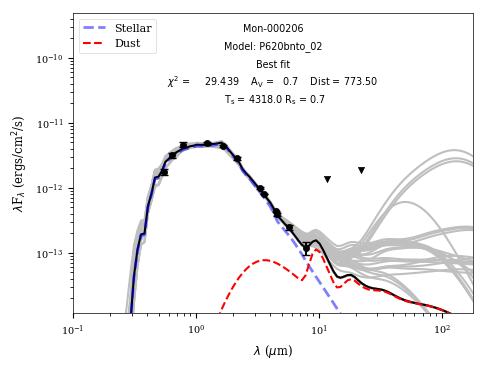}
\includegraphics[scale=0.29]{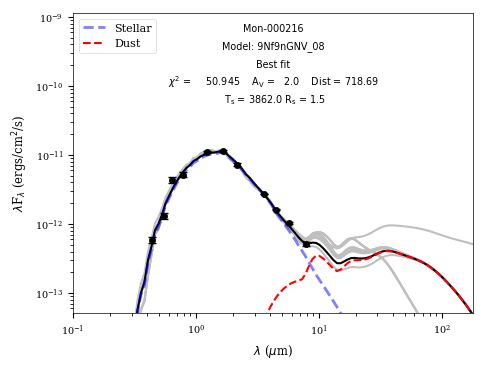}
\includegraphics[scale=0.29]{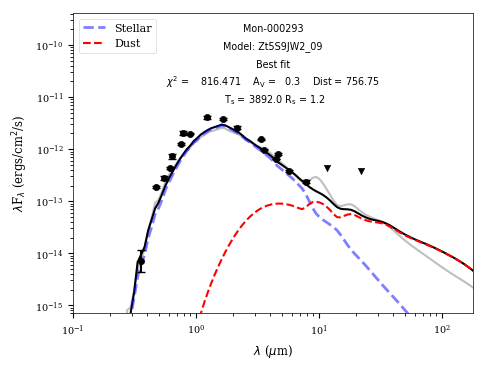}
\includegraphics[scale=0.29]{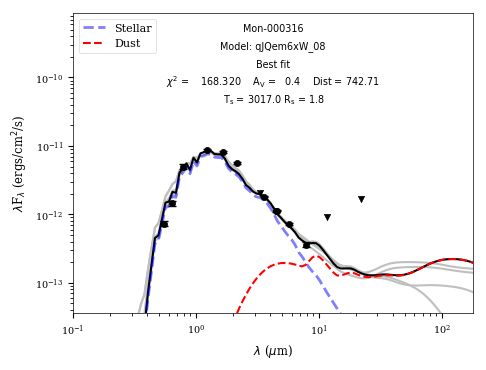}
\includegraphics[scale=0.29]{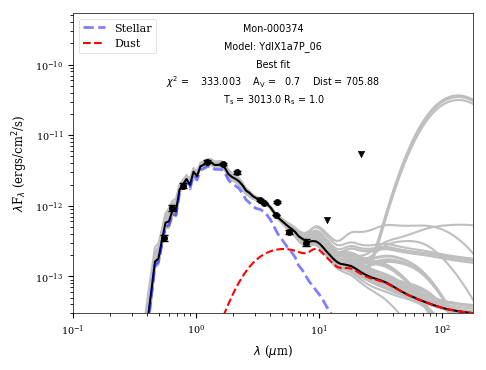}
\includegraphics[scale=0.29]{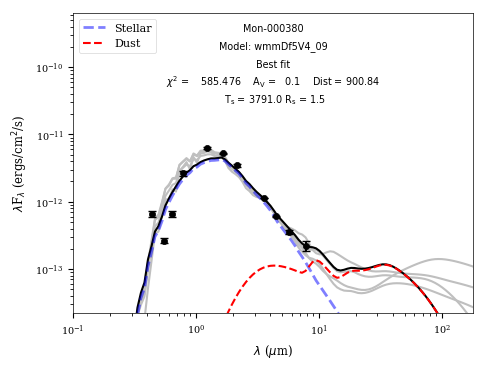}
\includegraphics[scale=0.29]{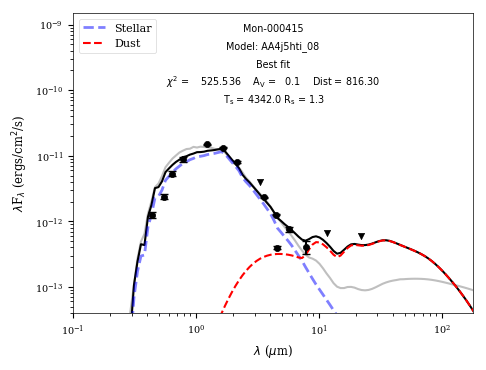}
\includegraphics[scale=0.29]{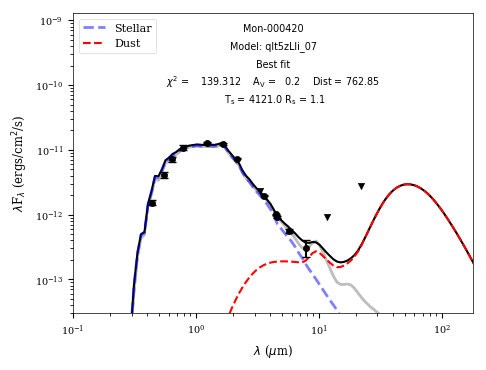}
\includegraphics[scale=0.29]{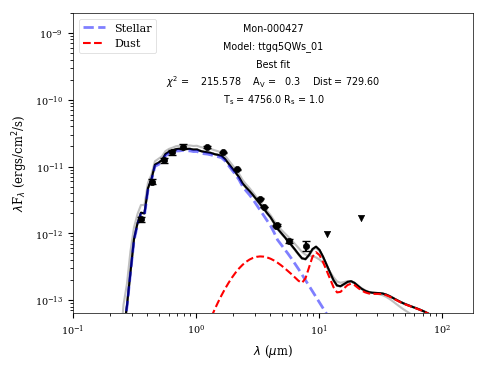}
\includegraphics[scale=0.29]{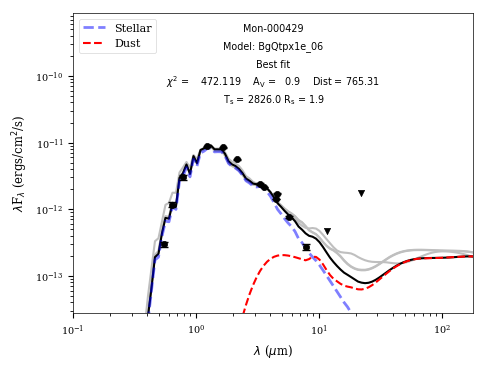}
\includegraphics[scale=0.29]{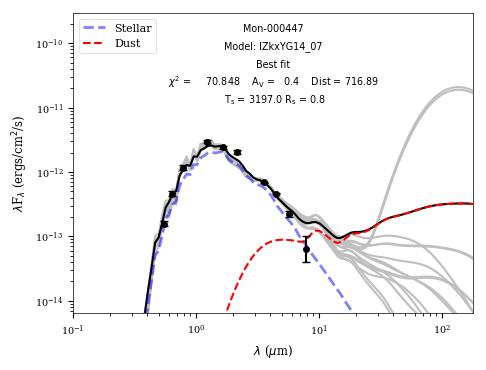}
\includegraphics[scale=0.29]{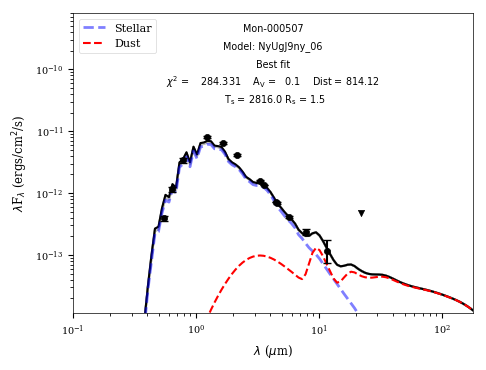}
\includegraphics[scale=0.29]{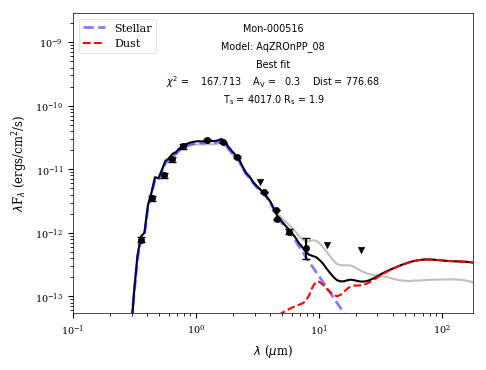}
\includegraphics[scale=0.29]{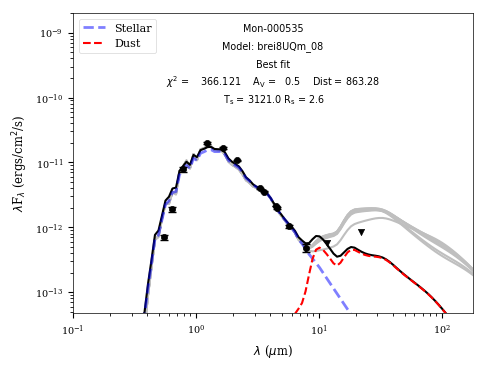}
\includegraphics[scale=0.29]{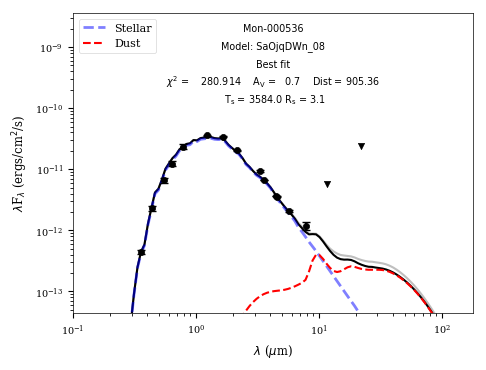}
\includegraphics[scale=0.29]{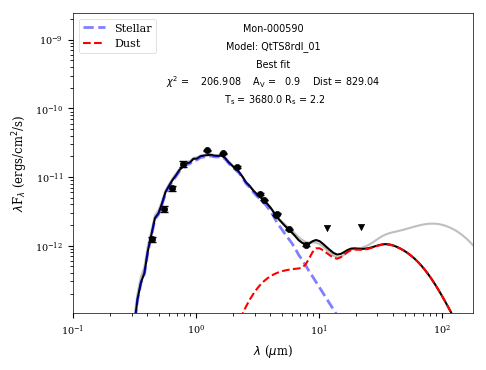}
\includegraphics[scale=0.29]{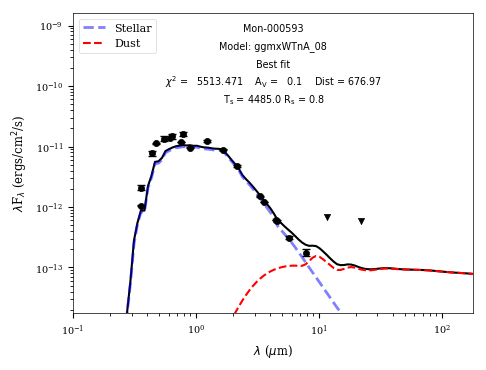}
\includegraphics[scale=0.29]{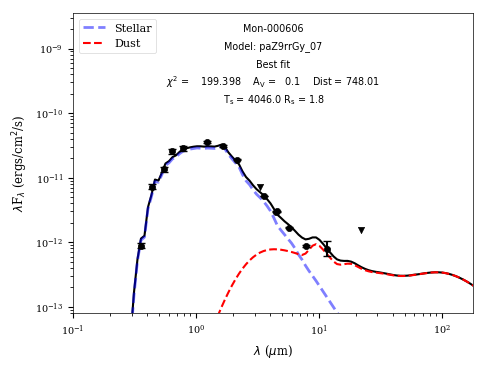}
\includegraphics[scale=0.29]{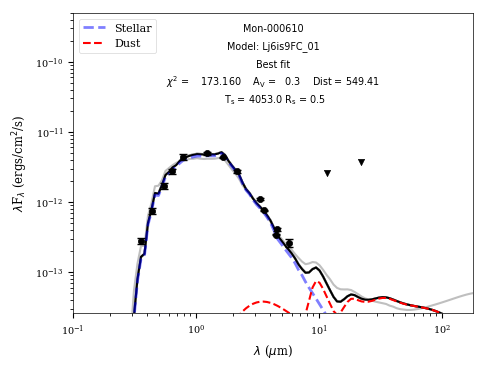}
\includegraphics[scale=0.29]{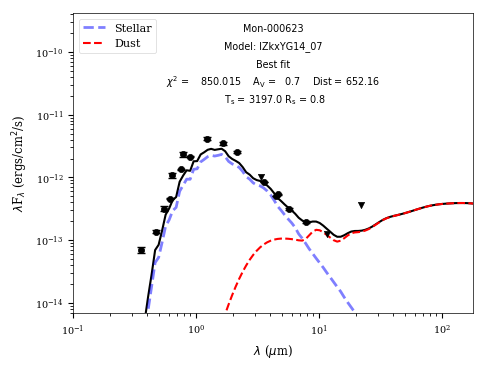}
\includegraphics[scale=0.29]{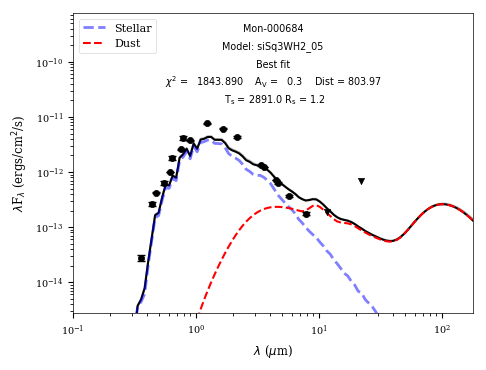}
\includegraphics[scale=0.29]{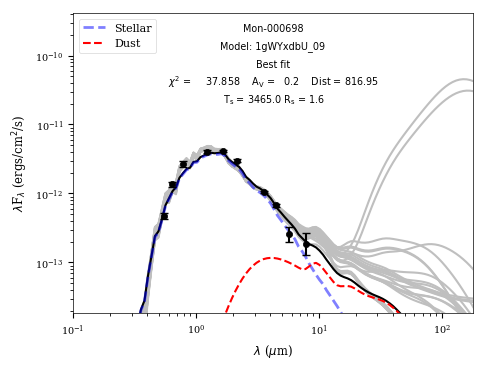}
\includegraphics[scale=0.29]{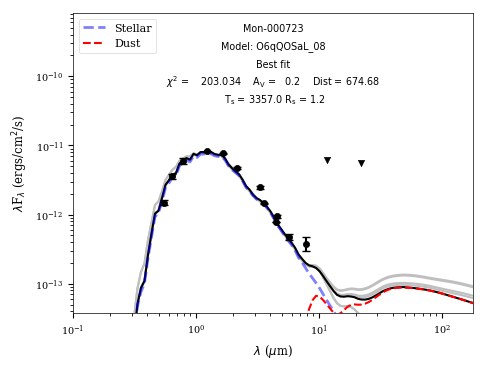}
\includegraphics[scale=0.29]{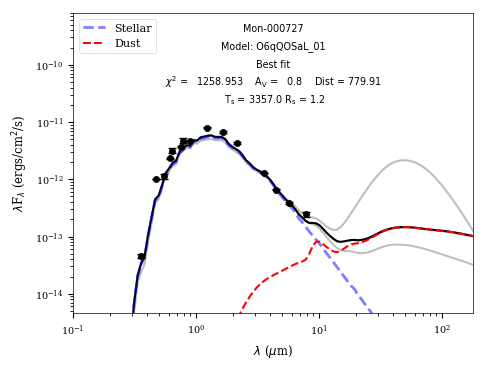}
\includegraphics[scale=0.29]{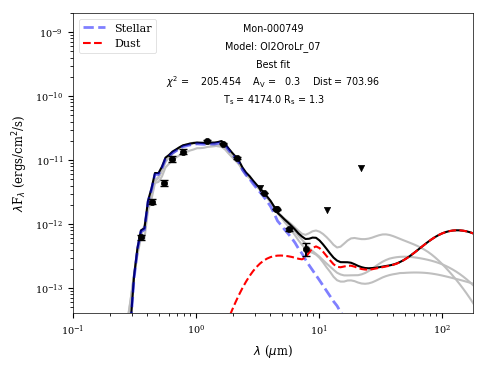}
\includegraphics[scale=0.29]{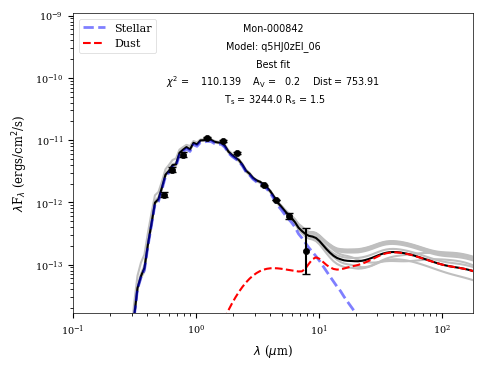}
\includegraphics[scale=0.29]{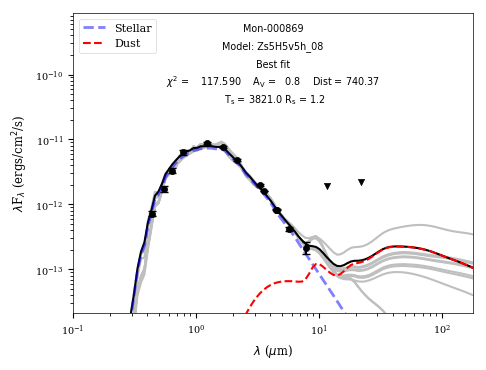}
\includegraphics[scale=0.29]{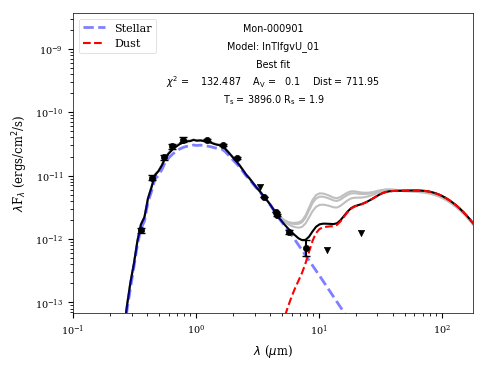}
\includegraphics[scale=0.29]{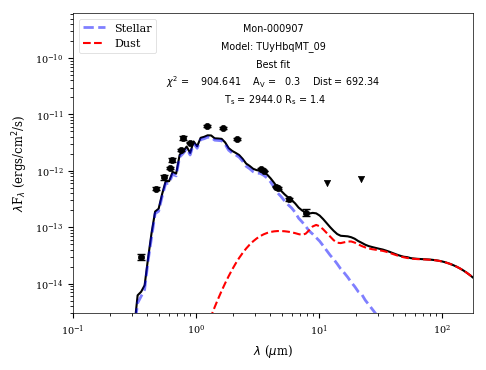}
\includegraphics[scale=0.29]{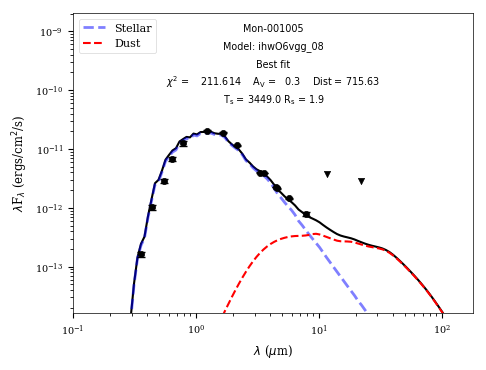}
\includegraphics[scale=0.29]{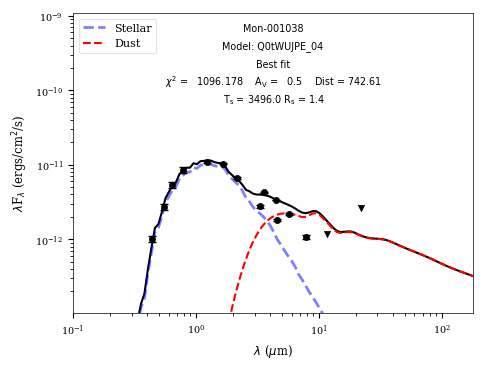}
\includegraphics[scale=0.29]{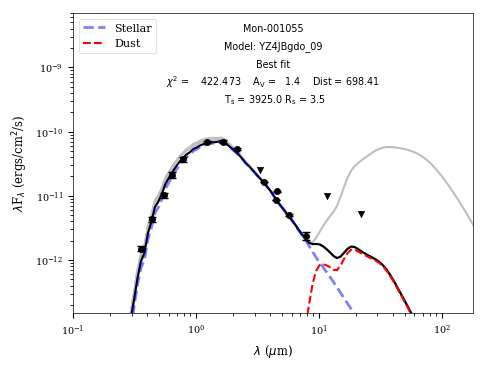}
\includegraphics[scale=0.29]{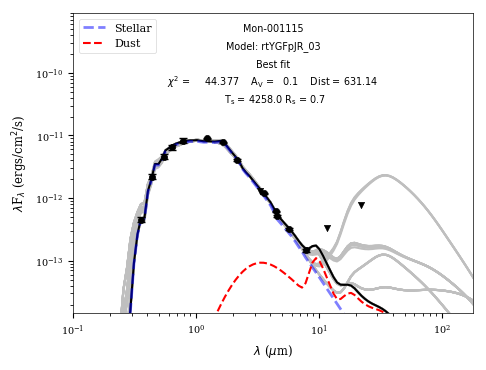}
\includegraphics[scale=0.29]{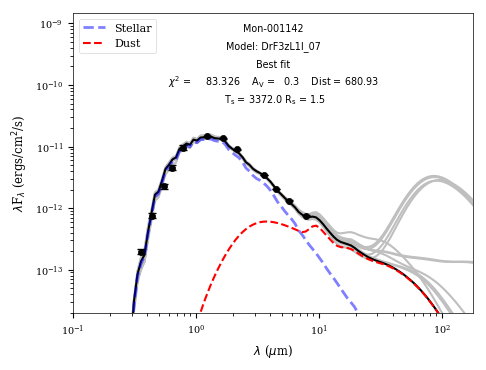}
\includegraphics[scale=0.29]{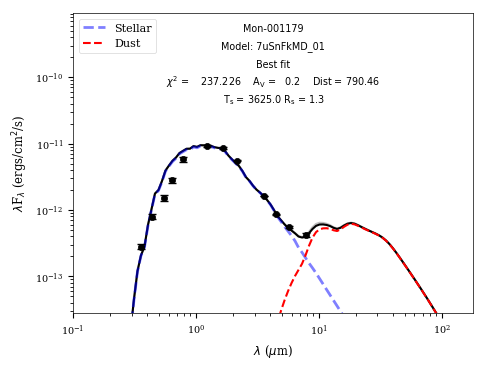}
\includegraphics[scale=0.29]{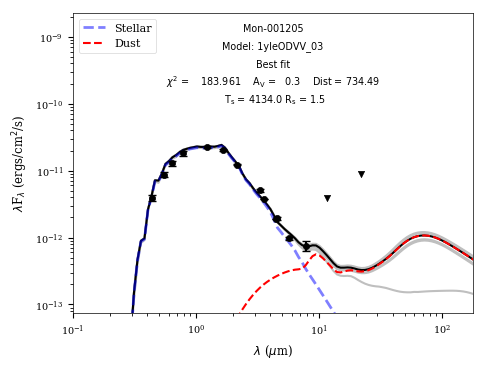}
\includegraphics[scale=0.29]{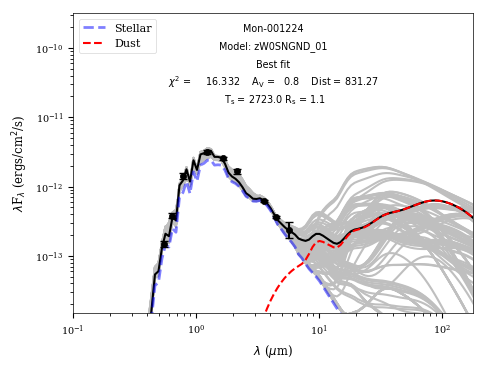}
\includegraphics[scale=0.29]{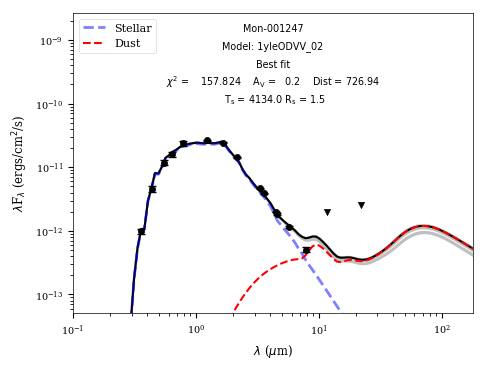}
\caption{\label{fig:Disk4}The same as Fig. \ref{fig:Disk1} but these systems 
were not observed by Spitzer at $24\,\mu\mathrm{m}$ and/or WISE 
at $22\,\mu\mathrm{m}$ or have only an upper limit of the flux at those
wavelengths (upside down triangles). }
\end{figure*}

\begin{figure*}
\includegraphics[scale=0.29]{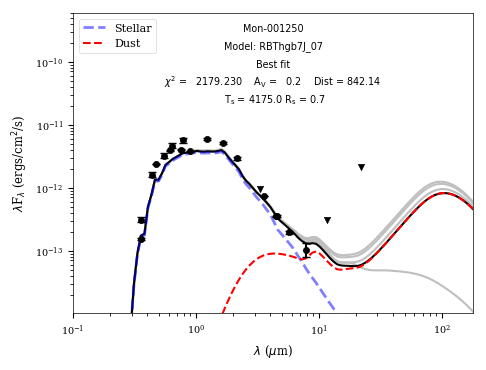}
\includegraphics[scale=0.29]{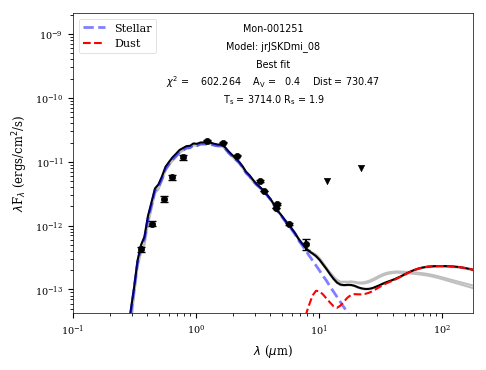}
\includegraphics[scale=0.29]{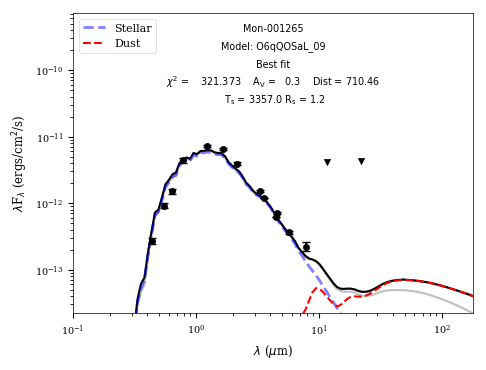}
\includegraphics[scale=0.29]{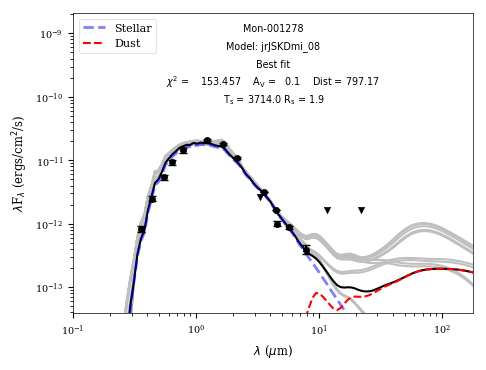}
\includegraphics[scale=0.29]{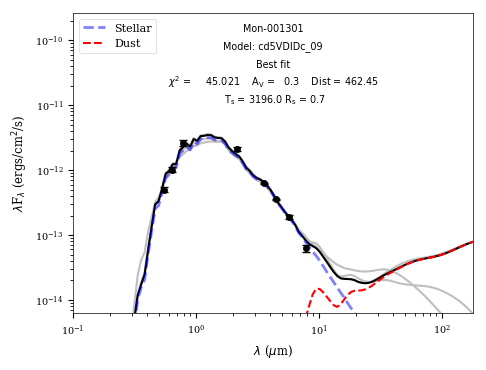}
\includegraphics[scale=0.29]{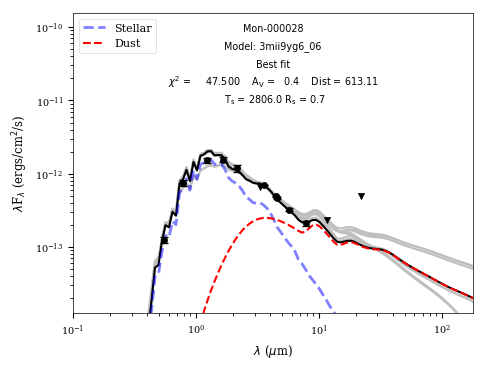}
\includegraphics[scale=0.29]{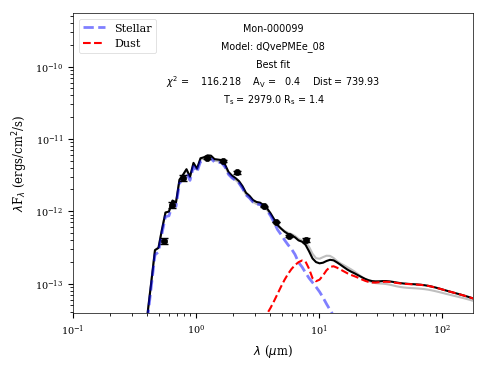}
\includegraphics[scale=0.29]{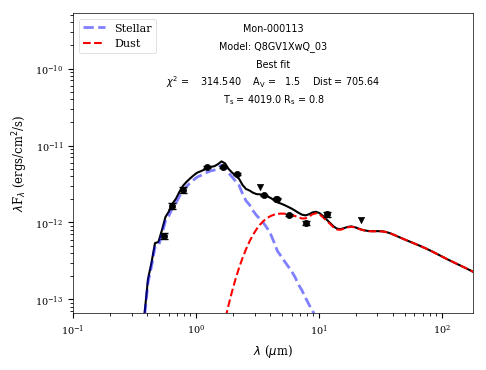}
\includegraphics[scale=0.29]{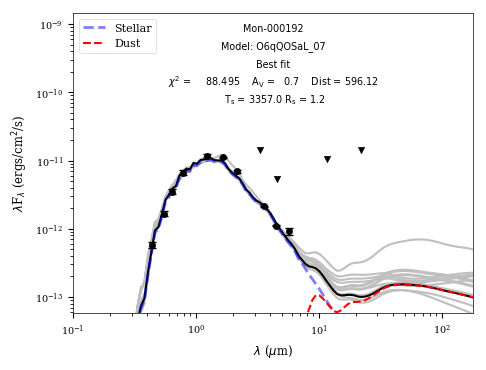}
\includegraphics[scale=0.29]{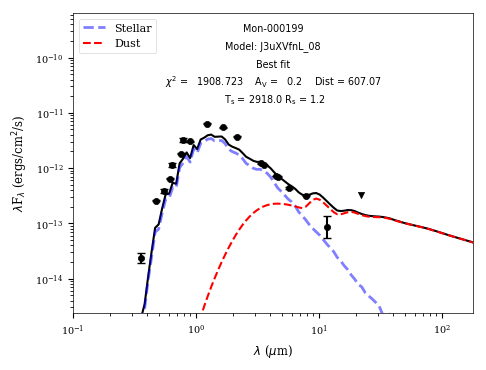}
\includegraphics[scale=0.29]{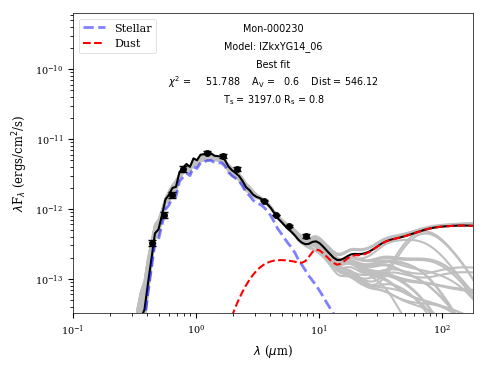}
\includegraphics[scale=0.29]{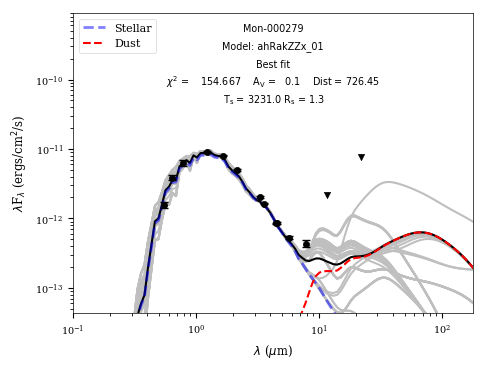}
\includegraphics[scale=0.29]{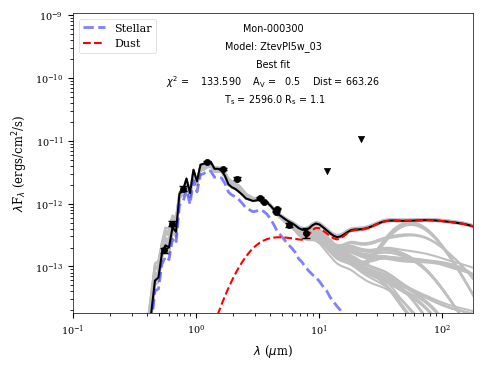}
\includegraphics[scale=0.29]{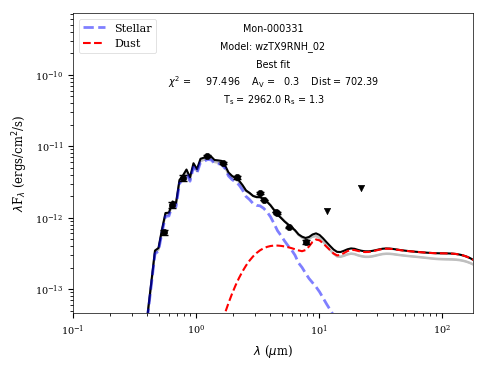}
\includegraphics[scale=0.29]{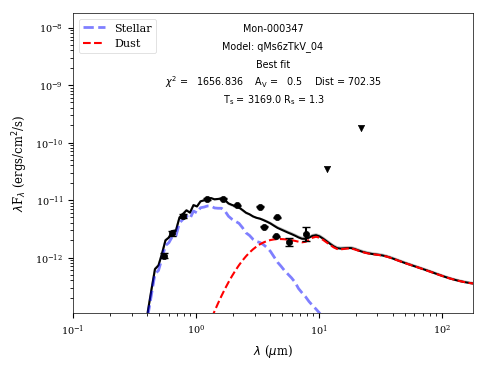}
\includegraphics[scale=0.29]{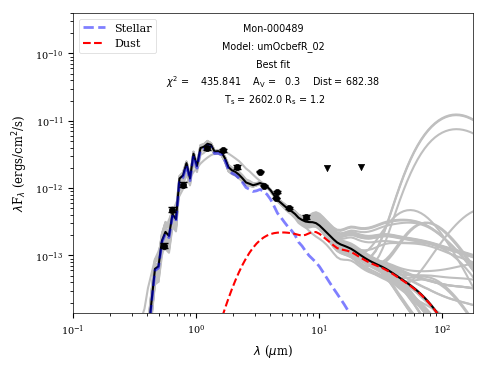}
\includegraphics[scale=0.29]{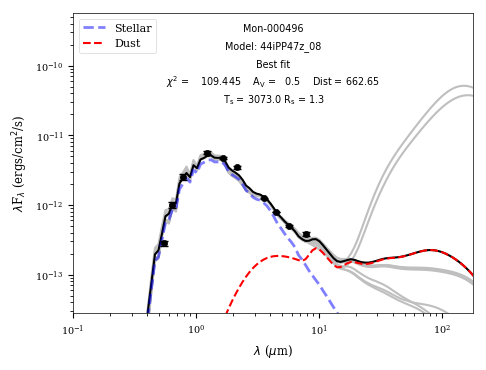}
\includegraphics[scale=0.29]{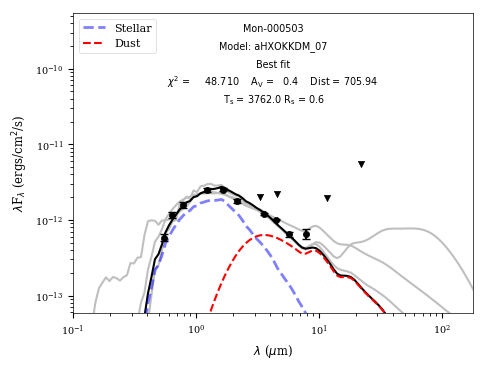}
\includegraphics[scale=0.29]{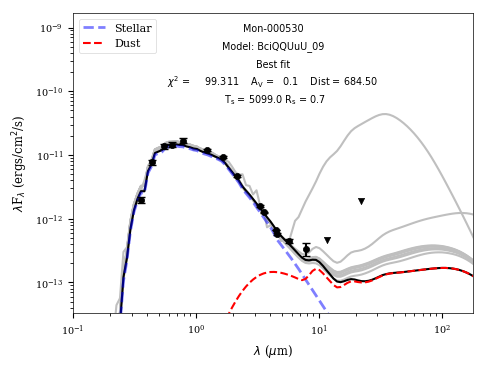}
\includegraphics[scale=0.29]{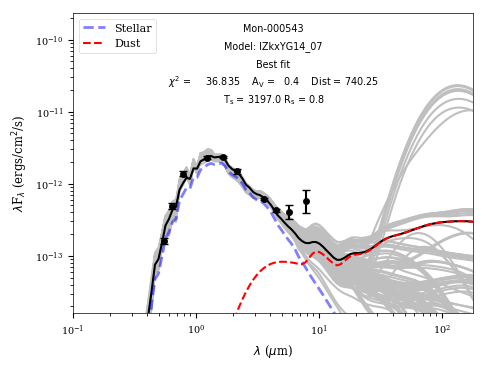}
\includegraphics[scale=0.29]{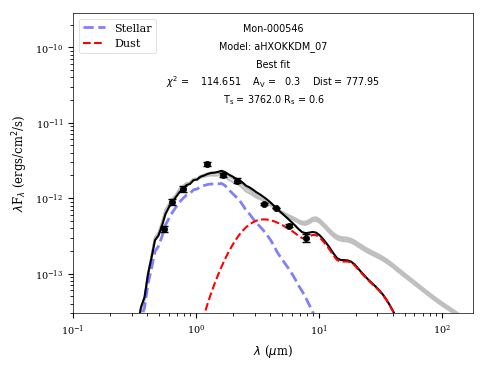}
\includegraphics[scale=0.29]{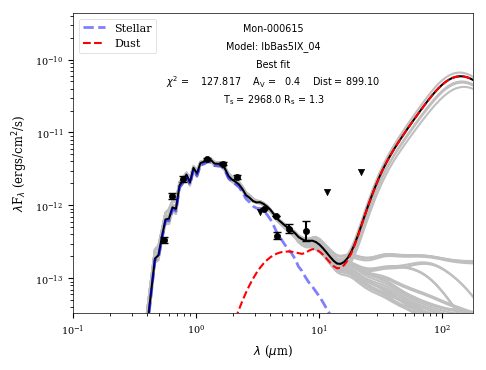}
\includegraphics[scale=0.29]{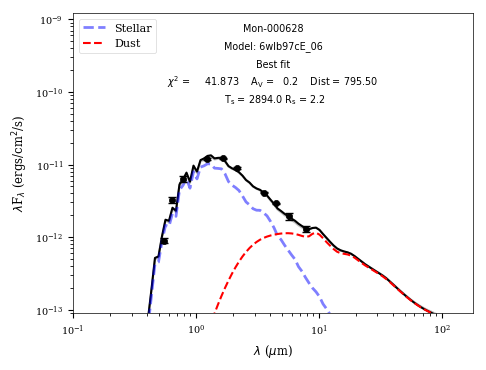}
\includegraphics[scale=0.29]{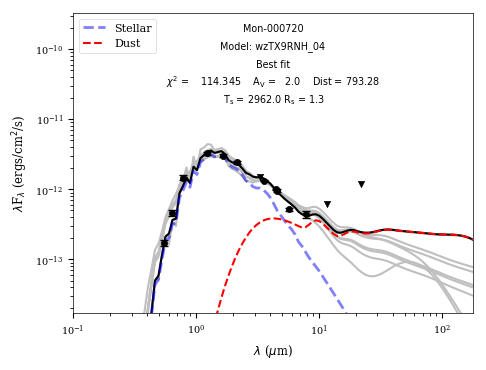}
\includegraphics[scale=0.29]{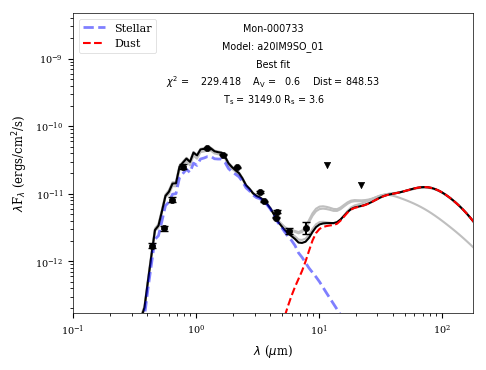}
\includegraphics[scale=0.29]{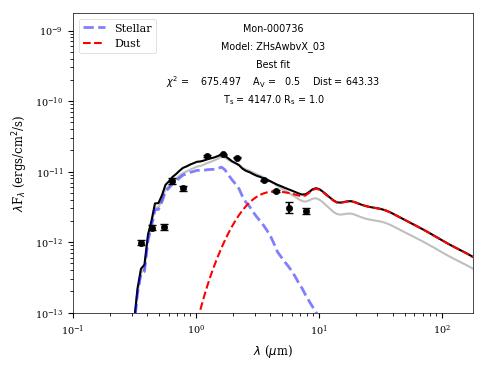}
\includegraphics[scale=0.29]{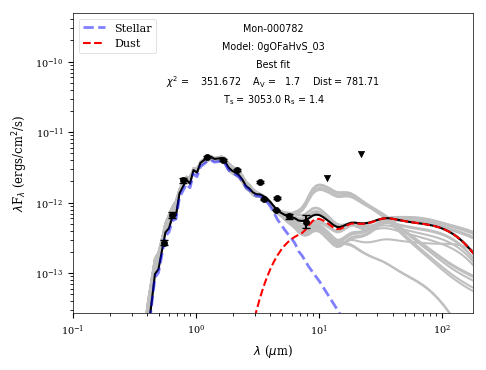}
\includegraphics[scale=0.29]{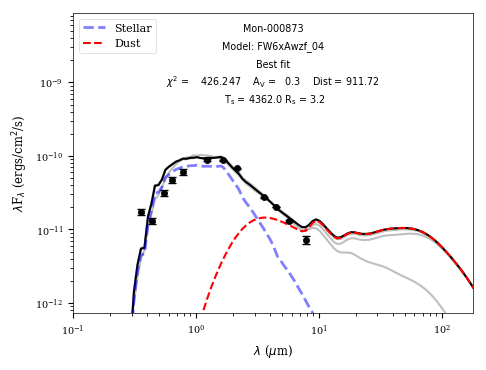}
\includegraphics[scale=0.29]{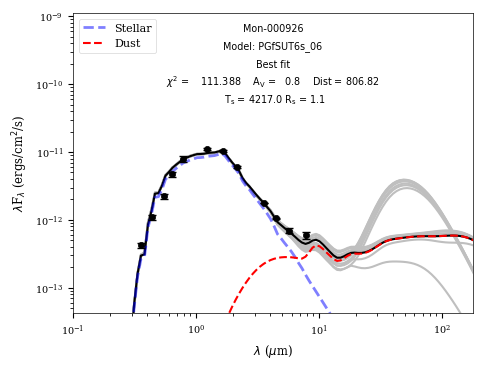}
\includegraphics[scale=0.29]{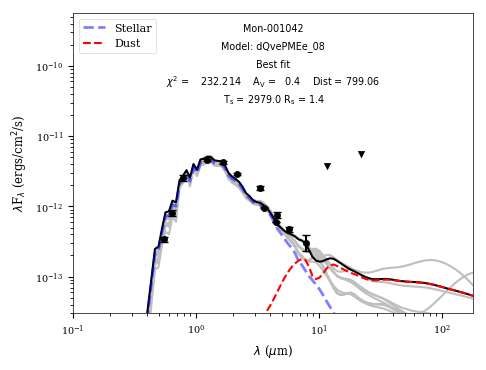}
\includegraphics[scale=0.29]{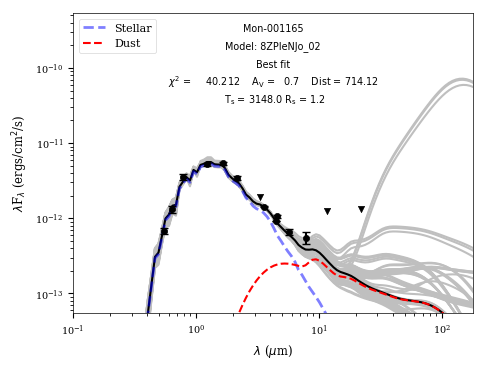}
\includegraphics[scale=0.29]{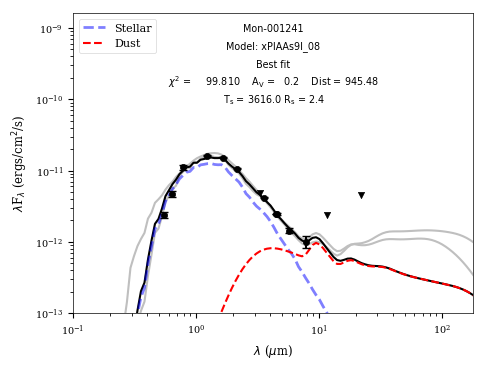}
\includegraphics[scale=0.29]{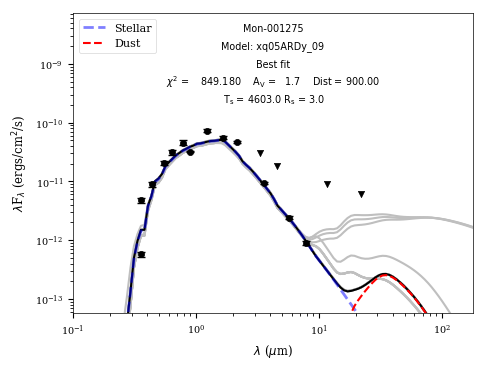}
\includegraphics[scale=0.29]{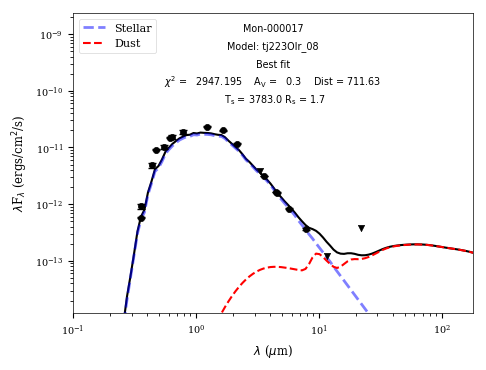}
\includegraphics[scale=0.29]{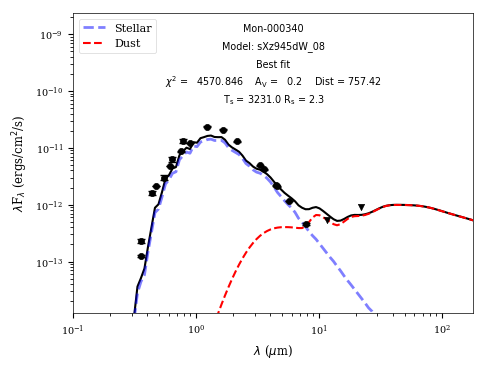}
\includegraphics[scale=0.29]{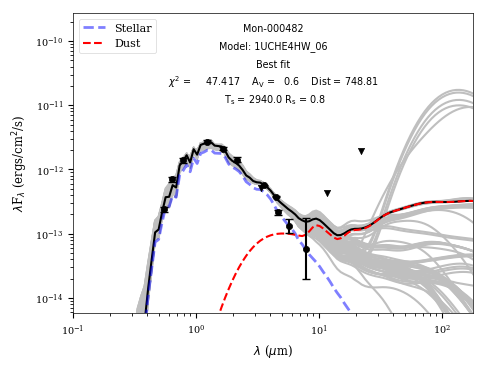}
\includegraphics[scale=0.29]{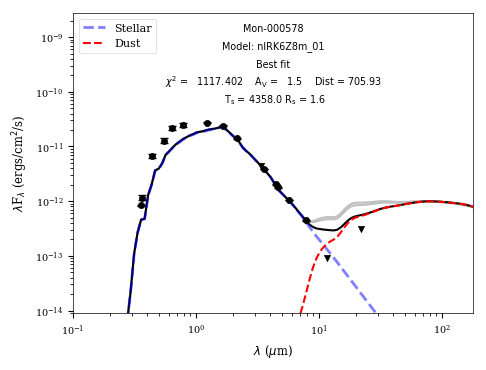}
\includegraphics[scale=0.29]{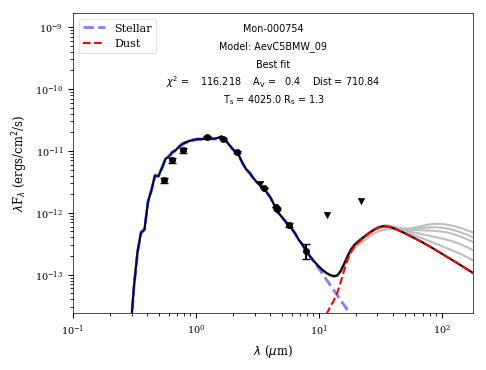}
\includegraphics[scale=0.29]{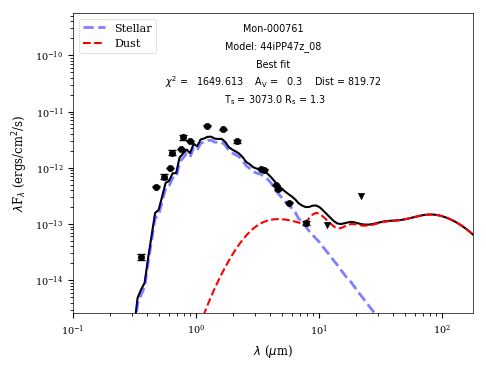}
\includegraphics[scale=0.29]{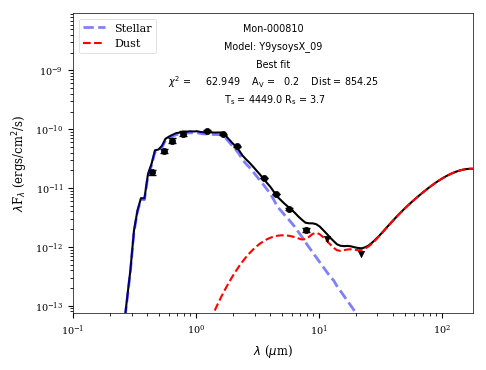}
\includegraphics[scale=0.29]{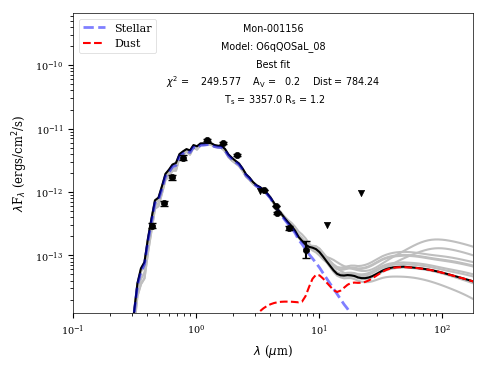}
\includegraphics[scale=0.29]{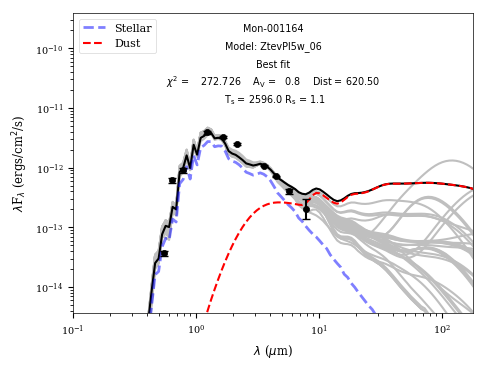}
\includegraphics[scale=0.29]{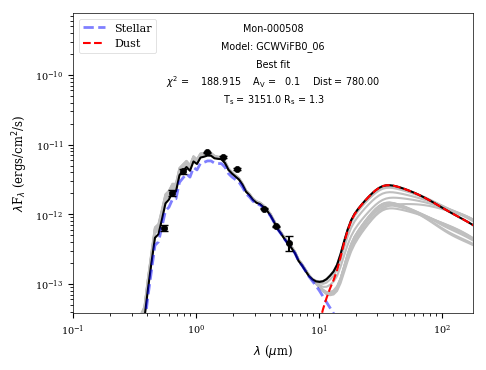}
\includegraphics[scale=0.29]{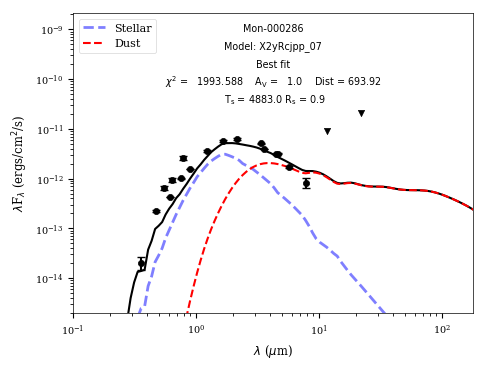}
\caption{\label{fig:Disk5} The same as Fig. \ref{fig:Disk4}.}
\end{figure*}

\clearpage

\begin{figure*}
\includegraphics[scale=0.29]{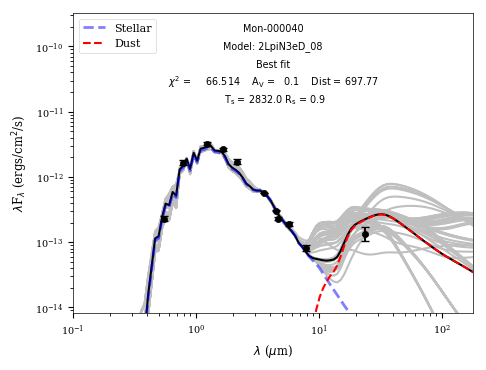}
\includegraphics[scale=0.29]{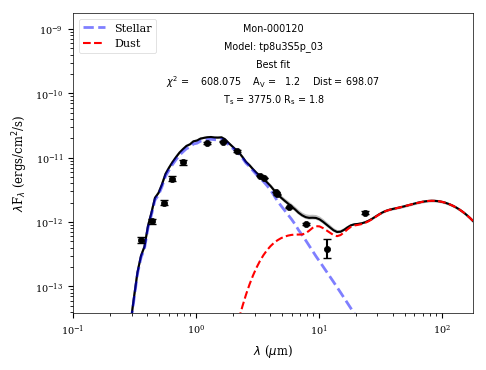}
\includegraphics[scale=0.29]{all_Mon-000122}
\includegraphics[scale=0.29]{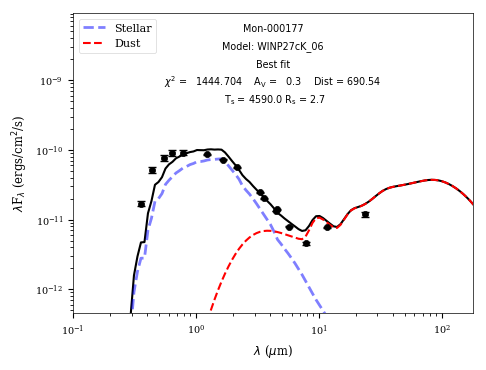}
\includegraphics[scale=0.29]{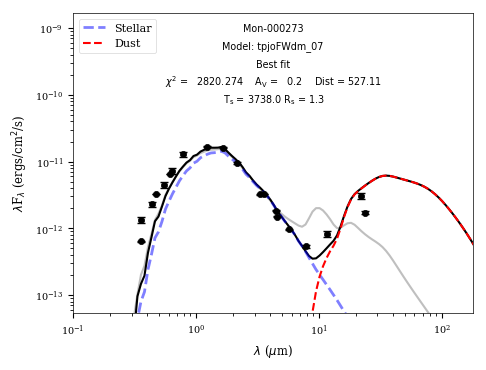}
\includegraphics[scale=0.29]{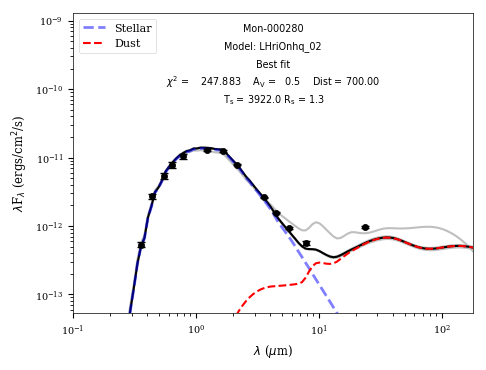}
\includegraphics[scale=0.29]{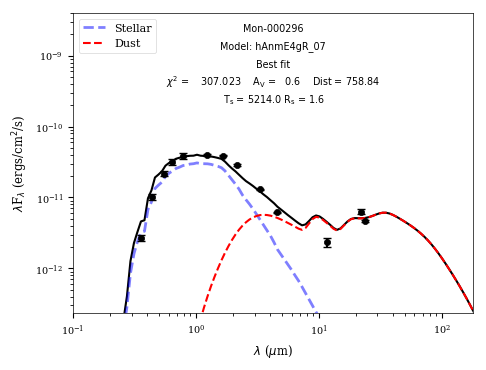}
\includegraphics[scale=0.29]{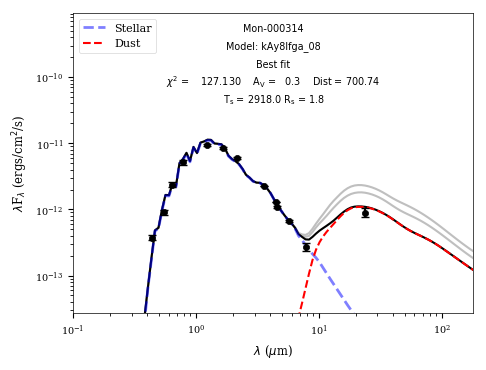}
\includegraphics[scale=0.29]{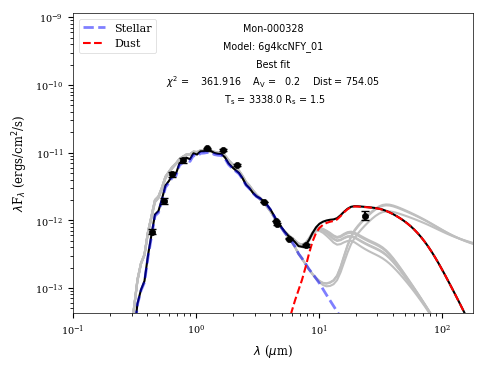}
\includegraphics[scale=0.29]{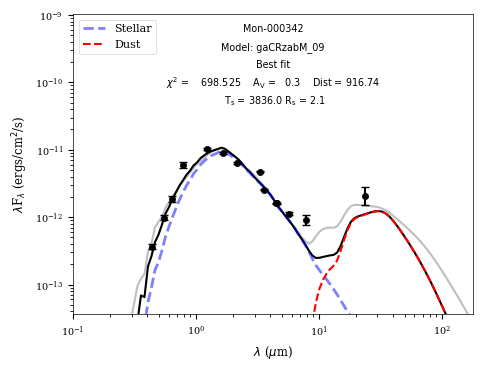}
\includegraphics[scale=0.29]{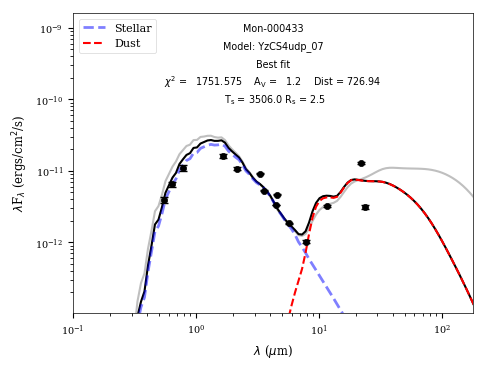}
\includegraphics[scale=0.29]{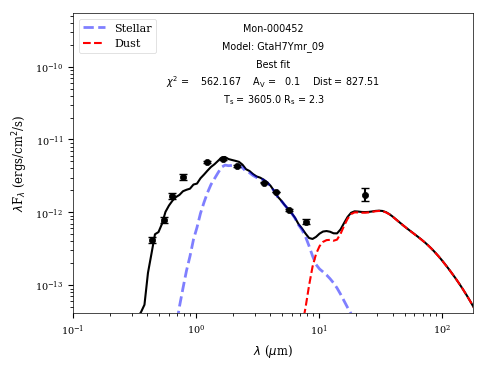}
\includegraphics[scale=0.29]{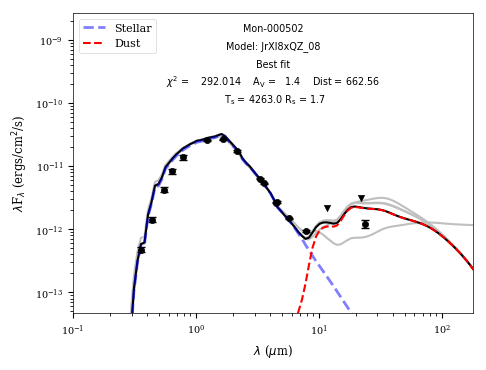}
\includegraphics[scale=0.29]{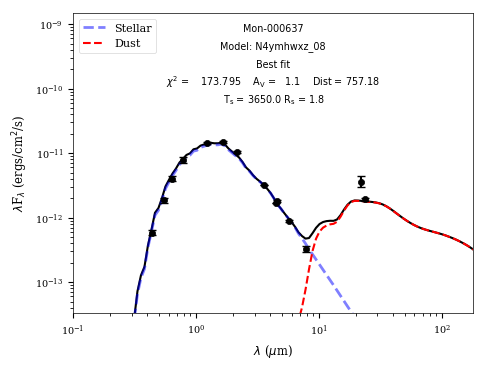}
\includegraphics[scale=0.29]{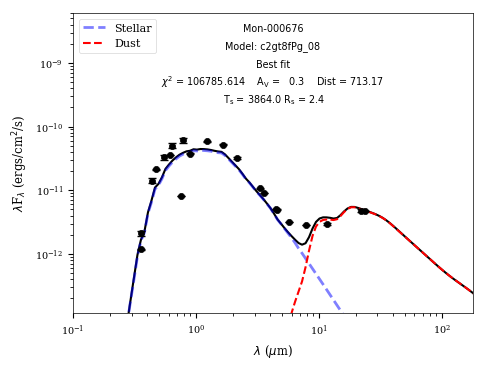}
\includegraphics[scale=0.29]{all_Mon-000771}
\includegraphics[scale=0.29]{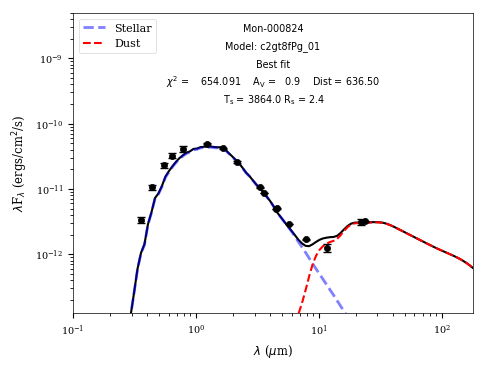}
\includegraphics[scale=0.29]{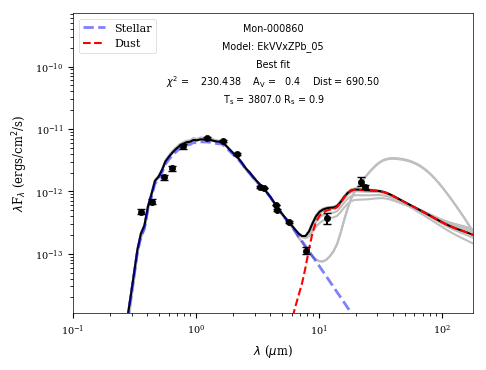}
\includegraphics[scale=0.29]{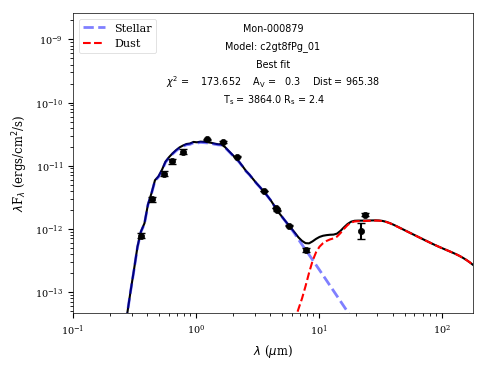}
\includegraphics[scale=0.29]{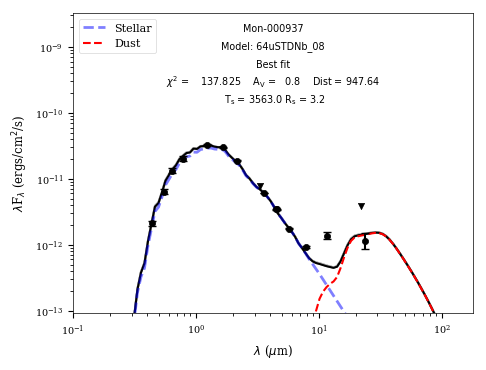}
\includegraphics[scale=0.29]{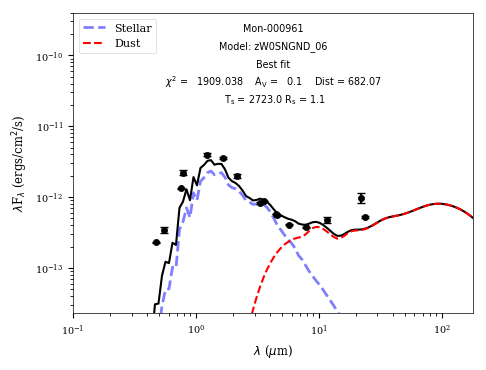}
\includegraphics[scale=0.29]{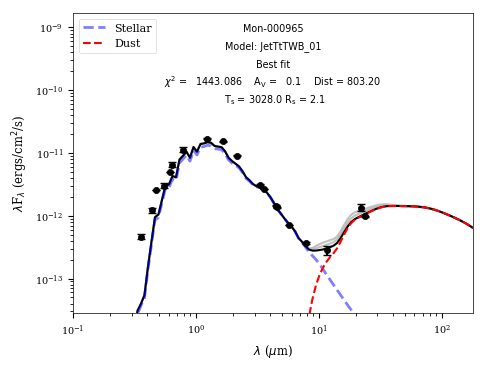}
\includegraphics[scale=0.29]{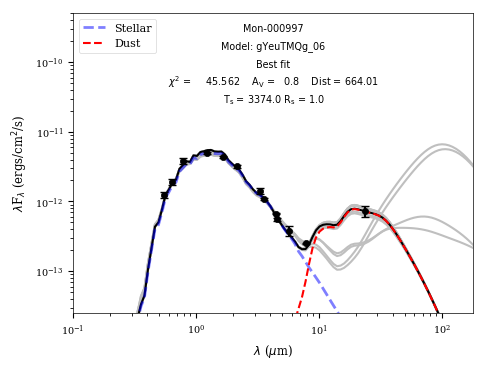}
\includegraphics[scale=0.29]{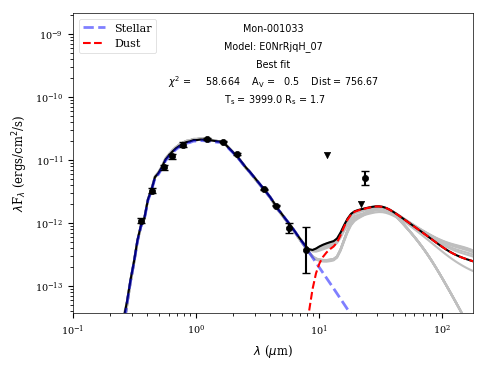}
\includegraphics[scale=0.29]{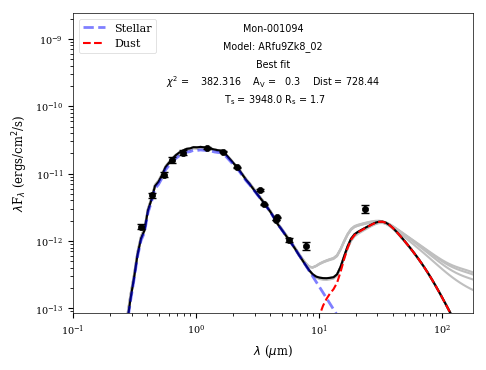}
\includegraphics[scale=0.29]{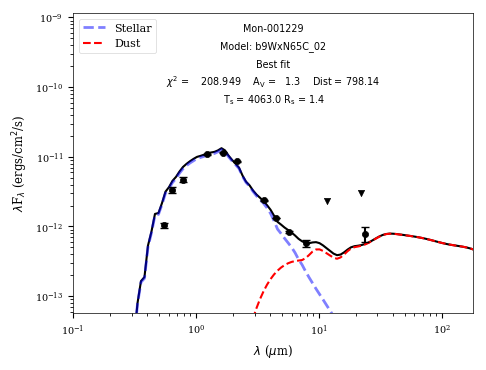}
\includegraphics[scale=0.29]{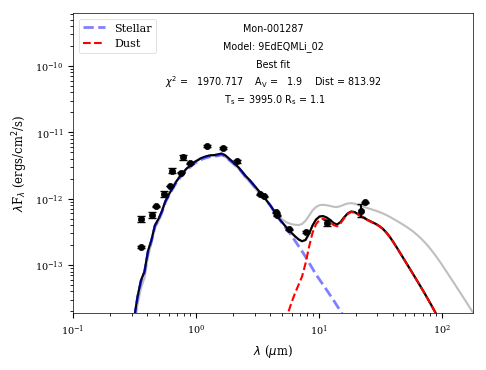}
\includegraphics[scale=0.29]{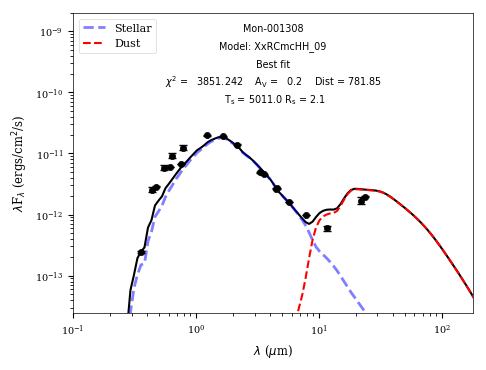}
\caption{\label{fig:TD} The same as Fig. \ref{fig:Disk1} but for systems classified as transition disk candidates.}
\end{figure*}


\begin{figure*}
\includegraphics[scale=0.29]{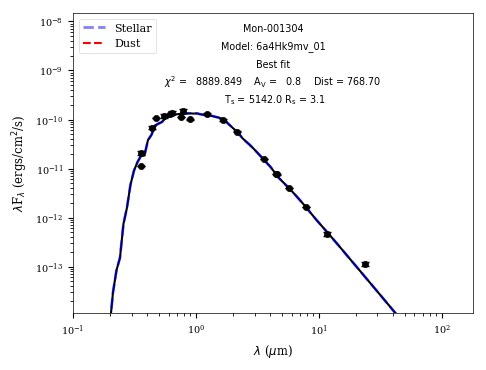}
\includegraphics[scale=0.29]{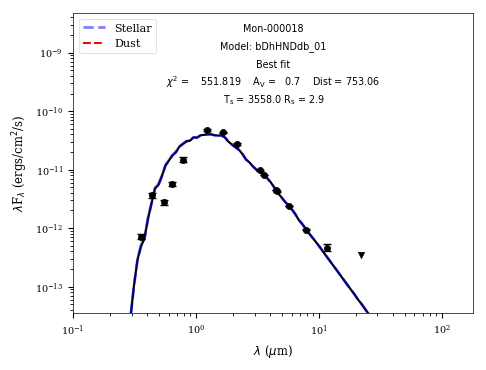}
\includegraphics[scale=0.29]{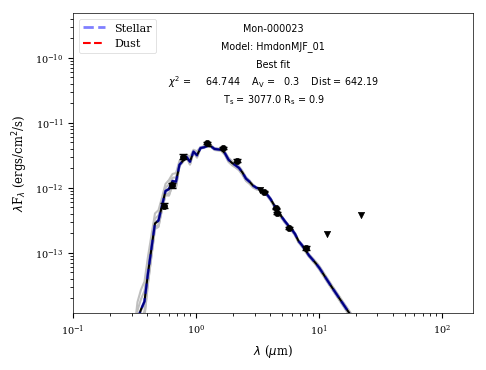}
\includegraphics[scale=0.29]{all_Mon-000029}
\includegraphics[scale=0.29]{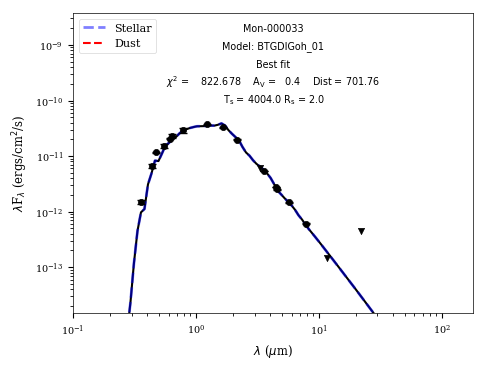}
\includegraphics[scale=0.29]{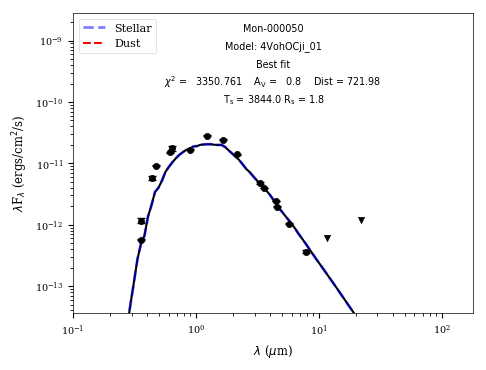}
\includegraphics[scale=0.29]{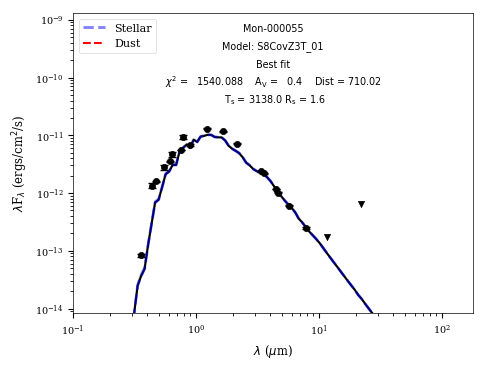}
\includegraphics[scale=0.29]{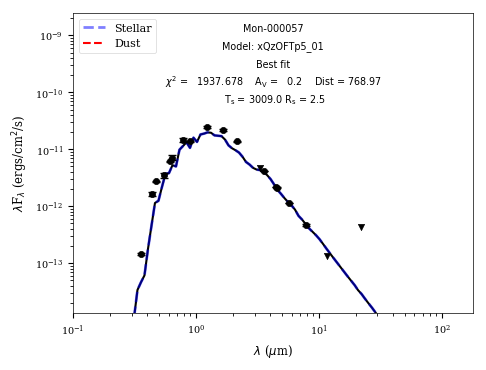}
\includegraphics[scale=0.29]{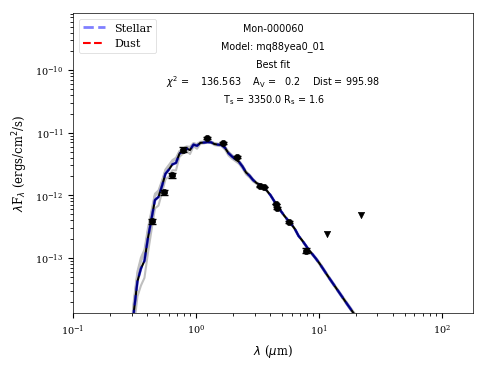}
\includegraphics[scale=0.29]{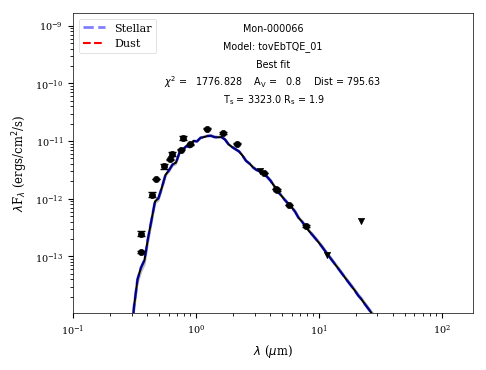}
\includegraphics[scale=0.29]{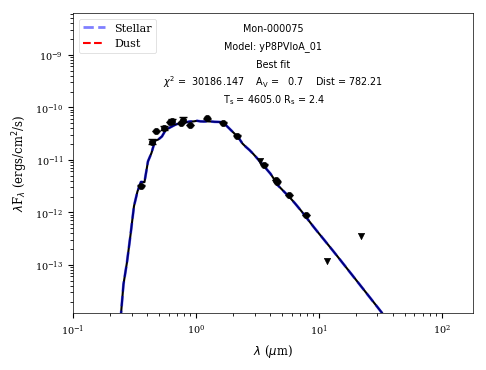}
\includegraphics[scale=0.29]{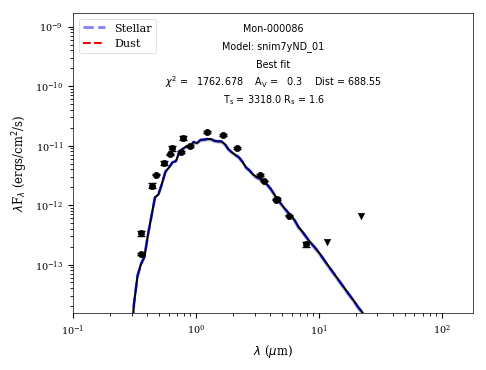}
\includegraphics[scale=0.29]{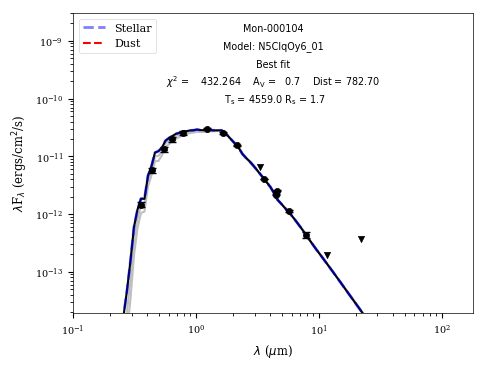}
\includegraphics[scale=0.29]{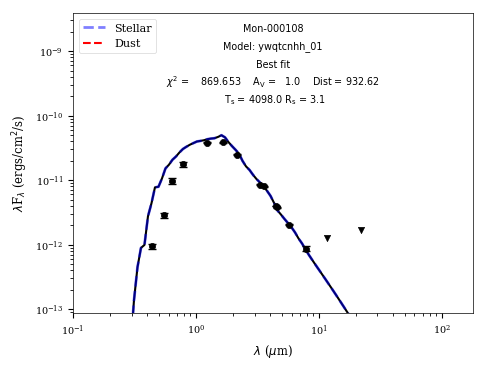}
\includegraphics[scale=0.29]{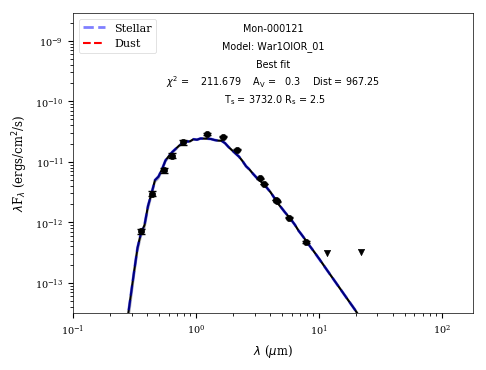}
\includegraphics[scale=0.29]{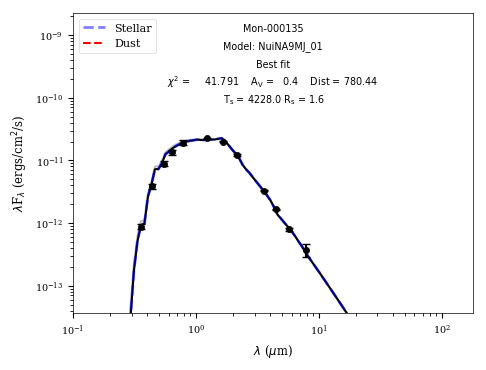}
\includegraphics[scale=0.29]{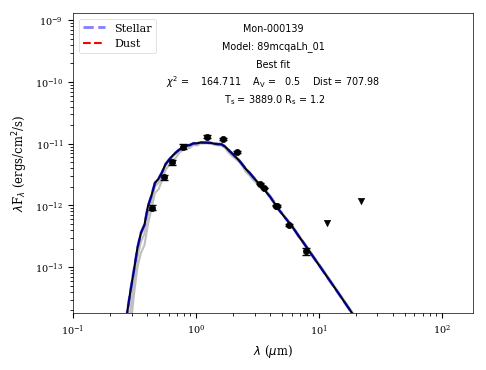}
\includegraphics[scale=0.29]{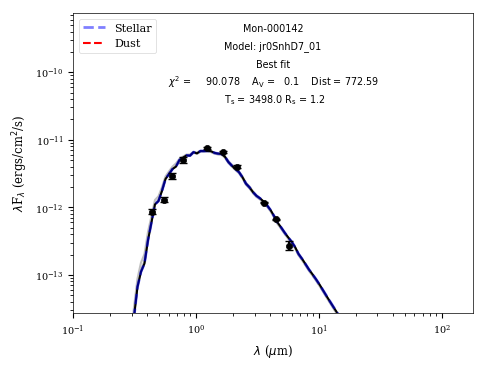}
\includegraphics[scale=0.29]{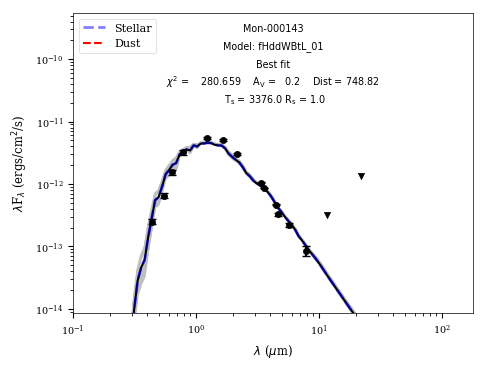}
\includegraphics[scale=0.29]{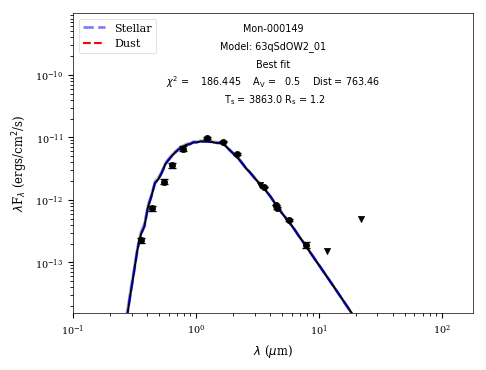}
\includegraphics[scale=0.29]{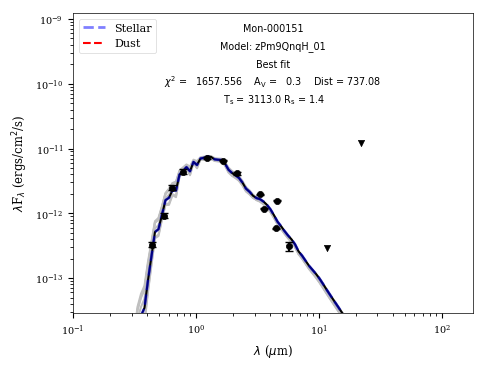}
\includegraphics[scale=0.29]{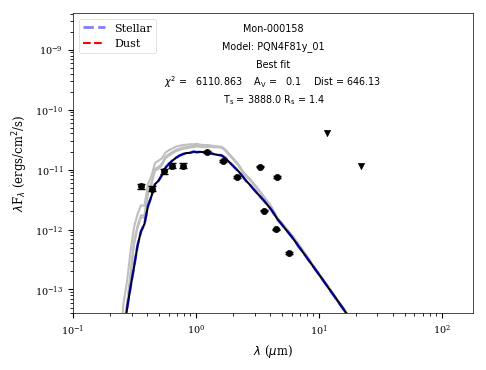}
\includegraphics[scale=0.29]{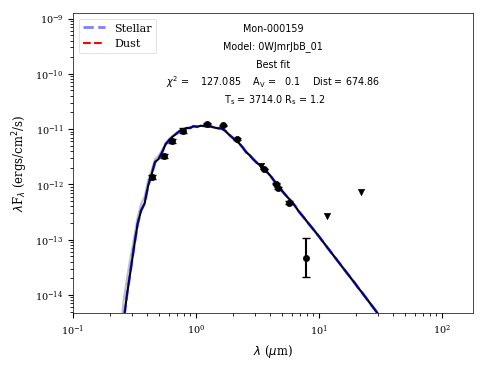}
\includegraphics[scale=0.29]{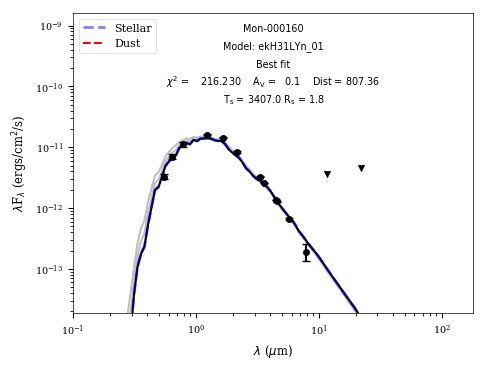}
\includegraphics[scale=0.29]{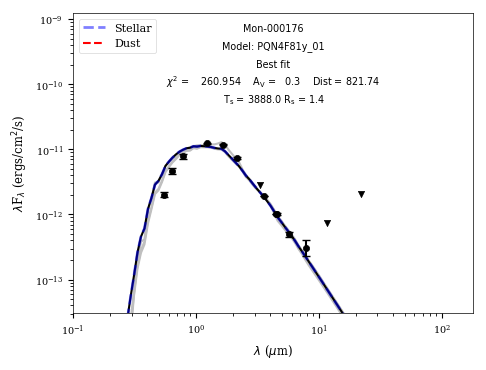}
\includegraphics[scale=0.29]{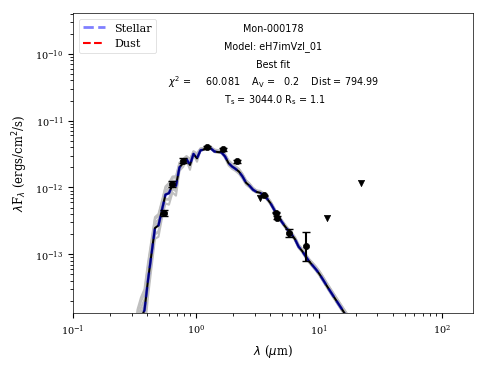}
\includegraphics[scale=0.29]{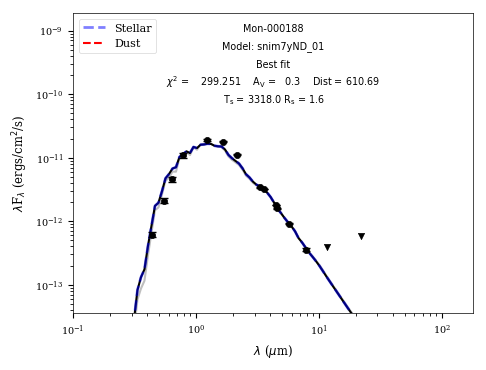}
\includegraphics[scale=0.29]{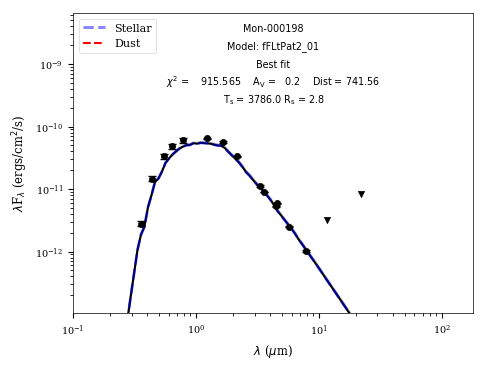}
\includegraphics[scale=0.29]{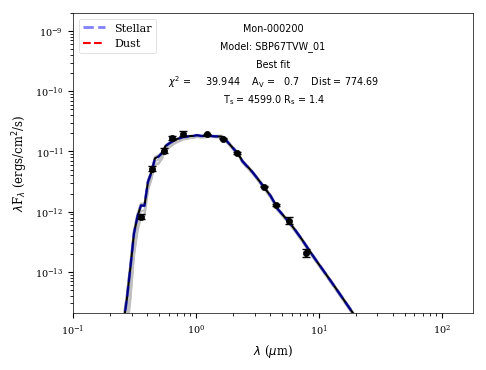}
\includegraphics[scale=0.29]{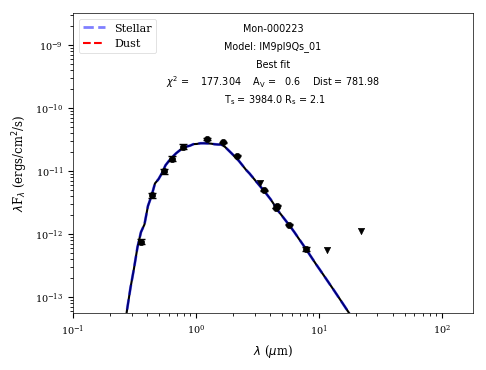}
\includegraphics[scale=0.29]{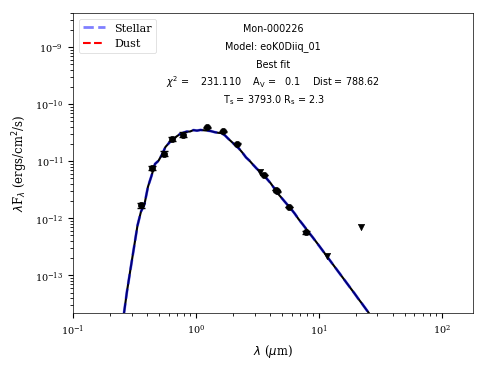}
\includegraphics[scale=0.29]{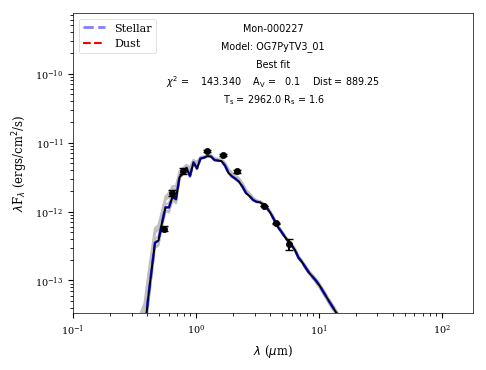}
\includegraphics[scale=0.29]{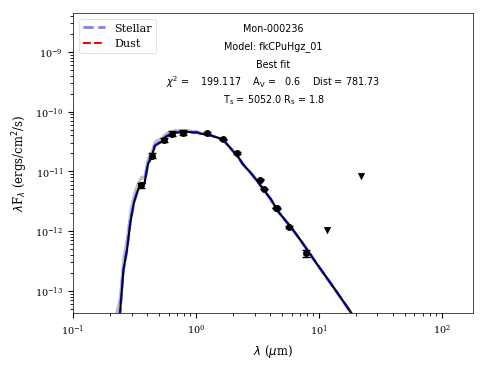}
\includegraphics[scale=0.29]{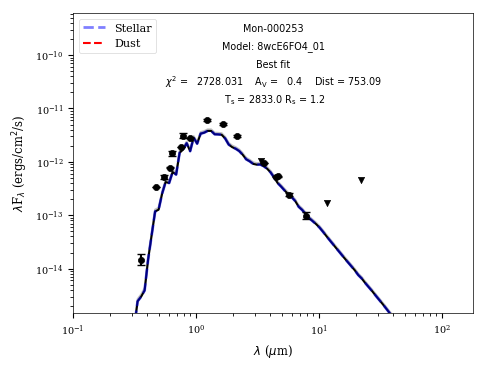}
\includegraphics[scale=0.29]{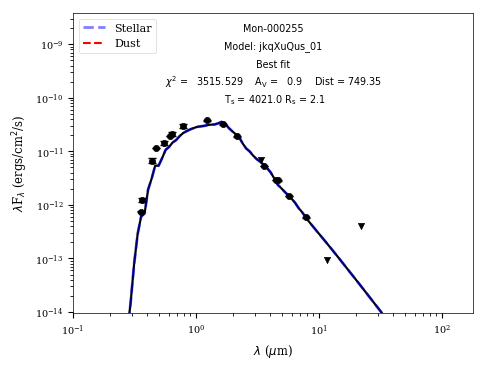}
\includegraphics[scale=0.29]{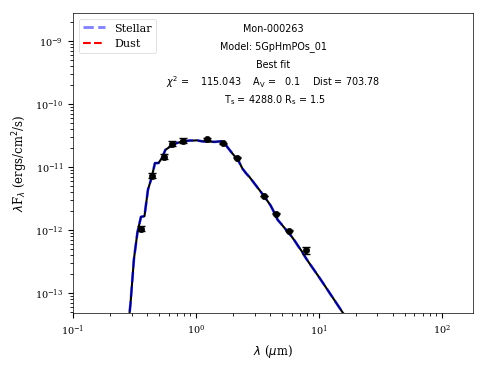}
\includegraphics[scale=0.29]{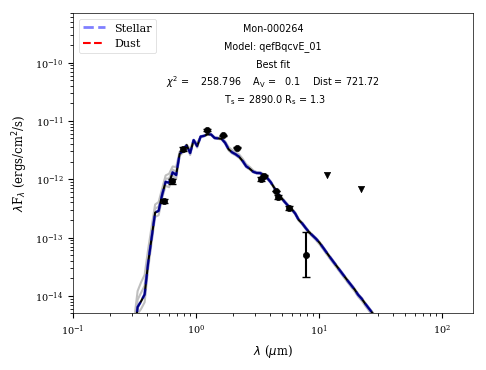}
\includegraphics[scale=0.29]{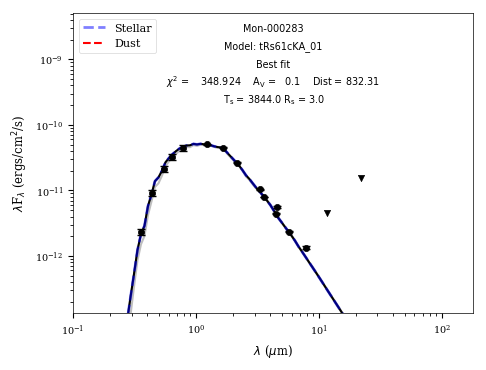}
\includegraphics[scale=0.29]{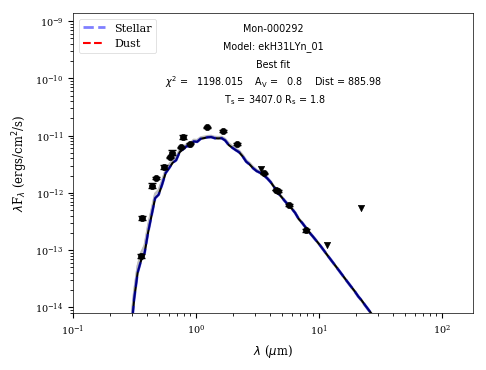}
\includegraphics[scale=0.29]{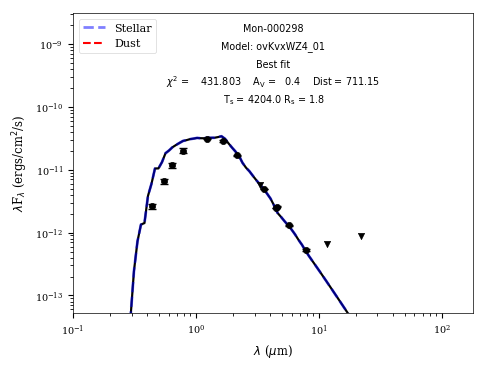}
\includegraphics[scale=0.29]{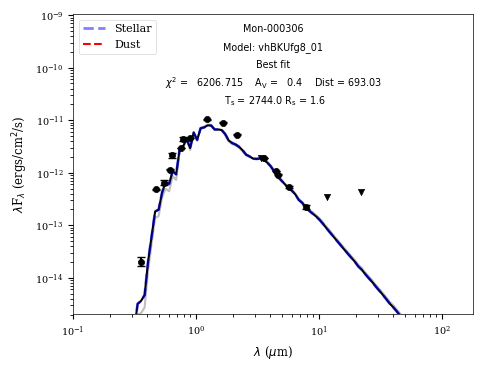}
\includegraphics[scale=0.29]{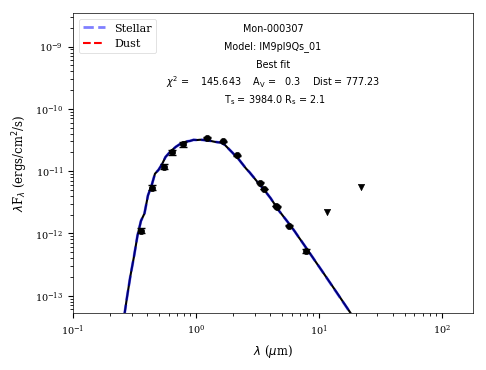}
\includegraphics[scale=0.29]{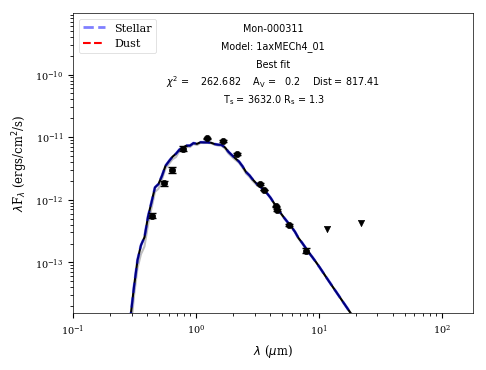}
\includegraphics[scale=0.29]{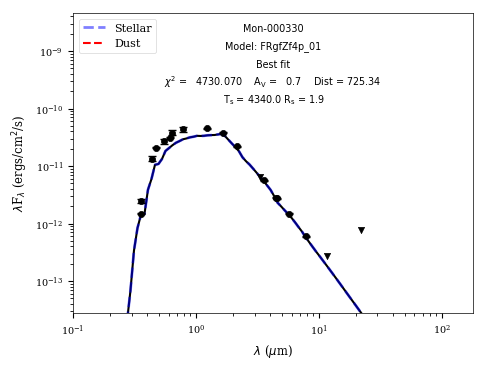}
\includegraphics[scale=0.29]{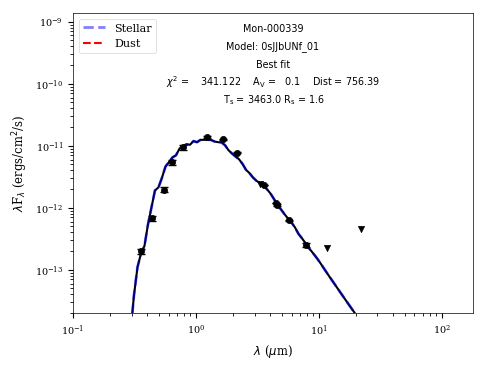}
\caption{\label{fig:Diskless} The same as Fig. \ref{fig:Disk1} but for stars classified as diskless.}
\end{figure*}

\begin{figure*}
\includegraphics[scale=0.29]{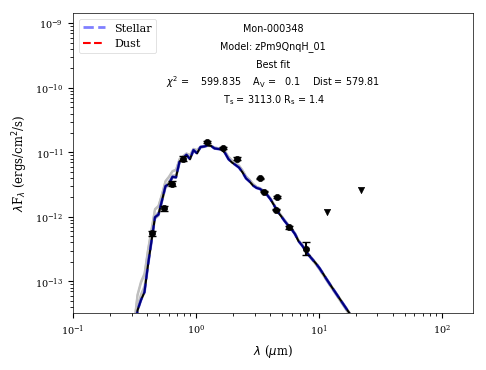}
\includegraphics[scale=0.29]{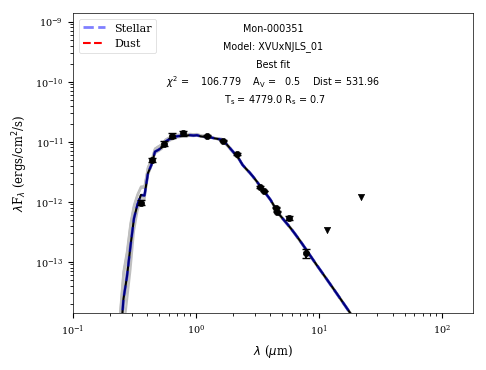}
\includegraphics[scale=0.29]{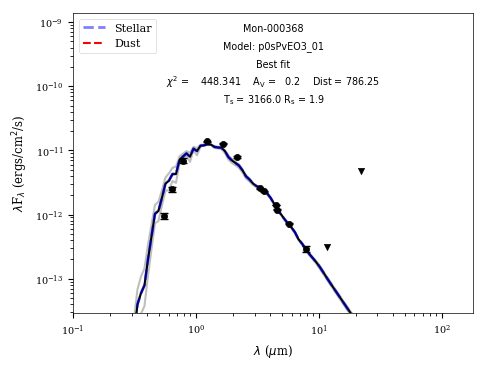}
\includegraphics[scale=0.29]{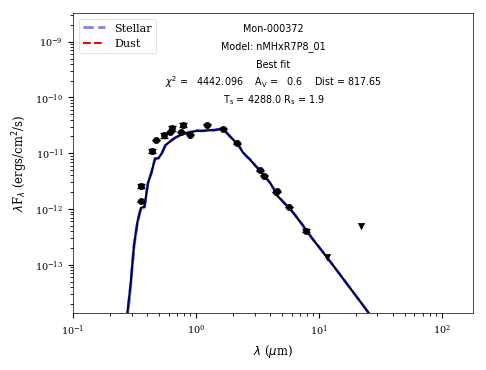}
\includegraphics[scale=0.29]{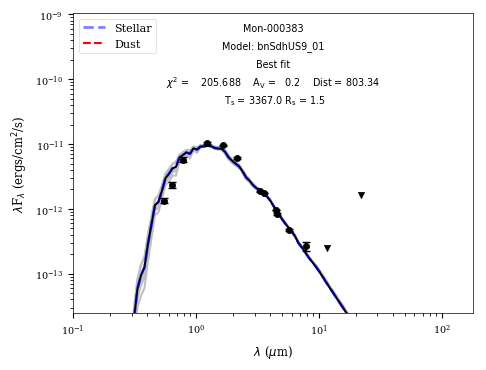}
\includegraphics[scale=0.29]{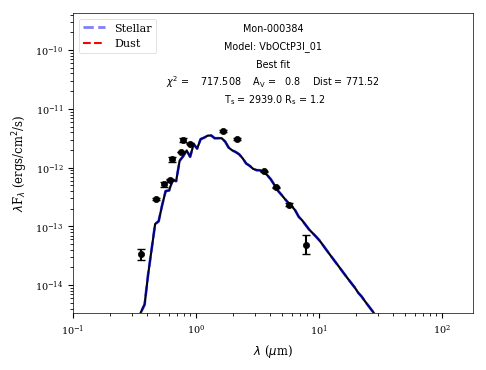}
\includegraphics[scale=0.29]{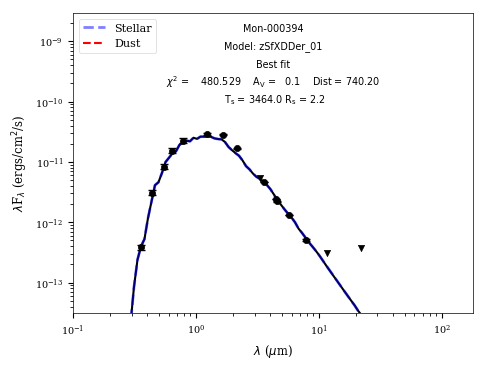}
\includegraphics[scale=0.29]{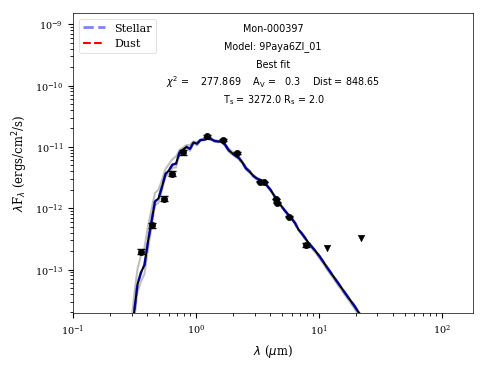}
\includegraphics[scale=0.29]{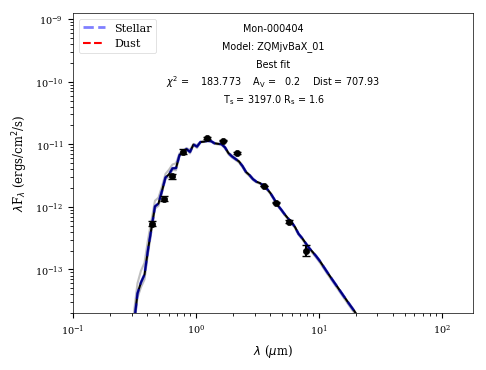}
\includegraphics[scale=0.29]{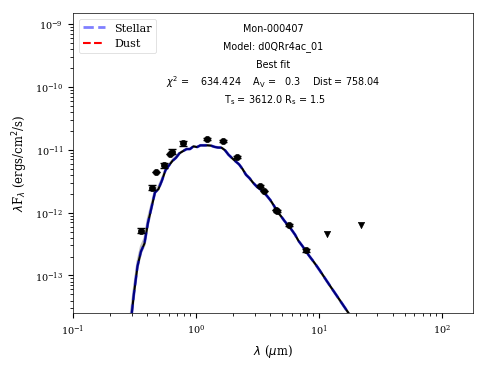}
\includegraphics[scale=0.29]{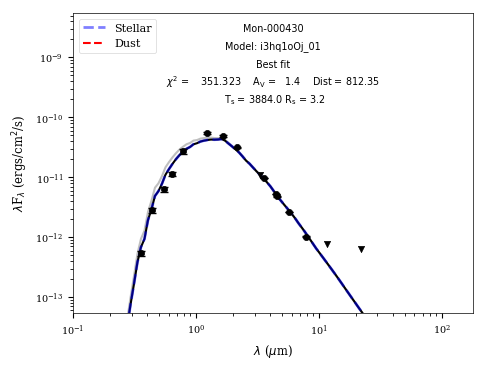}
\includegraphics[scale=0.29]{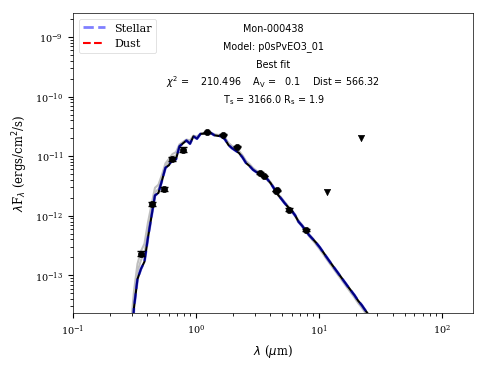}
\includegraphics[scale=0.29]{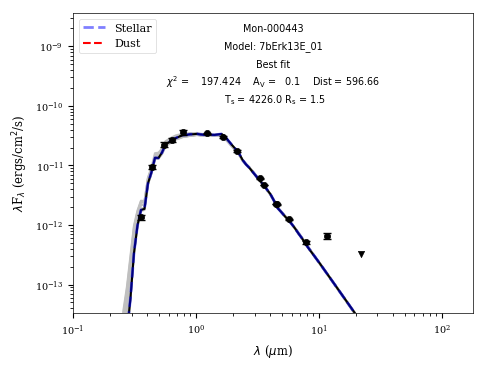}
\includegraphics[scale=0.29]{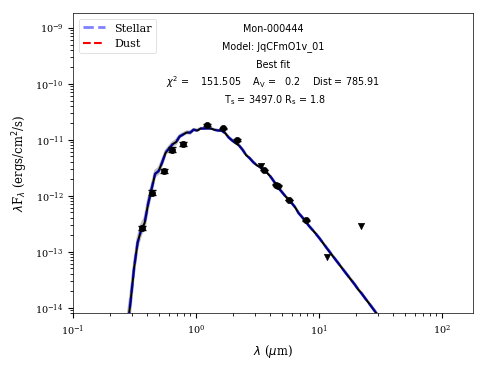}
\includegraphics[scale=0.29]{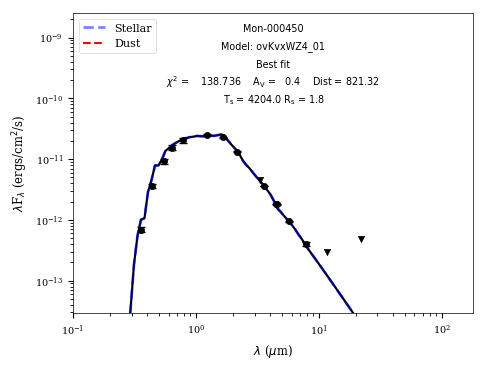}
\includegraphics[scale=0.29]{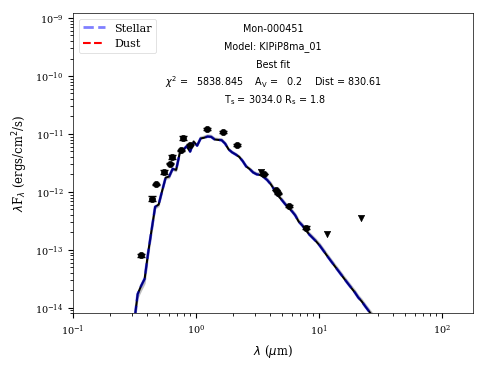}
\includegraphics[scale=0.29]{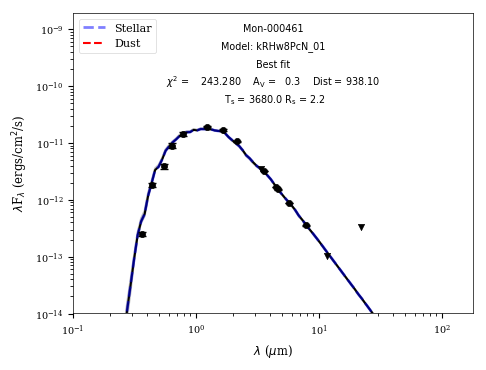}
\includegraphics[scale=0.29]{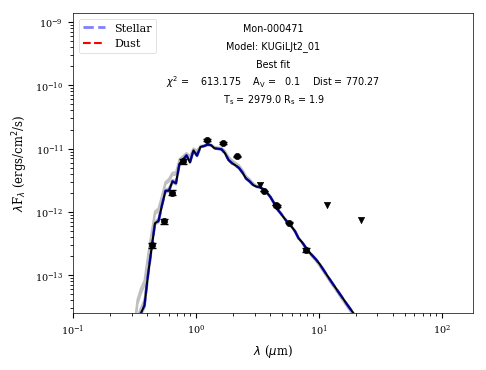}
\includegraphics[scale=0.29]{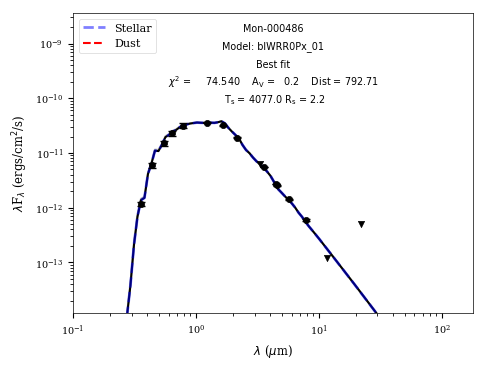}
\includegraphics[scale=0.29]{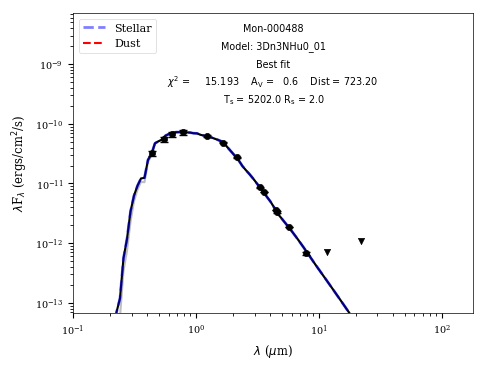}
\includegraphics[scale=0.29]{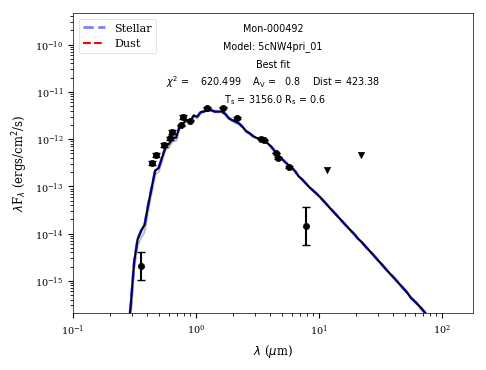}
\includegraphics[scale=0.29]{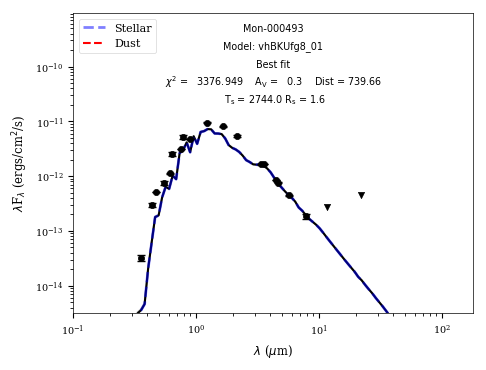}
\includegraphics[scale=0.29]{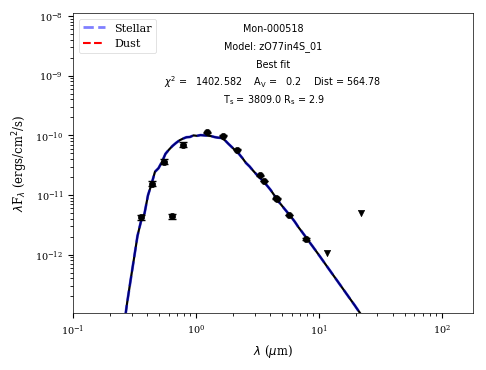}
\includegraphics[scale=0.29]{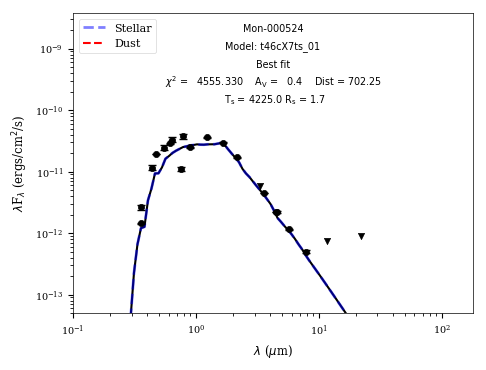}
\includegraphics[scale=0.29]{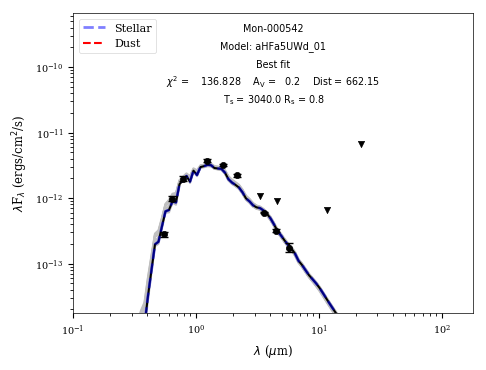}
\includegraphics[scale=0.29]{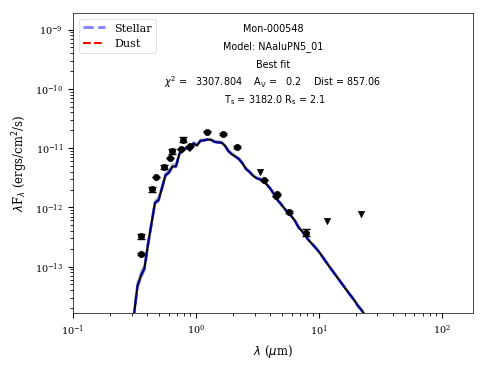}
\includegraphics[scale=0.29]{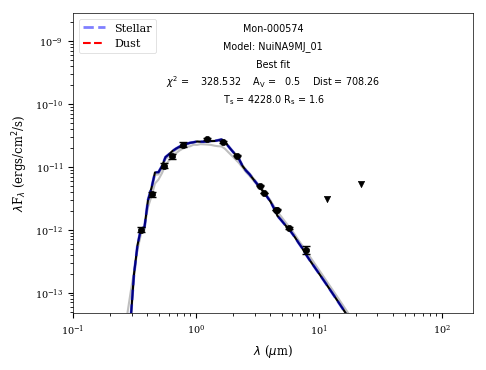}
\includegraphics[scale=0.29]{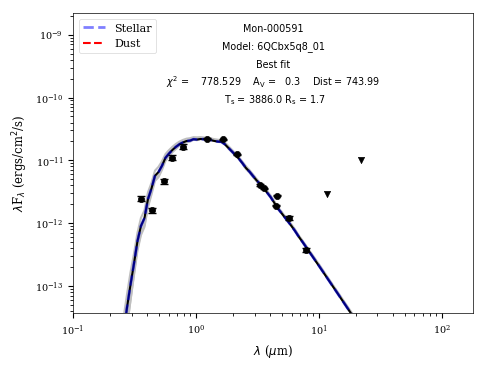}
\includegraphics[scale=0.29]{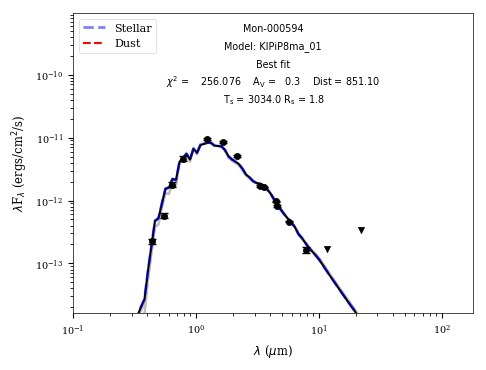}
\includegraphics[scale=0.29]{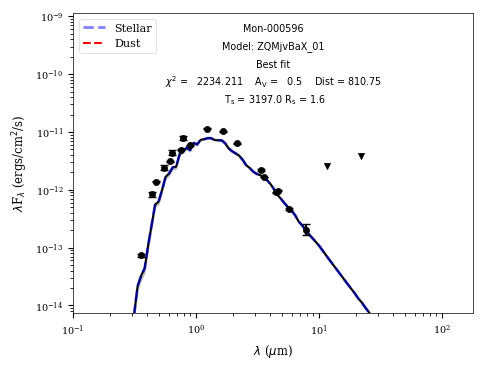}
\includegraphics[scale=0.29]{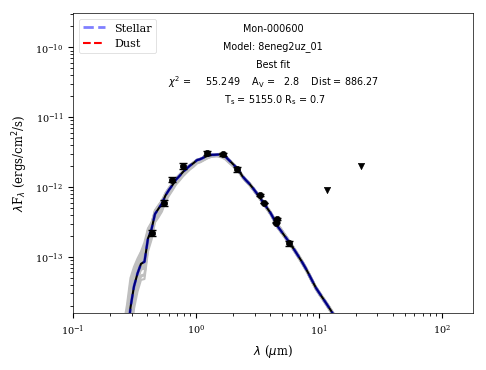}
\includegraphics[scale=0.29]{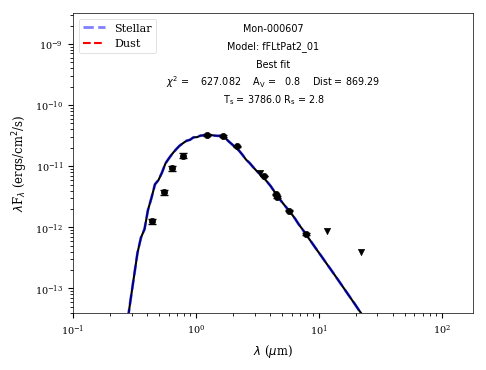}
\includegraphics[scale=0.29]{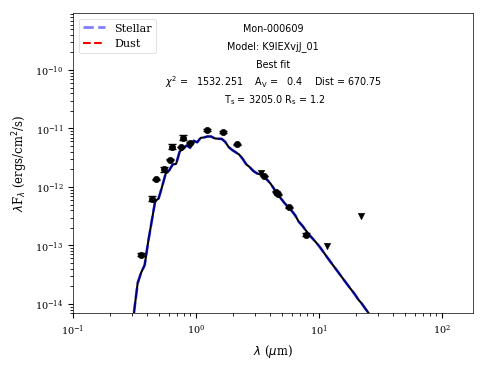}
\includegraphics[scale=0.29]{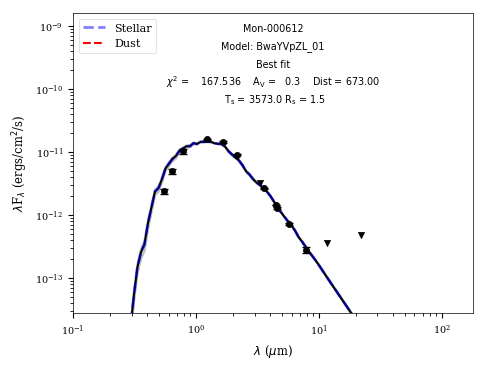}
\includegraphics[scale=0.29]{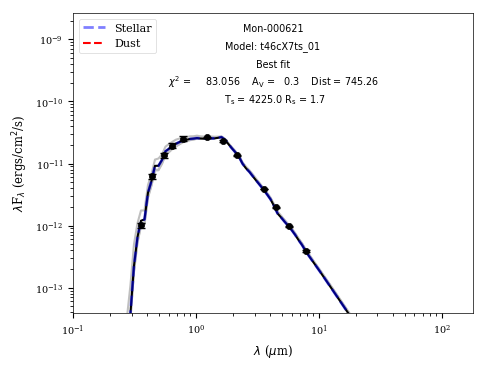}
\includegraphics[scale=0.29]{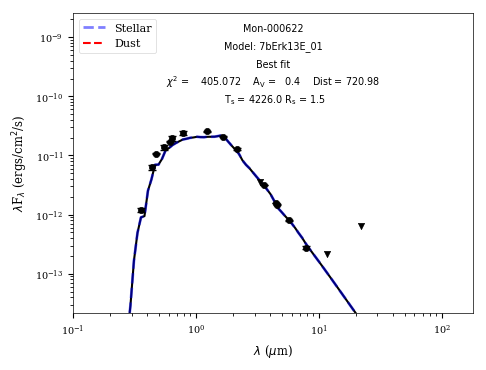}
\includegraphics[scale=0.29]{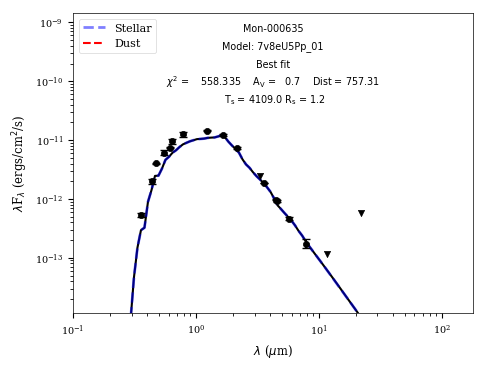}
\includegraphics[scale=0.29]{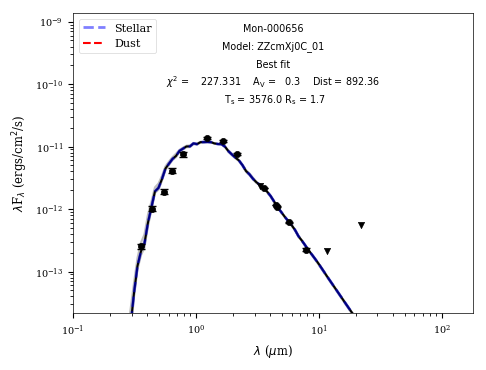}
\includegraphics[scale=0.29]{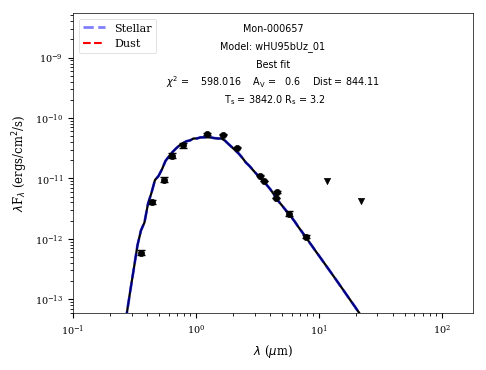}
\includegraphics[scale=0.29]{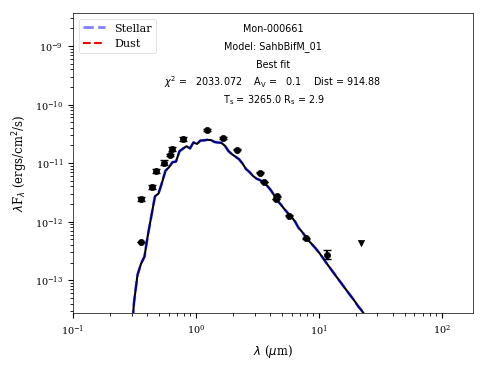}
\caption{\label{fig:Diskless2} The same as Fig. \ref{fig:Diskless}}
\end{figure*}

\begin{figure*}
\includegraphics[scale=0.29]{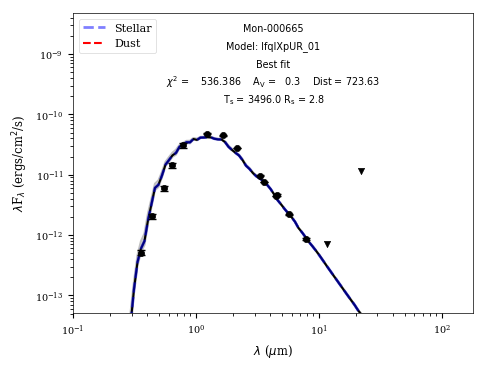}
\includegraphics[scale=0.29]{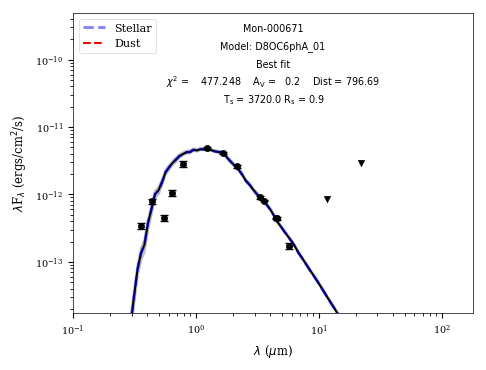}
\includegraphics[scale=0.29]{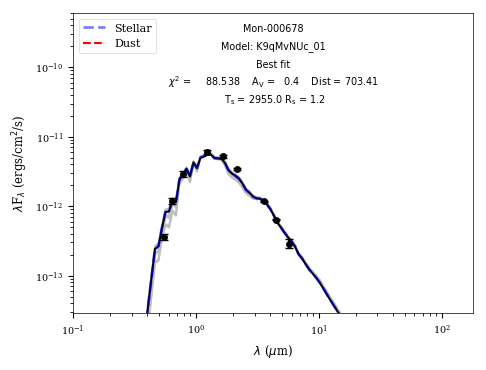}
\includegraphics[scale=0.29]{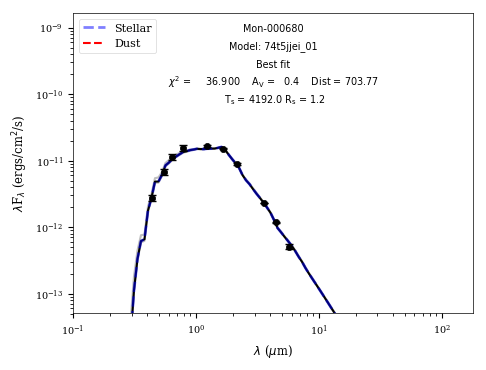}
\includegraphics[scale=0.29]{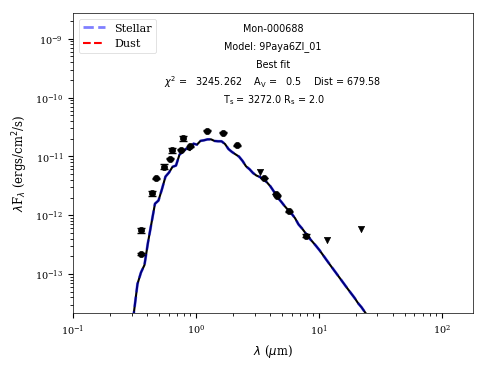}
\includegraphics[scale=0.29]{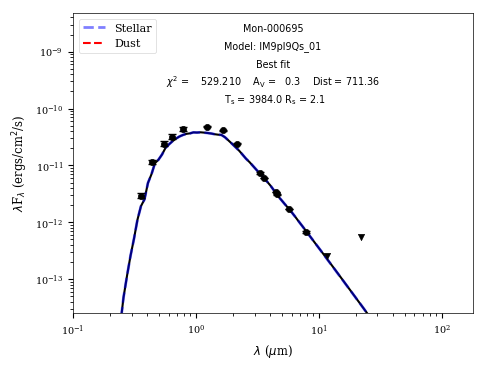}
\includegraphics[scale=0.29]{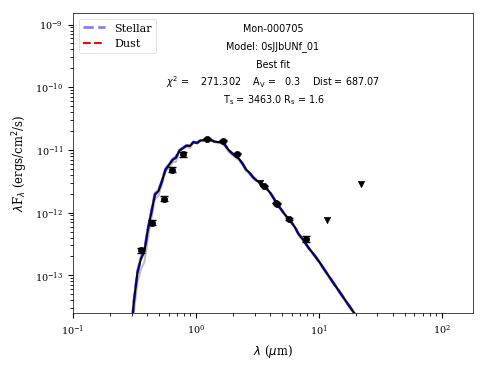}
\includegraphics[scale=0.29]{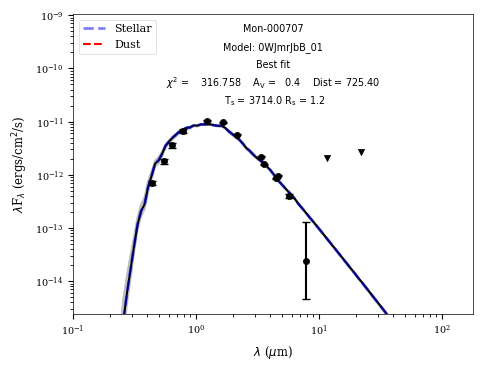}
\includegraphics[scale=0.29]{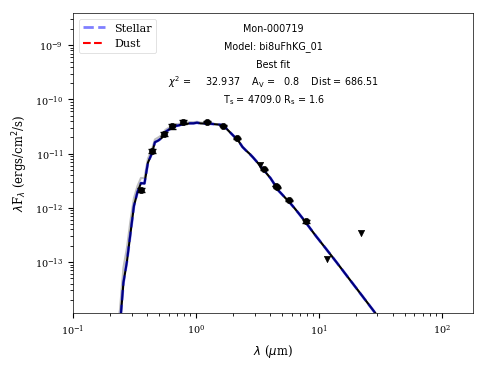}
\includegraphics[scale=0.29]{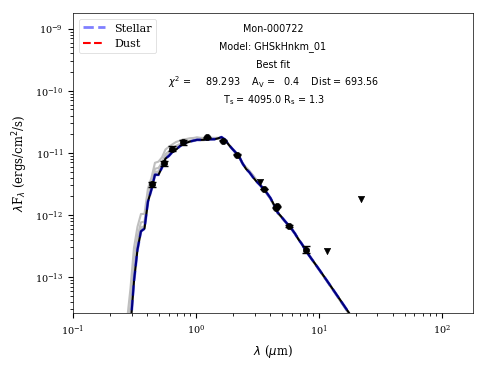}
\includegraphics[scale=0.29]{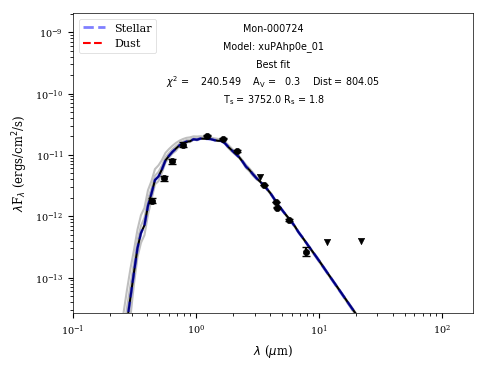}
\includegraphics[scale=0.29]{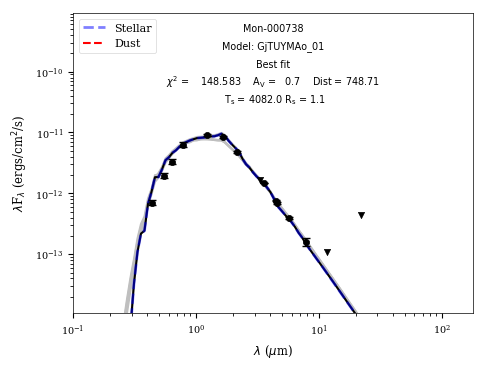}
\includegraphics[scale=0.29]{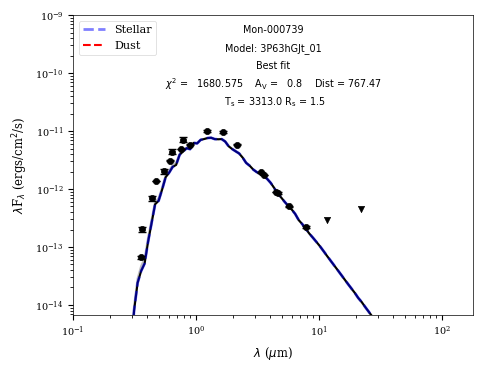}
\includegraphics[scale=0.29]{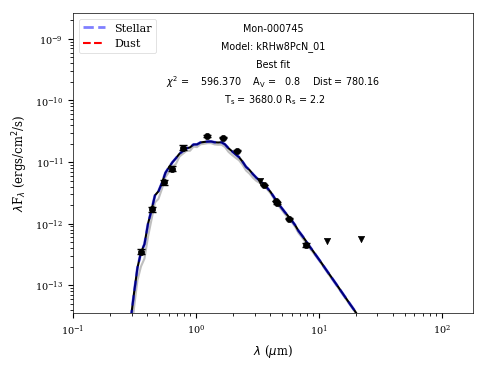}
\includegraphics[scale=0.29]{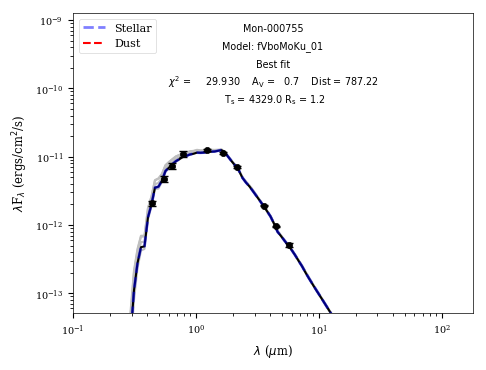}
\includegraphics[scale=0.29]{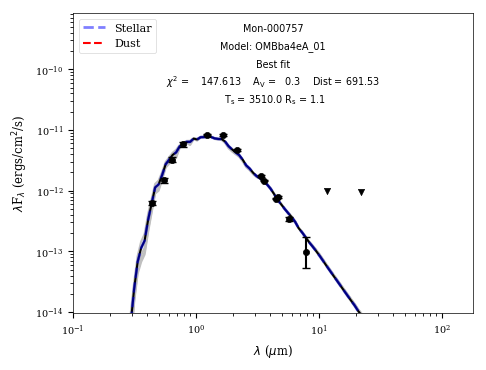}
\includegraphics[scale=0.29]{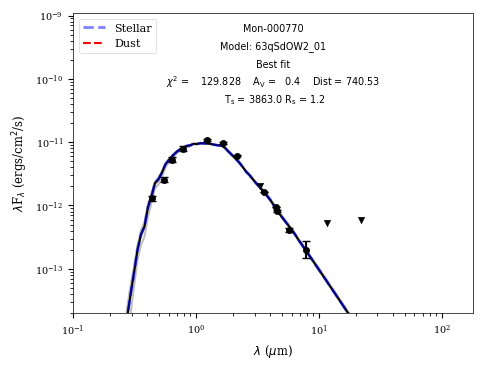}
\includegraphics[scale=0.29]{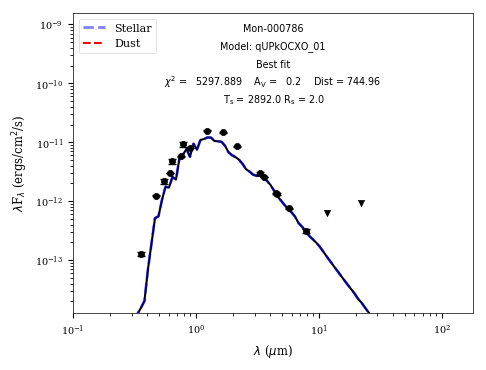}
\includegraphics[scale=0.29]{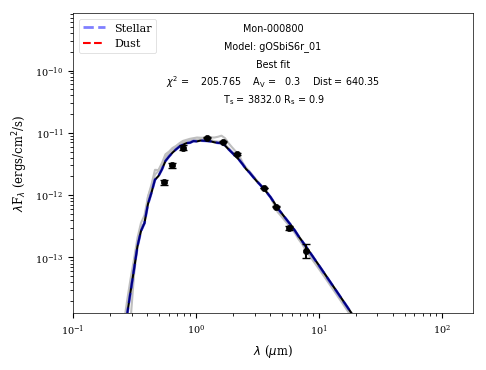}
\includegraphics[scale=0.29]{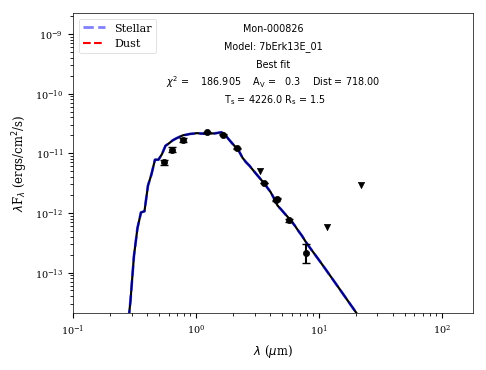}
\includegraphics[scale=0.29]{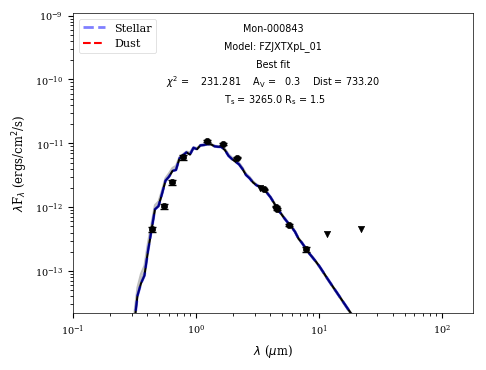}
\includegraphics[scale=0.29]{all_Mon-000881}
\includegraphics[scale=0.29]{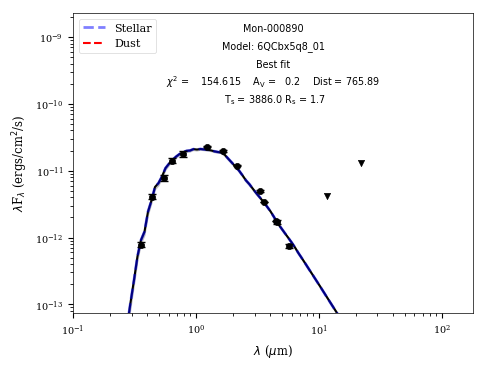}
\includegraphics[scale=0.29]{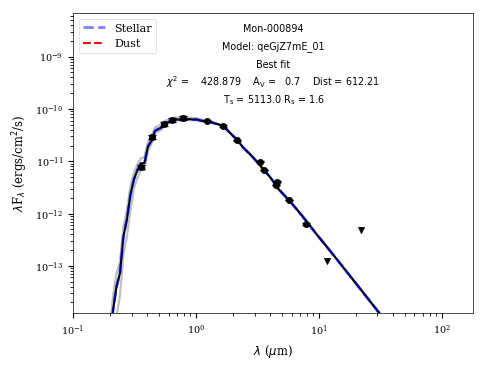}
\includegraphics[scale=0.29]{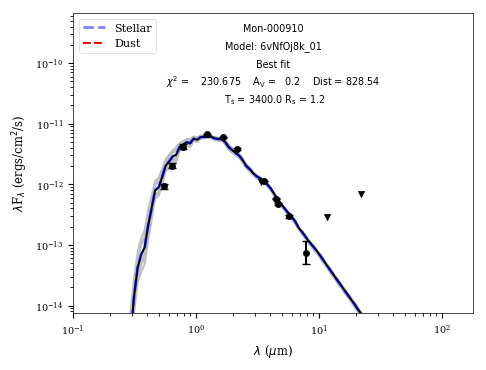}
\includegraphics[scale=0.29]{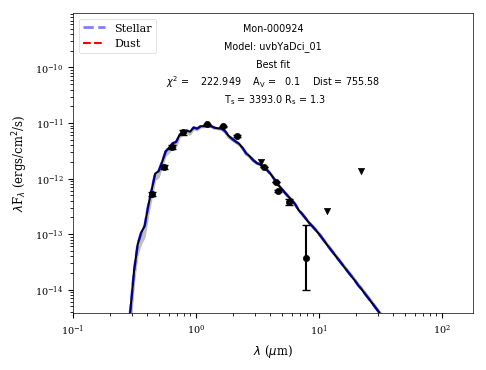}
\includegraphics[scale=0.29]{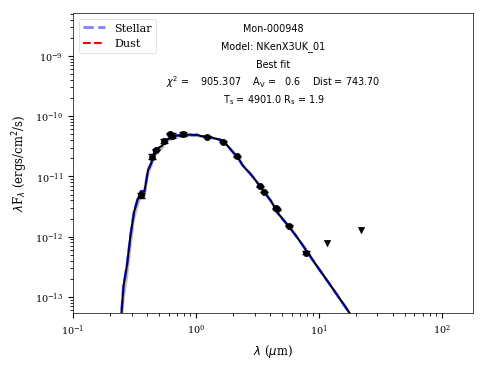}
\includegraphics[scale=0.29]{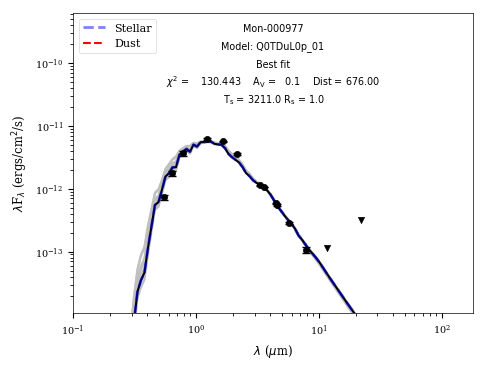}
\includegraphics[scale=0.29]{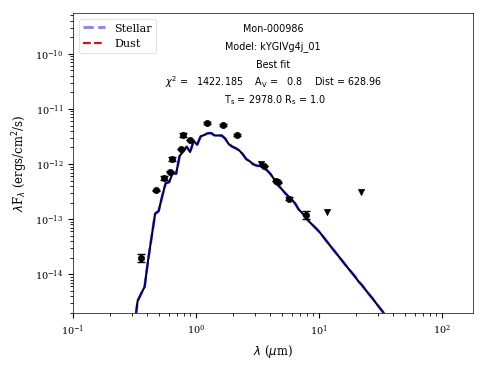}
\includegraphics[scale=0.29]{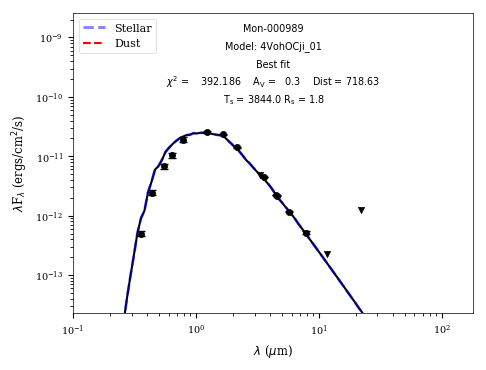}
\includegraphics[scale=0.29]{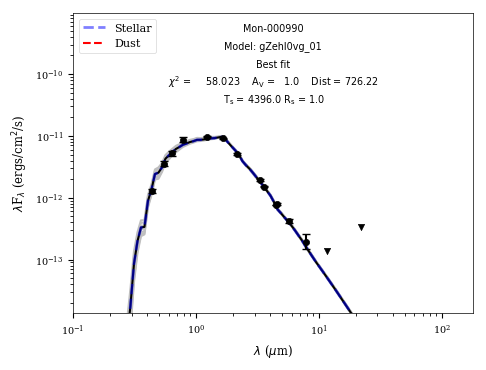}
\includegraphics[scale=0.29]{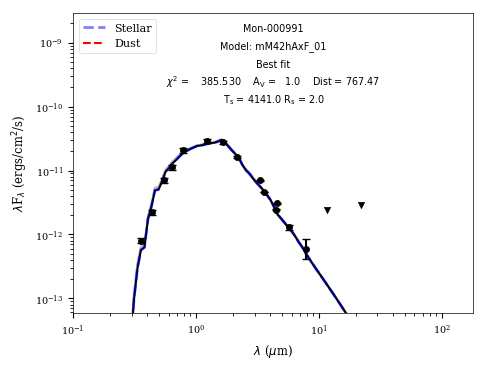}
\includegraphics[scale=0.29]{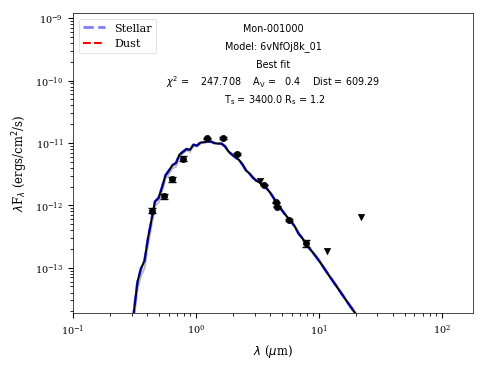}
\includegraphics[scale=0.29]{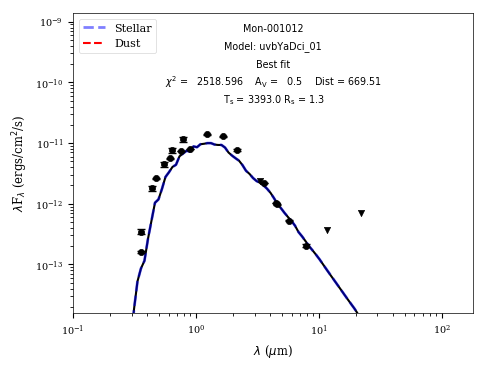}
\includegraphics[scale=0.29]{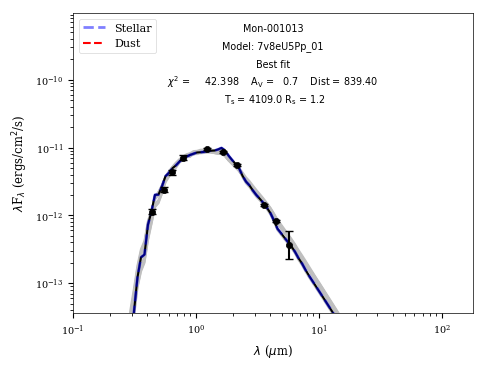}
\includegraphics[scale=0.29]{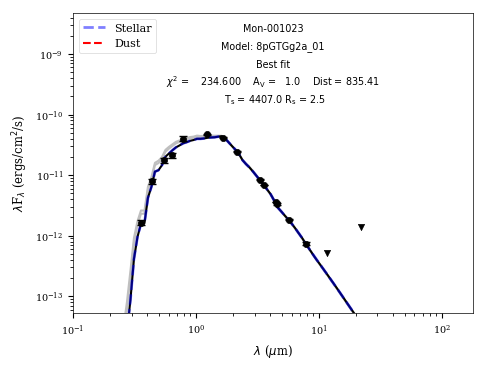}
\includegraphics[scale=0.29]{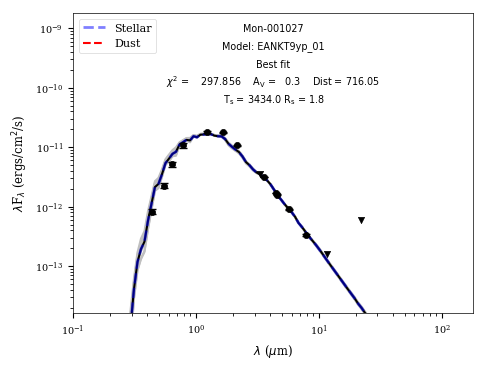}
\includegraphics[scale=0.29]{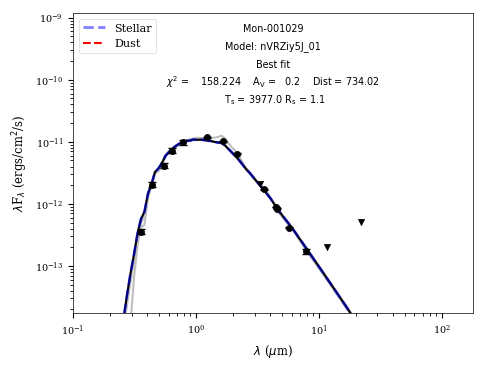}
\includegraphics[scale=0.29]{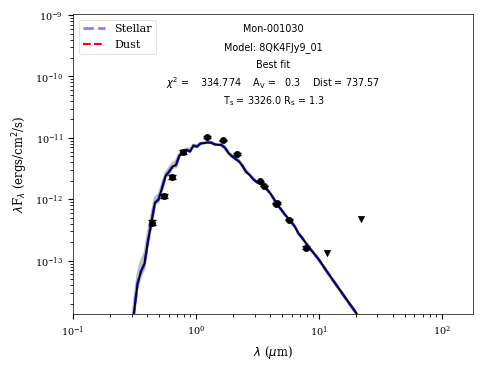}
\includegraphics[scale=0.29]{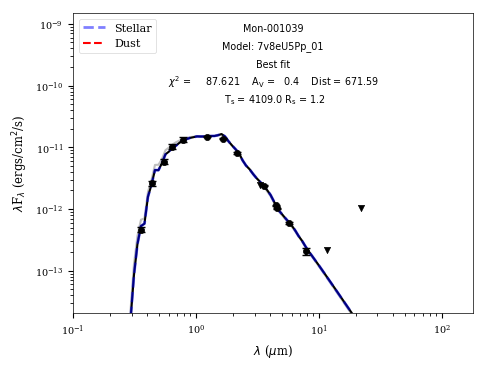}
\caption{\label{fig:Diskless3} The same as Fig. \ref{fig:Diskless}.}
\end{figure*}

\begin{figure*}
\includegraphics[scale=0.29]{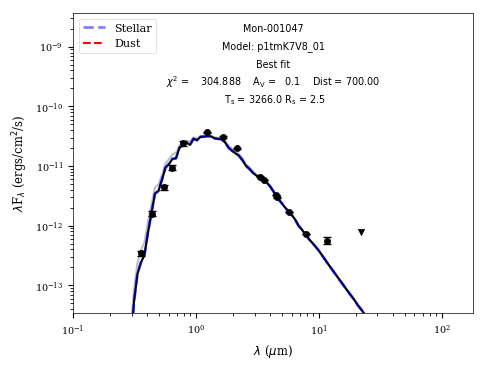}
\includegraphics[scale=0.29]{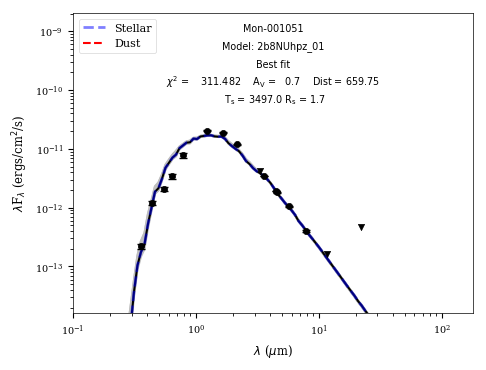}
\includegraphics[scale=0.29]{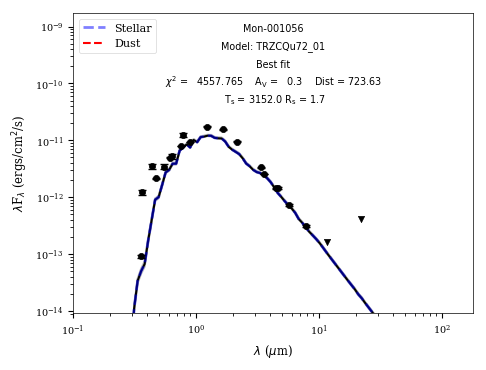}
\includegraphics[scale=0.29]{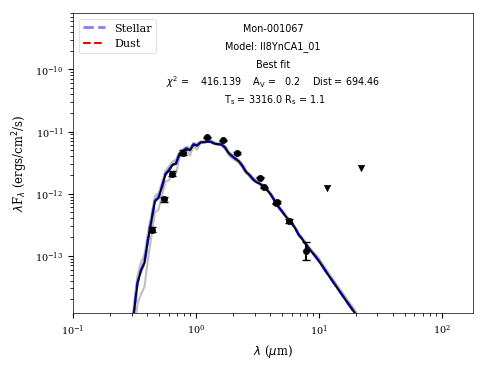}
\includegraphics[scale=0.29]{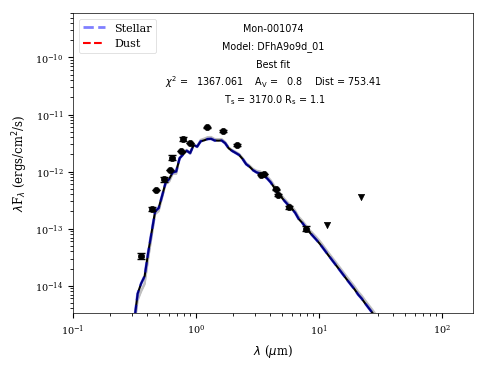}
\includegraphics[scale=0.29]{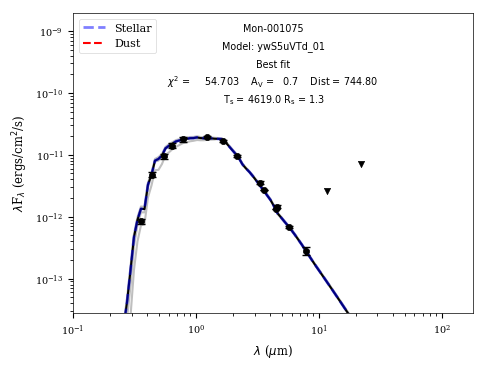}
\includegraphics[scale=0.29]{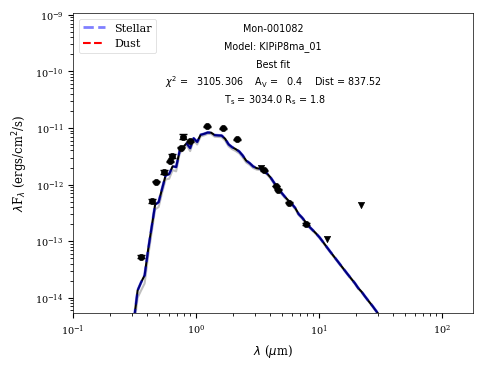}
\includegraphics[scale=0.29]{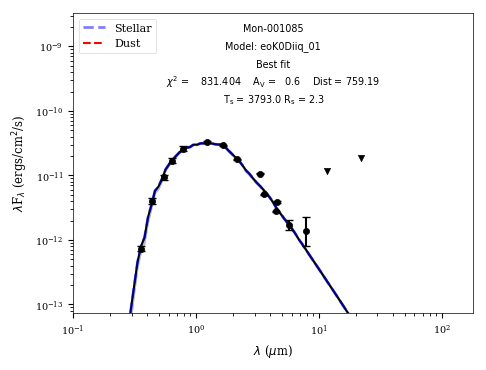}
\includegraphics[scale=0.29]{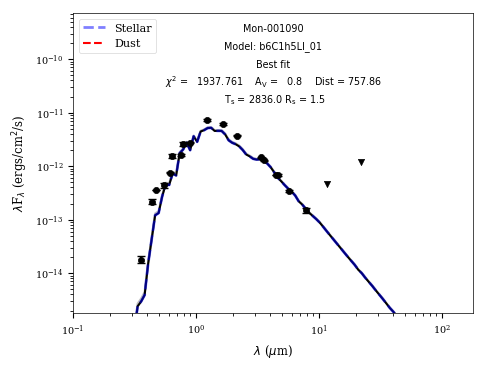}
\includegraphics[scale=0.29]{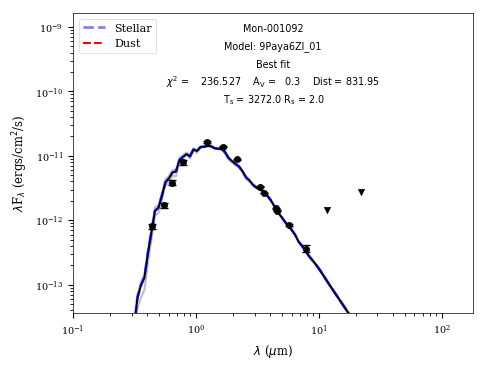}
\includegraphics[scale=0.29]{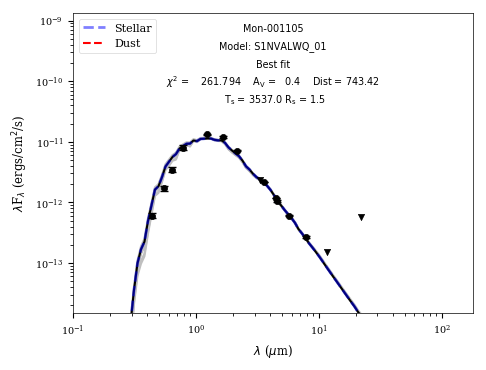}
\includegraphics[scale=0.29]{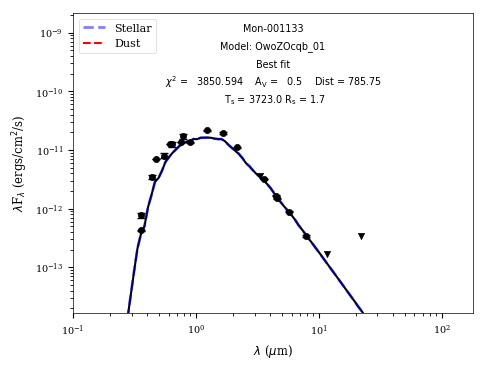}
\includegraphics[scale=0.29]{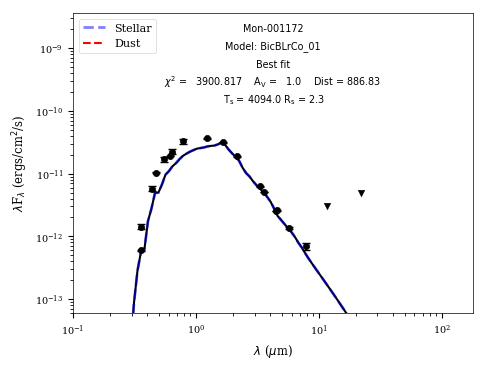}
\includegraphics[scale=0.29]{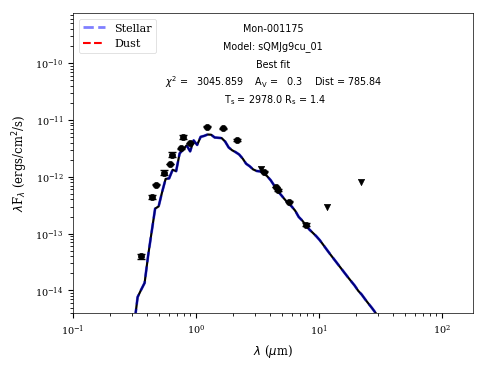}
\includegraphics[scale=0.29]{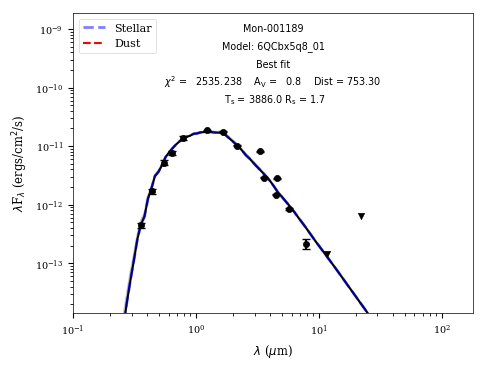}
\includegraphics[scale=0.29]{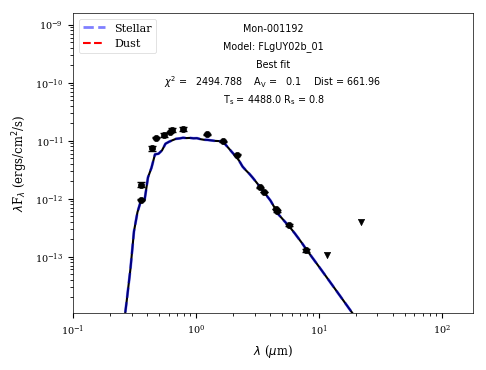}
\includegraphics[scale=0.29]{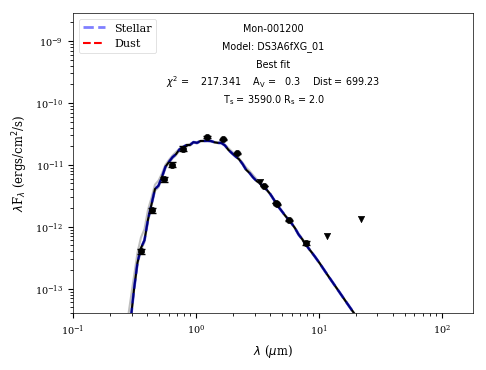}
\includegraphics[scale=0.29]{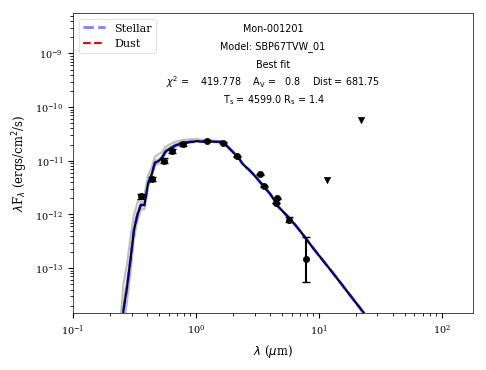}
\includegraphics[scale=0.29]{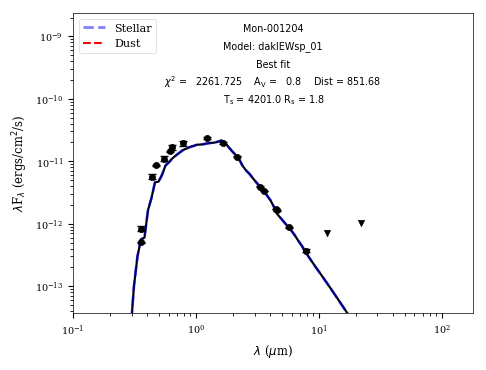}
\includegraphics[scale=0.29]{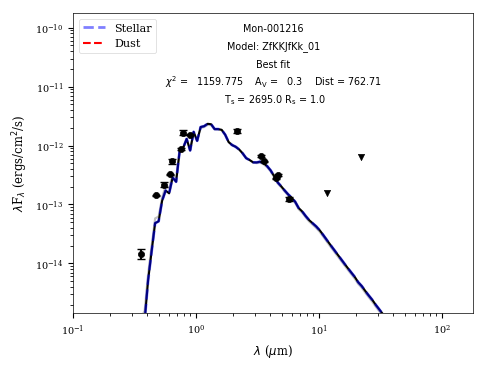}
\includegraphics[scale=0.29]{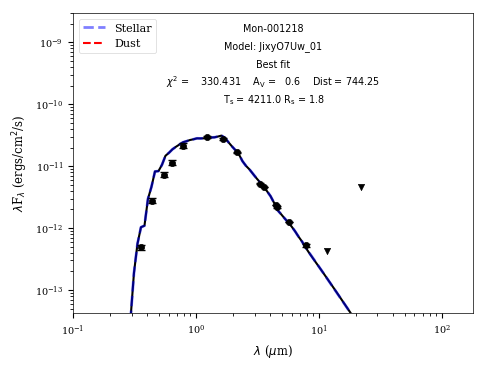}
\includegraphics[scale=0.29]{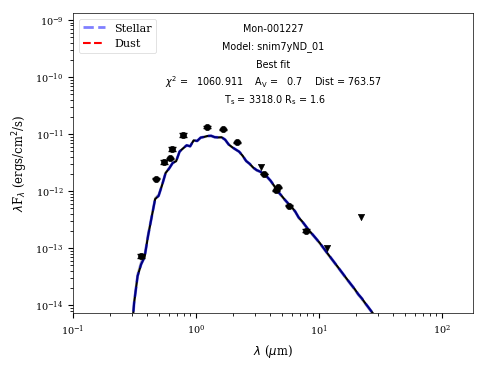}
\includegraphics[scale=0.29]{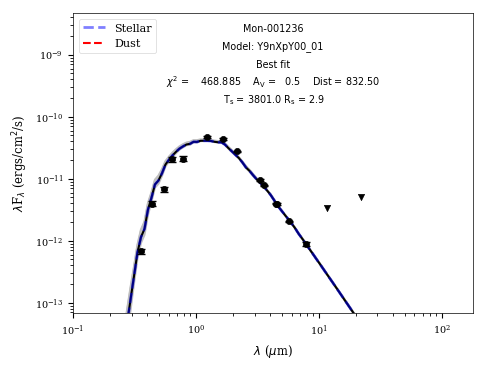}
\includegraphics[scale=0.29]{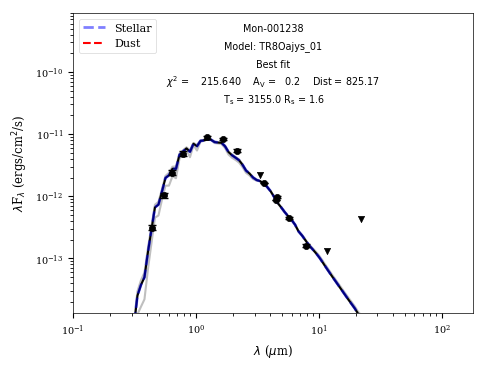}
\includegraphics[scale=0.29]{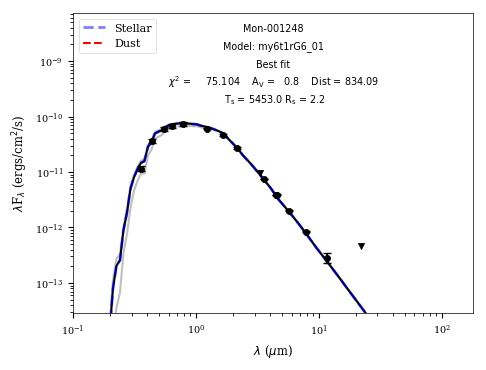}
\includegraphics[scale=0.29]{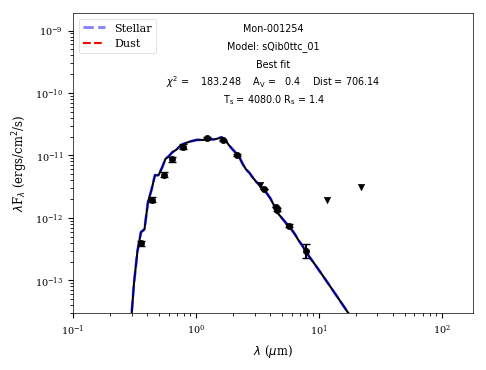}
\includegraphics[scale=0.29]{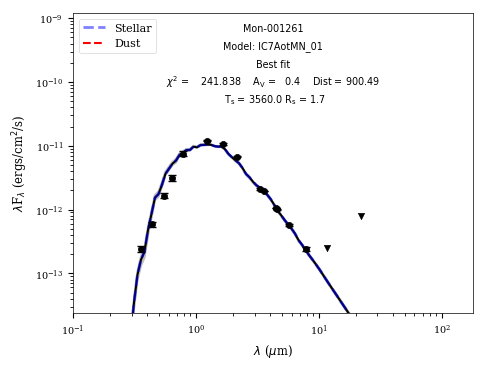}
\includegraphics[scale=0.29]{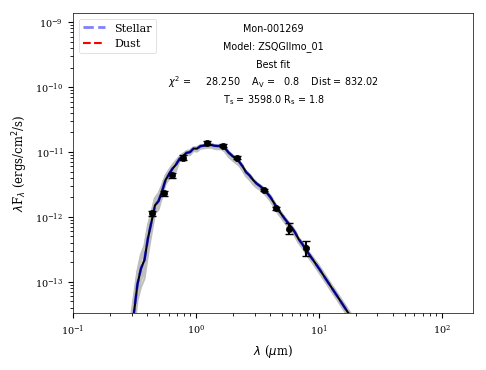}
\includegraphics[scale=0.29]{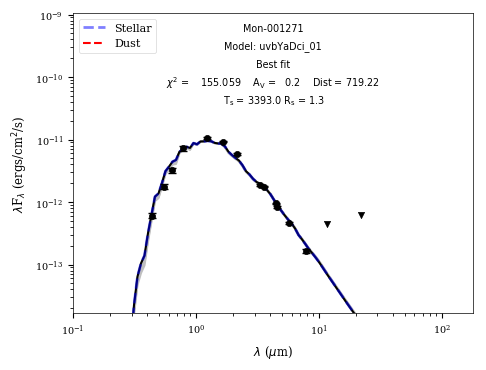}
\includegraphics[scale=0.29]{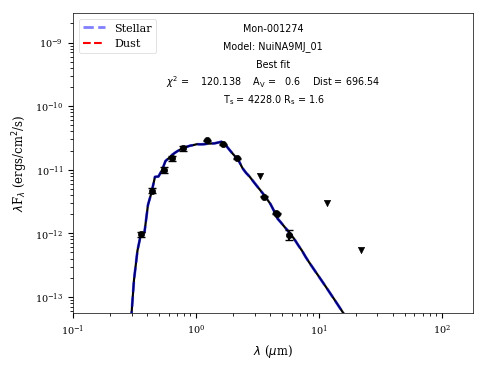}
\includegraphics[scale=0.29]{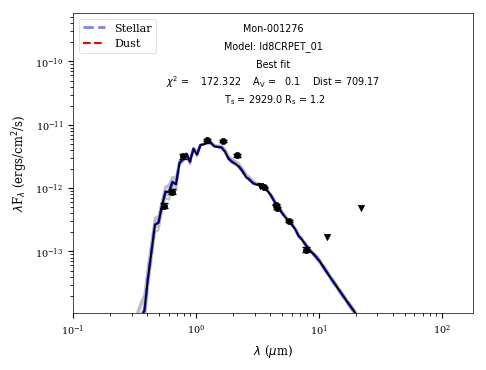}
\includegraphics[scale=0.29]{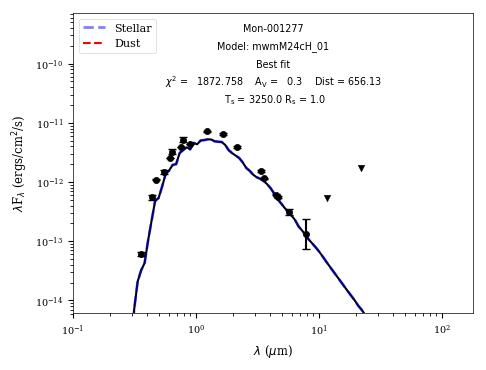}
\includegraphics[scale=0.29]{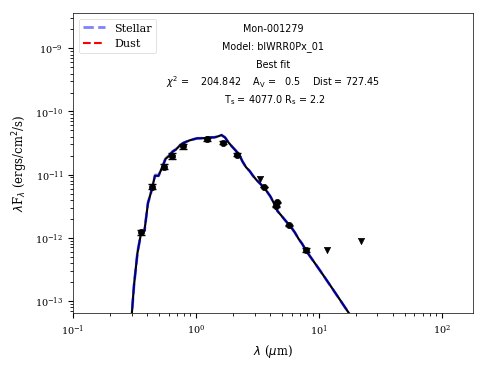}
\includegraphics[scale=0.29]{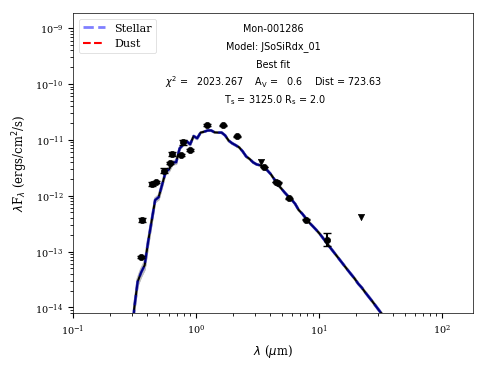}
\includegraphics[scale=0.29]{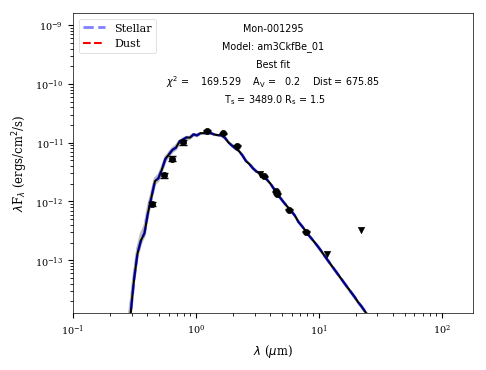}
\includegraphics[scale=0.29]{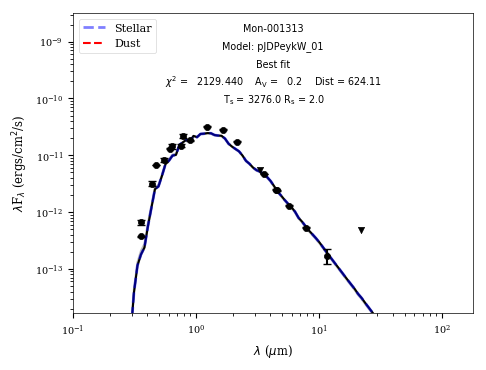}
\includegraphics[scale=0.29]{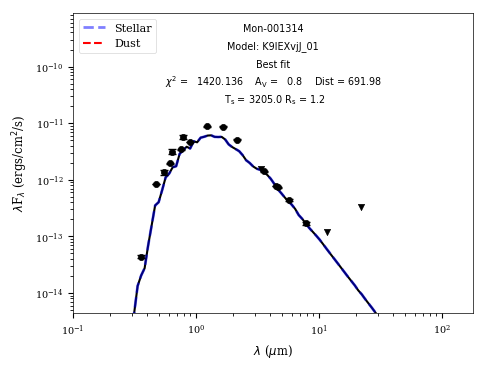}
\includegraphics[scale=0.29]{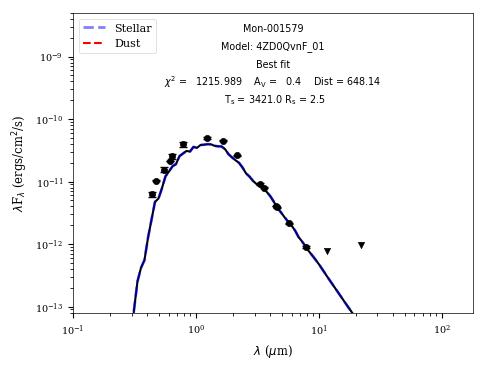}
\includegraphics[scale=0.29]{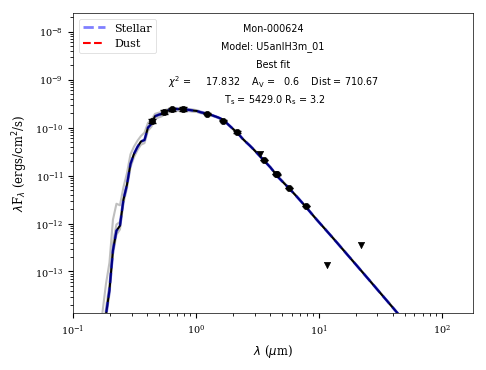}
\caption{\label{fig:Diskless4}The same as Fig. \ref{fig:Diskless}.}
 \end{figure*}

\end{appendix}

\end{document}